REPRESENTING BIG DATA AS NETWORKS:

NEW METHODS AND INSIGHTS

A Dissertation

Submitted to the Graduate School

of the University of Notre Dame

in Partial Fulfillment of the Requirements

for the Degree of

Doctor of Philosophy

by

Jian Xu

_______________________________

Nitesh V. Chawla, Director

Graduate Program in Computer Science and Engineering

Notre Dame, Indiana

July 2017

REPRESENTING BIG DATA AS NETWORKS:

NEW METHODS AND INSIGHTS

Abstract

by

Jian Xu

Our world produces massive data every day; they exist in diverse forms, from pairwise data and matrix to time series and trajectories. Meanwhile, we have access to the versatile toolkit of network analysis. Networks also have different forms; from simple networks to higher-order network, each representation has different capabilities in carrying information. For researchers who want to leverage the power of the network toolkit, and apply it beyond networks data to sequential data, diffusion data, and many more, the question is: *how to represent big data and networks?* This dissertation makes a first step to answering the question. It proposes the higher-order network, which is a critical piece for representing higher-order interaction data; it introduces a scalable algorithm for building the network, and visualization tools for interactive exploration. Finally, it presents broad applications of the higher-order network in the real-world.

Dedicated to those who strive to be better persons.

CONTENTS

FIGURES

TABLES

# ACKNOWLEDGMENTS

First, I would like to thank Professor Nitesh V. Chawla for supporting me through this rewarding journey. The most vividly, I remember the two "why not"s he said to me, one was in my first year of Ph.D. studies when my self-confidence was not yet established, he said "Why not submit your global shipping work to KDD? You have done some solid work." One was in my fourth year of Ph.D. when the HON paper had been rejected three times, he said "Why not submit it to Science Advances? I believe your work deserves to be seen by more people." It was his uncompromised support that turned the insecure me into a confident researcher. I would also like to thank Prof. David Lodge, Prof. Tijana Milenkovec, and Prof. Zoltán Toroczkai for serving on my thesis committee and providing me with insightful comments. It had always been a pleasure to work with Prof. Lodge on interdisciplinary challenges, and learn how he frame problems in the most concise way. It had always been a memorable lesson when explaining my HON poster to Prof. Milenkovec and was ruthlessly challenged, after which I learned from her how to back up my research. It had always been inspiring to see Prof. Toroczkai walking into my room and throwing random interesting questions at us, he is the living image of someone who is truly passionate about research.

I would also like to acknowledge my wonderful collaborators from across different disciplines, without whom this dissertation cannot have a breadth that reaches from biology to finance, from anomaly detection algorithm to visualization software: Zhi Da, Yuxiao Dong, John Drake, Erin Grey, Chao Huang, Reid Johnson, Reuben Keller, Karsten Steinhaeuser, Jun Tao, Chaoli Wang, Thanuka Wickramarathne, Xian Wu,

Mao Ye among others.

I would like to thank my mentors whom I have worked in person with and learned from: Prof. Zhi Da, who offered constructive suggestions to my career choices and made sure to follow up with me multiple times a year. I would also like to thank Prof. Bruno Ribeiro, whom I spent a summer with at Purdue University; that was the first time I had the first taste of and was seriously interested in academia. I would also like to thank Prof. Martin Rosvall, who had given me more insights in the lives of academia. Additionally, I would like to extend my appreciation to my supportive mentors from the industry: Francesco Calabrese and Fabio Pinelli at IBM Research, and Ananthram Swami at Army Research Lab.

I would like to thank my wonderful lab mates at iCeNSA, which I had always considered as my second family. My deepest thanks to Yuxiao Dong, who trained me to be a better researcher at work, and taught me to be a better person in life. And to other iCeNSA members I had the pleasure to share the room with, among whom many I had travelled with: Everaldo Aguiar, Daniel Barabási, Dipa Dasgupta, Louis Faust, Keith Feldman, Haoyun Feng, Chao Huang, Reid Johnson, Saurabh Nagrecha, Aastha Nigam, Yihui Ren, Pingjie Tang, Melinda Varga, Xian Wu and Yang Yang. And my thanks to other CSE and ND friends, Haipeng Cai, Siyuan Jiang, Yiji Zhang, Tian Jiang, among many others. And of course, to Jasmine Botello and Joyce Yeats, who made my life much easier and full of fun. My great appreciations to Barbara Mangione who taught me English and American culture, to Hannah Babbini and Cindi Fuja for bringing the best out of me in my last year of Ph.D., to Kevin McAward and Weiyang Xie for keeping me happy and healthy.

Finally, my deepest gratitude to my parents, who gave me unconditional love and support through this journey; and Ningxin Wang, who had been and will always be the highlight of everyday of mine.

CHAPTER 1

INTRODUCTION

In the world of data mining, there lies two gold mountains: One is the ubiquitous data that exist in diverse forms, from pairwise data and matrix to time series and trajectories. Another is the versatile network analysis toolkit that operates on different forms of networks from simple networks to higher-order networks.

For researchers who want to leverage the power of the network toolkit, and apply it to a wide range of data including sequential data, diffusion data, and many more, the question is: *how to represent data as networks*, to bridge the gap between the two mountains, and make the best of both worlds?

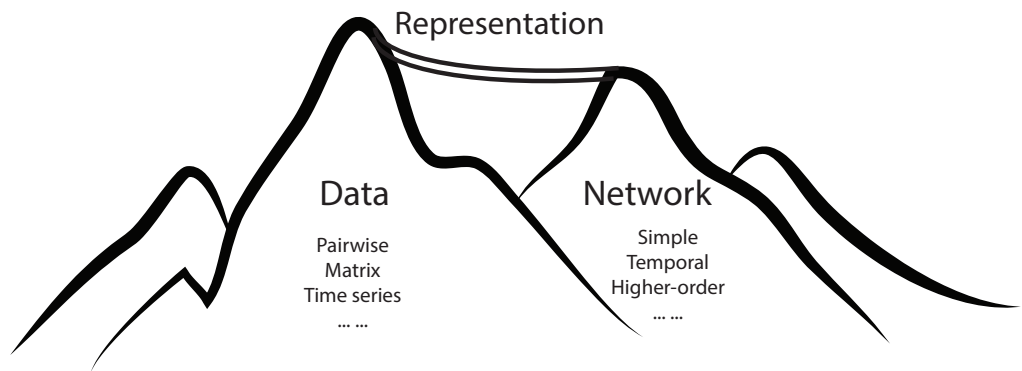

Figure 1.1. Bridging the gap between data and network.

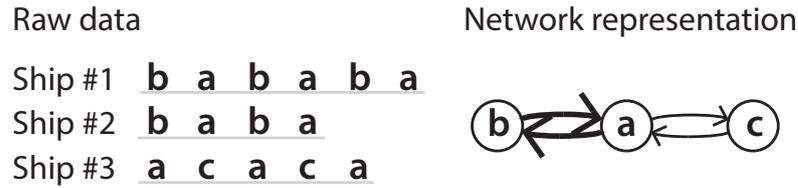

Figure 1.2. Conventional way of building network from sequential data.

When the network data is not readily available, representing other types of data as networks is the first step. The network then serves as the foundation for subsequent analyses. This representation step is critical: if the network loses important information in the raw data, the accuracy of subsequent analysis is undermined.

However, this important representation step — which determines the quality of subsequent analysis — is often overlooked. A common practice is to create a one-to-one mapping from entities in the raw data to nodes in the network, then count the number of interactions between entity pairs in the raw data, and take that as the edge weights in the network. Is that a universally appropriate way to represent any type of raw data as networks?

Suppose we have ship movement trajectory data as in Figure 1.2. The first two ship moves between port $a$ and $b$, and the third ship moves between $a$ and $c$. The conventional way to construct the network simply counts the traffic between pairs of ports, and take that as the edge weight, yielding a network on the right of Figure 1.2. However, the data suggests that if a ship goes between $a$ and $b$, the ship is likely to keep moving between $a$ and $b$ rather than going to $c$. But according to the structure of the conventional network, the ship may still escape the loop of $a$ and $b$ and move to $c$.

The example shows that the network representation is a non-trivial problem. Several questions remain:

- What data types exist? What properties do they have?

- What network representations exist? What information can they preserve?

- Which types of data can be represented as networks without losing information? Which types of data still do not have a good representation?

- If a new network representation exists, what can it preserve? How to construct it efficiently? How to evaluate its quality? How to visualize it? How do existing network analysis methods adapt to it? What are its influences on network analysis results? What are the real-world applications?

This dissertation gives a first answer to these questions.

## 1.1 Contributions and Organization

An overview of this dissertation is illustrated in Figure 1.3. The dissertation contains three parts: Part I discusses existing methods to represent data as networks, their limitations, and proposes the higher-order network representation; Part II discusses insights in real-world networks through the lens of the higher-order network; Part III focuses on diffusion processes on implicit social networks.

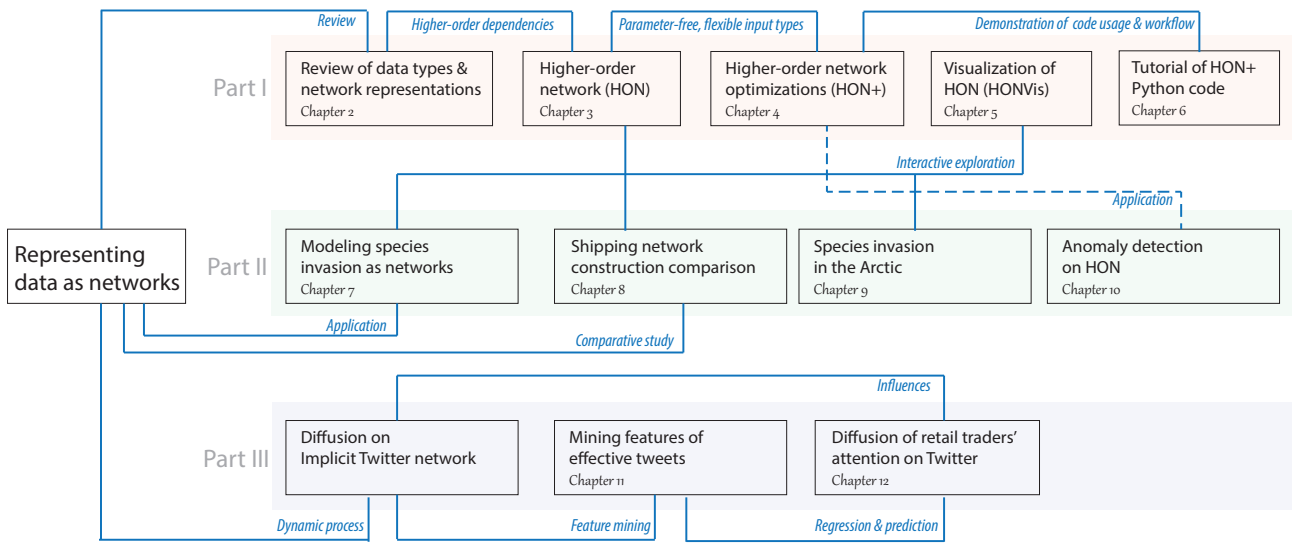

Figure 1.3. Organization of topics in the dissertation.

Specifically, Chapter 2, the review of data types & network representations, provides a comprehensive review of the most frequently seen types of raw data and their properties. Data can be categorized into two types: those that represent pairwise relationships, and those that represent higher-order relationships. Next, the chapter reviews network representations and analyze their capabilities in preserving information. Finally, the chapter provides a table illustrating the lossless and lossy network representations of the raw data, which calls for effective network representations of higher-order data.

Inspired by Chapter 2, Chapter 3, Higher-order Network (HON), shows that data derived from some complex systems show up to fifth-order dependencies, such that the oversimplification in the first-order network representation can later result in inaccurate network analysis results. The chapter then proposes the Higher-Order Network (HON) representation that can discover and embed variable orders of dependencies in a network representation. Through a comprehensive empirical evaluation and analysis, the chapter establishes several desirable characteristics of HON – accuracy, scalability, and direct compatibility with the existing suite of network analysis methods. The chapter illustrates the broad applicability of HON by using it as the input to a variety of tasks, such as random walking, clustering and ranking, where these methods yield more accurate results without modification.

To make the HON algorithm parameter free, scalable for big data, and can take more types of raw data, Chapter 4, higher-order network optimized for big data, proposes a parameter-free and fully scalable algorithm HON+. The chapter proposes a procedure that constructs observations of subsequences on demand, which is achieved by using an indexing cache with $\Theta(1)$ lookup time. The chapter provides the full pseudocode, and the complexity analysis for time and space. The new approach makes it possible to extract arbitrarily high orders of dependency. Finally, the chapter extends the input of HON from simple trajectory data to various other types of raw data including diffusion, time series, subsequence, temporal pairwise interaction, and heterogeneous data. This significantly extends the applications of HON.

To facilitate the visualization of HON, Chapter 5, Visualization of HON (HON-Vis), proposes a novel visual analytics framework for exploring higher-order dependencies, illustrated in the context of global ocean shipping. The framework leverages coordinated multiple views to reveal the network structure at three levels of detail (i.e., the global, local, and individual port levels). Users can quickly identify ports of interest at the global level and specify a port to investigate its higher-order nodes at the individual port level. Investigating a larger-scale impact is enabled through the exploration of HON at the local level. Using the global ocean shipping network data, the chapter demonstrates the effectiveness of HONVis with a real-world use case conducted by domain experts specializing in species invasion. Finally, the chapter discusses the generalizability of this framework to other real-world applications such as information diffusion in social networks and epidemic spreading through air transportation.

How exactly should one use the HON code, how to extend its input from sequential data to pairwise interactions, and how to do it efficiently? Chapter 6 provides a step-by-step walk-through of the non-trivial process of constructing HON based on a representative pairwise cellphone communications data. Specifically, the chapter proposes a time window-based approach to chain pairwise interactions into sequential data, and proposes an efficient algorithm with linear time complexity. The tutorial can serve as a reference for the pre-processing of data like social network direct messaging, wireless sensor network communication, and more.

As a practical application of network representation, Chapter 7 presents an approach for modeling species invasion through global shipping via creative use of computational techniques and multiple data sources, thus illustrating how data mining can be used for solving crucial, yet very complex problems towards social good. By modeling implicit species exchanges as a Species Flow Network (SFN), large-scale species flow dynamics are studied via a graph clustering approach that decomposes the SFN into clusters of ports and inter-cluster connections. The chapter then exploits this decomposition to discover crucial knowledge on how patterns in GSN affect aquatic invasions, and then illustrates how such knowledge can be used to devise effective and economical invasive species management strategies. By experimenting on actual GSN traffic data for years 1997-2006, the chapter has discovered crucial knowledge that can significantly aid the management authorities.

A question arises in the same context: how does different network representations of the same global shipping data affect network properties and analysis results? Chapter 8, shipping network construction comparison, conducts a comparative study. The chapter in itself does not aim to suggest which approach is "better" than the other. Rather, it is the first step to producing accurate and effective network representations that captures important information in the raw data, and it serves as a reference for researchers in related fields such as road or air traffic network analysis.

Climate change is melting the Arctic sea ice, opening up Arctic sea routes, and leads to the increased human activities in the Arctic region. This poses new challenges to species invasion to the Arctic. Chapter 9, species invasion in the Arctic, leverages network analysis and data mining techniques to assess, visualize and project ballast water mediated Arctic species introduction and diffusion risks. The chapter identifies high-risk connections between Arctic and non-Arctic ports that could be sources of invasive organisms, and critical links that facilitate species diffusion in the Arctic. Using higher-order network analysis, the chapter further distinguishes critical shipping routes that facilitate species dispersal within the Arctic. The decadal projections of current trends reveal the emergence of shipping hubs in the Arctic, and suggest that the cumulative risk of species introduction is increasing and becoming more concentrated at these emergent shipping hubs. The risk assessment and projection framework proposed in this chapter could inform risk-based assessment and management of ship-borne invasive species in the Arctic.

Another practical application that is based on dynamic network is anomaly detection, which conventionally relies on the first-order network representation. Chapter 10, anomaly detection on HON, proposes to replace the FON with HON in the dynamic network representation of raw trajectory data. The chapter shows that existing anomaly detection algorithms can then capture higher-order anomalies that may otherwise be ignored. A large-scale synthetic data with 11 billion movements was constructed to verify the effectiveness of HON in capturing variable orders of anomalies. The experiment on real-world taxi trajectory data demonstrates HON's ability in amplifying anomaly signals.

Information diffusion produces diffusion data derived from implicit social networks. In Chapter 11, mining features of effective tweets, the chapter presents a systematic review of tweet time, entities, composition, and user account features through the mining of 122 million engagements of 2.5 million original tweets. It is shown that the relationship between many features and tweeting effectiveness is non-linear; for example, tweets that use a few hashtags have higher effectiveness than using no or too many hashtags.

The mining of Twitter features serves as a foundation for analyzing the relationship between information diffusion on Twitter and retail trading, as in Chapter 12. Information diffusion in real time was collected by monitoring how the news is tweeted and retweeted on Twitter. It is found that news diffusion is highly correlated with intraday trading, especially for retail trading. News diffusion leads to a lower bid-ask spread and price pressure on the news day that is completely reverted the next day. The result is robust when the instrumental variables approach is employed. Results show that information diffusion via Twitter does not incorporate new information into the stock price. Rather, Twitter spreads stale news, albeit at a much higher speed than traditional media.

Overall, this dissertation makes a first step to answering the question "how to represent big data as networks", proposes the higher-order network which is a critical piece for representing higher-order interaction data, produces a collection of algorithms and tools for the task[1], and presents broad applications in the real-world.

---

[1] Code, video demo and paper available at http://www.HigherOrderNetwork.com

PART I

METHODS TO REPRESENT NETWORKS

# CHAPTER 2

# REVIEW OF DATA TYPES AND NETWORK REPRESENTATIONS

## 2.1 Overview

The world of data mining has two valuable resources. On one side, ubiquitous data exists in different formats, from pairwise data and matrix to time series and trajectories. On the other side, network representations range from simple networks to higher-order network, each representation have different capabilities in carrying information. The question is: how to make the best of both resources, and extend the power of the versatile network models to fully capture complex interactions in data?

**Intellectual merit:** In this chapter, we provide a comprehensive review of the most frequently seen types of raw data and their properties. We categorize the data into two types: those that represent pairwise relationships, and those that represent higher-order relationships. We also review network representations and analyze their capabilities in preserving information. Finally, we provide a table illustrating the lossless and lossy network representations of the raw data, which calls for effective network representations of higher-order data.

**Connections:** This chapter serves as the motivation of the higher-order network discussed throughout the rest of Part I.

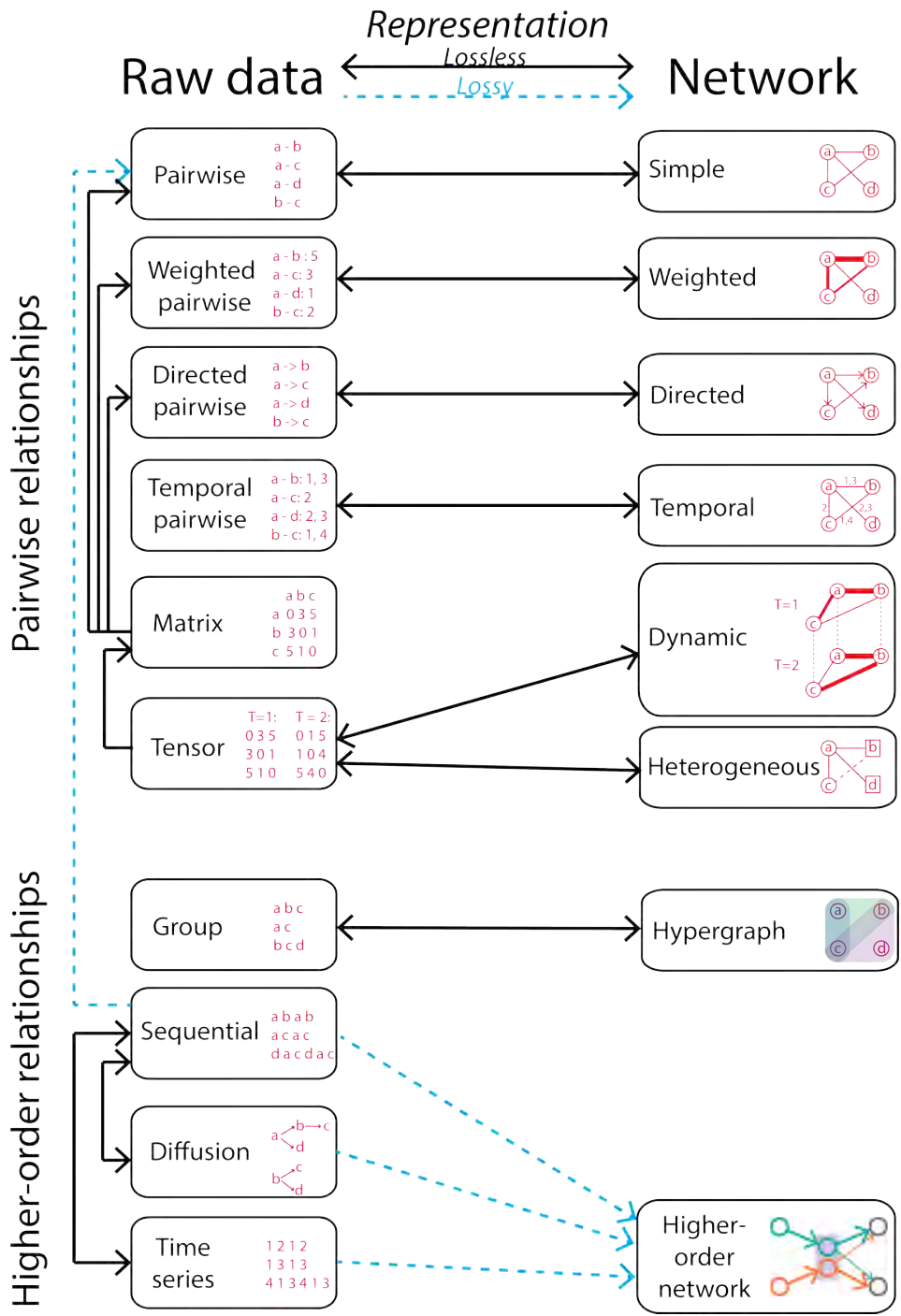

Figure 2.1. **Review of raw data types and network representations.**

## 2.2 Types of Raw Data

In this section, we review the most frequently seen types of raw data. It is not an attempt to cover all possible types of data; instead, this section reviews the types of data that are frequently collected in research tasks and later represented as networks. Hierarchical data types that are frequently represented as trees instead of networks are not discussed here. Compound data types can be seen as combinations of the data types to be discussed below.

### 2.2.1 Pairwise Data

The pairwise data is a collection of one-to-one relationships:

$$\mathbf{D} = set\{(E_i, E_j)\}$$

where $E_i$ and $E_j$ are entities in the entity set $\mathbf{E}$. Here we are assuming every pair of entities is unique in the set of relationships; $E_i$ and $E_j$ are also interchangeable.

A real-world example is the Facebook friendship network, in which every entity is a single person, and the pair of entities represent a friendship relationship between them:

$$\text{Alice} - \text{Bob}$$
$$\text{Alice} - \text{Carol}$$
$$\text{Bob} - \text{David}$$
$$......$$

Such types of data are prevailing in the real-world, from protein-protein interaction data to city-city airway connections.

### 2.2.2 Weighted Pairwise Data

This type of data is a natural extension of the pairwise data by *weighting* the pairs. Consider the phone call social network: five phone calls are made between Alice and Bob per week, three between Alice and Carol, and one between Bob and David. The data can be in the form of tuples as follows:

$$\text{Alice} - \text{Bob} : 5$$
$$\text{Alice} - \text{Carol} : 3$$
$$\text{Bob} - \text{David} : 1$$
$$\ldots\ldots$$

Such types of data are also ubiquitous, from social communications between people to road traffic between cities. All of them are in the form of $\mathbf{D} = set\{(E_i, E_j), W_{ij}\}$

### 2.2.3 Directed Pairwise Data

Another extension of the pairwise data is adding *directionality*. That is a major distinction between Facebook (where friendship are mutual) and Twitter (where one "follows" another and reciprocity is not guaranteed). For example, Alice may follow Lady Gaga on Twitter but Lady Gaga may not follow back to Alice. The data now becomes:

$$\text{Alice} \to \text{Bob}$$
$$\text{Alice} \to \text{Carol}$$
$$\text{Bob} \to \text{David}$$
$$\ldots\ldots$$

Such types of data can be seen in supply chains, global trades and so on. They are in the form of $\mathbf{D} = set\{<E_i, E_j>\}$, in which $E_i$ and $E_j$ are not interchangeable.

### 2.2.4 Temporal Pairwise Data

One more extension of the pairwise data is to specify the time stamps (or ranges) that the pairwise relationship is "activated". For example, WiFi connection records can be:

$$
\begin{aligned}
\text{Alice} - \text{Campus WiFi}: &\quad 9:00 - 13:00, 17:00 - 18:00 \\
\text{Alice} - \text{Starbucks WiFi}: &\quad 14:00 - 16:00 \\
\text{Bob} - \text{Campus WiFi}: &\quad 8:00 - 12:00 \\
\text{Carol} - \text{Campus WiFi}: &\quad 15:00 - 16:00 \\
&\quad \ldots \ldots
\end{aligned}
$$

If Bob's computer is unfortunately infected by a computer virus, the virus may infect Alice's computer via campus WiFi in the morning, and start to infect Starbucks visitors in the afternoon; Carol on the contrary is not affected. If the temporal information is not available, one cannot derive the accurate diffusion dynamics as mentioned above.

Temporal pairwise data is prevalent in communications, and exist in two major forms: the relationship is activated at discrete time stamps (such as text messaging), as $\mathbf{D} = set\{(E_i, E_j), t_1, t_2, \ldots\}$, or lasts for certain period (such as phone calls), as $\mathbf{D} = set\{(E_i, E_j), \Delta t_1, \Delta t_2, \ldots\}$.

### 2.2.5 Matrix Data

The matrix data can be seen as a special case of pairwise data: it explicitly lists the relationship for *every pair* of entities.

TABLE 2.1

EXAMPLE OF MATRIX DATA

|       | Alice | Bob | Carol | David |
|-------|-------|-----|-------|-------|
| Alice | 0     | 5   | 3     | 1     |
| Bob   | 5     | 0   | 4     | 2     |
| Carol | 3     | 4   | 0     | 2     |
| David | 1     | 2   | 2     | 0     |

The matrix data and pairwise data are mutually convertible, by enumerating the pairs of elements and assigning the corresponding values. When converting unweighted pairwise data, "exist" and "not exist" correspond to 1 and 0 in matrix data, respectively. When converting directed pairwise data, the resulting matrix data may not be symmetric. The matrix representation is suitable when the number of entities is not large, and the connections among entities is dense.

2.2.6   Tensor Data

Tensor, or *high-dimensional* matrix, is a generalization of the matrix data. A typical scenario when tensor data is used is to represent the temporal evolution of the relationships, such as "number of phone calls among people for every month". It can also be utilized add heterogeneous features such as geo-location, type of relationship, etc. Note that tensor data represents diverse types of *pairwise* relationships.

2.2.7   Group Data

When every observation include *arbitrary number of elements* instead of just two, the input can be seen as multiple (possibly overlapping) groups. For example:

$$\text{Alice, Bob, Carol} \in \text{CS department}$$

$$\text{Carol, David} \in \text{Math department}$$

$$\text{Alice, Carol, David} \in \text{Network Science Lab}$$

$$\ldots\ldots$$

Weights can also be added to group data, such as the number of times per month each group hold meetings. Besides social groups, group data can also be seen in store transactions, where each observation is a "basket" (group) of "commodities" (entities) bought together by a customer[1]. Formally, $\mathbf{D} = set\{(E_i, E_j, \ldots); weight_{i,j,\ldots}\}$. Note that the elements in each observation are interchangeable.

### 2.2.8 Sequential Data

Every observation in the sequential data also involves multiple entities. Unlike group data, in every observation of sequential data, the *order* of elements is now important, having the form of $\mathbf{D} = set\{S_a :< E_i, E_j, \cdots >\}$. A typical real-world scenario is human trajectory data:

$$\text{Alice : Lab} \to \text{Library} \to \text{Lab}$$

$$\text{Bob : Starbucks} \to \text{Lab} \to \text{Dining Hall} \to \text{Lab}$$

$$\text{Carol : Dorm} \to \text{Library} \to \text{Lab} \to \text{Dorm}$$

$$\ldots\ldots$$

Besides human trajectories, sequential data exist in road/air/railway trajectories as sequences of cities, user clickstreams as sequences of Web pages, and so on.

---

[1] Also a typical input for association rule mining.

2.2.9  Diffusion Data

An extension to the sequential data is to change every linear trajectory into *trees*, in the form of $\mathbf{D} = set\{T_a\}$ where each $T$ is a tree composed of entities selected from $\mathbf{E}$. The retweeting of Tweets, the propagation of epidemics, the spreading of rumors are all diffusion processes. The diffusion data is closely related to sequential data: (1) the elements in each observation have partial orders, with parent nodes preceding children nodes; (2) the diffusion data can be disassembled into multiple subsequences by following paths from root to leaves.

2.2.10  Time Series Data

The time series data, such as stock price data, can be seen as a special case of sequential data: the values are *real numbers* instead of arbitrary entities. Nevertheless, it is still possible to *categorize* real numbers in time series data into discrete states, and then treat as sequential data.

2.3  Representing Data as Networks

In this section, we review the diverse types of network representations, and how they can capture key features in the raw data. We will use the global shipping network as a real-world example to demonstrate what additional information can be incorporated in diverse network representations.

2.3.1  Simple Network

A simple network captures pairwise relationship between entities. The network $\mathbf{G}$ is composed of nodes $n \in \mathbf{N}$ and edges $e \in \mathbf{E}$, with $e = n_i, n_j$ capturing the pairwise relationship in the raw data. It can be constructed from simple pairwise data (Section 2.2.1) via a one-to-one mapping. For example, a simple global shipping network can tell which two ports in the world are connected.

Since the minimalistic simple network is trivial to build, it is perhaps the most frequently used network representation. However, the simple network preserves nothing more than pairwise relationships, if it is used as the basis for further analysis, any other potentially useful feature is ignored.

2.3.2 Weighted Network

The weighted network captures the weight of relationship as additional information. The network $\mathbf{G}$ is composed of nodes $n \in \mathbf{N}$ and edges $e \in \mathbf{E}$, with $e = n_i, n_j, w_{i,j}$ capturing the weighted pairwise relationship in the raw data. It is a lossless representation of the weighted pairwise data (Section 2.2.2) or the matrix data (Section 2.2.5). For example, the weighted global shipping network can represent the amount of traffic between two ports.

2.3.3 Directed Network

The directed network captures the directionality of relationship as additional information. The network $\mathbf{G}$ is composed of nodes $n \in \mathbf{N}$ and edges $e \in \mathbf{E}$, with $e = < n_i, n_j >$ capturing the directed pairwise relationship in the raw data. It is a lossless representation of the directed pairwise data (Section 2.2.3). For example, the weighted global shipping network can represent the amount of traffic between two ports.

2.3.4 Temporal Network

The temporal network captures the activation time (timestamps or durations) of relationship as additional information. The network $\mathbf{G}$ is composed of nodes $n \in \mathbf{N}$ and edges $e \in \mathbf{E}$, with $e = < n_i, n_j, T = < t_1, t_2, \cdots >$ capturing the discrete activation times of the relationship or $e = < n_i, n_j, T = < \Delta t_1, \Delta t_2, \cdots >$ It is a lossless representation of the temporal pairwise data (Section 2.2.4). For example, the changing coverage of sea ice in the Arctic region influences the availability of shipping routes. The temporal network can capture the changes in network topology.

2.3.5 Dynamic Network

The dynamic network captures the temporal evolution of relationships as additional information. It can be seen as a series of static networks, laid out on the time axis: $\mathbf{G_{dyn}} =< G_{t1}, G_{t2}, \cdots >$. It is a lossless representation of the tensor data (Section 2.2.6) if the third dimension in the tensor is time. For example, the dynamic global shipping network can represent the changing shipping traffic in different years.

An extension of dynamic network is multilayer network, which usually assumes some connection in different layers of networks. For example, there can be a network of ports, a network of ships, and a network of countries, and the entities in these different layers of networks may connect to each other. Nevertheless, such multilayer network is highly dependent on the research context.

2.3.6 Heterogeneous Network

The heterogeneous network assigns types to nodes and/or edges, therefore having the ability to storing metadata about pairwise relationships. The network $\mathbf{G}$ is composed of nodes $n \in \mathbf{N}$ and edges $e \in \mathbf{E}$, with $e = (n_i, n_j)$ having one or more attributes, and $n$ also having additional attributes. It is a lossless representation of the tensor data (Section 2.2.6). For example, in the global shipping heterogeneous network, the nodes can represent the type of ports (civilian or military), and the edges can represent the type of shipping routes (passenger or cargo).

2.3.7  Hypergraph

The hypergraph [31] introduces hyperedges that can connect multiple nodes, therefore it can capture relationships among multiple entities. The network **G** is composed of nodes $n \in \mathbf{N}$ and edges $e \in \mathbf{E}$, with $e = (n_i, n_j, \ldots)$ containing arbitrary number of entities. Note that the order of nodes in every hyperedge is interchangeable. Hypergraph is a lossless representation of the group data (Section 2.2.7). When representing the global shipping network, hypergraph can show the overlapping groups of ports different shipping companies have businesses at.

2.3.8  Higher-order Network

The higher-order network uses auxiliary higher-order nodes and edges to capture the previous steps. The network **G** is composed of higher-order nodes $n \in \mathbf{N}$ and higher-order edges $e \in \mathbf{E}$. Its edges $e = <n_i, n_j, w_{i,j}>$ are weighted and directed, and nodes $n = <E_t | E_{t-1}, E_{t-2}, \cdots>$ stores an *ordered* list of entities, with the first $E_t$ being the current physical entity, and the rest indicating the previous few steps.

There are several important distinctions between the higher-order network and the hypergraph: (1) in hypergraph, every node still represents a single entity, but in higher-order network, multiple nodes can represent the same entity; (2) in hypergraph, the order of entities in an edge is interchangeable, but in the higher-order network, the order of entities in a node cannot be changed; (3) the hypergraph is not directly compatible with most of network analysis methods, but the higher-order network can be treated as conventional networks and apply existing network analysis tools.

The higher-order network has several forms, notably the fixed-order network [178], and the variable-order network [229], which will be discussed in detail later. The higher-order network is a *lossy* representation of the raw sequential data (Section 2.2.8), the diffusion data (Section 2.2.9), and the time series data (Section 2.2.10). The qual-ity of representation (how much information is preserved from the raw data versus the size of network) is also influenced by the definition – these are the main focus of Part I. For example, the higher-order network of global shipping can highlight how a ship's previous steps can influence the ship's next step.

2.4 Representing Data as Networks

A representation is a re-organization of data. By representing data in an alternative format, one usually expects to:

- Reveal the main patterns of the raw data.
- Reduce storage space.
- Facilitate interpretation and visualization.
- Prepare for generalization, regression, or prediction.
- Enable access to data analysis tools in another domain.

Network-based representation has quickly emerged as the norm in representing rich interactions among the components of a complex system for analysis and modeling. It is critical for the network to truly represent the inherent phenomena in the complex system to avoid incorrect analysis results or conclusions. The question of *how to accurately represent the big data derived from these complex systems as networks*, although being the prerequisite of subsequent network analyses, does not receive equal attention as network analysis itself.

It is common to ignore the gap and represent the raw data as the simple network; nevertheless, given the aforementioned diverse types of raw data, and network representations that can capture different types of information, *how to best preserve the information in the raw data in network representations?* We aim to conduct a review of representation methods in this section as a first step to bridge the gap.

### 2.4.1 Lossless and Lossy Representations

While can be diverse representations for the same data, not all representations contain the same information. For example, for the same sequence of numbers $1, 2, 3, 5, 8$, one may perform linear regression and represent it as $x = 1.7n - 1.3$, or describe as $x_1 = 1, x_2 = 2, x_n = x_{n-1} + x_{n-2}$ for $n > 2$. Both representations have their values in different contexts; however, there is an important distinction between the two: *given the new representation, is it possible to reproduce the exact raw data?* Apparently, although the latter representation is slightly more complex, one can always reproduce the exact original series of numbers, we call it *lossless* representation. Whereas for the previous representation, the exact numbers in the raw data are lost forever: we call it *lossy* representation.

In general, a simple way to test if a representation is lossless is to check whether or not the original data can be restored given the representation. For diverse lossy representations, the quality representation depends on (1) how well the raw data can be restored, and (2) if the representation itself is efficient.

### 2.4.2 The Gap between Data and Network

We present the frequently seen data types, network types, and conversion methods in Figure 2.1 at the beginning of the chapter. We group data and network into two major categories: those that represent pairwise relationships on the upper half, and those representing higher-order relationships on the lower half. We use solid lines to denote lossless representations, and dotted lines to denote lossy representations.

From Figure 2.1, the most frequently seen networks (simple, weighted and directed) are all in the upper half, all corresponding to data with pairwise nature (tuple types and matrix types). However, for other important types of data, particularly the sequential type of data, there do not exist a lossless network representation.

Three options exist:

(1) Create a lossless representation. To preserve the full sequences in the sequential data, the lossless network representation (if exists) will use edges with the same length of the original sequence, which becomes impractical when the sequence is long.

(2) Fall back to pairwise relationship. One way to represent sequential data is first break down the higher-order relationships into pairwise relationships and create a pairwise data, then represented as simple networks. This higher-order to pairwise conversion, however, is lossy, that the resulting network will not incorporate any information about higher-order relationships.

(3) Find a network representation that can partially store higher-order relationships in the network structure. The network should ideally retain the advantage of network representation, incorporate the most important higher-order relationships; also, it should not induce significant overhead in the representation. The following chapters will discuss approaches in this direction, as concrete steps to fill the gap between higher-order relationship data and the network representation.

# CHAPTER 3

# HIGHER-ORDER NETWORK (HON): THE ORIGINAL ALGORITHM

## 3.1 Overview

To ensure the correctness of network analysis methods, the network (as the input) has to be a sufficiently accurate representation of the underlying data. However, when representing sequential data from complex systems such as global shipping traffic or web clickstream traffic as networks, the conventional network representation implicitly assuming the Markov property (first-order dependency) can quickly become limiting. That is, when movements are simulated on the network, the next movement depends only on the current node, failing to capture the fact that the movement may depend on multiple previous steps.

**Intellectual merit:** We show that data derived from some complex systems show up to fifth-order dependencies, such that the oversimplification in the first-order network representation can later result in inaccurate network analysis results. To that end, we propose the Higher-Order Network (HON) representation that can discover and embed variable orders of dependencies in a network representation. Through a comprehensive empirical evaluation and analysis, we establish several desirable characteristics of HON – accuracy, scalability, and direct compatibility with the existing suite of network analysis methods. We illustrate the broad applicability of HON by using it as the input to a variety of tasks, such as random walking, clustering and ranking, where these methods yield more accurate results without modification. Our approach brings the representative power of networks for handling the increasingly complex systems.

**Connections:** This research is an algorithm based on the discussions in Chapter 2. It is the foundation and the main contribution of this dissertation. It serves as the basis for the improved HON+ algorithm in Chapter 4, the underlying algorithm for the visualization software HoNVis in Chapter 5, and the algorithmic foundation for the global shipping & species invasion analysis in Chapter 8 and 9. It can be used as network features in Chapter 11 and 12.

**Work status:** This work is accomplished in collaboration with Prof. Nitesh Chawla and Prof. Thanuka Wickramarathne. It also receives helpful comments from Yuxiao Dong, Reid Johnson and Prof. David Lodge. It has been published at Science Advances[229].

3.2 Introduction

Today's systems are inherently complex, whether it is the billions of people on Facebook powering a global social network, the transportation networks powering the commute and the economy, or the interacting neurons powering the coherent activity in the brain. Complex systems such as these are made up of a number of interacting components that influence each other, and network-based representation has quickly emerged as the norm by which we represent the rich interactions among the components of such a complex system. These components are represented as nodes in the network, and the edges or links between these nodes represent the (ranges and strengths of) interactions. This conceptualization raises a fundamental question: Given the data, how should one construct the network representation such that it appropriately captures the interactions among the components of a complex system?

A common practice to construct the network from data (in a complex system) is to directly take the sum of pairwise connections in the sequential data as the edge weights in the network—for example, the sum of traffic between locations in an interval [17, 53, 67, 158], the sum of user traffic between two Web pages, and so on [23, 157, 204]. However, this direct conversion implicitly assumes the Markov property (first-order dependency) [141] and loses important information about dependencies in the raw data. For example, consider the shipping traffic network among ports, where the nodes are ports and the edges are a function of the pairwise shipping traffic between two ports. When interactions are simulated on the network, such as how the introduction of invasive species to ports is driven by the movements of ships via ballast water exchange, the next interaction (port-port species introduction) only depends on the current node (which port the ship is coming from), although, in fact, the interaction may be heavily influenced by the sequence of previous nodes (which ports the ship has visited before). Another example is user clickstreams on the Web, where nodes are Web pages and interactions are users navigating from one Web page to another. A users next page visit not only depends on the last page but also is influenced by the sequence of previous clicks. Thus, there are higher-order dependencies in networks and not just the first-order (Markovian) dependency, as captured in the common network representation. Here, we focus on deriving the network based on the specific set of interactions, namely, the interactions induced by movements among components of a complex system, wherein the sequence of movement patterns becomes pivotal in defining the interactions.

Let us again consider the process of constructing a network from the global shipping complex system by incorporating the movements from the ship trajectories (Figure 3.1A) [118, 228]. Conventionally [17, 23, 53, 67, 118, 157, 158, 204], a network is built by taking the number of trips between port pairs as edge weights (Figure 3.1B). When ship movements are simulated on this first-order network, according to the network structure where the edge Singapore Los Angeles and the edge Singapore Seattle have similar edge weights, a ship currently at Singapore has similar probabilities of going to Los Angeles or Seattle, no matter how it arrived at Singapore. In reality, the global shipping data indicate that a ships previous stops before arriving at Singapore influence the ships next movement: the ship is more likely to continue on to Los Angeles if it came from Shanghai and more likely to go to Seattle if it came from Tokyo. A first-order network representation fails to capture important information like this because, in every step, the flow of traffic on the network is simply aggregated and mixed. As a consequence, trajectories simulated on the first-order network do not follow true ship movement patterns. By contrast, by breaking down the node Singapore into Singapore—Tokyo and Singapore—Shanghai (Figure 3.1C), the higher-order network (HON) can better guide the movements simulated on the network. Because ships can translocate species along intermediate stops via partial ballast water exchanges (11), the ability to distinguish between these cases is critical for producing accurate species introduction probabilities for each port.

Such higher-order dependencies exist ubiquitously and are indispensable for modeling vehicle and human movements [178], email correspondence, article and Web browsing [52, 70, 201], conversations [210], stock market [115], and so on. Although higher-order dependencies have been studied in the field of time series [97, 115], information theory [194], frequent pattern mining [99], next-location prediction [152], variable-order Markov (VOM) [43, 198, 198], hidden Markov model [172], and Markov order estimation [6, 191, 218], they have focused on the stochastic process, rather than on how to represent higher-order dependencies in networks to adequately capture the intricate interactions in complex systems. In the field of network science, the frontier of addressing the higher-order dependencies still remains at the stage of assuming a fixed second order of dependency when constructing the network [100, 178, 187, 189, 190] or using multiplex networks [66], and there is neither a thorough discussion beyond second-order dependencies nor a systematic way of representing dependencies of variable orders in networks. Although there have also been efforts to incorporate HON structures for clustering [30], ranking [124], and so on, these approaches need to modify existing algorithms and are application-specific. As a result, these methods are not generalizable to broader applications, although we expect a network representation that is agnostic to the end-analysis methods (more discussions in Materials and Methods).

Here, we present a novel and generalizable process for extracting higher-order dependencies in the sequential data and constructing the HON that can represent dependencies of variable orders derived from the raw data. We demonstrate that HON is (i) more reflective of the underlying real-world phenomena (for example, when using HON instead of a first-order network to represent the global shipping data, the accuracy is doubled when simulating a ships next movement on the network and is higher by one magnitude when simulating three steps); (ii) efficient in scaling to higher orders, because auxiliary higher-order nodes and edges are added to a first-order network only where necessary; and (iii) consistent with the conventional network representation, allowing for a variety of existing network analysis methods and algorithms to run on HON without modification. These algorithms and methods produce considerably different and more accurate results on HON than on a first-order network, thus demonstrating the broad applications and potential influences of this novel network representation.

We analyze a variety of real-world data including global shipping transportation, clickstream Web browsing trajectories, and Weibo retweet information diffusions. We show that some of them have dependencies up to the fifth order, which the conventional first-order network representations or the fixed second-order network representations simply cannot capture, rendering the downstream network analysis tools, such as clustering and ranking, with limited and possibly erroneous information about the actual interactions in data. We also validate HONs ability to reveal higher-order dependencies on a synthetic data set, where we introduced dependencies of variable orders through a process completely independent of the construction of HON. We show that HON accurately identifies all the higher-order dependencies introduced.

## 3.3 Materials and Methods

### 3.3.1 The HON Representation

Conventionally, a network (also referred to as a graph) $G = (V, E)$ is represented with vertices or nodes $V$ as entities (for example, places, Web pages, etc.) and edges or links $E$ as connections between pairs of nodes (for example, traffic between cities, user traffic between Web pages, etc.). Edge weight $W(i \to j)$ is a number associated with an edge $i \to j$ representing the intensity of the connection, which is usually assigned as the (possibly weighted) sum of pairwise connections $i \to j$ (for example, the daily traffic from $i$ to $j$) in data [17, 23, 53, 67, 118, 157, 158, 204].

A wide range of network analysis methods, such as PageRank for ranking [161], MapEquation [177] and Walktrap [171] for clustering, and link prediction methods [15, 86] use random walking to simulate movements on networks (for example, ships traveling between ports, users clicking through Web pages, etc.). If the location of a random walker at time $t$ is denoted as a random variable $X_t$ where $X$ can take values from the node set $V$, then, conventionally [69, 86], the transition probability from node $i_t$ to the next step $i_{t+1}$ is proportional to the edge weight $W(i \rightarrow i_{t+1})$:

$$P(X_{t+1} = i_{t+1} | X_t = i_t) = \frac{W(i_t \rightarrow i_{t+1})}{\sum_j W(i_t \rightarrow j)} \tag{3.1}$$

This Markovian nature of random walking dictates that every movement simulated on the network is only dependent on the current node. In the conventional first-order network representation, every node maps to a unique entity or system component, so that every movement of a random walker is only dependent on a single entity (Singapore in Figure 3.1). Data with higher-order dependencies that involve more than two entities, such as "ships coming from Shanghai to Singapore are more likely to go to Los Angeles" in the global shipping data, cannot be modeled via the conventional first-order network representation. Thus, the simulation of movement performed on such networks will also fail to capture these higher-order patterns.

To represent higher-order dependencies in a network, we need to rethink the building blocks of a network: nodes and edges. Instead of using a node to represent a single entity (such as a port in the global shipping network), we break down the node into different higher-order nodes that carry different dependency relationships, where each node can now represent a series of entities. For example, in Figure 3.1C, Singapore is broken down into two nodes, Singapore *given* Tokyo as the previous step (represented as $Singapore|Tokyo$), and Singapore *given* Shanghai as the previous step (represented as $Singapore|Shanghai$). Consequently, the edges $Singapore|Shanghai \to LosAngeles$ and $Singapore|Shanghai \to Seattle$ can now involve three different ports as entities and carry different weights, thus representing second-order dependencies. Because the out-edges here are in the form of $i|h \to j$ instead of $i \to j$, a random walker's transition probability from node $i|h$ to node $j$ is:

$$P(X_{t+1} = j | X_t = (i|h)) = \frac{W(i|h \to j)}{\sum_k W(i|h \to k)} \qquad (3.2)$$

so that although a random walker's movement depends only on the current node, it now depends on multiple entities in the new network representation (as in Figure 3.1C), thus being able to simulate higher-order movement patterns in the data. This new representation is consistent with conventional networks and compatible with existing network analysis methods, because the data structure of HON is the same as the conventional network (the only change is the labeling of nodes). This makes it easy to use HON instead of the conventional first-order network as the input for network analysis methods, with no need to change the existing algorithms.

Rosvall et al. [178] consider a higher-order dependency, albeit with a fixed second-order assumption. They propose a network representation comprised of "physical nodes" and "memory nodes". As we will show with experiments, variable orders of dependencies can co-exist in the same data set, and be up to the fifth order in our data. So if the dependency is assumed as fixed second order, it could be redundant when first-order dependencies are sufficient, and could be insufficient when higher-order dependencies exist. In HON, every node can involve more than two entities and represent an arbitrarily high order of dependency, so variable orders of dependencies can co-exist in the same network representation, as shown in Figure 3.1D. For example, the fourth-order dependency relationship following the path of $Tianjin \to Busan \to Tokyo \to Singapore$ can now be represented as a fourth-order node $Singapore|Tokyo, Busan, Tianjin$; the second-order path $Shanghai \to Singapore$ is now a node $Singapore|Shanghai$; first-order relationships are now in a node $Singapore|\cdot$. Yet these nodes of variable orders all represent the same physical location Singapore. Compared with fixed order networks, we will show that our representation is compact in size by using variable orders and embedding higher-order dependencies only where necessary.

While the hypergraph [31] looks similar to HON in that its edges can connect to multiple nodes at the same time, it cannot directly represent dependencies. The reason is that dependencies are ordered relationships, but in a hypergraph the nodes connected by hyperedges are unordered, e.g., in the shipping example, an edge in a hypergraph may have the form of $set\{Tokyo, Busan, Tianjin\} \to set\{Singapore\}$, where Tokyo, Busan, and Tianjin are interchangeable and cannot represent the *path* of the ship before arriving at Singapore. On the contrary, the edges in HON have the form of $\{Tokyo|Busan, Tianjin\} \to \{Singapore|Tokyo, Busan, Tianjin\}$ where the entities in nodes are not interchangeable, thus HON can represent dependencies of arbitrary order.

### 3.3.2 The HON Construction Algorithm

The construction of the HON consists of two steps: rule extraction identifies higher-order dependencies that have sufficient support and can significantly alter a random walkers probability distribution of choosing the next step; then, network wiring adds these rules describing variable orders of dependencies into the conventional first-order network by adding higher-order nodes and rewiring edges. The data structure of the resulting network is consistent with the conventional network representation, so existing network analysis methods can be applied directly without being modified. We use global shipping traffic data as a working example to demonstrate the construction of HON, but it is generalizable to any sequential data.

3.3.2.1 Rule Extraction

The challenge of rule extraction is to identify the appropriate orders of dependencies in data; when building the first-order network, this step is often ignored by simply counting pairwise connections in the data to build the first-order network. We define a *path* as the movement from source node $A$ to target node $B$ with sufficient support (e.g. frequency $\geq 10$), though with nodes that differ from those in a conventional network: a node here can represent a sequence of entities, no longer necessarily a single entity. Then among those paths, given a source node $A$ containing a sequence of entities $[a_1, a_2, \ldots, a_k]$, if including an additional entity $a_0$ at the beginning of $A$ can significantly alter the normalized counts of movements (as probability distribution) to target nodes set $\{B\}$, it means $\{B\}$ has a higher-order dependency on $A_{ext} = [a_0, a_1, a_2, \ldots, a_k]$, and paths containing higher-order dependencies like $A_{ext} \to \{B\}$ are defined as rules. Then a rule like $Freq([a_0, a_1] \to a_2) = 50$ can map to an edge in the network in the form of $a_1 | a_0 \to a_2$ with edge weight 50. What are the expectations for the rule extraction process?

First, rules should represent dependencies that are significant. As in Figure 3.5 ③, if the probability distribution of a ship's next step from Singapore is significantly affected by knowing the ship came from Shanghai to Singapore, there is at least second-order dependency here. On the contrary, if the probability distribution of going to the next port is the same no matter how the ship reached Singapore, there is no evidence for second-order dependency (but third or higher-order dependencies may still exist, such as $g | f, d$ in Figure 3.5, and can be checked similarly).

Second, rules should have sufficient support. Only when some pattern happens sufficiently many times can it be considered as a "rule" or a "path" rather than some random event. Although this requirement of minimum support is not compulsory, not specifying a minimum support will result in a larger and more detailed network representation, and more infrequent routes are falsely considered as patterns, ulti-mately lowering the accuracy of the representation (see the discussions of parameters in Section 3.3.3).

Third, rules should be able to represent variable orders of dependencies. In real-world data such as the global shipping data, different paths can have different orders of dependencies, for example in Figure 3.1 the next step from Singapore is dependent on Tianjin through the fourth-order path $Tianjin \rightarrow Busan \rightarrow Tokyo \rightarrow Singapore$, as well as on Shanghai through the second-order path $Shanghai \rightarrow Singapore$. When variable orders can co-exist in the same data set, the rule extraction algorithm should not assign a fixed order to the data, but should be able to yield rules representing variable orders of dependencies.

Following the aforementioned three objectives of rule extraction, it is natural to *grow* rules incrementally: start with a first-order path, try to increase the order by including one more previous step, and check if the probability distribution for the next step changes significantly (Figure 3.5③). If the change is significant, the higher order is assumed, otherwise keep the old assumption of order. This rule growing process is iterated recursively until (a) the minimum support requirement is not met, or (b) the maximum order is exceeded.

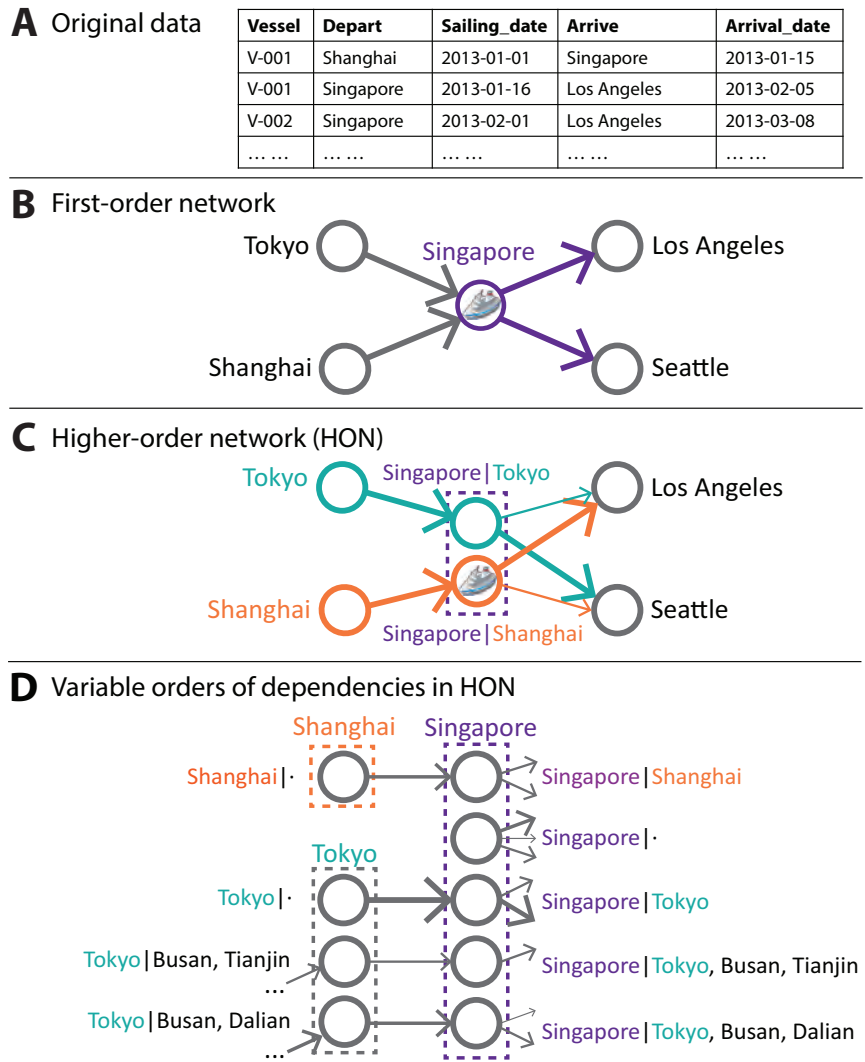

Figure 3.1. **Necessity of representing dependencies in networks.** (**A**) A global shipping data set, containing ship movements as sequential data. (**B**) A first-order network built by taking the number of trips between port pairs as edge weights. A ship currently at Singapore has similar probabilities of going to Los Angeles and Seattle, no matter where the ship came to Singapore from. (**C**) By breaking down the node Singapore, the ships next step from Singapore can depend on where the ship came to Singapore from and thus more accurately simulate movement patterns in the original data. (**D**) Variable orders of dependencies represented in HON. First-order to fourth-order dependencies are shown here and can easily extend to higher orders. Coming from different paths to Singapore, a ship will choose the next step differently.

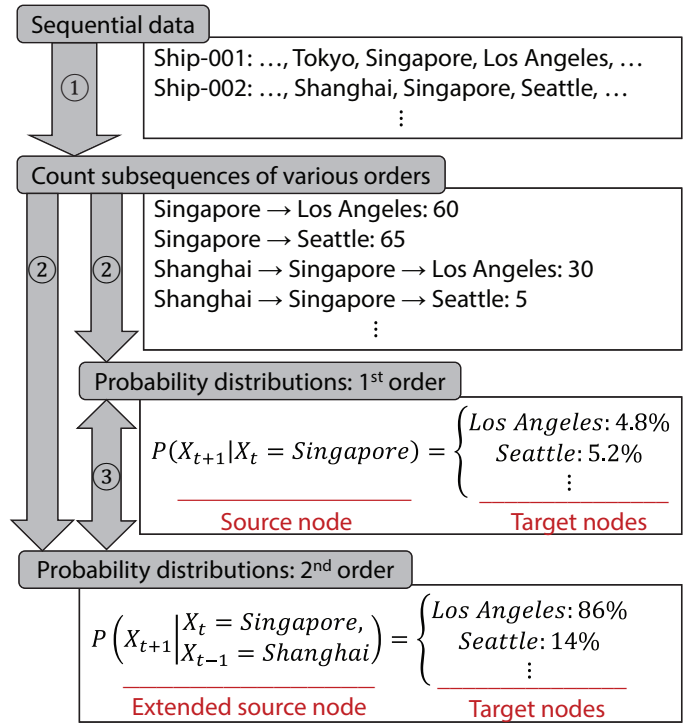

Figure 3.2. **Rule extraction example for the global shipping data.** Step 1: count the occurrences of subsequences from the first order to the maximum order, and keep those that meet the minimum support requirement. Step 2: given the source node representing a sequence of entities as the previous step(s), compute probability distributions for the next step. Step 3: given the original source node and an extended source node (extended by including an additional entity at the beginning of the entity sequence), compare the probability distributions of the next step. For example, when the current location is Singapore, knowing that a ship comes from Shanghai to Singapore (second order) significantly changes the probability distribution for the next step compared with not knowing where the ship came from (first order). So the second-order dependency is assumed here; then the probability distribution is compared with that of the third order, and so on, until the minimum support is not met or the maximum order is exceeded.

**Algorithm 1** Rule extraction. Given the sequential data $T$, outputs higher-order dependency rules $R$.
*Parameters*: $\mathcal{O}$: MaxOrder, $S$: MinSupport

 1: *global* counter $C \leftarrow \emptyset$
 2: *global* nested dictionary $D \leftarrow \emptyset$
 3: *global* nested dictionary $R \leftarrow \emptyset$
 4:
 5: **function** EXTRACTRULES($T$)
 6:     BUILDOBSERVATIONS($T$)
 7:     BUILDDISTRIBUTIONS()
 8:     GENERATEALLRULES()
 9:     **return** $R$
10:
11: **function** BUILDOBSERVATIONS($T$)
12:     **for** $t$ in $T$ **do**
13:         **for** $o$ from 2 to $\mathcal{O}$ **do**
14:             $SS \leftarrow$ ExtractSubSequences($t, o$)
15:             **for** $s$ in $SS$ **do**
16:                 $Target \leftarrow$ LastElement($s$)
17:                 $Source \leftarrow$ AllButLastElement($s$)
18:                 IncreaseCounter($C[Source][Target]$)
19:
20: **function** BUILDDISTRIBUTIONS()
21:     **for** $Source$ in $C$ **do**
22:         **for** $Target$ in $C[Source]$ **do**
23:             **if** $C[Source][Target] < S$ **then**
24:                 Remove(C[Source][Target])
25:         **for** $Target$ in $C[Source]$ **do**
26:             $D[Source][Target] \leftarrow C[Source][Target]$ / Sum($C[Source][*]$)
27:
28: **function** GENERATEALLRULES()
29:     **for** $Source$ in $D$ **do**
30:         **if** length($Source$)= 1 **then**
31:             ADDTORULES($Source$)
32:             EXTENDRULE($Source, Source, 1$)
33: *(To be continued on the next page.)*

**Algorithm 1** *(continued)*

```
34: function EXTENDRULE(Valid, Curr, Order)
35:     if Order ≥ 𝒪 then
36:         ADDTORULES(Valid)
37:     else
38:         Distr ← D[Valid]
39:         NewOrder ← Order + 1
40:         Extended ← all Source satisfying length(Source)= NewOrder and end with Curr
41:         if Extended = ∅ then
42:             ADDTORULES(Valid)
43:         else
44:             for ExtSource in Extended do
45:                 ExtDistr ← D[ExtSource]
```
46:             **if** $D_{KL}(ExtDistr||Distr) > \frac{NewOrder}{\log_2(1+\text{Sum}(C[ExtSource][*]))}$ **then**
```
47:                     EXTENDRULE(ExtSource, ExtSource, NewOrder)
48:                 else
49:                     EXTENDRULE(Valid, ExtSource, NewOrder)
50:
51: function ADDTORULES(Source)
52:     if length(Source) > 0 then
53:         R[Source] ← C[Source]
54:         PrevSource ← AllButLastElement(Source)
55:         ADDTORULES(PrevSource)
```

Algorithm 1 gives the pseudocode of rule extraction. The procedure consists of three major steps: BUILDOBSERVATIONS counts the frequencies of all subsequences from the second order to *MaxOrder* for every trajectory in T (Figure S1 ①); BUILDDISTRIBUTIONS first builds all paths by removing subsequences that appear less than *MinSupport* times, then estimates probability distributions of movements at every source node by normalizing the observed frequencies (Figure S1 ②). GENERATEALLRULES starts from the first order and tries to increase the order recursively, by including an additional entity at the beginning of the entity sequence of the source node and testing if the probability distribution for the next step changes significantly (Figure 3.5 ③).

The comparison between probability distributions is performed by the recursive function EXTENDRULE. $Curr$ is the current source node ($X_t = Singapore$ in Figure 3.5), which is to be extended into a higher-order source node $ExtSource$ by including the previous step ($X_t = Singapore, X_{t-1} = Shanghai$ in Figure 3.5) (line 40). $Valid$ is the last known source node from which a random walker has significantly different probability distributions towards the next step, i.e., the last assumed order for the path starting from node $Valid$ is length($Valid$). If at the extended source node, a random walker has a significantly different probability distribution of the next movement compared with that at node $Valid$, the extended source node will be marked as the new $Valid$ for the recursive growing of rule (line 47), otherwise the old $Valid$ is kept (line 49). The paths with $Valid$ as the source node have correct orders of dependencies, and will be added to the rules set $R$ whenever EXTENDRULE exceeds $MaxOrder$ (line 35) or the source node cannot be extended (line 41, true when no higher-order source node with the same last steps exists). When a higher-order rule (a path from Source) is added, all paths of the preceding steps of Source are added (line 54-55) to ensure the network wiring step can connect nodes with variable orders. For example, when paths from the source node $Singapore|Shanghai$ are added to $R$, the preceding step $Shanghai \rightarrow Singapore$ should also be added to $R$.

Specifically, to determine whether extending the source node from $Valid$ to $ExtSource$ significantly changes the probability distribution for the next step (line 46), we compute the Kullback-Leibler divergence [127] between the two distributions, since it is a widely-used and standard way of comparison probability distributions [29, 43, 198]. We consider the change is significant if the divergence satisfies

$$Divergence_{KL}(P_{ExtSource}||P_{Valid}) > \frac{Order_{ExtSource}}{log_2(1 + Support_{ExtSource})}$$

based on the following intuition, inspired by Ben-Gal 2005 [29] and Bühlmann 1999 [43]: for the two extended nodes showing the same divergence with regard to the original source node, (1) we are more inclined to prune the node with higher orders (the one with more previous steps embedded); and (2) we are less inclined to prune the node with higher support (the one with more trajectories going through). Instead of applying a universal threshold for all nodes that may have varying orders and supports, the threshold we adopt is self-adjustable for different nodes.

It is worth noting that Algorithm 1 also applies to data types other than trajectories such as diffusion data, which needs only one change of the ExtractSubSequences function such that it takes only the newest entity subsequence. In addition, although higher-order dependencies exist in many types of data, it is the type of data (vessel trajectory data / gene sequence data / language data / diffusion data ) that determines whether there are higher-order dependencies and how high the orders can be. Our proposed algorithm is backwards compatible with data that have no higher-order dependencies (such as the diffusion data): all rules extracted are first-order, thus the output will be a first-order network.

#### 3.3.2.2  Network Wiring

The remaining task is to convert the rules obtained from the last step into a graph representation. It is trivial for building conventional first-order networks because every rule is first-order and can directly map to an edge connecting two entities, but such direct conversion will not work when rules representing variable orders of dependencies co-exist. The reason is that during rule extraction, only the last entity of every path is taken as the target node, so that every edge points to a first-order node, which means higher-order nodes will not have in-edges. Rewiring is needed to ensure that higher-order nodes will have incoming edges, while preserving the sum of edge weights in the network. The detailed steps are illustrated as follows:

1. Converting all first-order rules into edges. This step is exactly the same as constructing a first-order network, where every first-order rule (a path from one entity to another) corresponds to a weighted edge. As illustrated in Figure 3.3A, $Shanghai \rightarrow Singapore$ is added to the network.

2. Converting higher-order rules. In this step, higher-order rules are converted to higher-order edges pointing out from higher-order nodes (the nodes are created if they do not already exist in the network). Figure 3.3B shows the conversion of rules $Singapore|Shanghai \rightarrow LosAngeles$ and $Singapore|Shanghai \rightarrow Seattle$, where the second-order node $Singapore|Shanghai$ is created and two edges pointing out from the node are added.

3. Rewiring in-edges for higher-order nodes. This step solves the problem that higher-order nodes have no incoming edges, by pointing existing edges to higher-order nodes. When adding the second-order node $Singapore|Shanghai$, a lower order rule and the corresponding edge $Shanghai \rightarrow Singapore$ are guaranteed to exist, because during rule extraction when a rule is added, all preceeding steps of the path are also added, as in ADDTORULES in Algorithm 1. As shown in Figure 3.3C, the edge from $Shanghai$ to $Singapore$ is redirected to $Singapore|Shanghai$. The rewiring step (instead of adding edges) also preserves the sum of edge weights. Converting higher-order rules (Step 2) and rewiring (Step 3) are repeated for all rules of first order, then second order, and likewise up to the maximum order, in order to guarantee that edges can connect to nodes with the highest possible orders. This step also implies that any two nodes that represent the same physical location will not have incoming edges from the same node.

4. Rewiring edges built from $Valid$ rules. After representing all rules as edges in HON, additional rewirings are needed for edges built from $Valid$ rules (refer to Algorithm 1). The reason is that the rule extraction step takes only the last entity of paths as targets, such that edges built from $Valid$ rules always point to first-order nodes. In Figure 3.3D, the node $Singapore|Shanghai$ was pointing to a first-order node $Seattle$. However, if a node of higher order $Seattle|Singapore$ already exists in the network, the edge $Singapore|Shanghai \rightarrow Seattle$ should point to $Seattle|Singapore$, otherwise the information about previous steps is lost. To preserve as much information as possible, the edges built from $Valid$ rules should point to nodes with the highest possible orders.

Following the above process, the algorithm for network wiring is given in Algorithm 2, along with more detailed explanations. Given a set of parameters, the result of HON is unique, so there is no optimization or greedy methods for the algorithm.

Algorithm 2 gives the pseudocode for converting rules extracted from data into a graph representation. In line 3, sorting rules by the length of key $Source$ is equivalent

**Algorithm 2** Network wiring. Given the higher-order dependency rules $R$, convert the rules of variable orders into edges, perform rewiring, and output the graph $G$ as HON.

```
 1: function BUILDNETWORK(R)
 2:     global nested dictionary G ← ∅
 3:     R ← Sort(R, ascending, by length of key Source)
 4:     for r in R do
 5:         G.add(r)
 6:         if length(r.Source) > 1 then
 7:             REWIRE(r)
 8:     REWIRETAILS()
 9:     return G
10:
11: function REWIRE(r)
12:     PrevSource ← AllButLastElement(r.Source)
13:     PrevTarget ← LastElement(r.Source)
14:     if (edge: (Source: PrevSource, Target: r.Source)) not in G then
15:         G.add(edge: (Source: PrevSource, Target: r.Source, Weight: (PrevSource → PrevTarget).Probability))
16:         G.remove(edge: (Source: PrevSource, Target: PrevTarget))
17:
18: function REWIRETAILS()
19:     ToAdd ← ∅; ToRemove ← ∅
20:     for r in R do
21:         if length(r.Target) = 1 then
22:             NewTarget ← concatenate(Source, Target)
23:             while length(NewTarget) > 1 do
24:                 if NewTarget in (all Sources of R) then
25:                     ToAdd.add(edge:(Source: r.Source, Target: NewTarget, Weight: r.Probability)
26:                     ToRemove.add(edge:(Source: r.Source, Target: r.Target))
27:                     Break
28:                 else
29:                     PopFirstElement(NewTarget)
30:     G ← (G ∪ ToAdd)\ToRemove
```

to sorting rules by ascending orders. This ensures that the `for` loop in line 4–7 converts all lower order rules before processing higher-order rules. For every order, line 5 converts rules to edges, and line 7 REWIRE(r) attempts rewiring if it is not the first order. Figure 3.3C illustrates the case $r = Singapore|Shanghai \rightarrow LosAngeles$ in line 11, indicating $PrevSource$ is $Shanghai$ and $PrevTarget$ is $Singapore|Shanghai$. The edge $Shanghai \rightarrow Singapore|Shanghai$ is not found in the network, so in line 15 and 16 $Shanghai \rightarrow Singapore|Shanghai$ replaces $Shanghai \rightarrow Singapore$ using the same edge weight.

After converting all rules in REWIRETAILS, edges created from $Valid$ rules are rewired to target nodes that have the highest possible orders. Figure 3.3D illustrates the following case: in line 21 $r = Singapore|Shanghai \rightarrow Seattle$, so in line 22 $NewTarget$ is assigned as $Seattle|Singapore, Shanghai$; assume $Seattle|Singapore, Shanghai$ is not found in all sources of $R$, so the lower order node $Seattle|Singapore$ is searched next; assume $Seattle|Singapore$ already exists in the graph, so $Singapore|Shanghai \rightarrow Seattle|Singapore$ replaces $Singapore|Shanghai \rightarrow Seattle$.

### 3.3.3 Parameter Discussion

#### 3.3.3.1 Minimum Support

Only when a subsequence occurs sufficiently many times (not below minimum support) can it be distinguished from noise and construed as a (non-trivial) path. While a minimum support is not compulsory, setting an appropriate minimum support can significantly reduce the network size and improve the accuracy of representation. As shown in Figure 3.4A, by increasing the minimum support from 1 to 10 (with a fixed maximum order of 5), the size of the network shrinks by 20 times while the accuracy of random walking simulation increases by -0.54%. By increasing the minimum support from 1 to 100, the accuracy first increases then decreases. The reason is that with low minimum support, some unusual subsequences that are noise are counted as paths; on the other hand, a high minimum support leaves out some true patterns that happen less frequently. The optimal minimum support (that can increase the accuracy of representation and greatly reduce the size of the network) may not be the same for different types of data, but can be found by parameter sweeping.

#### 3.3.3.2 Maximum Order

With a higher maximum order, the rule extraction algorithm can capture dependencies of higher orders, leading to higher accuracies of random walking simulations.

As shown in Figure 3.4B, when increasing the maximum order from 1 to 5 (with a fixed minimum support of 10), the accuracy of random walking simulation keeps increasing but converges at the maximum order of 5, and the same trend applies for the size of the network. The reason is that the majority of dependencies have lower orders while fewer dependencies have higher orders, which again justifies our approach of not assigning a fixed high order for the whole network. On the other hand, setting a high maximum order does not significantly increase the running time of building HON, because in the rule extraction algorithm, most subsequences of longer lengths do not satisfy the minimum support requirement and are not considered in the following steps. In brief, when building HON using the aforementioned algorithm, an order that is sufficiently high can be assumed as the maximum order (a maximum order of five is sufficient for most applications). With a higher maximum order, the rule extraction algorithm can capture dependencies of higher orders, leading to higher accuracies of random walking simulations.

### 3.3.4 Comparison with Related Methods

Although VOM models can be used on sequential data to learn a VOM tree [29, 43, 198] for predictions [28], our goal is to build a more accurate network representation that captures higher-order dependencies in the data. Although these two objectives are related, there are several key differences: (i) a VOM tree contains probabilities that are unnecessary (for example, nodes that are not leaves) for representing higher-order dependencies in a network; (ii) additional conditional probabilities are needed to connect nodes with different orders in HON, which are not guaranteed to exist in a pruned VOM tree; and (iii) VOM usually contains lots of unnecessary edges because of the "smoothing" process for the unobserved data, which is not desired for a network representation. Therefore, our work is not simply contained in a VOM implementation. The next section elaborates the differences and provides an empirical comparison between the HON and VOM.

Although a fixed $k^{th}$-order Markov model can be directly converted to a first-order model [178], the state space $S^k$ grows exponentially with the order. There has been plenty of research on Markov order estimation to determine the order $k$, such as using different information criteria [6, 191], cross-validation [52], and surrogate data [218], but these approaches produce a single global order for the model rather than variable orders, and no discussion was given to network representation. Other Markov-related works, such as hidden Markov model [172], frequent pattern mining [99], and next-location prediction [152] focus more on the stochastic process, rather than the network representation problem. For example, the hidden states in hidden Markov model do not represent clear dependency relationships like the higher-order nodes do in HON, and we are not learning a hidden layer that have "emission probabilities" to observations. From the network perspective, although there have been efforts to incorporate HON structures for clustering [30, 124], ranking [91], and so on, these methods are modifications of existing algorithms and are application-specific; instead, we embed higher-order dependencies into the network structure, so that the wide range of existing network analysis tools can be applied without modification.

3.3.5   Empirical Comparison with the Variable-order Markov (VOM) Model

In this section we first illustrate the differences between HON and VOM with an example, then provide an empirical comparison between HON and VOM. Note that because the "smoothing" process is not a compulsory step of VOM, we do not apply it in the following comparison, although "smoothing" is undesirable for network representation.

3.3.5.1 Example

In the context of global shipping, suppose ports $a, b, c, \ldots, h, i$ are connected as shown in Figure 3.5A. Port $f$ and $g$ are at the two ends of a canal. We assume that all ships coming from $d$ through the canal will go to $h$, and all ships coming from $e$ through the canal will go to $i$. A possible set of ship trajectories are listed in Figure 3.5B. Based on these trajectories, we can count the frequencies of subsequences, and compute the probability distributions of next steps given the previous ports visited (HON and VOM yield identical results). The subsequences of variable orders can naturally form a tree as shown in Figure 3.5C, where source nodes are in circles and target nodes (and the corresponding edge weights) are in the boxes below.

HON and VOM have different mechanisms of deciding which nodes to retain in the tree. In Figure 3.5C, the nodes kept by HON are denoted by red stars and nodes kept by VOM are denoted by purple triangles, showing a mismatch of the results. For HON, although $g|f$ does not show second order dependency (having the same probability distribution with $g$), $g|f,d$ shows third order dependency (having significantly different probability distribution compared with $g$), so $g|f,d$ is retained by HON. According to ADDTORULES of the "rule extraction" step (Algorithm 1), all preceding nodes are retained, including $f|d$ and $d$, such that the "network wiring" step already has exactly the nodes needed: there would be a path of $d \to f|d \to g|f,d$, as shown in Figure 3.5D. On the contrary, in the VOM construction process, after determining that $g|f,d$ is a higher-order node to be kept, VOM keeps $g|f$, and prunes $f|d$, despite that (1) $f|d$ is necessary for building the link to $g|f,d$ when constructing a network, and (2) $g|f$ is not necessary for building HON as it has the same probability distribution with $g$.

The eventual wiring of HON is shown in Figure 3.5D. Compared with the true connection in Figure 3.5A, HON not only keeps the first order links, but also adds higher-order nodes and edges for the two ports $f$ and $g$ in the canal, successfully capturing the pattern that "all ships coming from $d$ through the canal will go to $h$, and all ships coming from $e$ through the canal will go to $i$".

An additional difference between HON and VOM is how they determine the orders of rules. HON assumes the first order initially and compares with higher orders, while VOM "prunes" rules recursively from higher orders to lower orders, which as illustrated in Figure 3.5E, may prune higher-order nodes despite they have very different distributions than first-order nodes (e.g., $z|y,x,w$ compared with $z$), thus underestimating the orders of dependencies.

In brief, we have shown that VOM cannot be used directly to construct HON, given that VOM (1) retains unnecessary nodes for constructing HON, (2) prunes necessary nodes, and (3) has a pruning mechanism that may leave out certain higher-order dependencies.

3.3.5.2 Numerical Comparison

To show the differences of HON and VOM quantitatively, we apply both HON and VOM to the same global shipping data set, assume the same filtering for preprocessing ($MaxOrder = 5$ and $MinSupport = 10$), use the same distance measure (KL divergence), and for fair comparison, we use the same threshold $\frac{Order_{ExtSource}}{log_2(1+Support_{ExtSource})}$ for judging whether two distributions are significantly different. Table 3.2 gives the comparison of the number of rules extracted from both algorithms.

We can observe that the rules extracted by HON and VOM show considerable differences except for the first order, even though these two algorithms are given the same parameters. The different mechanisms of deciding which nodes to keep lead to the differences in the extracted rules. This further supports our claim that the rules extracted by VOM cannot be readily used for building HON, while the "rule extraction" process of HON has already prepared exactly the rules needed and only need to rewire some links.

## 3.4 Results

We start with an introduction with multiple real-world and synthetic data sets used in this study. Then we compare our proposed network representation with the conventional ones in terms of accuracy, scalability, and observations drawn from network analysis tools.

### 3.4.1 Data Sets

**Global shipping data.** This data made available by Lloyd's Maritime Intelligence Unit (LMIU) contains ship movement information such as `vessel_id`, `port_id`, `sail_date` and `arrival_date`. Our experiments are based on a recent LMIU data set that spans one year from May $1^{st}$, 2012 to April $30^{th}$, 2013, totaling $3,415,577$ individual voyages corresponding to $65,591$ ships that move among $4,108$ ports and regions globally. A minimum support of 10 is used to filter out noise in the data.

**Clickstream data.** This data made available by a media company contains logs of users clicking through web pages that belong to 50 news web sites owned by the company. Fields of interest include `user_ip`, `pagename` and `time`. Our experiments are based on the clickstream records that span two months from December $4^{th}$, 2012 to February $3^{rd}$, 2013, totaling $3,047,697$ page views made by $179,178$ distinct IP addresses on $45,257$ web pages. A minimum support of 5 is used to filter out noise in the data. Clickstreams that are likely to be created by crawlers (abnormally long clickstreams / clickstreams that frequently hit the error page) are omitted.

**Retweet data.** This data [233] records retweet history on Weibo (a Chinese microblogging website), with information about who retweets whose messages at what time. The data was crawled in 2012 and there are $23,755,810$ retweets recorded, involving $1,776,950$ users.

**Synthetic data.** We created a trajectory data set (data and code are available at https://github.com/xyjprc/hon) with known higher-order dependencies to verify the effectiveness of the rule extraction algorithm. In the context of shipping, we connect 100 ports as a 1010 grid, then generate trajectories of 100,000 ships moving among these ports. Each ship moves 100 steps, yielding 10,000,000 movements in total. Normally each ship has equal probabilities of going up/down/left/right on the grid in each step (with wrapping, e.g., going down at the bottom row will end up in the top row); we use additional higher-order rules to control the generation of ship movements. For example, a second-order rule can be defined as whenever a ship comes from Shanghai to Singapore, instead of randomly picking a neighboring port of Singapore for the next step, the ship has 70% chance of going to Los Angeles and 30% chance of going to Seattle. We predefine 10 second-order rules like this, and similarly 10 third-order rules, 10 fourth-order rules, and no other higher-order rules, so that movements that have variable orders of dependencies are generated. To test the rule extraction algorithm, we set the maximum order as five to see if the algorithm will incorrectly extract false rules beyond the fourth order which we did not define; we set minimum support as five for patterns to be considered as rules.

### 3.4.2 Higher-order Dependencies in Data Revealed by HON

First, we show that HON can correctly extract higher-order dependencies from synthetic data. The synthetic data set has 10,000,000 generated movements, based on the predefined 10 second-order dependencies, 10 third-order dependencies, and 10 fourth-order dependencies. On this synthetic data with known variable orders of dependencies, HON (i) correctly captures all 30 of the higher-order dependencies out of the 400 first-order dependencies, with variable orders (from second-order to fourth-order) of dependencies mixed in the same data set correctly identified; (ii) does not extract false dependencies beyond the fourth order even if a maximum order of five is allowed; and (iii) determines that all other dependencies are first-order, which reflects the fact that there is no other higher-order dependency in the data.

We then explore higher-order dependencies in real data: the global shipping data containing ship trajectories among ports, the clickstream data containing user browsing trajectories among Web pages, and the retweet data containing information diffusion paths among users. The global shipping data reveal variable orders of dependencies up to the fifth order, indicating that a ships movement can depend on up to five previous ports that it has visited. The clickstream data also show variable orders of dependencies up to the third order, indicating that the page a user will visit can depend on up to three pages that the user has visited before, matching the observation in another study on Web user browsing behaviors [52]. The fact that dependencies of variable orders up to the fifth order exist in real data further justifies our approach of representing variable and higher-order dependencies instead of imposing a fixed first or second order. On the contrary, the retweet data (recording information diffusion) show no higher-order dependency at all. The reason is that in diffusion processes, such as the diffusion of information and the propagation of epidemics, according to the classic spreading models [219], once a person A is infected, A will start to broadcast the information (or spread the disease) to all of its neighbors (A), irrespective of who infected A. Because of this Markovian nature of diffusion processes, all diffusion data only show first-order dependencies, and HON is identical to the first-order network. This also agrees with a previous finding that assuming second-order dependency has "marginal consequences for disease spreading" [178].

### 3.4.3 Improved Accuracy on Random Walking

Because random walking is a commonly used method to simulate movements on networks and is the foundation of many network analysis tools, such as PageRank for ranking, MapEquation and Walktrap for clustering, various link prediction algorithms, and so on, it is crucial that a naïve random walker (only aware of the current node and its out-edges) can simulate the movements in the network accurately. If different network representations are built for the same sequential data set (consisting of trajectories), how will the network structure affect the movements of random walkers? Do the random walkers produce trajectories more similar to the real ones when running on HON?

We take the global shipping data to explain the experimental procedures (the clickstream data have similar results). As illustrated in Figure 3.6A, for every trajectory of a ship, the last three locations are held for testing, and the others are used to construct the network. A first-order network (Figure 3.6B), a fixed second-order network [178], and a HON (Figure 3.6C) are constructed from the same data set, respectively. Given one of the networks, for every ship, a random walker simulates the ships movements on the network: it starts from the last location in the corresponding training trajectory and walks three steps. Then, the generated trajectories are compared with the ground truth in the testing set: a higher fraction of correct predictions means that the random walkers can simulate the ships movements better on the corresponding network. Random walking simulations in each network are repeated 1000 times, and the mean accuracies are reported. By comparing the accuracies of random walking, our intention here is not to solve a next-location prediction problem [152] or similar classification problems, but from a network perspective, we focus on improving the representative power of the network, as reflected by the accuracies of random walking simulations.

The comparison of results among the conventional first-order network, the fixed second-order network, and HONs with maximum orders of two to five is shown in Figure 3.6D. It is shown that random walkers running on the conventional first-order network have significantly lower accuracies compared with other networks. The rea-son is that the first-order network representation only accounts for pairwise connections and cannot capture higher-order dependencies in ships movement patterns. For example, a large proportion of ships are going back and forth between ports (for example, port a and port b in Figure 3.6A), which is naturally a higher-order dependency pattern because each ships next step is significantly affected by its previous steps. Such return patterns are captured by HON (Figure 3.6C) but not guaranteed in a first-order network (Figure 3.6B, where ships going from port a to port b may not return to port a). As shown in Table , the probability of a ship returning to the same port after two steps in a first-order network (10.7%) is substantially lower than that in HON (above 40%). From another perspective, in a first-order network, a random walker is given more choices every step and is more "uncertain" in making movements. Such "uncertainty" can be measured by the entropy rate [178, 194], defined as

$$H(X_{t+1}|X_t) = \sum_{i,j} \pi(i) p(i \to j) \log p(i \to j) \qquad (3.3)$$

where $\pi(i)$ is the stationary distribution at node $i$ and $p(i \to j)$ is the transition probability from node $i$ to node $j$, defined in Equation 3.1. The entropy rate measures the number of bits needed to describe every step of random walkingthe more bits needed, the higher the uncertainty. In Table 3.1, the first-order network has the highest entropy rate, indicating that every step of random walking is more uncertain because of the lack of knowledge of what the previous steps are, which leads to the low accuracy in the simulation of movements.

By assuming an order of two for the whole network, the accuracies on the fixed second-order network increase considerably as in Figure 3.6D, because the network structure can help the random walker remember its last two steps. Meanwhile, the accuracies on HON with a maximum order of two are comparable and slightly better than the fixed second-order network, because HON can capture second-order dependencies while avoiding the overfitting caused by splitting all first-order nodes into second-order nodes. Increasing the maximum order of HON can further improve the accuracy and lower the entropy rate; particularly, ship movements in bigger loops need more steps of memory and can only be captured with higher orders, as reflected in Table 3.1, where the probability of returning in three steps increases from 7.3 to 16.4% when increasing the maximum order from two to three in HON. By increasing the maximum order to five, HON can capture all dependencies below or equal to the fifth order, and the accuracy of simulating one step on HON doubles that of the conventional first-order network.

Furthermore, when simulating multiple steps, the advantage of using HON is even bigger. The reason is that in a first-order network, a random walker "forgets" where it came from after each step and has a higher chance of disobeying higher-order movement patterns. This error is amplified quickly in a few stepsthe accuracy of simulating three steps on the first-order network is almost zero. On the contrary, in HON, the higher-order nodes and edges can help the random walker remember where it came from and provide the corresponding probability distributions for the next step. As a consequence, the simulation of three steps on HON is one magnitude more accurate than on first-order network. This indicates that, when multiple steps are simulated (which is usually seen in methods such as PageRank and MapEquation that need multiple iterations), using HON (instead of the conventional first-order network) can help random walkers simulate movements more accurately; thus, the results of all random walking-based network analysis methods will be more reliable.

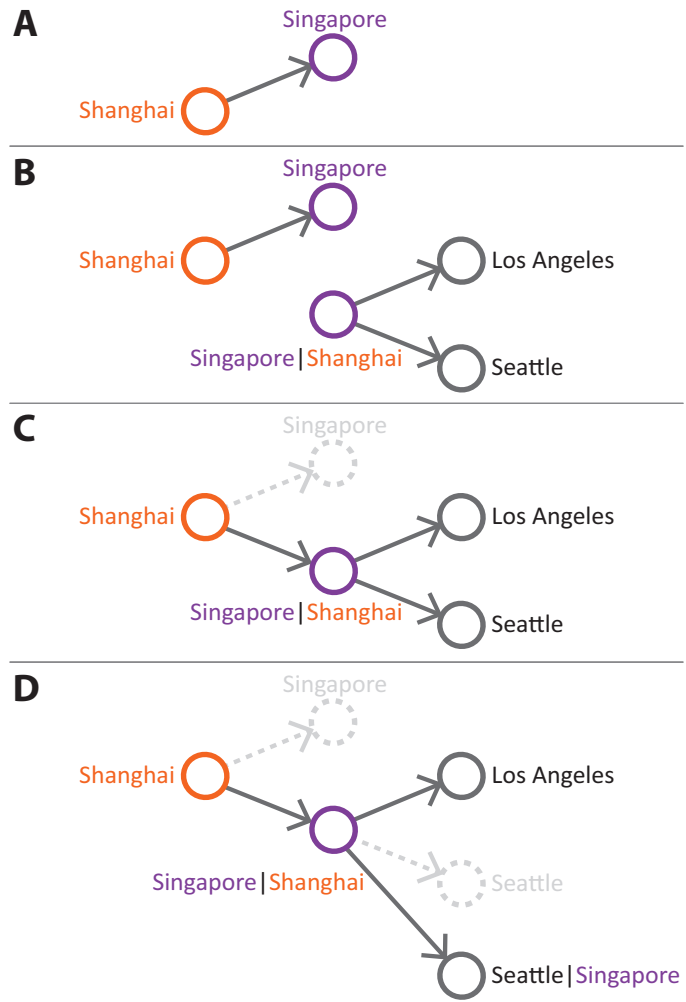

Figure 3.3. **Network wiring example for the global shipping data.** Figure shows how the dependency rules are represented as HON. (**A**) convert all first-order rules into edges; (**B**) convert higher-order rules, and add higher-order nodes when necessary, (**C**) rewire edges so that they point to newly added higher-order nodes (the edge weights are preserved); (**D**) rewire edges built from $Valid$ rules so that they point to nodes with the highest possible order.

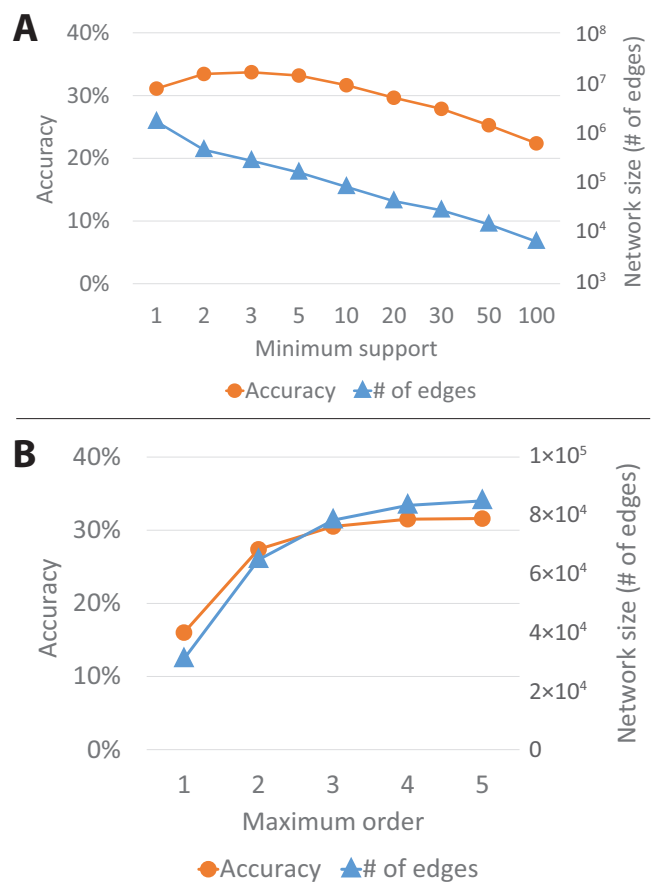

Figure 3.4. **Parameter senstivity of HON in terms of the accuracy and network size.** The global shipping data illustrated, and the accuracy is the percentage of correct predictions when using a random walker to predict the next step. (**A**) An appropriate minimum support can significantly reduce the network size and improve the accuracy of representation; (**B**) when increasing the maximum order, the accuracy of random walking simulation keeps improving but converges near the maximum order of 5.

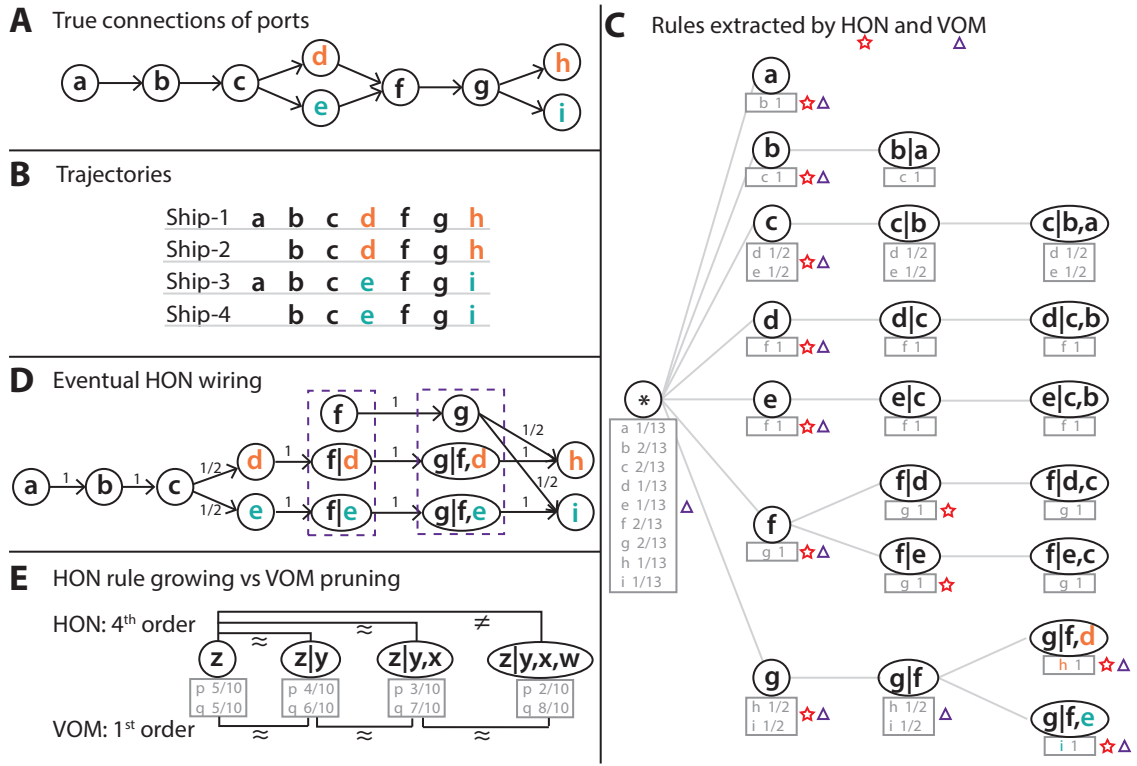

Figure 3.5. **Comparison between the HON and the VOM model** (**A**) In the context of global shipping, the true connection of ports. $f$ and $g$ are at the two sides of a canal. Ships coming from $d$ will go to $h$, and coming from $e$ will go to $i$. (**B**) Possible trajectories of ships. (**C**) Comparing the nodes retained by HON and VOM. VOM prunes nodes that are necessary for network representation while retaining nodes that are not necessary. (**D**) The eventual HON representation captures higher-order dependencies while retaining all first-order information. (**E**) HON "grows" rules from the first order, while VOM prunes rules from the highest order.

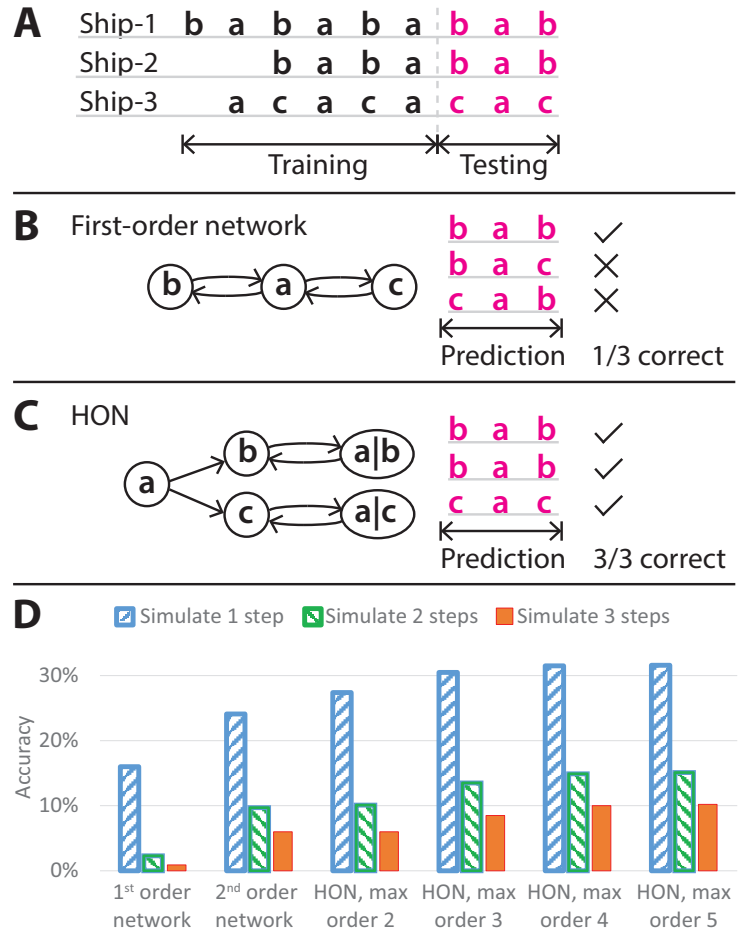

Figure 3.6. **Comparison of random walking accuracies.** (**A**) For the global shipping data composed of ships trajectories, hold the last three steps of each trajectory for testing and use the rest to build the network. (**B** and **C**) Given a generated shipping network, every ship is simulated by a random walker, which walks three steps from the last location in the corresponding training trajectory. The generated trajectories are compared with the ground truth, and the fraction of correct predictions is the random walking accuracy. (**D**) By using HON instead of the first-order network, the accuracy is doubled when simulating the next step and improved by one magnitude when simulating the next three steps. Note that error bars are too small to be seen (SDs on HONs are $0.11\% \pm 0.02\%$).

TABLE 3.1

COMPARING DIFFERENT NETWORK REPRESENTATIONS OF THE

SAME GLOBAL SHIPPING DATA

| Network representation | Number of edges | Number of nodes | Network density | Two step return | Three step return | Entropy rate (bits) | Clustering time (mins) | Ranking time (s) |
|---|---|---|---|---|---|---|---|---|
| Conventional first-order | 31,028 | 2,675 | $4.3 \times 10^{-3}$ | 10.7% | 1.5% | 3.44 | 4 | 1.3 |
| Fixed second-order | 116,611 | 19,182 | $3.2 \times 10^{-4}$ | 42.8% | 8.0% | 1.45 | 73 | 7.7 |
| HON, max order two | 64,914 | 17,235 | $2.2 \times 10^{-4}$ | 41.7% | 7.3% | 1.46 | 45 | 4.8 |
| HON, max order three | 78,415 | 26,577 | $1.1 \times 10^{-4}$ | 45.9% | 16.4% | 0.90 | 63 | 6.2 |
| HON, max order four | 83,480 | 30,631 | $8.9 \times 10^{-5}$ | 48.9% | 18.5% | 0.68 | 67 | 7.0 |
| HON, max order five | 85,025 | 31,854 | $8.4 \times 10^{-5}$ | 49.3% | 19.2% | 0.63 | 68 | 7.6 |

### 3.4.4  Effects on Clustering

One important family of network analysis methods is clustering, which identifies groups of nodes that are tightly connected. A variety of clustering algorithms such as MapEquation [177] and Walktrap [171] are based on random walking, following the intuition that random walkers are more likely to move within the same cluster rather than between different clusters. Since using HON instead of a first-order network alters the movement patterns of random walkers running upon, the compelling question becomes: how does HON affect the clustering results?

Consider an important real-world application of clustering: identifying regions wherein aquatic species invasions are likely to happen. Since the global shipping network is the dominant global vector for the unintentional translocation of non-native aquatic species [151] (species get translocated either during ballast water uptake/discharge, or by accumulating on the surfaces of ships [73]), identifying clusters of ports that are tightly coupled by frequent shipping can reveal ports that are likely to introduce non-native species to each other. The limitation of the existing approach [228] is that the clustering is based on a first-order network that only accounts for direct species flows, while in reality the species introduced to a port by a ship may also come from multiple previous ports at which the ship has stopped due to partial ballast water exchange and hull fouling. These indirect species introduction pathways driven by ship movements are already captured by HON and can influence the clustering result. As represented by the HON example in Figure 3.1C, following the most likely shipping route, species are more likely to be introduced to Los Angeles from Shanghai (via Singapore) rather than from Tokyo, so the clustering (driven by random walking) on HON prefers grouping Los Angeles with Shanghai rather than with Tokyo. In comparison, indirect species introduction pathways are ignored when performing clustering on a first-order network (Figure 3.1B), thus underestimating the risk of invasions via indirect shipping connections.

By clustering on HON, the overlap of different clusters is naturally revealed, highlighting ports that may be invaded by species from multiple regions. Since there can be multiple nodes representing the same physical location in HON (e.g., $Singapore|Tokyo$ and $Singapore|Shanghai$ both represent Singapore), and the ship movements through these nodes can be different, these higher-order nodes can belong to different clusters, so that Singapore as an international port belongs to multiple clusters, as one would expect.

The clustering results (using MapEquation) on a first-order network and HON are compared in Figure 3.7. For example, let us consider Malta, a European island country in the Mediterranean Sea. Malta has two ports: Valletta is a small port that mainly serves cruise ships in the Mediterranean, and Malta Freeport, on the contrary, is one of the busiest ports in Europe (many international shipping routes have a stop there). The clustering on the first-order network cannot tell the difference between the two ports and assigns both to the same Southern Europe cluster. On the contrary, the clustering on HON effectively separates Valletta and Malta Freeport by showing that Malta Freeport belongs to three additional clusters than Valletta, implying long-range shipping connections and species exchanges with ports all over the world. In summary, on HON, 45% of ports belong to more than one cluster, among which the Panama Canal belongs to six clusters, and 44 ports (1.7% of all) belong to as many as five clusters, including international ports such as New York, Shanghai, Hong Kong, Gibraltar, Hamburg, and so on, indicating challenges to the management of aquatic invasions, as well as opportunities for devising targeted management policies. These insights are gained by adopting HON as the network representation for the global shipping data, whereas the MapEquation algorithm is unmodified.

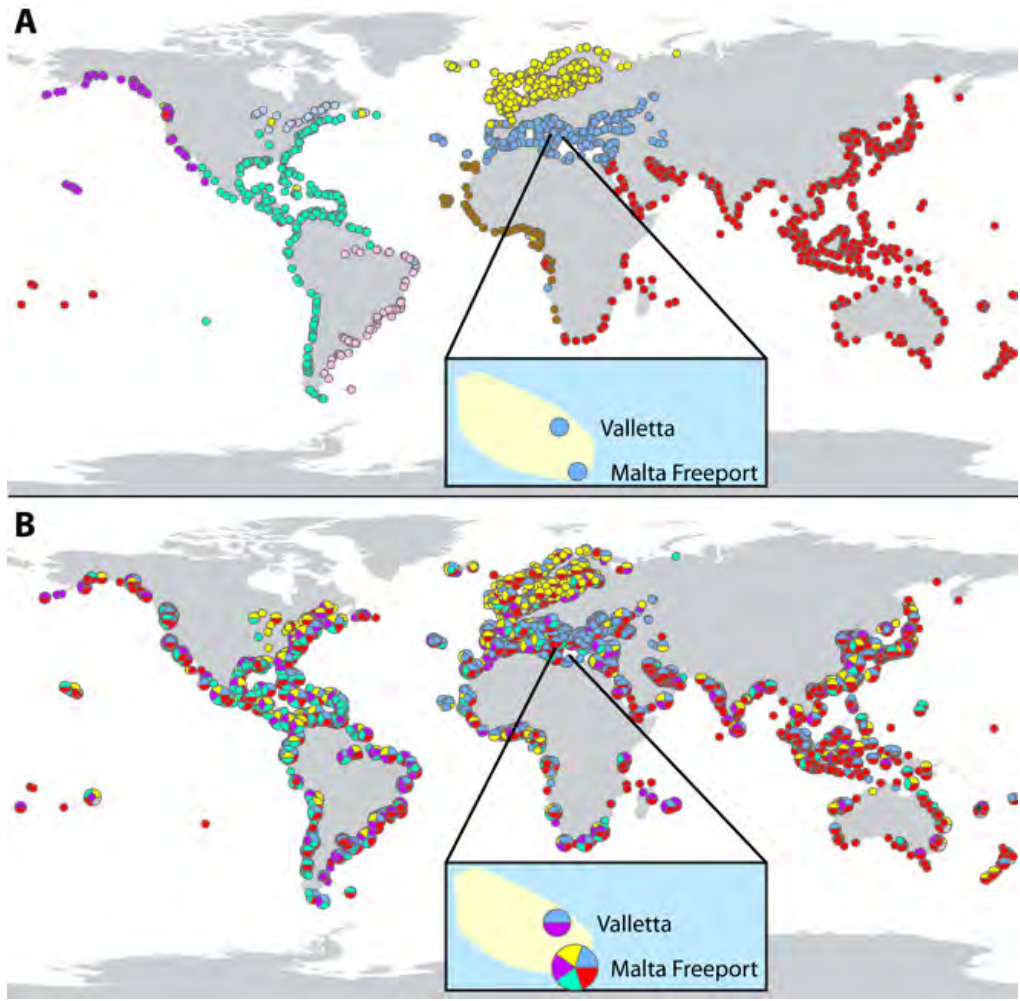

Figure 3.7. **Clustering of ports on different network representations of the global shipping data.** Ports tightly coupled by frequent shipping in a cluster are likely to introduce non-native species to each other. MapEquation [177] is used for clustering, and different colors represent different clusters. (**A**) Clustering on the first-order network. Although Valletta and Malta Freeport are local and international ports, respectively, the clustering result does not distinguish the two. (**B**) Clustering on HON. The overlapping clusters indicate how international ports (such as Malta Freeport) may suffer from species invasions from multiple sources.

### 3.4.5 Effects on Ranking

Another important family of network analysis methods is ranking. PageRank [161] is commonly used in assessing the importance of web pages by using random walkers (with random resets) to simulate users clicking through different pages, and pages with higher PageRank scores have higher chances of being visited. It has been shown that Web users are not Markovian[52], and PageRank on the conventional network representation fails to simulate real user traffic[143]. Because HON can help random walkers achieve higher accuracies in reproducing movement patterns, how can HON affect the PageRank scores, and why?

With the clickstream data, we can construct both a first-order network and a HON as the input for PageRank. In HON, the PageRank scores of multiple higher-order nodes representing the same Web page are summed up as the final score for the page. As shown in Figure 3.8, by using HON instead of the first-order network, 26% of the Web pages show more than 10% of relative changes in ranking; more than 90% of the Web pages lose PageRank scores, whereas the other pages show remarkable gains in scores. To have an idea of the changes, we list the Web pages that gain or lose the most scores by using HON as the input to PageRank, as shown in Table 3.2. Of the 15 Web pages that gain the most scores from HON, 6 are weather forecasts and 4 are obituaries, as one would expect considering that this data set is from Web sites of local newspapers and TVs. Of the 15 Web pages that lose the most scores, 3 are the lists of news personnel under the "about" page, which a normal reader will rarely visit, but are overvalued by ranking on the first-order network.

To further understand how the structural differences of HON and the first-order network lead to changes in PageRank scores, we choose Web pages that show significant changes in ranking and compare the corresponding subgraphs of the two network representations. A typical example is a pair of pages, *PHOTOS: January 17th snow - WDBJ7 / news* and *View/Upload your snow photos - WDBJ7 / news* these two pages gain 131 and 231% PageRank scores, respectively, on HON. In the first-order network representation, as shown in Figure 3.9A, where edge widths indicate the transition probabilities between Web pages, it appears that after viewing or uploading the snow photos, a user is very likely to go back to the WDBJ7 home page immediately. However, in reality, once a user views and uploads a photo, the user is likely to repeat this process to upload more photos while less likely to go back to the home page. This natural scenario is completely ignored in the first-order network but captured by HON, indicated by the strong loop between the two higher-order nodes (Figure 3.9B). The example also shows how the higher probability of returning after two (or more) steps on HON can affect the ranking results. Again, all these insights are gained by using HON instead of the conventional first-order network, without any change to the PageRank algorithm. Besides the ranking of Web pages, HON may also influence many other applications of ranking, such as citation ranking and key phrase extraction.

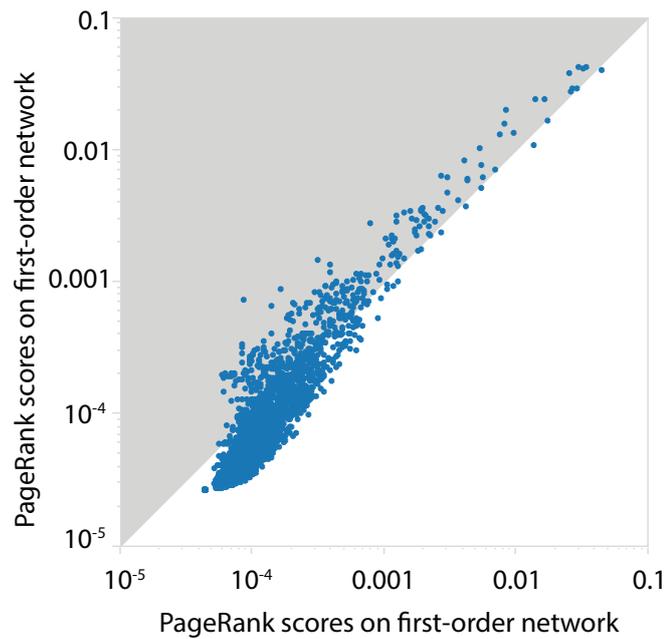

Figure 3.8. **Change of Web page rankings by using HON instead of first-order network.** PageRank [161] is used for ranking. Twenty-six percent of the pages show more than 10% of relative changes in ranking. More than 90% of the Web pages lose PageRank scores, whereas the other pages show remarkable gain in scores. Note that log-log scale is used in the figure, so a deviation from the diagonal indicates a significant change of the PageRank score.

### 3.4.6 Scalability of HON

We further show the scalability of HON, derived from its compact representation. In previous research (where a fixed second order is assumed for the network), from Table 3.1, it is shown that the network is considerably larger than the conventional first-order network, and assuming a fixed order beyond the second order becomes impractical because "higher-order Markov models are more complex" [178], due to combinatorial explosion. A network that is too large is computationally expensive to perform further analysis upon. On the contrary, although HON with maximum order of two has comparable accuracies in terms of random walking movement simulation, it has less nodes and about half the number of edges compared with a fixed second-order network, because it uses the first order whenever possible and embeds second-order dependencies only when necessary. Even when increasing the maximum order to five, HON still has less edges than the fixed second-order network, whereas all the useful dependencies up to the fifth order are incorporated in the network, resulting in considerably higher accuracies on random walking simulations.

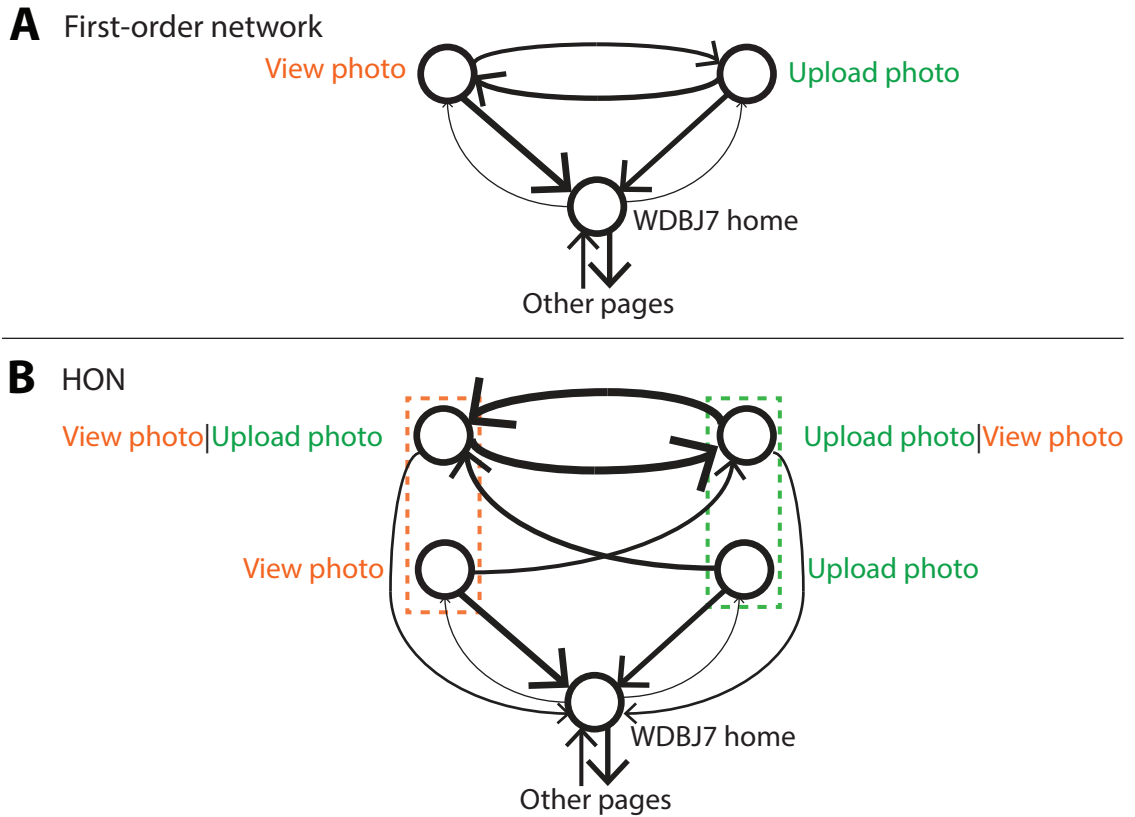

Figure 3.9. **Comparison of different network representations for the same clickstream data.** Edge widths indicate the transition probabilities. (**A**) First-order network representation, indicating that a user is likely to go back to the home page after viewing or uploading snow photos. (**B**) HON representation, which not only preserves the information in the first-order network but also uses higher-order nodes and edges to represent an additional scenario: once a user views and uploads a photo, the user is likely to repeat this process to upload more photos and is less likely to go back to the home page. Consequently, these photo viewing and uploading pages will receive higher PageRank scores [161] because the implicit random walkers of PageRank are more likely to be trapped in the loop of the higher-order nodes.

Another important advantage of HON over a fixed-order network is that network analysis algorithms can run faster on HON because of HONs compact representation. In addition, HON is sparser than the fixed-order representation, and many network toolkits are optimized for sparse networks. Table 3.1 shows the running time of two typical network analysis tasks: ranking (with PageRank [161]) and clustering (with MapEquation [177]). Compared with the fixed second-order network, these tasks run almost two times faster on HON with a maximum order of two and about the same speed on HON with a maximum order of five (which embeds more higher-order dependencies and is more accurate).

It is worth noting that the number of additional nodes/edges needed for HON (on top of a first-order network) is determined by the number of higher-order dependencies in the data and that additional size is affected neither by the size of the raw data nor by the density of the corresponding first-order network. For example, even if the first-order network representation of a data set is a complete graph with 1 million nodes, if 100 second-order dependencies exist in the data, HON needs only 100 additional auxiliary second-order nodes on top of the first-order network, rather than making the whole network the second order. Thus, the advantage of HON is being able to effectively represent higher-order dependencies, while being compact by trimming redundant higher-order connections.

## 3.5 Discussion

We have shown that for sequential data with higher-order dependencies, the conventional first-order network fails to represent such dependency patterns in the net-work structure, and the fixed second-order dependency can become limiting. If the network representation is not truly representative of the original data, then it will invariably lead to unreliable conclusions or insights from network analyses. We develop a new process for extracting higher-order dependencies in the raw data and for building a network (the HON) that can represent such higher-order dependencies. We demonstrate that our novel network representation is more accurate in representing the true movement patterns in data in comparison with the conventional first-order network or the fixed second-order network: for example, when using HON instead of a first-order network to represent the global shipping data, the accuracy is doubled when simulating a ships next movement on the network and is higher by one magnitude when simulating three steps, because the higher-order nodes and edges in HON can provide more detailed guidance for simulated movements. Besides improved accuracy, HON is more compact than fixed-order networks by embedding higher-order dependencies only when necessary, and thus, network analysis algorithms run faster on HON.

Furthermore, we show that using HON instead of conventional network representations can influence the results of network analysis methods that are based on random walking. For example, on HON, the clustering of ports takes indirect shipborne species introduction pathways into account and naturally produces overlapping clusters that indicate multiple sources of species invasion for international ports; the ranking of Web pages is corrected by incorporating the higher-order patterns of users browsing behaviors such as uploading multiple photos. Our work has the potential to influence a wide range of applications, such as improving PageRank for the task of unsupervised key phrase selection in language processing [147], because the proposed network representation is consistent with the input expected by various network analysis methods. Because nodes could be split into multiple ones in HON, it may require postprocessing to aggregate the results for interpretation. In the cur-rent method, the choice of parameters may influence the structure of the resulting network, so we provide parameter discussions.

TABLE 3.2

CHANGES OF PAGERANK SCORES BY USING HON INSTEAD OF A FIRST-ORDER NETWORK.

| Pages that gain PageRank scores | ΔPageRank |
|---|---|
| South Bend Tribune - Home. | +0.0119 |
| Hagerstown News / obituaries - Front. | +0.0115 |
| South Bend Tribune - Obits - 3rd Party. | +0.0112 |
| South Bend Tribune / sports / notredame - Front. | +0.0102 |
| Aberdeen News / news / obituaries - Front. | +0.0077 |
| WDBJ7 - Home. | +0.0075 |
| KY3 / weather - Front. | +0.0075 |
| Hagerstown News - Home. | +0.0072 |
| Daily American / lifestyle / obituaries - Front. | +0.0054 |
| WDBJ7 / weather / closings - Front. | +0.0048 |
| WSBT TV / weather - Front. | +0.0041 |
| Daily American - Home. | +0.0036 |
| WDBJ7 / weather / radar - Front. | +0.0036 |
| WDBJ7 / weather / 7-day-planner - Front. | +0.0031 |
| WDBJ7 / weather - Front. | +0.0019 |
| Pages that lose PageRank scores | ΔPageRank |
| KTUU - Home. | -0.0057 |
| KWCH - Home. | -0.0031 |
| Imperial Valley Press - Home. | -0.0011 |
| Hagerstown News / sports - Front. | -0.0005 |
| Imperial Valley Press / classifieds / topjobs - Front. | -0.0004 |
| Gaylord - Home. | -0.0004 |
| WDBJ7 / weather / web-cams - Front. | -0.0004 |
| KTUU / about / meetnewsteam - Front. | -0.0003 |
| Smithsburg man faces more charges ... - Hagerstown News / news - story. | -0.0003 |
| KWCH / about / station / newsteam - Front. | -0.0003 |
| South Bend Tribune / sports / highschoolsports - Front. | -0.0003 |
| Hagerstown News / opinion - Front. | -0.0002 |
| WDBJ7 / news / anchors-reporters - Front. | -0.0002 |
| Petoskey News / news / obituaries - Front. | -0.0002 |
| KWCH / news - Front. | -0.0002 |

CHAPTER 4

HIGHER-ORDER NETWORK PLUS (HON+): OPTIMIZED ALGORITHM FOR BIG DATA

4.1 Overview

The original higher-order network (HON) algorithm proposed in Chapter 3 requires two parameters: the maximum order to stop the growth of rules, and minimum support to determining if an observation is too trivial. The choice of parameters has significant influence on the resulting network, as discussed in the paper [229]. The challenge is that the determination of these parameters varies in different applications, and requires multiple trials or educated guesses to find the appropriate parameter.

**Intellectual merit:** In this chapter, we propose a *parameter-free algorithm* to construct HON. The original HON algorithm constructs observations of subsequences from the first order to the highest order in advance, which does not scale well for big data. In this chapter, we propose a procedure that constructs observations of subsequences *on demand*, which is achieved by using an indexing cache with $\Theta(1)$ lookup time. We provide the full pseudocode, and the complexity analysis for time and space. This new approach makes it possible to extract *arbitrarily high orders of dependency*. Finally, we extend the input of HON from simple trajectory data to various other types of raw data including *diffusion, time series, subsequence, temporal pairwise interaction, and heterogeneous data*. This significantly extends the applications of HON.

**Connections:** This work is an improved algorithm of HON in Chapter 3. It can apply to all applications of HON. The tutorial chapter 6 gives a step-by-step walkthrough on how to use the HON+ code made available online.

**Work Status:** The HON+ algorithm is developed by Jian Xu. The idea of computing the maximum divergence was inspired during the summer visit at Prof. Bruno Ribeiro's lab at Purdue University in 2016. The full software package (including source code) will be made available to the public at `www.HigherOrderNetwork.com`. The algorithm is in preparation for the Journal of Machine Learning Research.

## 4.2 HON+: Optimizing the HON Algorithm for Big Data

While HON embeds critical information for detecting higher-order anomalies, the original HON algorithm [229] has several limitations. First, the original algorithm requires multiple parameters, which varies for different data sets. Second, the original HON algorithm constructs observations of subsequences from the first order to the highest order in advance, which does not scale well for data with high orders of dependencies. Here we propose a fundamentally improved algorithm, HON+, which is *parameter-free*, and constructs observations of subsequences *on demand*.

### 4.2.1 Limitations of HON

The gist of the HON construction procedure is as follows:

- Input: sequential data (such as vehicle movements, flow of information, and so on).
- Rule extraction: extract dependency rules from sequential data, answering the questions "where do higher-order dependencies exist, and how high the orders are".
- Network wiring: connect nodes representing different orders of dependencies.
- Output: HON, which has data structure compatible with conventional networks, and can be used like the conventional network for analyses.

**Example.** Fig. 4.1 illustrates the dependency rule extraction step in the original HON algorithm. The original HON algorithm counts the subsequences of observations from first order to $MaxOrder$ (a required parameter, suppose $MaxOrder = 3$ in this example) in the raw data, then build distributions for the next steps given the current and previous steps. Finally test if knowing one more previous step significantly changes the distribution for the next step – if so, higher-order dependency exists for the path; this procedure (rule growing) is iterated recursively until *MaxOrder*. In this example, the probability distribution of the next steps from $C$ changes significantly if the previous step (coming to $C$ from $A$ or $B$) is known, but given more previous steps (coming to $C$ from $E \to A$ or $D \to B$) does not make a difference, demonstrating second-order dependencies.

**Complexity analysis.** The original HON algorithm, although intuitive and simple to implement, does not scale well for data with variable high orders of dependencies. For example, if the raw data has mostly lower-order dependencies but a few tenth-order dependencies, in order to extract the tenth-order dependencies, the algorithm has to build probability distributions and test for significant changes for all subsequences from first order to tenth order. Suppose the size of raw data is $L$, building observations and distributions up to $k^{th}$ order takes $\Theta(2L+3L+4L+\cdots+(k+1)L) = \Theta(k^2L)$ storage. The time complexity is $\Theta(Nk^2L)$: all observations will be traversed at least once; testing if adding a previous step significantly changes the probability distribution of the next step (if Kullback-Leibler divergence [127] is used) takes up to $\Theta(N)$ time where $N$ is the number of unique entities in the raw data.

### 4.2.2 Eliminating All Parameters

Starting from the first order $k = 1$, for each path $\mathcal{S} = [S_{t-k}, S_{t-(k-1)}, \ldots, S_t]$ of order $k$, the original HON algorithm initially assumes $k$ is the true order of dependency, which $\mathcal{S}$ has the distribution $D$ for the next step. The algorithm then adds one more previous step, namely extending $\mathcal{S}$ to $\mathcal{S}_{ext} = [S_{t-(k+1)}, S_{t-k}, S_{t-(k-1)}, \ldots, S_t]$, which has order $k + 1$ and distribution $D_{ext}$, and then tests if $D_{ext}$ is significantly different than that of $D$ The difference of the two distributions is measured with Kullback-Leibler divergence [127] as $\mathcal{D}_{KL}(D_{ext}||D)$, and compared with a dynamic threshold $\delta$ – if the divergence is larger than $\delta$, order $k + 1$ is assumed instead of $k$ for the path. The algorithm prefers lower orders than higher orders, unless higher orders have sufficient support (observations); therefore, the dynamic threshold $\delta$ is defined as $\delta = \frac{k_{ext}}{\log_2(1+Support_{\mathcal{S}_{ext}})}$. The whole is iterated recursively until $MaxOrder$.

The reason for having the $MaxOrder$ parameter in the original HON algorithm is to stop the rule growing process above from iterating indefinitely: without a constraint, even if the process has identified the true order $k_{true}$, the process will keep including previous steps indefinitely to test if the distribution of the next step with that of $k_{true}$. What if at some point $k' \geq k_{true}$, we can already determine that no matter how many more previous steps are included, the new distribution of the next step will never be significantly different than that of $k_{true}$?

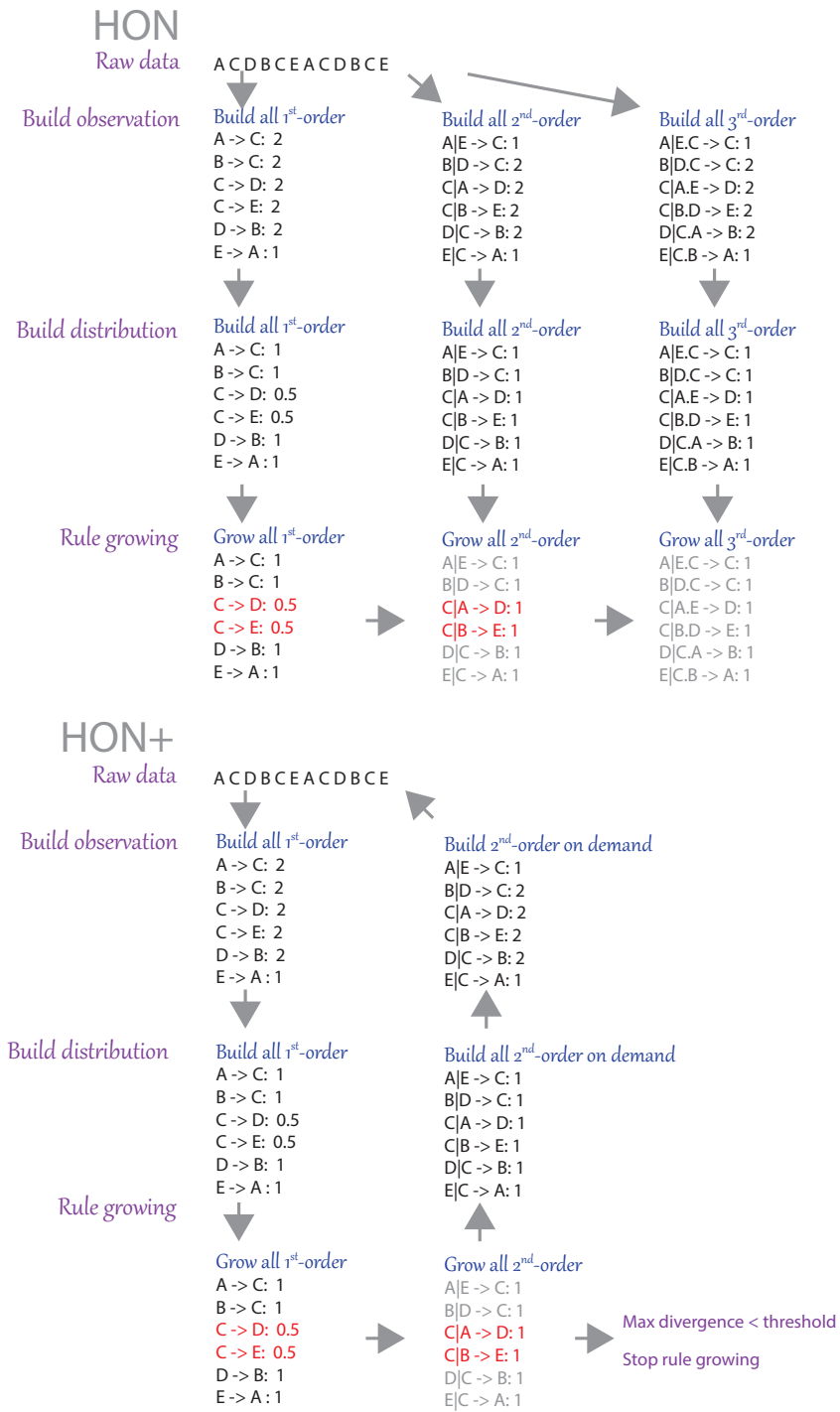

Figure 4.1. Comparison of the active observation construction in HON and the lazy observation construction in HON+.

**Lemma 4.1.** *The threshold* $\delta = \frac{k_{ext}}{\log_2(1+Support_{\mathcal{S}_{ext}})}$ *increases monotonically in the rule growing process when expanding $\mathcal{S}$ to $\mathcal{S}_{ext}$.*

*Proof.* The order of the extended sequence $\mathcal{S}_{ext}$, $k_{ext} = k_{true} + N$, increases monotonically with the inclusion of more previous steps. Every observation of $[S_{t-(k+1)}, S_{t-k}, \ldots, S_{t-1}, S_t]$ in the raw data can find a corresponding observation of $[S_{t-k}, \ldots, S_{t-1}, S_t]$, but not the other way around. Therefore, the support of $\mathcal{S}_{ext}$, $Support_{\mathcal{S}_{ext}}$, is equals to or smaller than that $Support_{\mathcal{S}}$ of the lower order $k = k_{ext} - 1$. As a result, the denominator decreases monotonically with the rule growing process. The overall threshold $\delta$ thus increases monotonically with the inclusion of more previous steps. $\square$

Given the next step distribution $D = [P_1, P_2, \ldots, P_N]$ of sequence $\mathcal{S}$, we can compute the maximum possible divergence:

$$\begin{aligned}
&max(D_{KL}(\mathcal{S}_{ext}||\mathcal{S})) \\
&= max(\sum_{i \in D} P_{ext}(i) \times \log_2 \frac{P_{ext}(i)}{P(i)}) \\
&= 1 \times \log_2 \frac{1}{min(P(i))} + 0 + 0 + \ldots \\
&= -\log_2(min(P(i)))
\end{aligned} \quad (4.1)$$

Therefore, at the rule extraction step, we can test if $-\log_2(min(P_{Distr}(i))) < \delta$; if it holds, then further increasing the order (adding more previous steps) will not yield significantly different distributions, so we can stop the rule growing process and take $k$ as the true order of dependency. An advantage of this parameter-free approach is that the algorithm can now extract *arbitrarily high order of dependency*, rather than terminating the rule growing process prematurely by the $MaxOrder$ threshold.

### 4.2.3 Scalability for Higher-orders

The original HON algorithm builds all observations and distributions up to $MaxOrder$ ahead of the rule growing process. We optimize this procedure through *lazy construction of observations and distributions*. Specifically, we do not count the occurrences of subsequences, nor calculate the distribution of the next steps, until the rule growing step explicitly asks for such information. This approach appears counter-intuitive, since the construction of observation requires traversal of the raw data, which every traversal has the complexity of $\Theta(L)$. However, given the following knowledge:

**Lemma 4.2.** *All observations of sequence $[S_{t-k-1}, S_{t-k}, \ldots, S_{t-1}, S_t]$ can be found exactly at the current and one preceding locations of all observations of sequence $S_{t-k}, \ldots, S_{t-1}, S_t]$ in the raw data.*

Instead of traversing the raw data every time when counting the occurrences of subsequences, we propose to use an *indexing cache* to store the locations of known observations, then use that information to narrow down higher-order subsequence lookups. The indexing cache is built for all first-order observations, recording their locations in the raw data. Then during the rule growing process, if $\mathcal{S}_{ext}$ has not been observed, *recursively* check if the lower-order observation is in the indexing cache, and use those cached indexes to perform fast lookup in the raw data. New observations from $\mathcal{S}_{ext}$ are then added to the indexing cache. This procedure guarantees the identification of observations of the previously unseen $\mathcal{S}_{ext}$, and the lookup time for each observation is $\Theta(1)$ when the indexing cache is implemented with hash tables.

**Example.** Recall that HON builds all observations and distributions from the first order to *MaxOrder* before rule growing (Fig. 4.1). Instead, HON+ first builds all first-order observations and distributions, then start from first-order test if the second order (adding one more previous step) significantly changes the probability distribution of the next step. Since second-order observations and distributions are not known yet, the algorithm utilizes the indexing cache to perform lookup from the raw data (which is $\Theta(1)$ for each lookup) and build the corresponding observations and distributions. Suppose HON+ determines that beyond the second order, the maximum possible divergence cannot possibly exceed the threshold $\delta$, HON+ will stop rule growing. Therefore, no unnecessary third-order observations, distributions, and rules will be generated.

**Complexity analysis.** The space complexity of HON is $\Theta(k^2 L)$, and is $\Theta(2R_1 + 3R_2 + \ldots)$ for HON+ (including observations, distributions, and the indexing cache), where $R_k$ is the actual number of higher-order dependency rules for order $k$. Since $R_k \leq L$, and in practice, when $k >> 2$, $R_k << L$ (higher-order rules are usually sparse), the overall space complexity of HON+ is significantly smaller than that of HON. The same applies to time complexity: while HON has $\Theta(Nk^2 L)$, HON+ has $\Theta(N(2R_1 + 3R_2 + \ldots))$.

In practice, when processing a real-world global ship movement data (with about 1.7 million records), the original HON algorithm becomes noticeably slow beyond $5^{th}$ order, and exhausts a personal computer's 16GB memory at around $6^{th}$ or $7^{th}$ order. On HON+, the algorithm can go well beyond $20^{th}$ order on the same computer. Theoretically, the HON+ algorithm can identify arbitrarily high orders of dependency (suppose it has sufficient support to make it significant at such high order).

The full pseudocode of HON+ is in Algorithm 3. The Python implementation is available at `http://www.HigherOrderNetwork.com`.

**Algorithm 3** HON+ rule extraction algorithm. Given the raw sequential data $T$, extracts arbitrarily high orders of dependencies, and output the dependency rules $R$. Optional parameters include $MaxOrder$, $MinSupport$, and $ThresholdMultiplier$

1: define *global C* as nested counter
2: define *global D,R* as nested dictionary
3: define *global SourceToExtSource, StartingPoints* as dictionary
4:
5: **function** EXTRACTRULES($T$, [$MaxOrder$, $MinSupport$, $ThresholdMultiplier = 1$])
6:     global $MaxOrder$, $MinSupport$, $Aggresiveness$
7:     BUILDFIRSTORDEROBSERVATIONS($T$)
8:     BUILDFIRSTORDERDISTRIBUTIONS($T$)
9:     GENERATEALLRULES($MaxOrder$, $T$)
10:
11: **function** BUILDFIRSTORDEROBSERVATIONS($T$)
12:     **for** $t$ in $T$ **do**
13:         **for** ($Source, Target$) in $t$ **do**
14:             $C[Source][Target] \mathrel{+}= 1$
15:             $IC$.add($Source$)
16:
17: **function** BUILDFIRSTORDERDISTRIBUTIONS($T$)
18:     **for** $Source$ in $C$ **do**
19:         **for** $Target$ in $C[Source]$ **do**
20:             **if** $C[Source][Target] < MinSupport$ **then**
21:                 $C[Source][Target] = 0$
22:         **for** $Target$ in $C[Source]$ **do**
23:             **if then**$C[Source][Target] > 0$
24:                 $D[Source][Target] = C[Source][Target]/(\sum C[Source][*])$
25:
26: **function** GENERATEALLRULES($MaxOrder$, $T$)
27:     **for** $Source$ in $D$ **do**
28:         ADDTORULES($Source$)
29:         EXTENDRULE($Source$, $Source$, $1$, $T$)
30:
31: **function** KLDTHRESHOLD($NewOrder, ExtSource$)
32:     **return** $ThresholdMultiplier \times NewOrder/log_2(1 + \sum C[ExtSource][*])$
33: **function** EXTENDRULE($Valid$, $Curr$, $order$, $T$)
34:     **if** $Order \leq MaxOrder$ **then**
35:         ADDTORULES($Source$)
36:     **else**
37:         $Distr = D[Valid]$
38:         **if** $-log_2(min(Distr[*].vals)) <$ KLDTHRESHOLD($order + 1$), $Curr$ **then**
39:             ADDTORULES($Valid$)
40:         **else**
41:             $NewOrder = order + 1$
42:             $Extended =$ EXTENDSOURCE($Curr$)
43:             **if** $Extended = \emptyset$ **then**
44:                 ADDTORULES($Valid$)
45:             **else**
46:                 **for** $ExtSource$ in $Extended$ **do**
47:                     $ExtDistr = D[ExtSource]$
48:                     $divergence =$ KLD($ExtDistr, Distr$)
49:                     **if** $divergence >$ KLDTHRESHOLD($NewOrder, ExtSource$) **then**
50:                         EXTENDRULE($ExtSource, ExtSource, NewOrder, T$)
51:                     **else**
52:                         EXTENDRULE($Valid, ExtSource, NewOrder, T$)

**Algorithm 3** *(continued)*

```
53: function ADDTORULES(Source):
54:     for order in [1..len(Source) + 1] do
55:         s = Source[0 : order]
56:         if not s in D or len(D[s]) == 0 then
57:             EXTENDSOURCE(s[1:])
58:         for t in C[s] do
59:             if C[s][t] > 0 then
60:                 R[s][t] = C[s][t]
61:
62: function EXTENDSOURCE(Curr)
63:     if Curr in SourceToExtSource then
64:         return SourceToExtSource[Curr]
65:     else
66:         EXTENDOBSERVATION(Curr)
67:         if Curr in SourceToExtSource then
68:             return SourceToExtsource[Curr]
69:         else
70:             return ∅
71:
72: function EXTENDOBSERVATION(Source)
73:     if length(Source) > 1 then
74:         if not Source[1 :] in ExtC or ExtC[Source] = ∅ then
75:             EXTENDOBSERVATION(Source[1 :])
76:     order = length(Source)
77:     define ExtC as nested counter
78:     for Tindex, index in StartingPoints[Source] do
79:         if index − 1 ≤ 0 and index + order < length(T[Tindex]) then
80:             ExtSource = T[Tindex][index − 1 : index + order]
81:             ExtC[ExtSource][Target]+ = 1
82:             StartingPoints[ExtSource].add((Tindex, index − 1))
83:     if ExtC = ∅ then
84:         return
85:     for S in ExtC do
86:         for t in ExtC[s] do
87:             if ExtC[s][t] < MinSupport then
88:                 ExtC[s][t] = 0
89:             C[s][t]+ = ExtC[s][t]
90:         CsSupport = ∑ ExtC[s][∗]
91:         for t in ExtC[s] do
92:             if ExtC[s][t] > 0 then
93:                 D[s][t] = ExtC[s][t]/CsSupport
94:                 SourceToExtSource[s[1 :]].add(s)
95:
96: function BUILDSOURCETOEXTSOURCE(order)
97:     for source in D do
98:         if len(source) = order then
99:             if len(source) > 1 then
100:                NewOrder = len(source)
101:                for starting in [1..len(source)] do
102:                    curr = source[starting :]
103:                    if not curr in SourceToExtSource then
104:                        SourceToExtSource[curr] = ∅
105:                    if not NewOrder in SourceToExtSource[curr] then
106:                        SourceToExtSource[curr][NewOrder] = ∅
107:                    SourceToExtSource[curr][NewOrder].add(source)
```

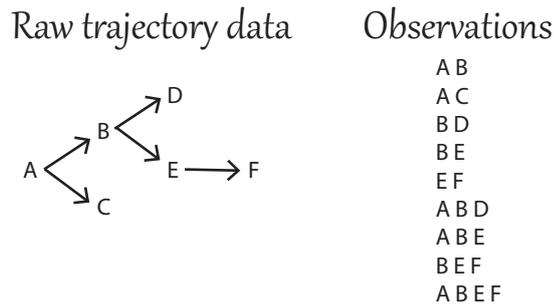

Figure 4.2. Converting diffusion data to entity sequences as the input for HON.

4.3 Flexible Input

4.3.1 Diffusion Data

The distinct property of diffusion data (also see Chapter 2 Section 2.2) is that for every diffusion process, the interactions form a tree, with the first "infected" entity being the root. Since there are no cycles in the tree, the tree can be decomposed into a collection of observed paths from the root to the leaves.

An example is given in Figure 4.2. Given the tree-shaped raw trajectory data on the left, entity pairs can be extracted, then triples, then quadruples. The resulting observations on the right can be directly taken as the *observations* for HON.

**Lemma 4.3.** *The observations thus constructed is a lossless representation of the raw trajectory data.*

*Proof.* Follow this procedure to reconstruct the tree from the observations data.

1. Start with the longest path, $ABEF$ in the example, use it as the initial tree $T$.

2. Remove all subsequences of $ABEF$, including $AB$, $BE$, $EF$, $ABE$ and $BEF$ from observations.

3. In the remaining observations, start with the longest path, $ABD$ in the example, find the overlap with the existing tree $T$, and add new branches when the entity subsequence diverges; in this example, the branch diverges at $D$, so a $B \to D$ branch is created.

4. Repeat (2) and (3) until all observations are exhausted. Return $T$.

□

With diffusion data, HON can capture patterns such as "information passed from A to B are more likely to go to E, and that passed from C to B are more likely to go to D.", which can be of interest in SMS communication, adoption of ideas, etc. Note that, diffusion processes such as disease propagation are first order by its nature.

This conversion process from diffusion data to observations can also be adapted to the HON+ algorithm for lazy observation construction. The only change is the indexing cache, which stores the location of observations in the tree $T$. A tree with reversed edge directionality can be built to facilitate the lookup of preceding (parent) entities.

### 4.3.2 Time Series Data

With discretization, time series data (also see Chapter 2 Section 2.2) can be converted to sequential data. An example is given in Figure 4.3. By discretizing the range of values into four categories $0-0.25, 0.25-0.5, 0.5-0.75, 0.75-1$, and assigning them as $A, B, C, D$, respectively, the time series data is converted to sequential data. An optional step is to remove duplicate consecutive entities to focus exclusively on the transitions between states (the HON algorithm itself does not prevent from taking such input, and can generate self-loops.) Finally, the observations of subsequences can be generated as usual from the discretized sequential data.

Various time series data have potentials for HON. For example, the Y axis can be stock prices changes, and we can learn patterns such as "price going up and up again will likely be followed by a price adjustment." Collective observations of multiple stock price change time series yields typical stock fluctuation patterns, and if such patterns change significantly, it signals anomalous patterns in price fluctuation.

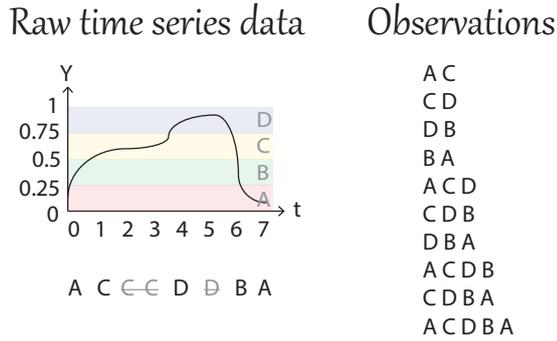

Figure 4.3. Discretizing time series data to entity sequences as the input for HON.

4.3.3 Subsequence Data

Certain applications have the subsequence observation directly. For example, dependency information derived from a grid of sensors. Those can be taken directly as the inputs of HON.

4.3.4 Pairwise Interaction Temporal Data

When temporal information is available, pairwise interactions can be chained together as multi-entity interactions. For example, if the raw data of phone call records have information such as "who called whom at what time", series of phone calls may form cascades of information.

An illustration is in Figure 4.4. Suppose phone calls within a time interval of 10 minutes are considered "related". The data shows that after $A$ called $B$, within 10 minutes $B$ called $E$, so there is a flow of information through the path $ABE$. However, $ABD$ does not exist since the gap of the two calls exceeds the interval of

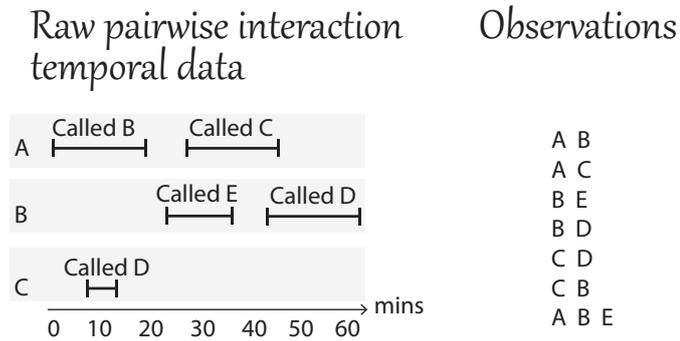

Figure 4.4. Chaining phone calls made within 10 minutes as sequential data.

10 minutes.

For this type of data, HON can capture information such as "$A$ is more likely to call $B$ if being called by $C$, more likely to call $D$ if being called by $E$." A detailed tutorial in Chapter 6 presents an efficient algorithm to chain pairwise interactions into sequential data, and demonstrates how to use the Python implementation of HON+ to analyze the sequential data.

4.3.5 Heterogeneous Data: Species Flow Higher-order Network (SF-HON)

In the original work that proposed the higher-order network [229], the algorithm constructs the network based on a sole source of data (e.g., ship trajectories), which cannot readily fit the needs of modeling species flows that is a function of multiple factors. In this work, we propose to extend the HON construction algorithm such that it can (1) take arbitrary number of data sources as input, (2) have customized aggregation functions, and (3) have customized thresholds.

We present a side-by-side comparison of the original algorithm and the extended algorithm used in this work; Figure 4.5 highlights the key differences in network construction. We extend HON in the species invasion context, and call it Species Flow Higher-order Network (SF-HON).

The original HON algorithm can take a single source of event sequence (illustrated as ship trajectories) as the input, whereas SF-HON not only takes ship trajectories but also ship types, trip durations and ballast water discharge for every ship movement as the input. Computing the influence per trip is trivial in HON, where every ship movement are treated the same and counted as "one" from source port to target port; in SF-HON however, the influence every trip can be different due to variations in trip duration, ballast discharge, ship type and so on, therefore we compute the risk of invasion $P_{i \to j}^{(t)}$ separately for every trip.

When aggregating the influence through a given pathway, in HON it is simply counting the number of trips observed through the pathway; in SF-HON we cannot simply add up the probabilities of invasions for different trips, instead, we take the joint probability assuming different trips are independent. Finally, as the HON algorithm needs a parameter "minimum support" as the terminating condition for higher-order rule extraction, in HON the minimum support is a positive integer, that pathways with less than the specified trips through them will be discarded; in SF-HON we extend minimum support to probabilities, that pathways with aggregated probability of species invasion less than the specified threshold will be discarded.

## 4.4 Discussion

In this chapter, we have proposed a parameter-free algorithm to construct HON, a procedure that constructs observations of subsequences on demand, and ways to extend the input of HON from simple trajectory data to various other types of raw data including diffusion, time series, subsequence, temporal pairwise interaction, and heterogeneous data. The extension to the original HON algorithm enables the extraction of arbitrarily high orders of dependency, and greatly reduces the time and space requirement. It significantly broadens the applications of HON, and opens the gate to new possibilities to representing big data.

| | HON for ship movements | HON for species flow (SF-HON) |
|---|---|---|
| Input | Ship 1: Port 1 —1 trip→ Port 3 —1 trip→ Port 4 --→<br>Ship 2: Port 2 —1 trip→ Port 3 —1 trip→ Port 5 --→ | Ship 1: Type $k_1$, Port 1 —$D_{t1}\,\Delta t_{t1}$→ Port 3 —$D_{t2}\,\Delta t_{t2}$→ Port 4 --→<br>Ship 2: Type $k_2$, Port 2 —$D_{t1}\,\Delta t_{t1}$→ Port 3 —$D_{t2}\,\Delta t_{t2}$→ Port 5 --→ |
| Influence per trip | Ship 1: Port 1 —1 trip→ Port 3 | Ship 1: Type $k_1$, Port 1 $\xrightarrow{P^{(1)}_{1\to 3} = (1-e^{-\lambda D^{(1)}_{1\to 3}})e^{-u\Delta^{(1)}_{1\to 3}}}$ Port 3 |
| Counting subsequences | Port 3 → Port 4, 30 trips = 1 + 1 + ...<br>Port 3 → Port 5, 10 trips = 1 + 1 + ...<br>Port 1 → Port 3 → Port 4, 8 trips = 1 + 1 + ... | Port 3 → Port 4, $0.6 = P_{3\to 4} = 1 - \prod_i (1 - P^{(i)}_{3\to 4})$<br>Port 3 → Port 5, $0.4 = P_{3\to 5} = 1 - \prod_i (1 - P^{(i)}_{3\to 5})$<br>Port 1 → Port 3 → Port 4, $0.1 = P_{1\to 3\to 4} = 1 - \prod_i (1 - P^{(i)}_{1\to 3\to 4})$ |
| Normalization | Port 3 —0.75→ Port 4<br>Port 3 —0.25→ Port 5 | Port 3 —0.6→ Port 4<br>Port 3 —0.4→ Port 5 |
| Rule extraction terminating condition | Minimum support = 10<br>Port 3 → Port 4, 30 trips > 10<br>Port 1 → Port 3 → Port 4, 8 trips < 10 | Minimum support = 0.2<br>Port 3 → Port 4, 0.6 > 0.2<br>Port 1 → Port 3 → Port 4, 0.1 < 0.2 |

Figure 4.5. A comparison of the original HON construction algorithm that takes a single source of data (left) and the extended algorithm used in this work that can build SF-HON from multiple sources of data (right).

The next step will be parallelizing the implementation of the rule extraction process. Meanwhile, further extend the potential types of inputs of HON based on the inspirations from domain applications, such biological and financial data.

CHAPTER 5

VISUALIZING AND EXPLORING HIGHER-ORDER NETWORKS

5.1 Overview

Unlike the conventional first-order network (FON), the higher-order network (HON) provides a more accurate description of transitions by creating additional nodes to encode higher-order dependencies. However, there exists no visualization and exploration tool for the HON. For applications such as the development of strategies to control species invasion through global shipping which is known to exhibit higher-order dependencies, the existing FON visualization tools are limited.

**Intellectual merit:** In this paper, we present HONVis, a novel visual analytics framework for exploring higher-order dependencies of the global ocean shipping network. Our framework leverages coordinated multiple views to reveal the network structure at three levels of detail (i.e., the global, local, and individual port levels). Users can quickly identify ports of interest at the global level and specify a port to investigate its higher-order nodes at the individual port level. Investigating a larger-scale impact is enabled through the exploration of HON at the local level. Using the global ocean shipping network data, we demonstrate the effectiveness of our approach with a real-world use case conducted by domain experts specializing in species invasion. Finally, we discuss the generalizability of this framework to other real-world applications such as information diffusion in social networks and epidemic spreading through air transportation.

**Connections:** This software framework is a direct application of HON in Chapter 3. It also serves as a handy tool for exploring the global shipping network and species invasion in Chapter 8, 9. The anomaly detection procedure in Chapter 10 can serve as a module of HONVis.

5.2   Introduction

Modern day systems are complex, whether they are movements of hundreds of thousands of ships to form a global shipping network [112], powering the transportation and economy while inadvertently translocating invasive species; interactions of billions of people on social networks, facilitating the diffusion of information; or complex metabolic systems representing rich cellular interactions.

The complex systems are often represented as networks, where the components of the system are represented as nodes and the interactions among them are represented as edges or links. This network based representation facilitates subsequent analysis and visualization. For example, the global shipping activities are usually represented as a global shipping network, with ports as nodes, and the amount of traffic between port pairs as edge weights [118]. Traditionally, creating networks from such ship movement data has followed the port-to-port movement of a ship, and ignores the historic trajectory of the ship. This becomes extremely limiting as it has been observed that ship movements actually depend on up to *five* previously visited ports [229]; other types of interaction data from communication to transportation often exhibit *higher-order dependencies* [52, 178]. Therefore, when representing data derived from these complex systems, conventional network representations that implicitly assume the Markov property (i.e., *first-order dependency*) can quickly become limiting, undermining subsequent network analysis that relies on the network representation.

To address this problem, prior work has proposed the use of *higher-order network* (HON) to discover higher-order dependencies and embed conditional transition probabilities into a network representation [229]. For the global shipping network example, instead of mapping every port to a single node, each higher-order node in HON encodes not only the current step (the port that a ship currently stays) but also a sequence of previous steps (the ports that a ship visited before arriving at the cur-rent port). Therefore, the transitions among nodes in a HON are now conditional, and are able to reproduce complex ship movement patterns more accurately from the raw data. HON features direct compatibility with the existing suite of network analysis methods, such as random walking, clustering, and ranking, thus serving as a powerful tool for modeling the increasingly complex systems.

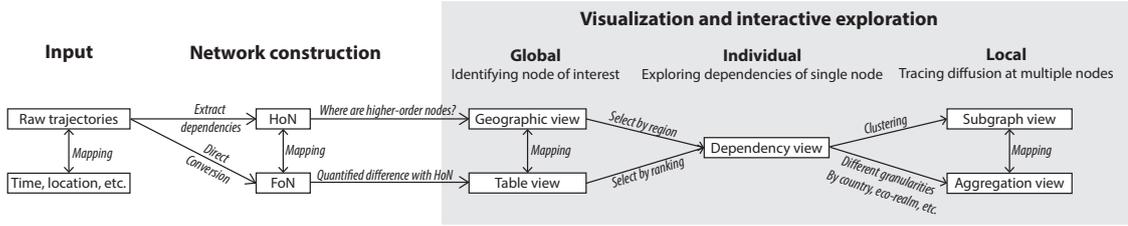

Figure 5.1. The framework of HONVis design. FON and HON are converted and extracted from the raw trajectory data, from which we identify nodes of interest. Five linked views are designed to enable the interrogation of single and multiple nodes.

HON is the correct way of representing complex systems that defy the first-order dependency assumptions. Despite the importance of HON and its applicability to network analysis, there has not yet been a visualization tool that can handle the richness of the HON representation. In this work, we team up with two domain experts in network science and marine ecology and develop a visual analytics framework, named HONVis, to facilitate the exploration and understanding of HON. The global shipping network, being an important application of HON for the study of invasive species, is used as a case study and for illustration throughout this paper, although the general approach we take can be applied to other types of HONs. We focus on the formation and impact of higher-order nodes, e.g., why a higher-order node exists in a HON and how the species may propagate from a port to other ports given the previous steps? Specific to the shipping network case study, we aim to answer these questions through a three-step exploration process: 1) *global identifica-* *tion of ports of interest*, 2) *detailed observation of the connections of an individual port*, and 3) *tracing the propagation of invasive species from port to port through shipping*. Accordingly, we lay out the design of HONVis in Figure 5.1. The input data are converted to the FON and dependencies are extracted to construct the HON. From these network representations, we identify nodes of interest. The visualization includes five coordinated views: *geographic view* and *table view* show information related to a single node; *dependency view*, *subgraph view*, and *aggregation view* show connections among multiple nodes. Together these five views enable users to explore higher-order nodes and their dependencies, allowing insights to be gained from this comprehensive system.

5.3  Related Work

**HON Visualization.** HON visualization is sporadic in the literature. Blaas et al. [33] proposed to visualize higher-order transitions by connecting nodes using higher-order curves. By following a smooth curve from one end to the other, one can identify which nodes are associated with higher-order transitions and what are the orders of the nodes. Rosvall et al. [178] grouped higher-order nodes by their current nodes and drew directed edges between connected nodes. The higher-order nodes representing the same physical locations are placed in one circle to build the correspondence. This approach, although intuitive, does not scale beyond the second order, nor when more than a dozen of higher-order nodes representing the same location coexist. HON is also used as an analysis tool in unsteady flow visualization for better workload distribution. Zhang et al. [234] employed high-order dependencies to estimate the destinations of particles given their previous locations, providing more accurate information about which data blocks to load at the next step.

**Visual Path Analysis and Graph Comparison.** For first-order networks, quite a few works have been presented on visual path analysis and graph compar-ison. We refer readers to survey papers [8, 10, 27, 220, 222] for a comprehensive overview. Bodesinsky et al. [37] proposed an interface with coordinated multiple views to explore sequences of events. The event view visualizes each sequence of events as horizontally aligned bars. Event patterns are summarized in a pattern overview from which users can query a certain pattern to highlight the recurring instances in the event view. Partl et al. [163] designed Pathfinder to analyze paths in multivariate graphs. A node-link diagram visualizes the paths between queried nodes, and a ranked list shows the attributes associated with the nodes. Wongsuphasawat et al. [226] presented LifeFlow to study and compare multiple event sequences. Each sequence is displayed in a horizontal bar, and the events in different sequences are aligned vertically for easy comparison. Detailed event information for each sequence is displayed as a list.

**Movement Data Visualization.** Spatiotemporal movement data (e.g., traffic and trajectory) are often encoded as conventional graphs, where each node represents a location and an edge represents the traffic volume between two locations without distinguishing their previous locations. Guo [94] used the location-to-location graph to visualize population migration in the United States. The spatial regions are partitioned to form hierarchies and support node aggregation at the regional level. The flow is clustered based on the associated variables, such as the number of migrants for different ages and income levels. von Landesberger et al. [221] presented the MobilityGraph to visualize mass mobility. They also grouped the regions for clearer observation. To obtain a common movement pattern, a temporal clustering is performed based on the graph's feature vectors generated in different time spans. However, both works do not consider higher-order dependencies and therefore, they are not able to answer the questions such as how many migrants in Chicago who came from Los Angles would finally move to New York City.

5.4  Design Rationales

We invited two domain experts from the NSF Coastal SEES collaborative research project, who specialize in data mining and the application of marine ecology, and hold positions in R1 universities. They had spent five years on the modeling of species invasion, and had published works in interdisciplinary journals and top conferences in the domain. The experts noticed that indirect species flow through shipping exhibit higher-order dependencies [229], but had been using the FON for visualization and control strategy development due to the lack of tools for the HON. In this section we review the background of species invasion research, then identify requirements to guide our design of HONVis.

5.4.1  Application Background

The ever-increasing human activities unintentionally facilitate the transportation of species, which may outcompete native species and cause substantial environmental and economical harm. The annual damage and control costs of invasive species in the United States is estimated to be more than 120 billion US dollars [169]. The global shipping network is the dominant vector for the unintentional introduction of invasive species [151]: species "hitchhike" on ships from port to port in ballast water or via hull fouling [73]. Understanding the global shipping network is crucial for devising species control management strategies. The data mining community has recently produced promising observations on the global shipping network [228]. For example, several clusters of ports which are loosely connected to each other are revealed in the global shipping network. Targeted species management strategies can be devised toward the loose connections among the clusters to prevent or slow down the species propagation from one cluster to another.

However, even the state-of-the-art research still faces unresolved challenges. For example, the recent network approach by Xu et al. [228] uses the FON to model the species flow between port pairs; it is unclear from the FON how species may propagate after multiple steps, and it is impossible to know which port or pathway plays an important role connecting different clusters, eco-regions, or countries. Therefore, the ability to explore the process interactively is important for the development of species management strategies.

Meanwhile, the FON that was used to model and visualize global shipping is an *oversimplification* of higher-order dependencies that exist ubiquitously in ship movements and species flows [229]. In the iconic work of Kaluza et al. [118], a global cargo ship network was built by taking the number of trips between port pairs as edge weights, while multi-step ship movement patterns were ignored. From the visualization of the FON thus built, one cannot tell if ships coming from Shanghai to Singapore are more likely to visit Los Angeles or Seattle. Such higher-order dependencies in networks are crucial for accurately modeling the flow of ships and species. However, no such a tool currently exists.

Finally, although it has been shown how ship types, ship sizes, geographical locations and seasonality can influence the structure of the first-order global shipping and species flow patterns [228], there has been no discussion on how such factors influence higher-order shipping patterns. It is unknown whether higher-order movement patterns are mainly formed by oil tankers, or located at estuaries, or appear mainly in winters. Such information can provide insight in revealing the driving forces behind the formation of higher-order dependencies in ship movements, and aid the development of invasive species management strategies.

5.4.2  Design Requirements

Given the gap between the demand to visualize higher-order dependencies in global shipping and the lack of HON visualization tools, we identify key requirements for our visual analytics system.

**R1. Create a mapping between the HON and FON, and quantify the differences.** The experts expect to see geographical locations of ports and their connections on a map, in order to select ports at places of interest; the experts want to know if higher-order dependencies are more likely to exist in certain geographical locations (e.g., canals and straits). Additionally, the experts expect to learn how do the FON and HON representations compare with each other in terms of network properties such as port centralities.

While the HON contains richer information, the FON has the simplicity of one-to-one mapping from nodes to geographical locations on a map. To combine the advantage of both representations, we map the structure of HON back to the FON when visualizing it on the map, and assign scalar values to the corresponding nodes and edges in the FON for comparison. The comparison can be defined in multiple ways depending on the exploration goal. By default, we quantify the difference of the transition probabilities between the HON and FON. The difference can also be quantified by comparing the network analysis results. For example, domain experts are interested in the nodes with the largest PageRank [161], which effectively simulate the flow of invasive species; the PageRank differences can help to identify ports with underestimated risks in FON. In brief, mapping the difference or important values to FON provides clearer observation on the map view and allows users to effectively identify and select the regions of interest for further exploration.

**R2. Provide aggregation view of the higher-order nodes.** The experts would like to explore port connections at different granularities, such as connections among countries, continents, eco-regions, eco-realms, etc. Therefore, the higher-order nodes should be aggregated and visualized for high-level knowledge discovery. For example, it should provide information such as how many nodes with the highest order exist in an eco-realm (to reveal geographical distribution of higher-order dependencies), how many pathways incorporated in higher-order nodes navigate through multiple eco-realms (to identify non-indigenous species diffusion pathways), and so on. The level of aggregation should be flexible so that users can observe the connections at different granularities, such as countries, continents, eco-regions, eco-realms, temperature and salinity ranges, etc.

**R3. Visualize higher-order dependencies associated with a given port.** The experts first want to know that given a port, *how* do the previous steps change a ship's choice of the next step. For example, ships currently at Singapore may have equal probabilities of going to Los Angeles and Seattle. The experts wonder if ships coming from certain ports to Singapore will make them more likely go to Los Angeles, and how much the difference is. Meanwhile, the experts want to know if certain features correlate with the existence of higher-order dependencies. For example, are higher-order dependencies mainly associated with certain types of ships (such as oil tankers), or certain geographical locations (such as canals)?

Therefore, when a port of interest is designated, a subgraph of HON containing all higher-order nodes and edges associated with the port of interest should be generated. The transition probabilities from different higher-order nodes to the next node should be represented, in order to show how the previous ports a ship has visited may influence the ship's next step. Additionally, the attributes of ships corresponding to the transitions should be shown, such that users may discover certain higher-order movement patterns exclusively associated with certain types of ships, particular months, and so on. For example, if the ships moving between two ports are mostly passenger ships, the ship is likely to return to the previous port, since passenger ships are likely to move between two ports instead of among multiple ports. Therefore, we should encode these attributes associated with transitions, so that once transitions of interest are identified, users can observe the corresponding attributes.

**R4. Visualize and expand a subgraph.** In the context of invasive species studies, the experts hope to see if higher-order dependencies are evenly distributed in the network or only exist in certain groups of tightly connected ports. The experts also expect to visualize and expand a subgraph of invaded ports to understand how invasive species propagate from a given port. The expansion should be performed forward or backward to cover more nodes along paths of interest. This allows interactive exploration and facilitates case studies on studying the species flow along certain shipping pathways. To understand the influence of these paths to the entire network, such as which are the important pathways that connect different clusters of ports, visual connections should be established between the subgraph and the entire network.

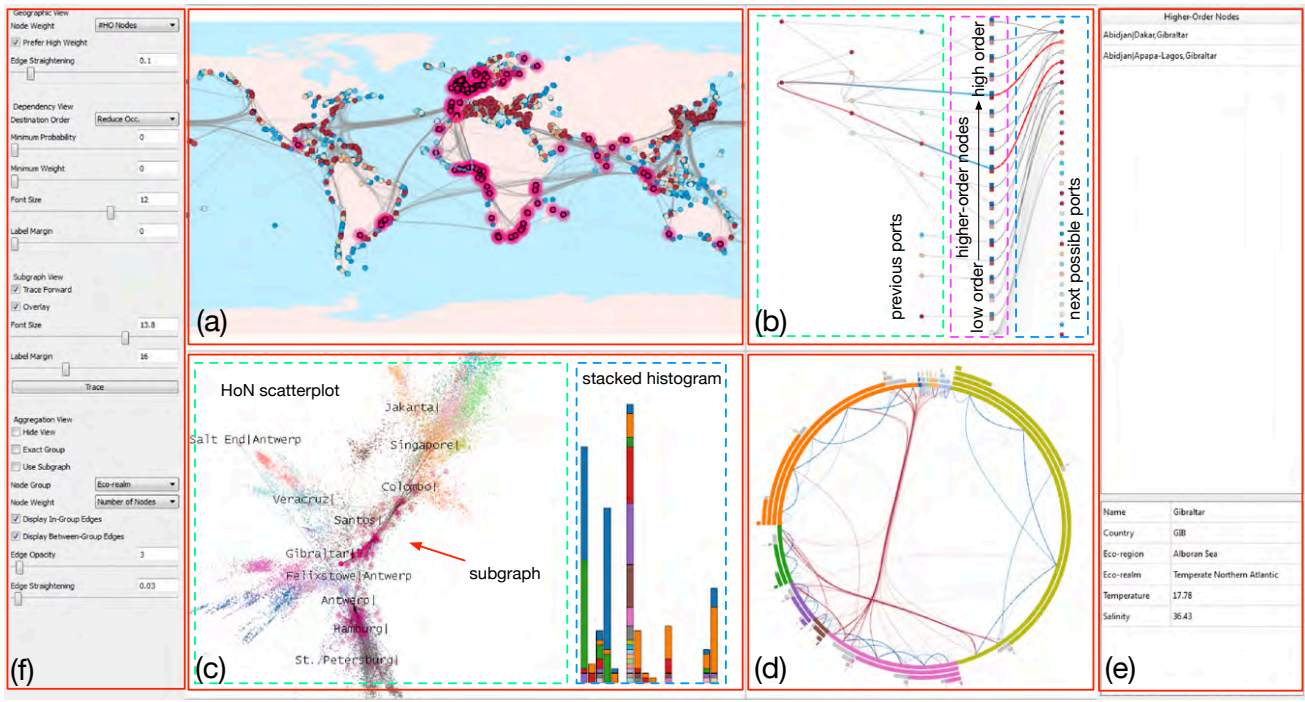

Figure 5.2. The overview of HONVis, our visual analytics system for exploring the global shipping higher-order network. (a) Geographic view. (b) Dependency view. (c) Subgraph view. (d) Aggregation view. (e) Table view. (f) Parameter panel.

## 5.5 System Description

We design five coordinated views to meet the design requirements stated in Section 5.4. The five views of our HONVis are: 1) a *geographic view* where the geographical locations of ports and the connections among them are displayed; 2) a *dependency view* that shows all the higher-order nodes associated with a given port, as well as the previously visited ports and the next possible ports to visit (Section 5.5.1); 3) a *subgraph view* that compares a user-generated subgraph with the graph showing the entire HON (Section 5.5.2); 4) an *aggregation view* that visualizes higher-order dependencies under a certain aggregation criterion (Section 5.5.3); and 5) a *table view* that displays the detailed text information of a port or the current user exploration status. Users can hide the aggregation view to leave more vertical space for the dependency view. All these five views are linked together through brushing and linking. Labels of higher-order nodes/ports will be shown in the dependency view and subgraph view, since they cannot be inferred from the respective layout.

With HONVis, a typical user workflow is as follows. Users start from the geographic view and aggregation view. In the geographic view, they identify through visual encoding (red to gray to blue for high to medium to low), the ports with more higher-order nodes or ports whose rankings change the most in the HON compared to the FON. In the aggregation view, both the current nodes and their previous steps are aggregated according to a given criterion. For example, when the entire network is aggregated at the eco-realm level, users can efficiently identify the higher-order nodes whose previous steps contain ports in other eco-realms, suggesting non-indigenous species introduction pathways. Users may then specify a port for individual port investigation: all higher-order nodes containing pathways leading to the given port will be visualized in the dependency view, showing how ships or species coming from different pathways to the current port will have different probability distributions of choosing the next port. Assuming a potentially invasive species in the current port, users can also trace species diffusion in the subgraph view, and understand how the species may propagate to different clusters of ports. Starting from the higher-order nodes directly related to the specified port, users can expand the subgraph of invaded ports by tracing forward or backward and including the nodes visited. This stepwise expansion gradually fills the gap between the one-step neighborhood of the selected port and the entire global shipping network, which helps users evaluate the impact of a port or a higher-order node at a larger scale. After each user operation, we use animated transition to emphasize the changes in other views, indicating where to explore in the next step. In the following, we describe the dependency view, subgraph view, and aggregation view. The other two views (refer to Figure 5.2 (a) and (e)) are omitted as their design and roles are straightforward.

### 5.5.1 Dependency View

Given a set of higher-order nodes, the dependency view shows the connections among previously visited ports and next possible ports to visit. It corresponds to the design requirements **R1** and **R3**. The higher-order nodes being investigated can be the higher-order nodes associated with a port selected in the geographic view, or multiple higher-order nodes contained in an aggregated node selected in the aggregation view. The transitions between the higher-order nodes and their next possible ports can be filtered by the probabilities or the number of ships associated with the transitions. This produces a compact visualization allowing the more important transitions to be observed clearly. A set of attributes is assigned to the ports, providing visual hints to guide the exploration. These attributes include computed ones (e.g., PageRank in the FON, aggregated PageRank in the HON, and the number of associated higher-order nodes) and the geographical properties (e.g., temperature, salinity, and eco-realm).

**Higher-Order Nodes.** Each higher-order node is displayed as a rectangle, as shown in Figure 5.2 (b). Each rectangle is divided into two boxes: the upper and lower boxes. The upper box indicates the entropy of transition probabilities starting from the higher-order node, where blue/white corresponds to low/high entropy (*low* entropy corresponds to *high* certainty). The lower box indicates the KLD of the transition probability distributions of the higher-order node and its corresponding first-order node, where red/white corresponds to high/low KLD. These two properties are of particular interest, since the first one represents the *certainty* of the next port to visit given the higher-order dependency and the second one represents the *difference* between the higher-order node and its corresponding first-order node. Therefore, distinct higher-order patterns significantly different from first-order ones show a combination of blue and red boxes and can be identified at a glance. In Figure 5.2 (b), we observe considerable blue/red combinations, indicating higher-order dependencies of potential interest that are not captured in the FON. Higher-order nodes with high entropy or low KLD values, though less interesting by themselves, are indispensable for bridging the connection of other higher-order nodes.

If the number of higher-order nodes is large, we only display the lower KLD boxes of nodes, since KLD is the deciding factor for extracting higher-order dependencies and is more relevant to the formation of higher-order nodes. The higher-order nodes are lined up according to their current ports and orders: the nodes with the same current port are contiguous and the node with highest/lowest order is placed at the top/bottom of that contiguous space.

**Previous Ports.** We display the previous ports as circles to the left side of the higher-order nodes, as shown in Figure 5.2 (b). For each higher-order node, we draw a smooth high-order Catmull-Rom spline to connect its corresponding ports in the visit order for clear observation, as suggested by Blaas et al. [33]. The curves exhibit color transition from red to blue, indicating the visit order of ports (i.e., red indicates the port visited first and blue indicates the current port).

We determine the layout of the previous ports using a simple heuristic: their $x$-coordinates are determined by their earliest appearance in any higher-order nodes; and their $y$-coordinates are determined by the average $y$-coordinates of the higher-order nodes containing them. The ports that are placed at the same locations are moved vertically to resolve the conflict. In Figure 5.2 (b), we find that the ports are aligned from left to right in their visit order for most higher-order nodes. The ports associated with individual second-order nodes are mostly placed at the lower part of the dependency view and the ports associated with more higher-order nodes are mostly placed at the upper part. More sophisticated algorithms exist for drawing directed graphs, but they tend to increase the horizontal span in order to better preserve the order of nodes, which may not be ideal in our scenario given the limited screen space.

**Next Possible Ports.** We display the next possible ports as circles to the right side of the higher-order nodes, as shown in Figure 5.2 (b). The opacity of an edge connecting a higher-order node and a next possible port indicates the corresponding transition probability. In Figure 5.2 (b), since most edges associated with higher-order nodes are dark, their next steps to take are fairly certain. Furthermore, the edges associated with the first-order node at the bottom share similar light colors, which indicates that the next possible ports will be visited with similar probabilities.

The next possible ports can be lined up to reduce edge crossing or reflect a user-specified property. To reduce edge crossing, we first estimate the $y$-coordinate of a port using the average $y$-coordinates of the higher-order nodes connecting to that port weighted by their respective transition probabilities. Thus, a port will be placed closer to the higher-order nodes that are more likely to transit to it. Then, all ports are evenly spaced to span the entire screen space along a vertical line, preserving their estimated $y$-coordinates. Users can also arrange the ports according to an associated property. This facilitates the identification of transitions related to certain characteristics (e.g., high temperature or a certain eco-realm).

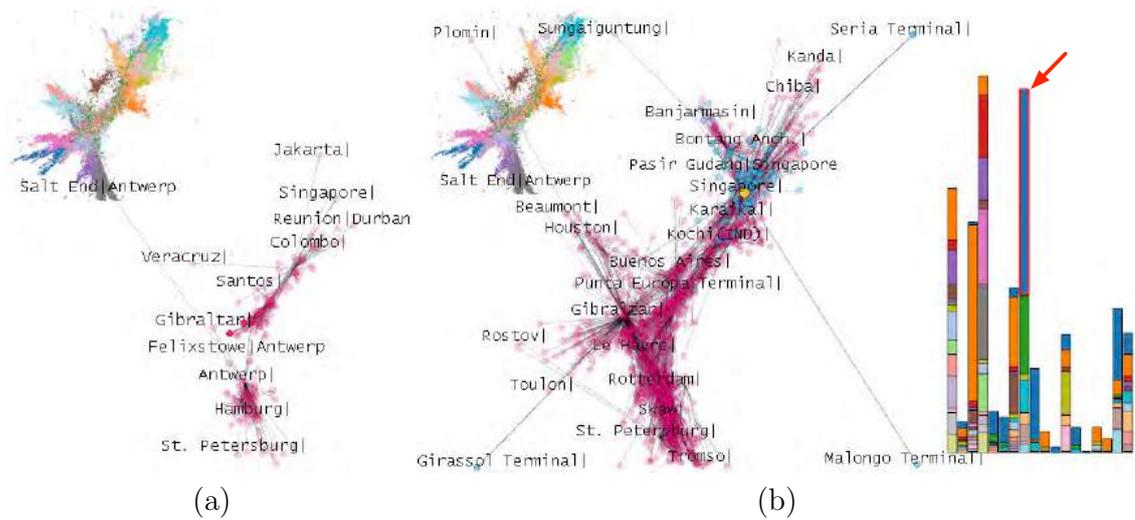

Figure 5.3. The subgraph view. (a) HON scatterplot and subgraph. (b) HON scatterplot, subgraph expanded from the subgraph shown in (a), and stacked histogram showing node contribution.

**Interaction.** Users can select a previous port in the dependency view for investigation. The curves associated with that port will maintain their colors, while the other curves will become gray. In the table view, we display the names of the higher-order nodes containing the selected port and the information of this port. In the subgraph view, the subgraph will be updated as well, so that users can study the propagation pattern given that port as a previous node. Users can further select a set of next possible ports. To provide detailed information, we display two histograms of ship types and temporal activities of the transitions between the selected higher-order nodes and the next possible ports.

### 5.5.2 Subgraph View

The subgraph view visualizes a subgraph of the HON in the context of the entire network, corresponding to the design requirement **R2**. It shows the topological proximity of ports, and allows users to expand the subgraph of invaded ports to explore how the invasive species will propagate over the network. The entire HON is described by a layout of the network using ForceAtlas2 [114]. Meanwhile, the structural organizations of HON also influence the propagation dynamics. For example, the global shipping network is naturally organized into multiple communities; in each community the ports are tightly coupled by shipping traffic. Once a given species is introduced to a community, the species will propagate through the whole community shortly. Therefore, locating the *entry points* and *pathways* to communities is essential to devising species control strategies. We apply the widely-used Louvain method [36] for community detection, using edge weights and the default resolution of 1.0. Note that higher-order nodes representing the same port could belong to different communities, which naturally yield overlapping clusters and indicate how certain ports may be susceptible to multiple sources of species invasion.

We visualize the entire HON using scatterplot, where each point represents a node in HON, colored by the community of that node. The edges in the HON are ignored for clutter reduction. The subgraph is then displayed on top of the scatterplot. Each node in the subgraph is drawn as a semi-transparent circle, whose center is placed at the corresponding point in the scatterplot. The transparency of a circle indicates the probability of the corresponding node being reached during the expansion of the subgraph. An edge in the subgraph is drawn as a straight line with transparency indicating the corresponding transition probability. In Figure 5.2 (c), the subgraph expanded from the two higher-order nodes selected in the dependency view is displayed on top of the HON scatterplot. We can see that the subgraph mostly covers the lower right branch and the lower middle region of the network. As an option, users can choose not to overlay the subgraph and the HON scatterplot. In that case, the HON scatterplot will be displayed in the top-left corner of the subgraph view, as shown in Figure 5.3 (a). Without the overlay, the nodes in the subgraph can be observed more clearly, but the covered regions can only be roughly interpreted.

**Subgraph Expansion.** Subgraph expansion is performed by tracing from the nodes in the current subgraph and including the nodes reached during the tracing. Users can trace backward to find out through which nodes the subgraph can be reached or trace forward to explore the nodes that will be reached from the nodes in the current subgraph. The subgraph expansion procedure starts from a set of higher-order nodes selected in the other views. The initial probability of reaching a node is proportional to the number of ships leaving/arriving that node when tracing forward/backward. After each tracing step, the probability of reaching a node $n_i$ will be updated to $\sum_{n_j \in N(n_i)} p(n_j) p(e_{ji})$, where $N(n_i)$ is the set of nodes from which $n_i$ will be reached, $p(n_j)$ is the probability of $n_j$ being reached, and $p(e_{ji})$ is the transition probability from $n_j$ to $n_i$. The expansion can be observed in both the HON scatterplot and the geographic map, where the ports associated with any node in the subgraph is highlighted. A tracing step is only performed when users click the "Trace" button in the parameter panel. This allows users to observe the propagation pattern in a stepwise manner.

**Identification of Contributing Nodes.** By *contributing nodes*, we mean the nodes that lead to the coverage of a certain community or certain regions in the HON. The contribution of a node $n$ to a community $c$ is measured by the number of nodes in $c$ that are reached directly through $n$ for the first time. The total contribution of a node $n$ is the summation of its contributions to all communities. We choose to visualize twenty nodes with the highest total contributions using a stacked histogram. Each bin in the histogram corresponds to the coverage of one community. The bars with the same color correspond to the same contributing node. In Figure 5.3 (a) and (b), we show the subgraph before and after a critical tracing step. After that tracing step, the subgraph propagates to the upper part of the HON. We can see that many nodes in the 8-th community are covered after this step, as indicated by the red arrow in Figure 5.3 (b). The node corresponding to the blue bars contributes most to the coverage of that community, as the blue bar in the 8-th bin is the tallest. By clicking on that blue bar, the contributing node is highlighted in yellow and the nodes reached from it are highlighted in blue in the subgraph. This indicates that the contributing node is an important transit point for the ships to propagate into the 8-th community. By identifying such nodes, domain experts can devise targeted species control strategies at certain critical ports to maximize the effectiveness and minimize the cost.

### 5.5.3 Aggregation View

The aggregation view provides an overview of the higher-order dependencies among groups of ports and their connections, corresponding to the design requirement **R2**. It also serves as a convenient interface to select the higher-order nodes with desired properties, e.g., the fifth-order nodes that contain ports in different eco-realms. The aggregation can be performed on the entire HON or synchronized with the subgraph under expansion based on port grouping. The aggregated node corresponding to an original higher-order node is determined by converting each port associated with the higher-order node to the group containing that port. Formally, denoting a $k$-th-order node as a sequence of ports $\mathbf{n_i} = [p_{i_0}|p_{i_1}, \ldots, p_{i_{k-1}}]$, where $p_{i_0}$ is the current port and $p_{i_1}, \ldots, p_{i_{k-1}}$ are the previously visited ports, and the group of a port $p$ as $G(p)$, the aggregated node corresponds to node $\mathbf{n_i}$ can be written as

$$A(\mathbf{n_i}) = [G(p_{i_0})|G(p_{i_1}), \ldots, G(p_{i_{k-1}})]. \tag{5.1}$$

The edges are aggregated accordingly by summing up the weights of edges corresponding to the same pair of aggregated nodes.

We group the ports according to their eco-realms. This means that the higher-order nodes containing sequences of ports are aggregated into the higher-order nodes containing sequences of eco-realms. The edges are aggregated to show the number of ships moving among the eco-realms. Twelve groups of ports (i.e., eleven marine eco-realms and one group containing all freshwater ports) are considered. Unlike the original nodes, where two consecutive ports are always distinct, an aggregated node may contain two consecutive appearances of the same eco-realm, meaning that the ships move from one port to another in the same eco-realm. This will be effective for domain experts to distinguish the higher-order dependencies inside each eco-realm and among the eco-realms, which is critically important to the study of species invasion.

**Coarse Grouping Aggregation.** In some cases, the aggregation technique with the above *exact grouping* may not be necessary. For example, users may be interested in the higher-order nodes whose previous steps contain ports in other eco-realms without caring exactly what those eco-realms are. In other words, it suffices to distinguish the ports in the same eco-realm as the current port and the ports in different eco-realms. To accommodate this need, we further design an aggregation scheme with *coarse grouping*. With coarse grouping, the aggregated nodes can still be generated using Equation (5.1) but with a slightly different grouping function $G(p)$. Unlike the exact grouping function that always maps a port to a group, the coarse grouping function either maps a port to the group representing the eco-realm of the current port, or to a special status indicating that the port is in a different eco-realm. For example, the node [Singapore|Port Klang, Shanghai] will be aggregated into [Central Indo-Pacific|Central Indo-Pacific, Temperate Northern Pacific] with exact grouping but [Central Indo-Pacific|Central Indo-Pacific, Different Eco-

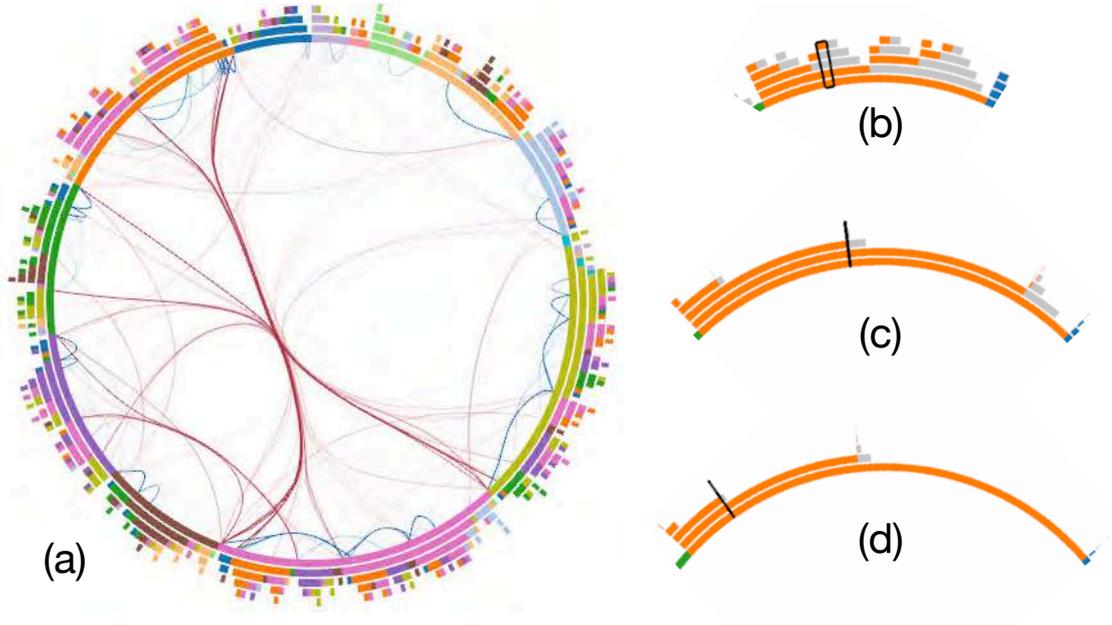

Figure 5.4. The aggregation view. (a) Exact grouping using eco-realms. (b) to (d) The eco-realm of "Temperate Northern Pacific" with coarse grouping. (b) Uniform node weight. (c) Nodes are weighted by the number of original nodes. (d) Nodes are weighted by the number of ships. The same aggregated node is highlighted in black in (b) to (d).

realm] with coarse grouping. In our experiment, the number of aggregated nodes reduces from 396 to 180 with coarse grouping, allowing users to focus more on the between-group dependencies. Users can switch between exact grouping and coarse grouping depending on their needs.

**Network Layout.** We show the aggregation view using the circular layout, where the nodes are aligned on a circle and their connections are displayed inside the circle. The edges among nodes belonging to the same current group are colored in blue, while the edges among nodes belonging to different current groups are colored in brown. We bundle the edges for visual clarity using the force-directed edge bundling algorithm [103]. An aggregated node covers a sector of the circle, as highlighted by the black rectangle in Figure 5.4 (b). The number of layers in the sector represents the node's order, and the color of a layer represents the group of ports (i.e., eco-realm). The groups of ports are visited in the order from the outermost layer to the innermost layer (i.e., the current group is in the innermost layer). The gray color is reserved for the special group "different eco-realm". For example, the aggregated node highlighted in Figure 5.4 (b) exhibits five layers, from outermost to innermost, colored in orange, gray, gray, orange, and orange, respectively. This indicates that the node is fifth-order and the ships visited different eco-realms two and three steps before. The nodes are ordered according to their corresponding sequences of groups. That is, the nodes belonging to the same current group occupy a consecutive sector at the innermost layer, and then the nodes belonging to the same previous group are organized consecutively at the second inner layer, and so on. In Figure 5.4 (b), we can see that the nodes corresponding to the orange group are placed together. The second inner layer shows orange on the left side and gray on the right side, indicating that the nodes with the same previous group are on the left side and the nodes with different previous groups are on the right side.

The arc length of the sector is decided by the weight of the corresponding node. We provide three types of node weights. Figure 5.4 (b) shows the orange group with the uniform weight, where each node occupies the same arc length so that different nodes can be distinguished more easily. In Figure 5.4 (c), the aggregated nodes are weighted by the number of original nodes contained in them. We can observe from the arc lengths that most higher-order nodes exist among ports in the same eco-realm. In Figure 5.4 (d), the aggregated nodes are weighted by the number of ships related to each node. We observe that about half of the sector shows higher-order dependencies, within which a large proportion of ships travel within the same eco-realm, while a small proportion may bring in invasive species from other eco-realms, suggesting targeted control opportunities. A complete picture of the aggregation view with coarse grouping can be found in Figure 5.2 (d). Figure 5.4 (a) shows the aggregation view with exact grouping. Although it provides more details, it is more difficult to interpret as an overview due to its complexity.

## 5.6 Case Study on Species Invasion via Global Shipping Network

We worked in person with two domain experts in network science and marine ecology, and in this section we record the experts' workflow and observations when they first used HONVis to explore the global shipping network. We then show how HONVis reveals novel patterns at the global scale, which are valuable for decision-makers to devise effective species control strategies.

### 5.6.1 Data

Diverse types of data were used for this case study. The global ship movement data are made available by the Lloyd's List Intelligence, which contains more than two thirds of active ships globally (measured in dead weight tonnages). The raw data contain 3,415,577 individual ship voyages corresponding to 65,591 ships that move among 4,108 ports globally between May 1, 2012 and April 30, 2013. The data also contain metadata of ships, such as ship type, voyage start and end time, ship size, as well as metadata of ports such as coordinates and country. The environmental conditions (temperature and salinity) of ports are obtained from the Global Ports Database [120] and the World Ocean Atlas [11, 136]. The eco-region information comes from Marine Ecoregions of the World [205] and Freshwater Ecoregions of the World [2]. Ports (and associated ship movements) that have corresponding coordinates, eco-region and environmental conditions are retained for analysis.

### 5.6.2 Domain Experts' Workflow and Insights

**Locating Ports with Higher-Order Dependencies (R1).** The experts wanted to investigate potential species invasions from South America to Europe via global shipping, and evaluate the influence of higher-order movement patterns of shipping. As shown in Figure 5.5 (a), the experts first used the geographic view to zoom in to South America. To identify ports through which ships demonstrate higher-order movement patterns, the experts chose to color the ports by the number of higher-order dependencies, and focused on ports shown in red (the ones that demonstrate the most higher-order dependencies). The number of candidate ports is thus reduced from hundreds to tens. The experts then simply clicked on the area of interest, and in the table view (Figure 5.5 (b)), the ports in the area were sorted by the number of higher-order dependencies. The experts clicked through the top ports to highlight shipping paths from those ports, and quickly identified Salvador in Brazil, which shows a direct connection in the bundle from South America to Europe.

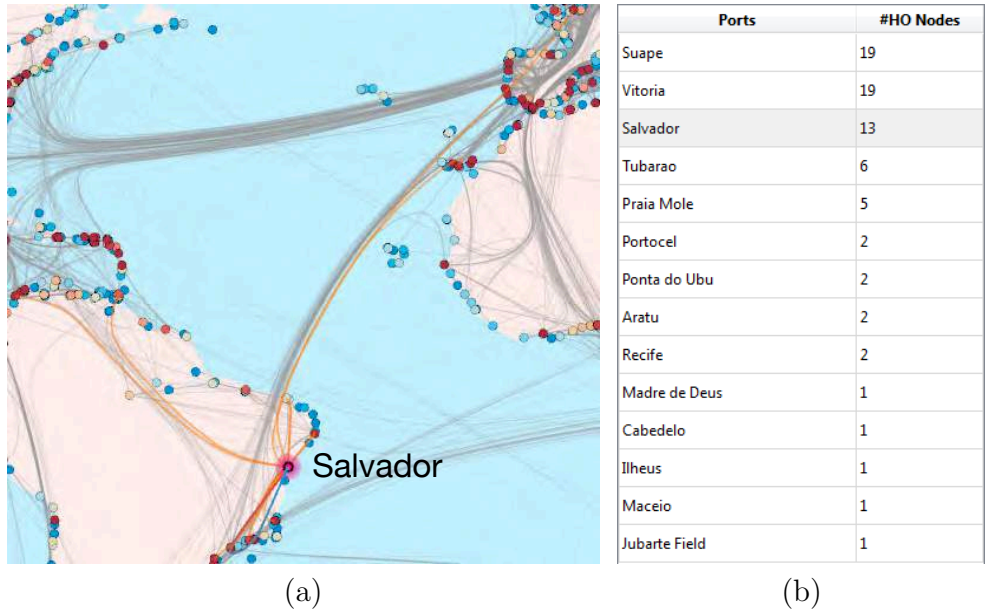

Figure 5.5. Identifying a port of interest. (a) The port Salvador in Brazil is highlighted with a magenta halo in the geographic view. (b) The nearby ports are listed in the table view ordered by their numbers of associated higher-order nodes.

**Exploring Higher-Order Dependencies (R3).** The experts then evaluated

how the movement pattern from Salvador is influenced by from where the ships came to Salvador. After selecting Salvador in the table view, all its higher-order dependencies are displayed in the dependency view (Figure 5.6 (c)). At a glance, the experts knew that without knowing a ship's previous locations, the ship's next step from Salvador is uncertain. This is revealed by both the weak connections (dimmed visually) from the first-order node [Salvador|] (highlighted by the blue arrow in Figure 5.6 (c)) to all 16 potential destination ports on the right, and the white entropy box of [Salvador|] (high entropy indicating low certainty). A quick drag-and-drop selection of the destination ports reveals that the ships from Salvador are mainly container carriers (UCC), and shipping at Salvador remains active throughout the year (Figure 5.6 (b)).

Following the link from Rio de Janeiro to the second-order node [Salvador|Rio de Janeiro] (highlighted by the red arrow in Figure 5.6 (c)), the experts discovered that knowing ships came from Rio de Janeiro to Salvador does not significantly influence the ships' choices for the next step, indicated by the light red KLD box (meaning low difference compared with the distribution from the first-order node), and the light blue entropy box (indicating low certainty). Essentially, this implies that *the second order is insufficient in capturing the complex dependencies in this case.* It is likely that Rio de Janeiro, being the second largest city of Brazil, has a port so versatile and provides limited information in narrowing down complex ship movement patterns. The reason that the second-order node [Salvador|Rio de Janeiro] is included in HON is that it bridges connections from other essential higher-order nodes.

The experts then proceeded to explore dependencies beyond the second order. By selecting the fourth-order path Salvador → Santos → Rio de Janeiro → Salvador, as highlighted in Figure 5.6 (c), the experts observed a *loop*, that if a ship has been observed following the loop at least once, the ship will keep following the loop for sure. The dark blue entropy box and dark red KLD box at port [Salvador|Rio de Janeiro,

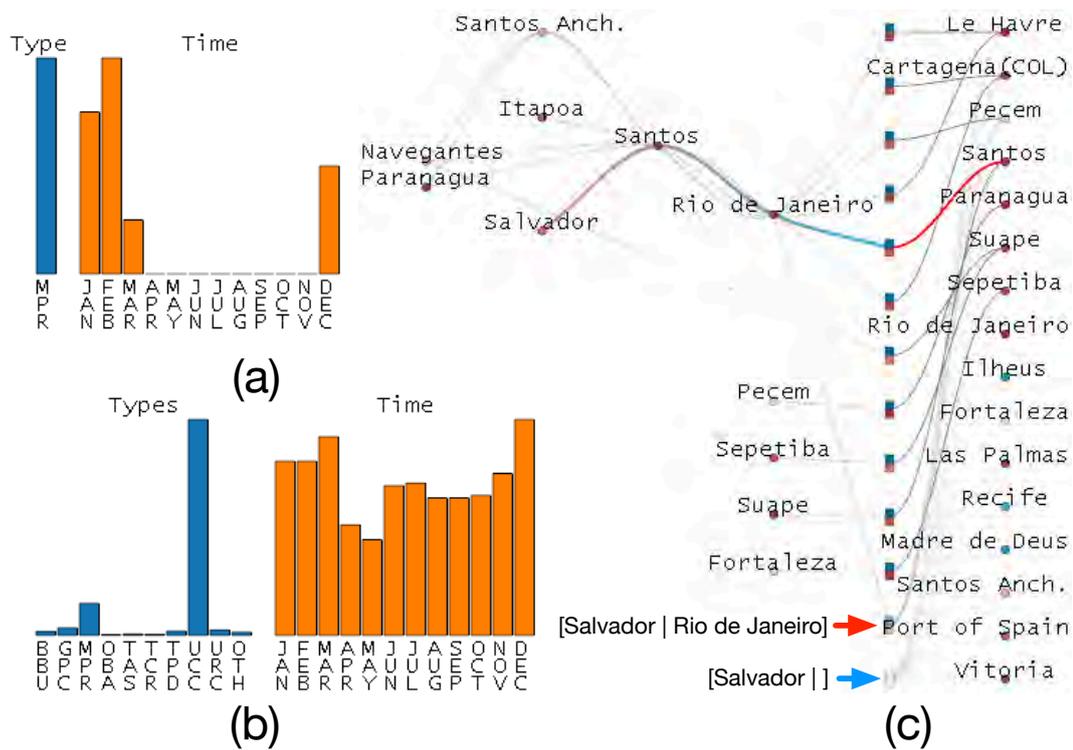

Figure 5.6. The higher-order dependencies related to Salvador. (a) Histograms of ship types and temporal activities of fourth-order movement patterns from Salvador. (b) Histograms of ship types and temporal activities for all ships from Salvador. (c) Higher-order dependencies related to Salvador in the dependency view.

Santos, Salvador] indicate that this pattern displays high certainty and is significantly different than the first-order movement pattern. Moreover, the bar charts (Figure 5.6 (a)) in the dependency view show that ships following this fourth-order pattern are exclusively cruise ships (MPR) and are only active in the summer (December to March in the South Hemisphere), *revealing the underlying reason behind this higher-order dependency.*

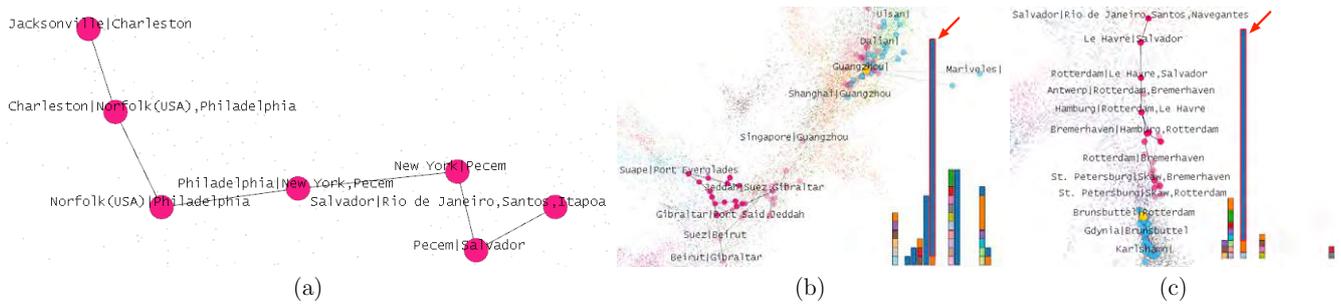

Figure 5.7. (a) Tracing how the species may propagate from Salvador in a stepwise manner. (b) The propagation eventually influences multiple ports in East Asia, which are far away from Salvador. (c) Another direction of the propagation covers multiple ports in Northwest Europe.

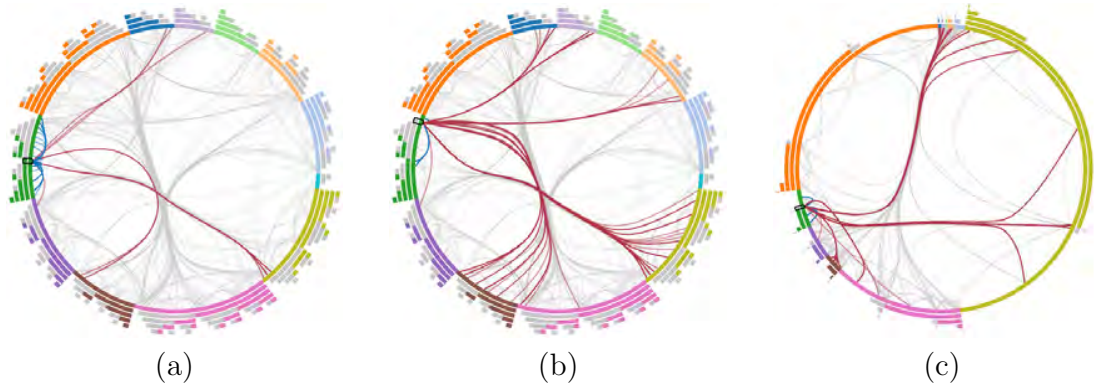

Figure 5.8. Investigating higher-order dependencies at different granularities. (a) Studying a sector which both the current and previous ports are in the Tropical Atlantic eco-realm. (b) Studying a sector which the current ports are in the Tropical Atlantic eco-realm, but the previous ports are not. (c) Changing the view in (b) from uniform node weight to weighted by the number of ships.

**Exploring the Influence of Higher-Order Dependencies in Propagation (R4).** The experts further explored how higher-order dependencies influence the propagation of invasive species via shipping. Specifically, knowing that the ships came from Itapoa or Navengates before sailing through Santos and Rio de Janeiro to Salvador, the experts wanted to figure out how the species propagate differently. The experts first selected the fourth-order pathway Itapoa → Santos → Rio de Janeiro → Salvador in the dependency view, and the corresponding node [Salvador|Rio de Janeiro, Santos, Itapoa] is automatically selected in the subgraph view. The experts clicked "Trace" button to see how the species may propagate from the given port in a stepwise manner. As shown in Figure 5.7 (a), the species first went to Pecem in Brazil and then to New York City in USA. After that, with high certainty, the species were propagating toward the blue cluster on the left, which mainly consists of ports in Northeast America. After tracing a few more steps, the possible diffusion diverged. A branch kept propagating in Northeast America with high certainty. More interest-ingly, the species may influence multiple ports in East Asia, represented as the green cluster at the top-right corner as shown in Figure 5.7 (b), which was topologically far from the initial port Salvador on the lower left. The experts noticed the new spike in the stacked histogram, consisting mainly of a single color (blue). This indicates that a port is making significant contribution to the massive dispersion of species in that cluster. The experts clicked on the dominating blue bar of that spike, and the subgraph view reveals that Guangzhou was the port that facilitated the potential massive spread of invasive species in East Asia. *Knowing that Guangzhou is the entry point to species spreading in that region is vital when developing targeted invasive species control strategies* to prevent Brazilian species from invading East Asia. Tracing back, Guangzhou was invaded by ships sailing from Gibraltar through the Mediterranean Sea, then through the Suez Canal to the Red Sea, passing Jeddah and finally to Guangzhou. These ports on the shipping path also deserve close monitoring.

On the contrary, when the experts selected the pathway Navengantes → Santos Arch → Santos → Rio de Janeiro → Salvador in the dependency view, with high certainty the species will propagate toward the gray cluster at the bottom as shown in Figure 5.7 (c), which mainly consists of ports in Northwest Europe. The port leading to the mass diffusion in the cluster was Brunsbuttel. Through the interactive exploration and comparison, the experts gained a comprehensive understanding on how the higher-order dependencies may influence the subsequent propagation.

**Exploring Higher-Order Dependencies at Different Granularities (R2).** Finally, the experts wanted to explore the connections at a higher level: the *eco-realm* is the largest biogeographic division of the sea [205]; species coming from other eco-realms are more likely to be non-indigenous and will incur invasions. The question is: how do the connections differ whether the previous port was also in the Tropical Atlantic eco-realm (which Salvador is in) or was in a different eco-realm? The experts first chose to color the ports in the geographic view with eco-realms. Tropical Atlantic was colored dark green. Then the experts shifted to the aggregation view, and chose the sector which both the current and previous ports are in the Tropical Atlantic eco-realm. The sector is denoted by two layers of dark green, as shown in Figure 5.8 (a). The aggregation view reveals ampler and stronger intra-eco-realm connections as denoted by blue links, compared with inter-eco-realm connections as denoted by brown links (mainly connections to Temperate Southern Africa, Temperate Northern Atlantic, and Temperate South America). By cross-checking with ship types in the dependency view, the experts found out that variable types of ships exist for this case.

On the contrary, when the experts chose the sector which the current ports are in Tropical Atlantic but the previous ports are not (the sector denoted by the innermost layer as dark green and the outer layer as gray, as shown in Figure 5.8 (b)), the inter-eco-realm connections are stronger, including additional connections to Tropical Eastern Pacific and Central Indo-Pacific. Meanwhile, the dependency view suggested that these inter-eco-realm navigation patterns were exclusively made by container carriers. The experts came to the preliminary conclusion that *ships coming from different eco-realms were more likely to keep traveling among eco-realms, posing higher risks of bringing in non-indigenous species.* Furthermore, in terms of species management strategies for specific types of ships, *container carriers posed the highest risk for the introduction of non-indigenous species.*

Last, the experts changed the widths of sectors from uniform to the number of ships, as shown in Figure 5.8 (c), which gives an intuitive overview of the composition of all higher-order dependencies. The experts noticed that although ships coming from other eco-realms to Tropical Atlantic have higher chances of keeping with the inter-eco-realm voyages, the number of inter-eco-realm trips was much less than that of intra-eco-realm trips. The fact that the more risky inter-eco-realm voyages were the minorities suggested that *targeted species control policies only need to focus on a* *small fraction of ships and routes.*

**Insights Revealed by HONVis at the Global Scale.** HONVis not only enables interactive exploration as shown in the above use case, but also reveals the influence of higher-order dependencies at the global scale. For example, one observation was for ports in the Arctic. The change of climate had been melting the Arctic sea ice at an alarming speed and opening up Arctic shipping routes [82]. Therefore, there are growing concerns on threats to the valuable resources in the Arctic posed by invasive species via the unprecedented growth of shipping. The PageRank algorithm naturally simulates the flow of species hitchhiking onto ships, with random resets accounting for the changing or unobserved shipping activities. The PageRank score of each port indicates the relative risk that species will end up to the port in multiple steps. The PageRank risk estimation on the FON marks multiple ports in the Arctic as high risk, but as HON can improve the result of PageRank running upon, the estimated risks for Arctic ports were in fact overwhelmingly overestimated on the FON. This is indicated by the ports in blue as shown in Figure 5.9. For example, the PageRank score of Murmansk, a major Arctic port in Russia, was $4.52 \times 10^{-4}$ on the FON, but only $1.57 \times 10^{-4}$ on the HON. The dependency view suggested that by using the HON, traffic from hub ports such as Rotterdam to the Arctic ports is more likely to go back immediately to those hub ports rather than moving randomly among Arctic ports. Thus the relative flow of species in the Arctic is smaller on HON. The information on the overestimation of risks made possible by HONVis is important for policy makers.

5.7  Conclusions and Future Work

We have presented HONVis, a visual analytics framework for visualizing and exploring higher-order networks. We focus on the global shipping network and work closely with domain experts in network science and marine ecology to compile the task list and define design requirements. Our HONVis design leverages five linked views to enable users to explore the HON at different levels of detail and investigate higher-order dependencies among higher-order nodes. By directly contrasting the HON and its FON counterpart and visualizing higher-order dependencies, we tackle the key challenges in visualizing higher-order dependencies in networks, which is a milestone in pushing the understanding of the formation and impact of higher-order dependencies. The efficacy of HONVis is demonstrated through results gathered by two domain experts who use the system to investigate species invasion in the global shipping network. Several critical insights that can only be obtained with the use of HONVis are reported.

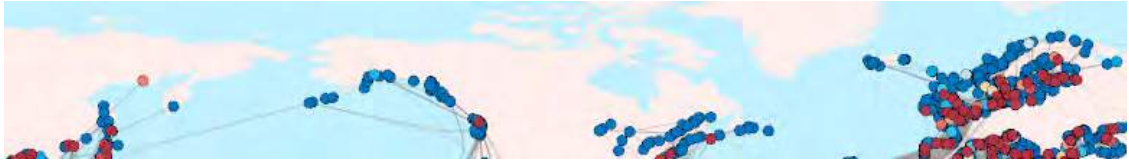

Figure 5.9. Comparison of PageRank risk simulation on the FON and the HON. Blue ports are risks overestimated on the FON and red ports are risks underestimated on the FON.

We acknowledge the limitations of the current version of HONVis, including the lack of effective visual hints to aid the users in navigating through the different views, and the challenge of labeling when the data are large. We advocate the idea of automatically producing statistics of all possible dependency structures (such as large loops) and aiding in the identification of principal patterns, which is a non-trivial task given the computational complexity.

Besides the application in global shipping and species invasion, the framework of HONVis can be generalizable to other types of HONs, which we plan to implement in the near future. For example, given that air transportation exhibits higher-order dependencies [178], HONVis can help to explore epidemic outbreak scenarios through domestic and international travels, by substituting ships with airplanes and invasive species with contagious diseases. Similarly, HONVis can also help to explore information diffusion patterns through phone call or online activities in social networks, by treating phone call or retweet cascades as ship trajectories.

# CHAPTER 6

# TUTORIAL: CONSTRUCTING HON FROM PAIRWISE INTERACTIONS DATA

## 6.1 Overview

How exactly should one use the HON code, how to extend its input from sequential data to pairwise interactions (such as phone calls between people and direct messaging between social network users), and how to do it efficiently?

**Intellectual merits:** Given a representative pairwise cellphone communications data, this tutorial chapter provides a step-by-step walk-through of the non-trivial process of constructing HON. Specifically, it proposes a time window-based approach to chain pairwise interactions into sequential data, and proposes an efficient algorithm with linear time complexity. The tutorial can serve as a reference for the pre-processing of data like social network direct messaging, wireless sensor network communication, and more.

**Connections:** This work is based on the discussion of data types in Chapter 2, and directly uses the HON+ algorithm and software implementation in Chapter 4.

**Work status:** This work is based on discussions with Mandana Saebi. It is being compiled as a Jupyter Notebook to be shared online.

## 6.2 Data understanding

The data set we will use throughout this tutorial is the anonymized cell phone call records in a 24-hour window in an European country. Assuming a typical Unix/Linux/Mac environment, getting the number of records:

```
wc -l data.txt
```
```
9364462 data.txt
```

For the 9.36 million records, each line is a pairwise phone call record in the format of:

`[Caller],[Callee],[Duration],[StartTime],[EndTime]`

Recall that the global shipping data has the format of

`[ShipID],[PortID],[ArrivalTime],[DepartureTime]`

The two data sets both look like pairwise data, but one important distinction exists. In the global shipping data, we know the `ShipID` associated with `PortID`, so that we can chain the ports together using the `ShipID` to obtain *sequences of ports for each ship*. In the phone call data, however, we do not know the implicit `[Information]` that can be used to chain phone calls together to build *sequences of cell phone users for each piece of information*.

Although phone calls may exhibit higher-order dependencies like "$A$ is more likely to call $B$ if being called by $C$, but more likely to call $D$ if being called by $E$.", each record of the raw interaction data involves only two entities (cell phone users), which does not contain any higher-order dependencies. Therefore, such pairwise data cannot be used directly as the input for HON; it is necessary to find the implicit `[Information]` to chain pairwise calls into sequences as the input for HON.

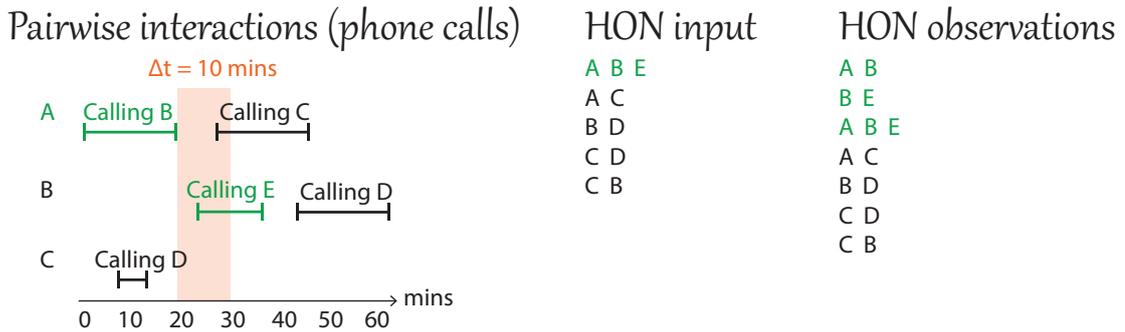

Figure 6.1. Chaining phone calls made within 10 minutes as sequential data.

## 6.3 Chaining Pairwise Data as Sequential Data

### 6.3.1 Main Idea

Here we make an important assumption: phone calls made within a time interval of 10 minutes are considered "related", and can be chained together. Figure 6.1 illustrates the chaining process: after $A$ finished the phone call with $B$, $B$ called $E$ within 10 minutes, so we suppose the two phone calls are about the same piece of `information`; in other words, there is a flow of information from $A$ through $B$ to $E$. Therefore, we can chain the two pairwise interactions $A \to B$ and $B \to E$ as $A \to B \to E$.

In contrast, the path $A \to B \to D$ does not exist in this example, since the gap between the two calls (the ending time of the $A \to B$ call and the starting time of the $B \to D$ call) exceeds the interval of 10 minutes. Also, $A \to C \to D$ does not exist, since the call $A \to C$ happened after $C \to D$.

The chaining algorithm takes the raw pairwise interactions as the input, and outputs the sequential data as the input for HON. The algorithm should only retain the longest sequence (e.g., $A \to B \to E$), instead of its subsequences (e.g., $A \to B$ or $B \to E$). The reason is that HON will automatically count all subsequences as the observations, as illustrated in Figure 6.1. If we include both $A \to B \to E$, $A \to B$, and $B \to E$, then in HON's observation construction process, $A \to B$ and $B \to E$ will be counted twice each, which is incorrect.

For the same reason, the chaining process should be as inclusive as possible. For example, if $A \to B \to D$, $A \to C \to D$ and $A \to B \to C \to D$ are all possible sequences, we only keep the most inclusive $A \to B \to C \to D$.

Formally, the data $\mathcal{D}$ is comprised of pairwise interaction $p = Caller \to Callee$ starting at time $t_s(p)$ and ending at $t_e(p)$. Assuming a time window $\Delta$, the chaining algorithm builds sequences of calls $S = [p^{(1)}, p^{(2)}, \ldots, p^{(k)}]$, satisfying:

- Temporal order of interactions, i.e., $t_e(p^{(k-1)}) \leq t_s(p^{(k)})$
- Temporal proximity of interactions, i.e., $t_s(p^{(k)}) - t_e(p^{(k-1)}) \leq \Delta$
- Maximum length, i.e., does not exist $S' = S_1 S_2$ or $S_2 S_1$
- Inclusiveness, i.e., does not exist $p^{(k')}$ satisfying $0 \leq t_s(p^{(k')}) - t_e(p^{(k-1)}) \leq \Delta$ and $0 \leq t_s(p^{(k+1)}) - t_e(p^{(k')}) \leq \Delta$

The resulting set of chains $\mathcal{S} = S_1, S_2, \ldots, S_n$ is the output of the chaining process, and serves as the input of the HON algorithm.

6.3.2 Naïve Chaining Algorithm

A direct transformation from the aforementioned idea to an algorithm is given in Algorithm 4. This method, despite being intuitive, has several issues. First, the two nested for loops makes the time complexity $\Theta(N^2)$, which is not scalable for data with more than 10,000 records. More importantly, this naïve approach does not guarantee the "maximum length" and "inclusiveness" properties discussed before. Some sequences thus identified can be subsequences of others, which will lead to duplicate counting during the construction of HON.

**Algorithm 4** Naïve chaining algorithm. Time complexity $\Theta(N^2)$

1: **for** pairwise interaction $p_i$ in $\mathcal{P}$ **do**
2:     **for** pairwise interaction $p_j$ in $\mathcal{P}$ **do**
3:         **if** $0 \leq t_s(p^{(k)}) - t_e(p^{(k-1)}) \geq \Delta$ **then**
4:             Chain $p_i$ and $p_j$

6.3.3 Optimized Chaining Algorithm with Linear Time Complexity

How can we reduce the time complexity, and perform all the necessary chaining with a single pass of the pairwise interactions? Suppose we have a existing pool of chains, and a new pairwise interaction $p_k$. We test if $p_k$ can be attached to the chains in the pool. If it can, we update the chain with the new "tail". If it cannot, we add $p_k$ to the pool as a chain with length of two. Then we proceed to testing $p_{k+1}$.

An illustration is given in Figure 6.2. Suppose we are testing the pair $A \to C$. We check if there are chains in the pool that terminates in the time window of $[t_s - \Delta, t_s]$. In this case, although the existing chain $A \to B \to C \to D$ ends in the time window, the ending element $D$ does not match the starting element $A$; therefore, these two cannot be chained. Next we proceed to the pair of $D \to B$, and we see that it satisfies all chaining requirements, so that $D \to B$ will be appended to $A \to B \to C \to D$.

A compelling questions is: given a pairwise interaction $p$ that lasts in $[t_s, t_e]$, how can we efficiently lookup candidate chains from pool? We notice that for an existing chain, all we care about are the ending element of the chain and the ending time of the chain. Here we use a nested dictionary, *chains*, with the first key being the ending time, and the second key being ending person. Since dictionaries are implemented using hash table, each lookup is $\Theta(1)$. For interaction $p = i \to j$, all we need to do is to check if there are chains in the pool ending in $i$, which is done in constant time; if so, check from $t_s - \Delta$ to $t_s$ if there are chains ending in the time window, which is done in $\Theta(\Delta)$. Since the length of time window is a small constant, we have reduced the lookup time of candidate chains to constant for each $p$. Therefore, the overall

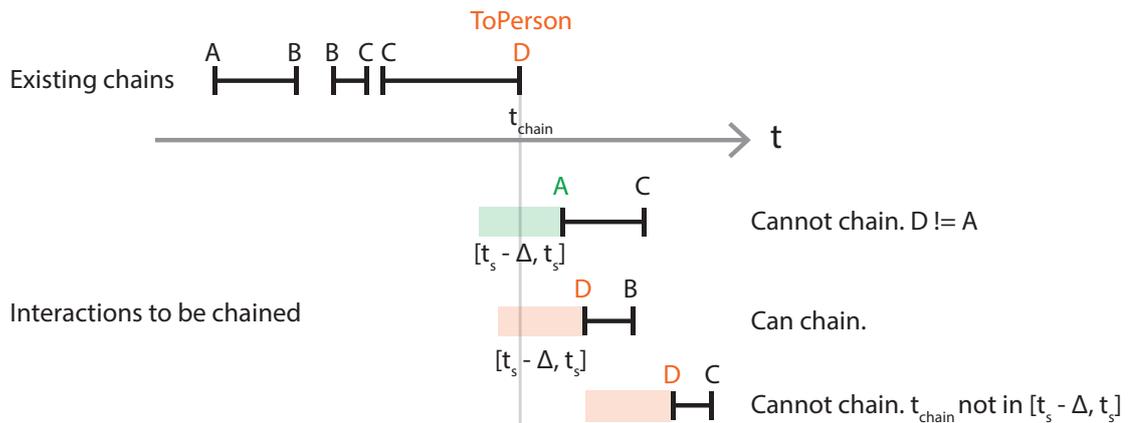

Figure 6.2. Illustration of the optimized chaining algorithm. The green shade is the range of $TryStartTime$.

time complexity is reduced to $\Theta(\Delta N)$. The algorithm is given in Algorithm 5.

6.4  Using the HON+ Code

First, download the latest HON+ implementation from the HON website `http://www.HigherOrderNetwork.com` (Figure 6.3), on which we share the code and explain algorithm details with videos and animations. The latest HON+ code contains both the Python and Common Lisp implementations; we will use the Python version (pyHON) in the following tutorial.

Suppose the sequential data has been prepared using Algorithm 5 as the input of HON. Note that each line should use the following format:

[SequencdID] [Person_1] [Person_2] [Person_3] ...

The pyHON package contains the following files: `main.py` is the main function that controls file I/O and optional parameters; `BuildRulesFastParameterFree.py` is the parameter-free HON+ implementation for building rules and outputting the higher-order dependencies, which we will use for illustration in this tutorial; `BuildNetwork.py` performs the network wiring step and outputting the final HON.

**Algorithm 5** Optimized Chaining Algorithm. Input: pairwise interactions $\mathcal{P}$ sorted by starting time. Output: interactions chained into sequences that can be used as the input of HON. Time complexity $\Theta(\Delta N)$ where $\Delta$ the width of time window and is a constant.

1: define $chains[EndPerson][EndTime]$ as nested dictionary
2: **for** $p = [FromPerson \to ToPerson]$ in $\mathcal{P}$ **do**
3:     **if** $FromPerson$ in $chains$ **then**
4:         **for** $TryStartTime$ in $[t_s(p^{(i)} - \Delta, t_s(p^{(i)}]$ **do**
5:             **if** $TryStartTime$ in $chains[FromPerson]$ **then**
6:                 $NewChain = chains[FromPerson][TryStartTime] \to ToPerson$
7:                 $chains$.remove($chains[FromPerson][TryStartTime]$)
8:                 $chains[ToPerson][EndTime]$.add($NewChain$)
9:                 continue
10:     $chains[ToPerson][EndTime]$.add($p$)

**File I/O:** The input/output file names can be set by changing the values of `InputFileName, OutputRulesFile, OutputNetworkFile`.

**MaxOrder:** By default, the parameter $MaxOrder = 99$ does not limit the order of dependency rules. If the algorithm keeps increasing the order and taking too long, one can impose a smaller $MaxOrder$.

**MinSupport:** By default, the parameter $MinSupport = 1$ does not impose a minimum support requirement. However, if the raw data contains significant noise and observations that happened only once are not of interest, one may impose a higher minimum support value.

**LastStepsHoldOutForTesting:** By default, the parameter LastStepsHoldOutForTesting = 0 uses all inputs for training the network. However, to test the representational power of the network, one may choose to hold out the last few steps in the input sequence for testing, and use random walkers to recover the testing steps.

**MinimumLengthForTraining:** By default, the parameter MinimumLengthForTraining = 1 uses all sequential data for training. Sometimes if the sequences are too short and are considered trivial (e.g., shipping sequences with less than five stops), they can be filtered out using this parameter.

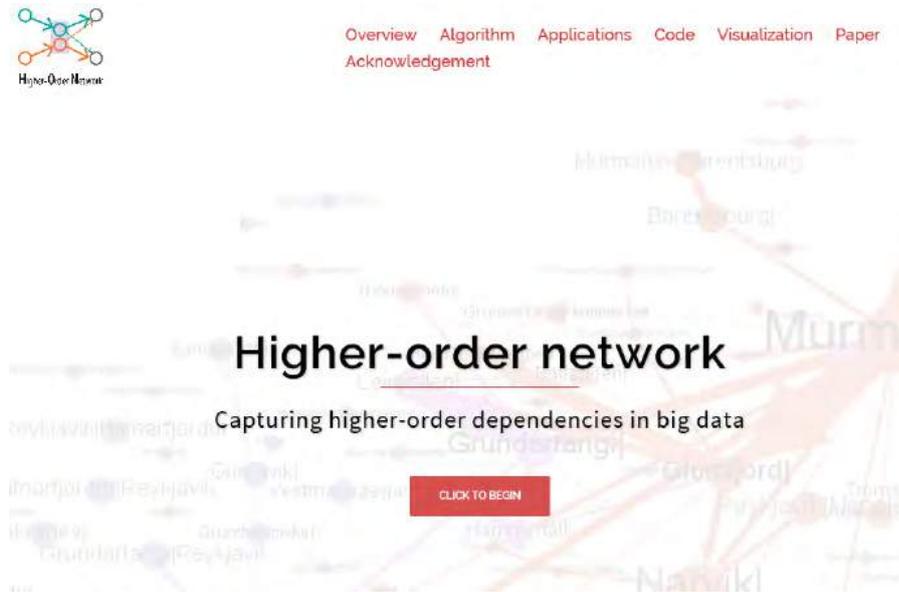

Figure 6.3. The higher-order network website which contains code, papers, video demos and more.

**InputFileDeliminator:** If the input file is deliminated by comma or other symbols, it can be set with this parameter.

**Verbose:** Outputting computation progress to the terminal. Default `True`.

**ThresholdMultiplier:** Positive float number. Directly controls the KL-Divergence threshold that determines to what extent two distributions are considered "significantly" different. Indirectly controls how many higher-order dependencies are produced. Default 1.

**Execution:** simply run `main.py` with Python3 or Python2.

PART II

INSIGHTS IN REAL-WORLD APPLICATIONS

# CHAPTER 7

# MODELING SPECIES INVASION AS NETWORKS

## 7.1 Overview

The unintentional transport of invasive species (i.e., nonnative and harmful species that adversely affect habitats and native species) through the Global Shipping Network (GSN) causes substantial losses to social and economic welfare (e.g., annual losses due to ship-borne invasions in the Laurentian Great Lakes is estimated to be as high as USD 800 million). Despite the huge negative impacts, management of such invasions remains challenging because of the complex processes that lead to species transport and establishment. Numerous difficulties associated with quantitative risk assessments (e.g., inadequate characterizations of invasion processes, lack of crucial data, large uncertainties associated with available data, etc.) have hampered the usefulness of such estimates in the task of supporting the authorities who are battling to manage invasions with limited resources.

**Intellectual merit:** We present here an approach for addressing the problem at hand via creative use of computational techniques and multiple data sources, thus illustrating how data mining can be used for solving crucial, yet very complex problems towards social good. By modeling implicit species exchanges as a network that we refer to as the Species Flow Network (SFN), large-scale species flow dynamics are studied via a graph clustering approach that decomposes the SFN into clusters of ports and inter-cluster connections. We then exploit this decomposition to discover crucial knowledge on how patterns in GSN affect aquatic invasions, and then illustrate how such knowledge can be used to devise effective and economical invasive species management strategies. By experimenting on actual GSN traffic data for years 1997-2006, we have discovered crucial knowledge that can significantly aid the management authorities.

**Connections:** This work can be extended from the network construction discussions in Chapter 8, the SF-HON proposed in Chapter 4, the analysis and projection methods in the Arctic region in Chapter 9, and visualized by HONVis in Chapter 5.

**Work status:** This work is accomplished in collaboration with the CoastalSEES research group funded by NSF. It has been published at KDD 2014[228].

## 7.2 Introduction

Networks of human trade and travel transport different plants, animals and pathogens across the globe.

> *"Human activity such as trade, travel and tourism have all increased substantially, increasing the speed and volume of species movement to unprecedented levels. Invasive species are often unintended hitchhikers on cargo and other trade conveyances, (Page 4)."*

*Invasive species* (i.e., non-native species that adversely affect habitats and bioregions) are among the top three drivers of global environmental change. Such invasive species include both plants and animals, and cause substantial economic and environmental harm by outcompeting or preying on native species. For instance, the impacts of *aquatic invasions* include increased diseases in humans (e.g., *cholera*) and aquaculture species (e.g., fish virus), losses of wildcaught fisheries (e.g, *comb jelly* invasion of the Black Sea), and losses of other ecosystem services. From an economic perspective, the estimated annual damage and control costs in the U.S. alone amount to more than USD 120 billion [169]. These species are introduced via the networks of human trade and travel, and analyzing these networks can illuminate potential man-agement strategies, regulatory policies, incentive structures, and risks from changing climate.

**Ship-borne invasive species problem.** The Global Shipping Network (GSN) is the dominant global vector for unintentional translocation of non-native aquatic species [151]: species get translocated via *ballast water* (during ballast water uptake/discharge) and *biofouling* (i.e., the accumulation of microorganisms, plants, algae, or animals) on the surfaces of ships [73]. To reduce invasion risks, authorities (e.g., International Maritime Organization (IMO)) have proposed standards for the maximum density of organisms that can be discharged in ships' ballast water. These standards are based on the premise that reducing the concentration of live organisms in ballast tanks will reduce the number of invasions, but the extent to which this approach will actually reduce the invasion risk is unknown. In addition, this approach does not address invasions via biofouling, nor does it consider many poorly known, but likely significant, biological and ecological factors that influence invasion risk. Moreover, the problem cannot be understood at a regional level, because ship-borne species can arrive from anywhere via the GSN. *With all these uncertainties in place, decisions are still needed to be made about the most efficient ways to target limited management resources.*

**Significance of the problem.** In the few coastal areas with good invasive species monitoring, increased shipping connecting an expanding network of global ports is correlated with an accelerating accumulation curve of established species (e.g., San Francisco Bay), and is estimated to be responsible for 69% of known aquatic invasions [151]. Although only a portion of species transported via GSN become invasive (i.e., spread, become abundant, and cause harm), environmental and economic damages from these species are often large and increase over time [96, 119]. For instance, the annual loss to the US Great Lakes regional economy due to ship-borne aquatic invasive species may be as high as USD 800 million [179]. However, GSN undoubtedly provides enormous benefits to the US economy, and is also responsible for approximately 90% of global trade. Furthermore, global trade patterns are optimized based on economic and trade considerations, but not necessarily to safeguard against aquatic invasions. Therefore, imposing expensive and cumbersome regulations on the shipping industry could cause serious adverse effects to a country's economy and trade relationships.

**Motivation and Goals.** It is clear that a thorough understanding of ship-borne invasion risks in terms of overall data about trade patterns, ports, vessel types, etc. is necessary to devise practices and policies that are feasible, effective and capable of bringing to bear the net long-term benefits to human welfare. With this motivation, our goal is to develop computational and data-driven frameworks that can inform invasive species management policies and practices.

### 7.2.1 Data Mining for Social Good

Ship-borne invasions are a result of a complex interplay of ship traffic, ballast uptake/discharge dynamics, species survival during transport, various environmental/biological variables, etc [227]. Incorporating these complexities into a quantitative risk assessment framework is extremely difficult, since the majority of the governing relationships are poorly quantified. The few studies that have attempted to quantify invasion risks via probabilistic approaches [120, 193] have relied on multiple simplifying assumptions. Moreover, usefulness of these approaches towards development of efficient invasion management policies is further hampered by the inability to incorporate crucial invasion mechanisms (e.g., "stepping-stone" process) into risk assessment. However, numerous streams of data that capture vessel movement pat-terns, ballast uptake/discharge and other environmental/biological factors (that affect species transport and establishment) are increasingly being collected by several agencies for research/commercial purposes. Therefore, one can creatively combine domain expertise and computational data analysis to understand the underlying patterns of ship-borne invasions in order to develop a sufficient understanding towards the development of effective and economical management strategies. *Our work is in fact a multi-disciplinary attempt towards utilizing this data to create insights and knowledge that can eventually lead to decision-making tools for policy makers.*

**Data.** We now introduce the numerous data sources utilized for the research.

(i) **Vessel movement data:** made available by Lloyd's Maritime Intelligence Unit (LMIU) contains travel information for vessels such as `portID`, `sail_date` and `arrival_date`, along with vessel metadata, such as `vessel_type` and `DWT` (Dead Weight Tonnage), etc. This information can be readily used to build a *network* to represent species flow paths and patterns among ports. Our experiments are based on LMIU data that spans four (4) two-years-long periods starting $1^{st}$ of May 1997, 1999, 2002 and 2005, totaling $6,889,748$ individual voyages corresponding to a total of $50,487$ vessels of various types that move among a total of $5,545$ ports and regions. However, none of the existing vessel movement datasets (including LMIU) provide explicit ballast water exchange amounts (or even whether a vessel discharged ballast water).

(ii) **Ballast discharge data:** made available by the National Ballast Information Clearinghouse (NBIC) contains the `date` and `discharge_volume` of all ships visiting U.S. ports from Jan. 2004 to present. As suggested in [193], NBIC data can be used to estimate an average ballast discharge based on `vessel_type` and `DWT` using a linear regression model.

(iii) **Ecoregion data:** are available via Marine Ecoregions of the World [205]

and the Freshwater Ecoregions of the World [1], where *ecoregions* are defined by species composition and shared evolutionary history [205], and are thereby capable of providing an index of native ranges. Therefore, these definitions can be used for more realistic and qualitative invasion risk analysis, in comparison to, for example, geographic distance as used in [193].

(iv) **Environmental data:** on port temperature and salinity (i.e., the two most crucial variables for identifying survivability of species in non-native coastal environments) are available via Global Ports Database (GPD) [120]. These estimates can be used for calculating species establishment risk based on environmental similarity; the missing values in GPD can also be supplemented via estimates from the World Ocean Atlas 2009 [11, 136] when necessary.

**Problem Statement.** Given the complexity of the problem and lack of relationships that are required for robust risk assessment, we set forth to extract knowledge on large-scale patterns of GSN in order to obtain better insight towards ship-borne invasions of non-indigenous species. Furthermore, we will illustrate how such knowledge can be used to derive efficient invasion management strategies.

**Framework.** Our method is devised to tackle the limitations due to lack of data and governing relationships that are required for quantitative risk assessment. Towards this, we take the following approach: **(i)** a network that represents the general species flow tendency among ports is built; then, utilizing a graph clustering method [177] that operates on the basis of flow-dynamics, **(ii)** a map of the species flow network, i.e., a cogent representation that extracts the main structure of flow while retaining information about relationships among modules (of main structure), is built; finally, using this map that summarizes the species flow dynamics in terms of *clusters* (or groups) of ports and highlights *inter-cluster* (i.e., between clusters) and *intra-cluster* (i.e., within cluster) relationships, **(iii)** the impact of GSN dynamics on aquatic invasions is studied in conjunction with ecological and environmental aspects that govern species establishment.

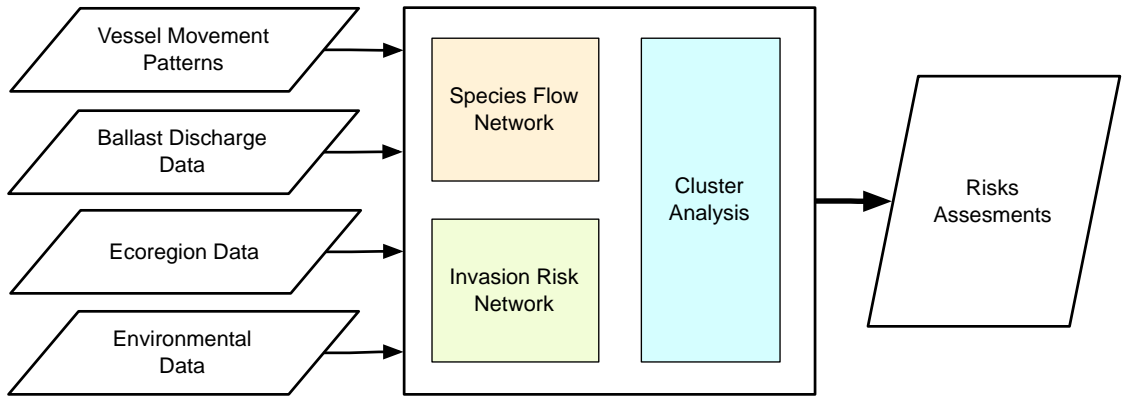

Figure 7.1. A **concept diagram** illustrating the integration of multiple data sources, modeling and data mining techniques for extracting useful knowledge.

### 7.2.2 Contributions and Broader Impact

We provide a data-driven foundation for more effective and efficient risk assessment and management by modeling the spread of aquatic non-indigenous species through the GSN, which is the most important vector of aquatic invasions. We have discovered vital information on patterns of GSN that can inform management strategies and regulatory policies. In a potential deployed configuration (see Figure 7.2), the discovered knowledge can efficiently be used to analyze the invasion risks with respect to changing climate, policy and infrastructure. Understanding the structure of the component networks and the dynamic interactions between the different networks is crucial to the design of policies that could cost-effectively reduce invasions.

This paper is organized as follows: Section 7.3 presents the formulation of species flow networks using LMIU and NBIC data, graph clustering approach for understand-ing the large-scale dynamics of GSN, and the main discoveries; Section 7.4 presents invasion risk assessment that incorporates ecoregion definitions and environmental conditions via a unique graphical approach; Section 7.5 presents how the emerging knowledge and methods can be potentially deployed towards development of species management strategies; and finally, Section 7.6 contains the concluding remarks.

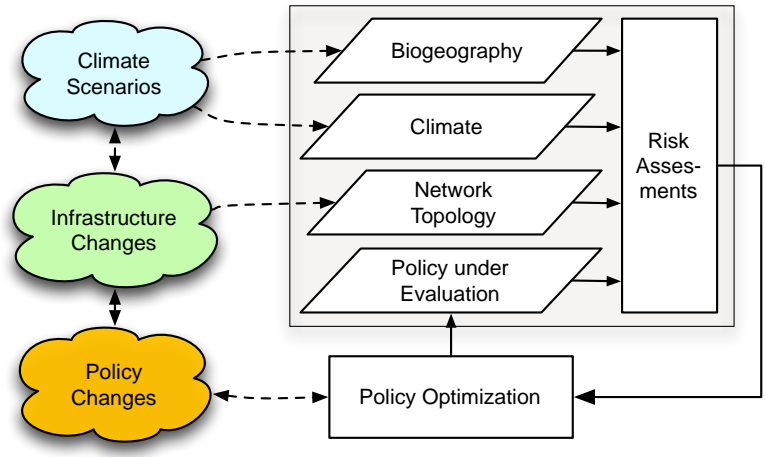

Figure 7.2. **Use of discovered knowledge in a potential deployed setting** for invasion risk assessment with respect to changing climate, policy and infrastructure.

7.3 Species Flow Analysis

The basic idea behind our work is to find patterns of species flow in order to identify ports and shipping routes for which interventions would be the most effective in stopping invasions through the entire network. Such knowledge can then be further leveraged with auxiliary information (e.g., vessel types) in order to inform management strategies in a targeted manner. Since GSN naturally forms a graph, LMIU and NBIC data can be utilized to build a network to represent species flow among ports (see Figure 7.3). This network can then be analyzed via graph mining

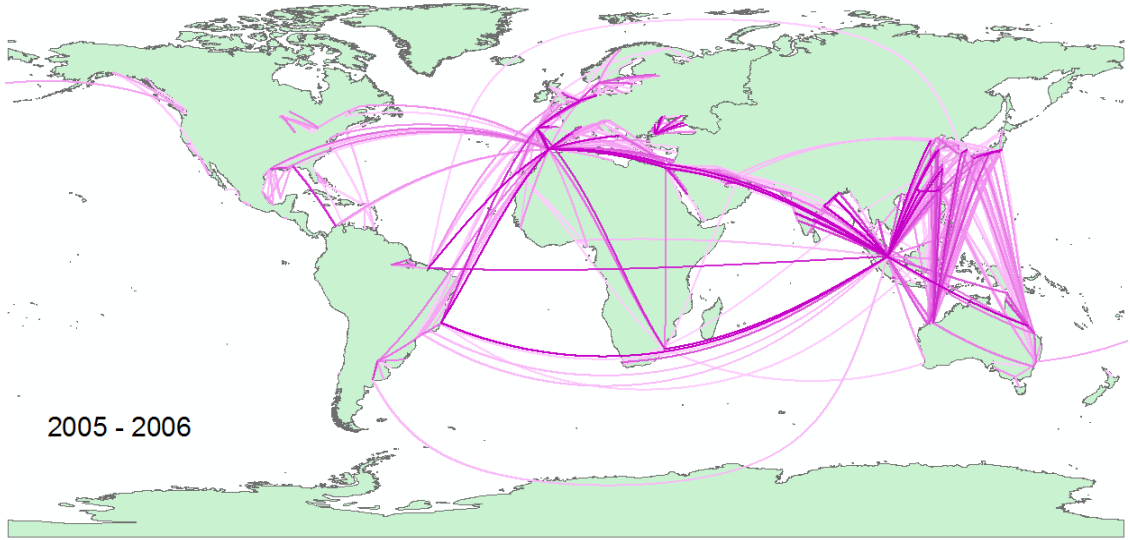

Figure 7.3. **Species flow between ports corresponding to vessel movements given in the LMIU 2005–2006 dataset.** The edges represent the aggregated species flow between ports, where the color intensity is proportional to the magnitude of flow. Approximately 2300 paths with the highest species flow are shown.

or network science techniques to extract relevant insights.

### 7.3.1 Species Flow Network (SFN)

Let $\mathcal{G} \equiv (\mathcal{N}, \mathcal{E})$ be a directed weighted graph, where $\mathcal{N} \equiv \{n_1, \ldots, n_n\}$ and $\mathcal{E} \subset \mathcal{N} \times \mathcal{N}$ denote the set of nodes and edges of $\mathcal{G}$, respectively. Let the nodes in $\mathcal{N}$ correspond to ports visited by vessels in the GSN and the *weight* of the directed edge $e_{ij} \in \mathcal{E}$ given by $w_{ij} \in (0,1]$ represents the *total probability of species flow* corresponding to all vessels traveling from port $n_i$ to $n_j$ (without intermediate stopovers), for all $n_i, n_j \in \mathcal{N}$.

**Species Flow Estimation.** Estimation of exact amounts of species exchanged between ports is extremely difficult. However, as proposed in [193], vessel movement and ballast discharge data can be leveraged to estimate the likelihood of species exchange. We now briefly explain how LMIU and NBIC datasets are used to estimate species flow (i.e., the edge weights of $\mathcal{G}$) and refer the interested reader to [193] for a comprehensive discussion on probabilistic species flow modeling including details on model assumptions, development and validation.

**(a) Calculation of edge weights:** Consider a vessel $v$ traveling from port $n_i$ to $n_j$ in $\Delta t_{ij}^{(v)}$ time (without intermediate stopovers), during which the species in ballast water may die at a *mortality rate* of $\mu$ (is set to a constant average of $0.02/day$ for all routes $r$ and vessel types). In addition, let $D_{ij}^{(v)}$, $\rho_{ij}^{(v)} \in [0,1]$ and $\lambda$ denote the amount of ballast water discharged by vessel $v$ at $n_j$, the efficacy of ballast water management for $v$ for the route $n_i \rightarrow n_j$, and the characteristic constant of discharge, respectively. Then, the probability of vessel $v$ introducing species from $n_i$ to $n_j$ (without intermediate stopovers) is given by:

$$p_{ij}^{(v)} = \rho_{ij}^{(v)} \left(1 - e^{-\lambda D_{ij}^{(v)}}\right) e^{-\mu \Delta t_{ij}^{(v)}}; \quad (7.1)$$

the weight of the edge $e_{ij}$ is taken to be the total probability of species introduction for all vessels traveling from $n_i$ to $n_j$, and is given by:

$$w_{ij} = 1 - \prod_{\substack{r \in DB \\ r = v : n_i \rightarrow n_j}} (1 - p_{ij}^{(v)}), \quad (7.2)$$

where the product is taken over all routes $r$ in LMIU database $DB$ s.t. a vessel $v$ travels from port $n_i$ to $n_j$.

**(b) Estimation of ballast discharge:** Information available on ballast discharge are incomplete, where estimation of exact quantities exchanged for each and every ship route $r$ is impossible for most ports of the world: **(i)** ballast discharges in ports are not recorded globally, and are known to differ significantly by port and ship type; **(ii)** vessels may have intermediate stopovers, thus exchanging and mixing ballast water with existing water in ballast tanks; and **(iii)** data are largely unavailable for offshore discharges. Therefore, in order to mitigate the above difficulties, ballast discharge is estimated based on linear regression models on `DWT` per `vessel_type` as in [193]. Specifically, linear regression models on `DWT` for vessels of type `Bulk Dry`, `General Cargo`, `Ro-Ro Cargo`, `Chemical`, `Liquified Gas Tankers`, `Oil Tankers`, `Passenger Vessels`, `Refrigerated Cargo`, `Container Ships` and `Unknown/Other`). Furthermore, the relationship of ballast discharge amount to the likelihood of species introduction is not well defined. For estimation of (7.1), $\lambda$ is chosen s.t. $p_{ij}^{(v)} = 0.80$ for a ballast discharge of $500,000\,m^3$, when $\rho_{ij}^{(v)} = 1$ and $\Delta t_{ij}^{(v)} = 0$, i.e., a discharge volume of $500,000\,m^3$ has a probability of 0.8 of introducing species if the vessel travels with zero mortalities and has no ballast management strategies in place.

**Characteristics of the SFN.** Summary of characteristics for SFNs generated for the four available years of data are shown in Table 7.1. The *path length* of a network identifies the number of stops required to reach a given port from another. An *average path length* of three (3) is observed in all four SFNs. This is perhaps mainly due to the presence of *hubs* (i.e., ports that are connected to many other ports) in GSN (e.g., Singapore). The *in-/out-degree* of a node is defined as the number of other nodes connected to/by it. Therefore, *average degree* in SFN describes the average number of direct pathways of species introduction. Furthermore, as empirical evaluations for *power law degree distribution* [55] suggest that SFNs fall under the category of *scale-free networks* [21], for degree $\geq$139.

TABLE 7.1

CHARACTERISTICS OF SPECIES FLOW NETWORKS

| Feature | 97–98 | 99–00 | 02–03 | 05–06 |
| --- | ---: | ---: | ---: | ---: |
| Number of nodes | 3971 | 4045 | 4264 | 4250 |
| Number of edges | 150479 | 150150 | 143560 | 145199 |
| Average path length | 2.987 | 2.998 | 3.018 | 3.041 |
| Average in/out- degree | 37.9 | 37.1 | 33.7 | 34.2 |
| Diameter | 8 | 7 | 7 | 9 |
| Density | 0.010 | 0.009 | 0.008 | 0.008 |

TABLE 7.2

PORTS THAT REMAINED IN THE SAME CLUSTER FOR THE

DURATION OF 1997–2006.

| Pacific  %TP=28.33%, #P=818 | | | Mediterranean  %TP=15.61%, #P=513 | | | W. European  %TP=15.37%, #P=1117 | | | E. North_America  %TP=9.31%, #P=363 | | | Indian_Ocean  %TP=6.12%, #P=137 | | | South_America  %TP=3.41%, #P=80 | | |
|---|---|---|---|---|---|---|---|---|---|---|---|---|---|---|---|---|---|
| Port name | %TF | %CF | Port name | %TF | %CF | Port name | %TF | %CF | Port name | %TF | %CF | Port name | %TF | %CF | Port name | %TF | %CF |
| Singapore | 2.82 | 9.96 | Gibraltar | 2.56 | 16.37 | Rotterdam | 0.87 | 5.68 | Houston | 0.52 | 5.57 | Jebel Ali | 0.25 | 4.07 | Santos | 0.42 | 12.37 |
| Hong Kong | 0.68 | 2.41 | Tarifa | 0.86 | 5.54 | Skaw | 0.60 | 3.93 | New Orleans | 0.37 | 3.94 | Ras Tanura | 0.22 | 3.67 | Tubarao | 0.33 | 9.70 |
| Kaohsiung | 0.58 | 2.05 | Port Said | 0.84 | 5.38 | Antwerp | 0.55 | 3.59 | New York | 0.35 | 3.80 | Mumbai | 0.20 | 3.29 | San Lorenzo* | 0.33 | 9.57 |
| Port Hedland | 0.52 | 1.83 | Suez | 0.48 | 3.09 | Brunsbuttel | 0.44 | 2.85 | Baltimore | 0.23 | 2.42 | Juaymah Term. | 0.19 | 3.12 | Paranagua | 0.21 | 6.11 |
| Busan | 0.50 | 1.76 | Barcelona | 0.29 | 1.83 | Hamburg | 0.42 | 2.76 | Port Arthur | 0.21 | 2.28 | Kharg Is. | 0.18 | 2.91 | Rio de Janeiro | 0.15 | 4.45 |
| Hay point | 0.49 | 1.72 | Venice | 0.24 | 1.52 | Amsterdam | 0.31 | 2.02 | Santa Marta | 0.20 | 2.17 | Jubail | 0.17 | 2.76 | Bahia Blanca | 0.15 | 4.31 |
| Newcastle** | 0.48 | 1.71 | Genoa | 0.23 | 1.47 | Immingham | 0.28 | 1.83 | Tampa | 0.20 | 2.16 | New Mangalore | 0.15 | 2.50 | Rosario | 0.14 | 4.07 |
| Gladstone | 0.47 | 1.67 | Piraeus | 0.22 | 1.39 | St. Petersburg | 0.27 | 1.73 | Port Everglades | 0.20 | 2.13 | Mesaieed | 0.13 | 2.08 | Sepetiba | 0.12 | 3.60 |
| Nagoya | 0.46 | 1.61 | Leghorn | 0.21 | 1.32 | Tees | 0.22 | 1.41 | Mobile | 0.19 | 2.04 | Bandar Abbas | 0.12 | 2.03 | Rio Grande*** | 0.12 | 3.59 |
| Incheon | 0.45 | 1.60 | Augusta | 0.20 | 1.26 | Zeebrugge | 0.21 | 1.36 | Savannah | 0.18 | 1.95 | Jebel Dhanna Term. | 0.12 | 1.95 | Praia Mole | 0.12 | 3.50 |

Ports corresponding to highest %TF:=percentage flow w.r.t. total flow and %CF:=percentage flow w.r.t. flow within cluster are shown for six major clusters; for each cluster, the aggregated %TF:=percentage flow in the cluster w.r.t. total flow and number of ports in the cluster are given in the first row of table. Here, San Lorenzo*:=San Lorenzo, Argentina; Newcastle**:=Newcastle, Australia; Rio Grande***:=Rio Grande, Brazil.

### 7.3.2 Clustering Analysis of SFN

Complex networks are efficient abstractions for highly complex systems that consist of numerous, often complex underlying patterns and relationships. However, these abstractions still remain too complex to derive useful inferences. Therefore, a decomposition that represents such complex networks via *modules* and their interactions [90, 162, 183] can be very useful in understanding the underlying patterns. We utilize a graph clustering approach in order to simplify the underlying flow dynamics of SFN. The clusters can capture the ship movement activity among ports leading to a better identification of risk corridors, which can then be used to estimate invasion risk based on ecological and environmental conditions.

**Choice of Clustering Method.** For the task at hand, we are interested in understanding how the structure of SFN relates to species flow across the network. Therefore, among many alternatives, *MapEquation* [177]—a graph clustering method that attempts to decompose the network with respect to *flow-dynamics* (in comparison to optimization of *modularity*)—is used. The basic principle of operation behind MapEquation-based clustering stems from the notions of information theory, which states the fact that a data stream can be compressed by a *code* that exploits regularities in the process that generates the stream [194]. Therefore, a group of nodes among which information flows quickly and easily can be aggregated and described as a single well connected module; the links between modules capture the avenues of information flow between those modules. MapEquation identifies clusters by optimizing the entropy corresponding to intra- and inter-cluster in a *recursive* manner—the clusters identified cannot be further refined or partitioned.

**Clusters of Ports based on Species Flow.** Clustering analysis of SFN reveals several clusters of ports. These clusters represent groups of ports among which the species exchange is relatively higher; if inter-cluster pathways are controlled, species flow would be combinatorially reduced. Once such clusters are identified, species flow characteristics within clusters can be analyzed in conjunction with ecological and environmental data for invasion risk assessment. While clustering is derived based on species flow dynamics, geographical orientation of the major (in terms of aggregated flow) clusters is also intuitive (see Figure 7.4). A few major clusters correspond to a significant proportion of total species flow among ports (see Table 7.2 for major ports that are in 6 of these major clusters). For instance, in 2005–2006, six (6) major clusters (out of 64 in total), viz., the clusters of `Pacific`, `Mediterranean`, `Western_Europe`, `Eastern_North_America`, `Indian_Ocean` and `South_America` contained 68.6% of total ports and corresponded to 76.3% of the total species flow.

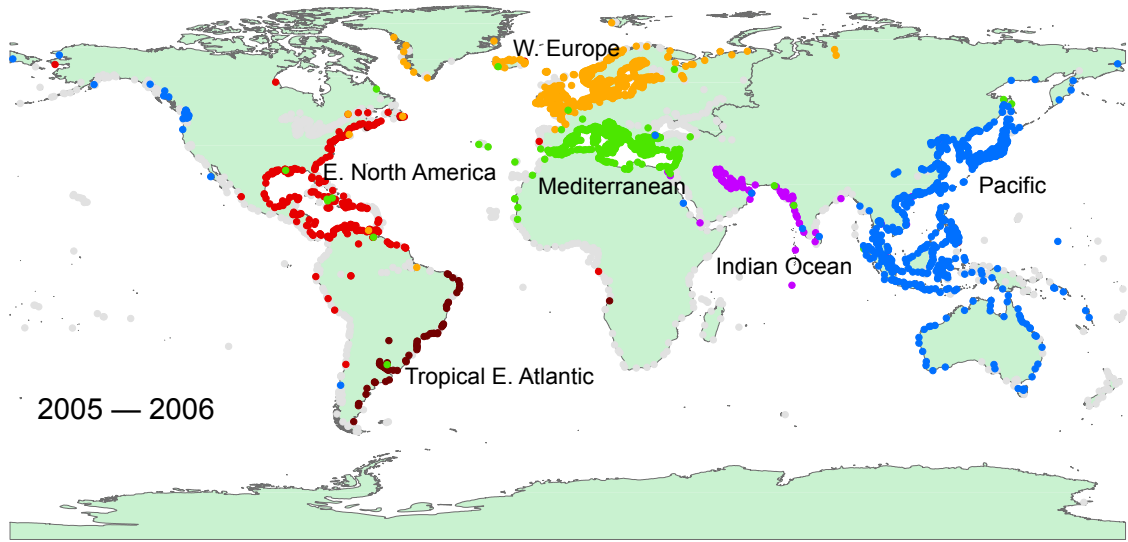

Figure 7.4. **The Six Major clusters of SFN during 2005–2006.** Color of dots correspond to that in Figure 7.5, and white dots are not included in any of the six major clusters. Major clusters remain largely unchanged for the duration of 1997–2006, and contain a significant proportion of total species flow between ports.

**Cluster Consistency.** From a deployment perspective, perhaps the most crucial contribution of our analysis is that these major clusters continue to exist over the duration studied; for a given cluster, while some ports leave/join over time, the vast majority of the ports continue to remain in the same cluster (see Figure 7.5). This provides a solid foundation for devising management strategies targeting clusters and inter-cluster connections to efficiently control species flow. Furthermore, evolution of clusters (or how the clustering patterns change over time) can reveal important information on how changes in vessel movement (and ballast discharge) patterns affect species flow dynamics. For instance, the exchange of the order of the two clusters `Mediterranean` and `Western_Europe` from 2002–2003 to 2005–2006 indicates a relative increase of species exchange among ports that belong to these clusters during 2005–2006, which can be attributed to the merger of a significant proportion of ports belonging to `Mediterranean` cluster with `South European Atlantic Shelf` cluster to form the `Tropical_East_Atlantic` cluster in 2005–2006. Another example is the formation of a new smaller cluster (the eighth in Figure 7.5) in 1999–2000 by 21 ports in California and Hawaii, including ports such as San Francisco, Los Angeles and San Diego that previously belonged to the `Pacific` cluster in 1997–1998. Such changes can reveal large-scale trends that may be very useful in devising long term management strategies.

7.4  Invasion Risk Analysis

Quantification of invasion risk is a challenging problem because of the complex interactions between species and their abiotic and biotic environment [227]. Here, we shift our attention from inter-cluster species flow to *intra-cluster* (i.e., ports within a cluster) NIS invasion risk in order to gain insight into the plausibility of invasions in terms of environmental similarity. Previous studies have assumed that the invasion risk is proportional to *Euclidean* distance between annual averages of temperature and salinity [85, 120, 193]. However, this assumption may not be valid for invasive species that often exhibit broad environmental tolerances [71, 92]. We take an approach that is based on biogeographic patterns, and empirically observed temperature and salinity tolerances for ranking invasion risks.

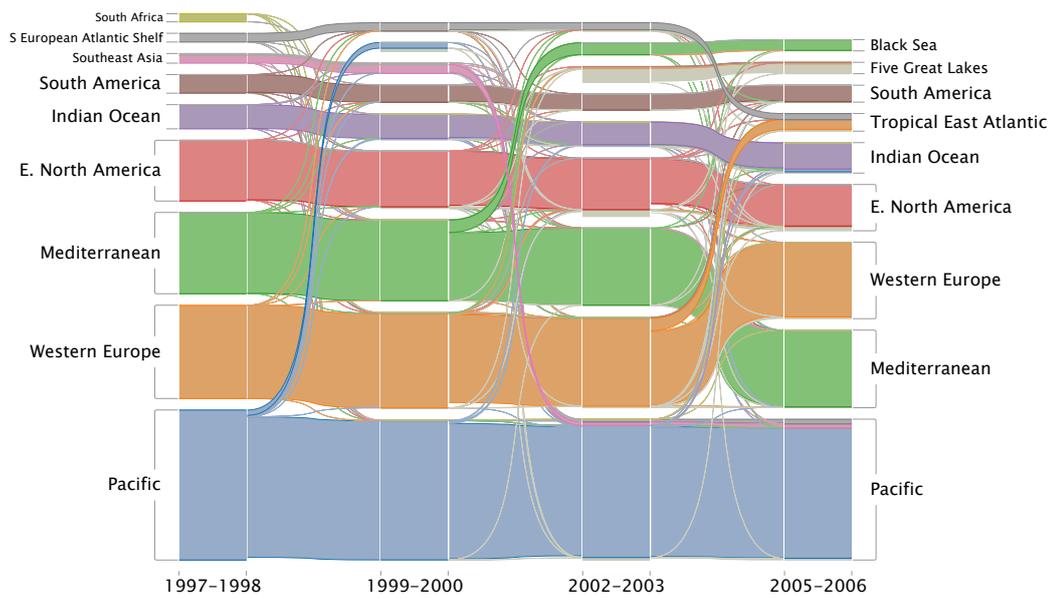

Figure 7.5. **Illustration of evolution of major clusters during the period of 1997–2006.** The clusters in *alluvial diagram* [177] are ranked by aggregated flow within the cluster. Here, the columns 1997, 1999, 2002 and 2005 represent the major clusters of SFN generated from LMIU datasets for 1997–1998, 1999–2000, 2002–2003 and 2005–2006, respectively.

### 7.4.1 Invasion Risk Modeling

For an exchange of species to become an invasion, the introduced species must be: **(i)** non-indigenous, viz., movements between non-contiguous ecoregions; and **(ii)** able to survive and establish in its new environment. Invasion risk between port environments can then be ranked by considering a species *assemblage* (or a collection) that contains "generalist" and "specialist" species. We have taken this approach to counteract the lack of relationships or data to calculate or estimate exact invasion risks. Based on temperature and salinity tolerance levels (empirically-estimated long term thermal tolerances of marine invertebrate taxa [175]), we define invasion risk in terms of number of species groups that can tolerate the given conditions. Specifically, six (6) different species tolerance groups based on two (2) temperature and three (3) salinity tolerance levels are considered (see Table 7.3). Here, salinity tolerance levels were set to capture species types that are completely intolerant to salinity (i.e., freshwater species), those that are restricted to marine waters (i.e., low tolerance), and species that can survive in a wide range of salinities (i.e., high tolerance). Risk between any two ports is then quantified as an index created by overlapping the species tolerance groups as shown in Figure 7.6.

### 7.4.2 Environmental Similarity Network

Quantification of port-wise invasion risk is both difficult and not very useful in terms of species control and management. In order to gain insight into what ports

TABLE 7.3

GROUPING BASED ON ENVIRONMENTAL TOLERANCE

| Species Tolerance Group | Tolerance Levels | |
|---|---|---|
| | $\Delta T\,(^\circ C)$ | $\Delta S\,(ppt)$ |
| Tolerance Group 1 | [0, 2.9] | [0, 0.2] |
| Tolerance Group 2 | [0, 2.9] | [0, 2.0] |
| Tolerance Group 3 | [0, 2.9] | [0, 12] |
| Tolerance Group 4 | [0, 9.7] | [0, 0.2] |
| Tolerance Group 5 | [0, 9.7] | [0, 2.0] |
| Tolerance Group 6 | [0, 9.7] | [0, 12] |

are at risk based on species flow and environmental similarity, we utilize a graphical representation that is referred to as the *Invasion Risk Network (IRN)*. An IRN is built for every *major* cluster in SFN to intuitively represent the invasion risk based on how easily species can establish in the new environment (based on environmental tolerance, see Table 7.3). Note that IRN is an undirected weighted graph, since environmental match is symmetric, and the risk level (based on number of tolerance groups at risk) can vary between port pairs, respectively.

7.4.3  Clustering Analysis of IRN

With the edges representing the invasion risk between ports, clustering can help detect groups of ports that have similar environmental conditions (while belonging to different ecoregions). The basic idea here is to exploit the fact that the invasion risk between groups of ports that are very dissimilar (e.g., fresh-water ports and marine ports) is lower than ports within the same group (with relatively similar conditions). Clustering analysis (Section 7.3.2) can therefore again be utilized on IRN to identify groups of ports that are similar in terms of invasion risk. The clusters detected here are sub-clusters of SFN clusters (that are based on species flow dynamics); therefore, if two ports are in the same cluster of IRN, then it is very likely for an invasion to occur between these two ports. Furthermore, if adequate species flow control is not in place, given the frequent species exchanges and high chances of species establishment within ports in an IRN cluster, an invasion to one single port will immediately put all the other ports in the IRN sub-cluster at risk of an invasion. Therefore, IRN clusters identify groups of ports that would most benefit from some form of species flow control to avoid invasions.

**Note:** Clustering based on flow-dynamics (simulating random walks) is used here for identifying ports with similar environmental conditions, since an approach only considering pair-wise distance is not capable of capturing the stepping-stone effect.

## 7.5 Emergent Species Flow Control Strategies

Clustering analysis on SFN has discovered a consistent pattern of port groupings in terms of potential species exchanges. One can easily identify such regions at risk, by overlaying the ecoregions to the IRN clusters above (see Figure 7.9(a)). This allows one to consider the four factors of vessel movement, ballast discharge, environmental conditions, and ecoregion in a unique but an intuitive fashion. With this knowledge in place, species exchange among ports can be efficiently controlled by management strategies that target high species exchange pathways in order to isolate ports and clusters of ports.

### 7.5.1 Managing Inter-cluster Exchanges

Consider Figure 7.8 which illustrates the inter- and intra-cluster species flow among major clusters. While we observe some changes in inter-cluster connections,

TABLE 7.4

HIGHEST INTER-CLUSTER FLOW FOR PACIFIC CLUSTER IN

2005–2006

| From Port | To Port |
|---|---|
| Singapore | Port Said |
| Singapore | Richards Bay |
| Mormugao | Singapore |
| Suez | Singapore |
| Paradip | Singapore |
| Visakhapatnam | Singapore |
| Tubarao | Singapore |
| Chennai | Singapore |
| Ponta da Madeira | Singapore |

major clusters and species exchange pathways are virtually consistent over time. Therefore, by limiting species flow on inter-cluster connections, species exchanges among ports could be restricted to ports within clusters. This will combinatorially reduce the species introduction pathways. For instance, consider the `Pacific` cluster in year 2005–2006. There are 37,596 inter-cluster connections, where Table 7.5.1 tabulates the strongest connections.

Singapore alone contributes to approximately 26% of total inter-cluster flow from/to `Pacific` cluster that contains 818 ports (see Figure 7.9 for an illustration of invasion risk with respect to Singapore). Here, via targeted ballast management on inter-cluster connections to/from Singapore and a few other "influential" ports, inter-cluster flow from/to `Pacific` cluster can be significantly reduced (see Figure 7.9(b)).

TABLE 7.5

MAJOR INTER-CLUSTER CONTRIBUTORS IN 2005–2006

| Cluster | Port |
|---|---|
| Pacific | Singapore |
| Mediterranean | Gibraltar |
| W. Euro | Rotterdam |
| E. N. America | New York |
| Indian Ocean | Mormugao |
| S. America | Tubarao |
| Great Lakes | Seven Islands |
| Black Sea | Istanbul |
| W. N. America | Long Beach |

Table 7.5 lists ports corresponding to the highest inter-cluster flow in major clusters for 2005–2006. Any practices that reduced species movements through these ports would potentially reduce a large proportion of inter-cluster species flow. Increases in species surveillance in these ports would strengthen the foundation for geographic allocation of risk management efforts. Finally, increased surveillance of ports would provide a baseline against which to measure the effectiveness of future risk reduction efforts–a baseline that is now largely absent globally (Costello et al. 2007)

### 7.5.2 Targeting Hubs for Species Flow Control

Average path length of three (3) that is observed on SFN indicates that species could be translocated between any two given ports within two (2) stopovers on average. This indicates a generally high risk of invasions in the absence of risk reduction practices. In order to understand the impact of targeted ballast management on average path length, a test scenario based on a hypothetical SFN—$\widehat{\text{SFN}}$ is derived as follows: **(i)** choose an SFN (SFN corresponding to 2005–2006 LMIU dataset was chosen for our experiment); then, **(ii)** identify 20% of all ports with the highest degree (see Table 7.6); and, finally **(iii)** generate $\widehat{\text{SFN}}$ by removing all edges to/from the above ports; this corresponds to ballast management with 100% efficiency, i.e., zero (0) species flow from/to these ports. *Then, the average path length increases to 6.4 indicating that it will be at least twice as difficult for species to be translocated from one port to another.* Furthermore, higher average path length also implies, **(i)** longer travel times (hence, very lower chance of survival for species during the voyage) and **(ii)** increased number of intermediate stop-overs (which is likely to dilute ballast water and expose organisms to multiple shocks).

### 7.5.3 Vessel Type-Based Management Strategies

The exact amount of species relocated by a vessel depends on many factors: ballast size, average duration per trip, frequently visited ports, etc. Furthermore, vessel types we observe in GSN are often chosen for specific tasks (e.g., oil transportation, vehicle transportation, etc.) and these vessels often have their respective frequent ports/routes. Therefore, we investigate the relationship of vessel types to inter- and intra-cluster species flow in order to understand existing patterns that can be helpful in devising species management strategies (based on the 2005-2006 LMIU dataset).

TABLE 7.6

PORTS WITH DEGREE ¿ 1000 IN 2005–2006 THAT ACT AS "HUBS" IN SFN

| Port name | Degree | Important pathways (connected ports) |
|---|---|---|
| Gibraltar | 1882 | Cape Finisterre, Tubarao |
| Dover Strait* | 1747 | Cape Finisterre, Rotterdam, Tubarao |
| Singapore | 1569 | Mormugao, Tubarao |
| Cape Finisterre | 1387 | Gibraltar, Rotterdam, Tubarao |
| Panama Canal* | 1275 | New Orleans |
| Tarifa | 1224 | Gibraltar, Cape Finisterre |
| Rotterdam | 1126 | Cape Finisterre, Dover Strait |

* indicates locations in LMIU database, but do not correspond to actual ports; connected ports are listed in decreasing order of degree.

(i) **Frequent inter-cluster travelers:** While not being the most active vessel in the GSN, `container carriers` correspond to 57,909, or equivalently 24% of all inter-cluster trips in 2005-2006. Among the most frequently seen vessel types, `bulkers`, `crude oil tankers`, `refrigerated general cargo ships` and `combined bulk and oil carriers` tend to travel inter-cluster for over 25% of the time. Furthermore, among the vessel types that do not travel frequently, some vessel types tend to travel inter-cluster in a majority of their trips (e.g., `wood-chip carriers`: 40.4%, `livestock carriers`: 34.3%, `semi-sub HL vessels`: 37.4% and `barge container carriers`: 55.7%).

(ii) **Frequent intra-cluster travelers:** Among the most frequently travelled vessel types, `passenger carriers` tend to stay within clusters for 97.6% of their trips, thus imposing only a very minimal risk in terms of inter-cluster species translo-cation. Similarly, `barge ships` also stay within the cluster for 98.1% of total trips.

### 7.5.4 Impact of Environmental Conditions

With proper species control on inter-cluster connections, species flow can be confined to clusters. Even though ports within a cluster have higher species exchanges among them, if the environmental conditions are significantly different, invasions are less likely to occur. On the other hand, for ports in the same IRN cluster (hence, environmental conditions are very similar), if proper species flow control is not in place, invasions will be nearly unavoidable. For instance, in the `Pacific` cluster, we observe that *Hong Kong, Qingdao* and *Kaohsiung* have higher species exchanges among them; and, following clustering analysis of IRN, we also notice that Hong Kong and Kaohsiung are in the same IRN cluster. Therefore, invasions are very likely to occur in between these two ports. On the other hand, Qingdao is in a different sub-cluster to Hong Kong, indicating these two ports have significantly different environmental conditions—invasions are less likely to happen between these two ports, even with high species exchanges.

### 7.6 Concluding Remarks

Aquatic invasions via the GSN are a result of a complex interplay of ship traffic, ballast uptake/discharge dynamics, species survival during transport and numerous environmental/biological variables. The inherent complexity of the invasive species problem has made risk assessment very difficult, and thereby has severely hampered the effectiveness of species management efforts. To that end, we have developed an approach for more effective and efficient risk assessment and management by modeling the spread of aquatic non-indigenous species through the GSN, which is the most important vector of aquatic invasions. Knowledge about the patterns of GSN, within the context of species flow and invasion risk, appropriate risk assessments can be generated to help inform management strategies and regulatory policies.

In a management context, the discovered knowledge could efficiently be used to analyze the invasion risks with respect to changing climate, policy and infrastructure. Understanding the structure of the component networks and the dynamic interactions between the different networks is crucial to the design of policies that could cost-effectively reduce invasions. The analyses outlined and performed in this paper could also be used to geographically prioritize species surveillance efforts using traditional organism sampling methods (e.g., water samples, nets) and/or newer genetic approaches [38]. Furthermore, our work illustrates the value of creative use of data mining for social good via the application to a significant societal problem.

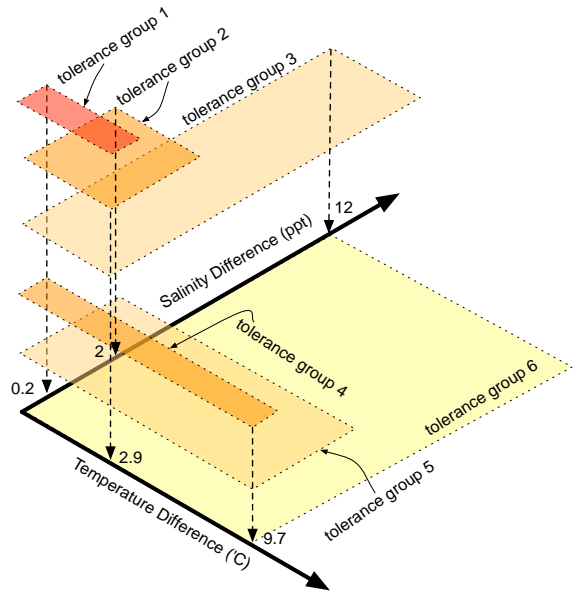

(a) Species tolerance groups

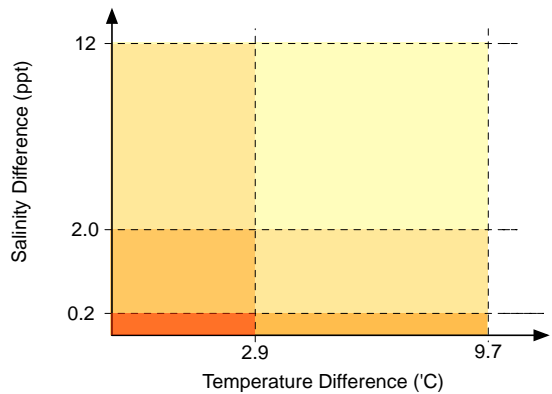

(b) Risk level definition

Figure 7.6. **Illustration of *risk level* definition based on species tolerance groups and between-port environmental differences.** Sub-figure (a): identifies six (6) different species groups that categorizes the risk of survival relative to given difference in temperature and salinity based on two (2) temperate tolerance levels (high = can survive up to $9.7°C$ and low = can survive up to $2.9°C$ temperature difference) and three (3) salinity tolerance levels (zero = $0.2ppt$, low = $2.0ppt$ and high = $12.0ppt$ tolerance). Sub-figure (b): definition of risk level, defined based on number of species groups as identified in (a); the colors are generated by overlapping the layers and later enhanced for clarity and ease of distinction. In this setting, risk level ranges from 0 to 6.

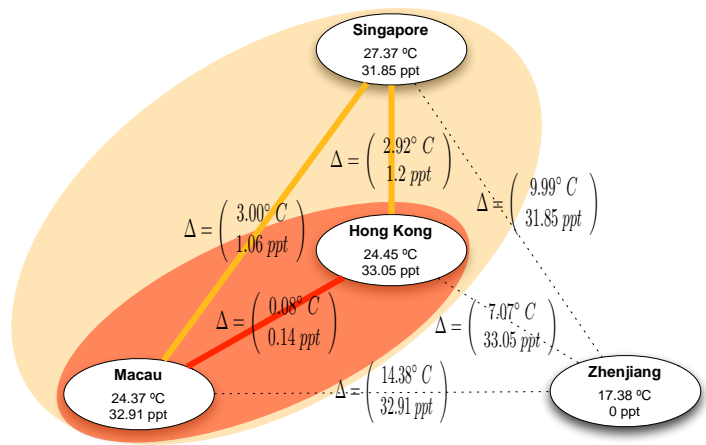

Figure 7.7. **Illustrating the generation of Invasion Risk Network (IRN).** The IRN is an undirected graph where nodes and edges are given by the ports visited in the GSN and invasion risk level, respectively. Shown here as examples are four ports along with annual average temperature and salinity, and pair-wise salinity and temperature differences. Edges drawn in solid lines represent the risk level between ports as defined in Figure 7.6; dotted-lines show zero (0) risk edges; colored-patches are used to show the overlap of species tolerance groups shared by a port-pair.

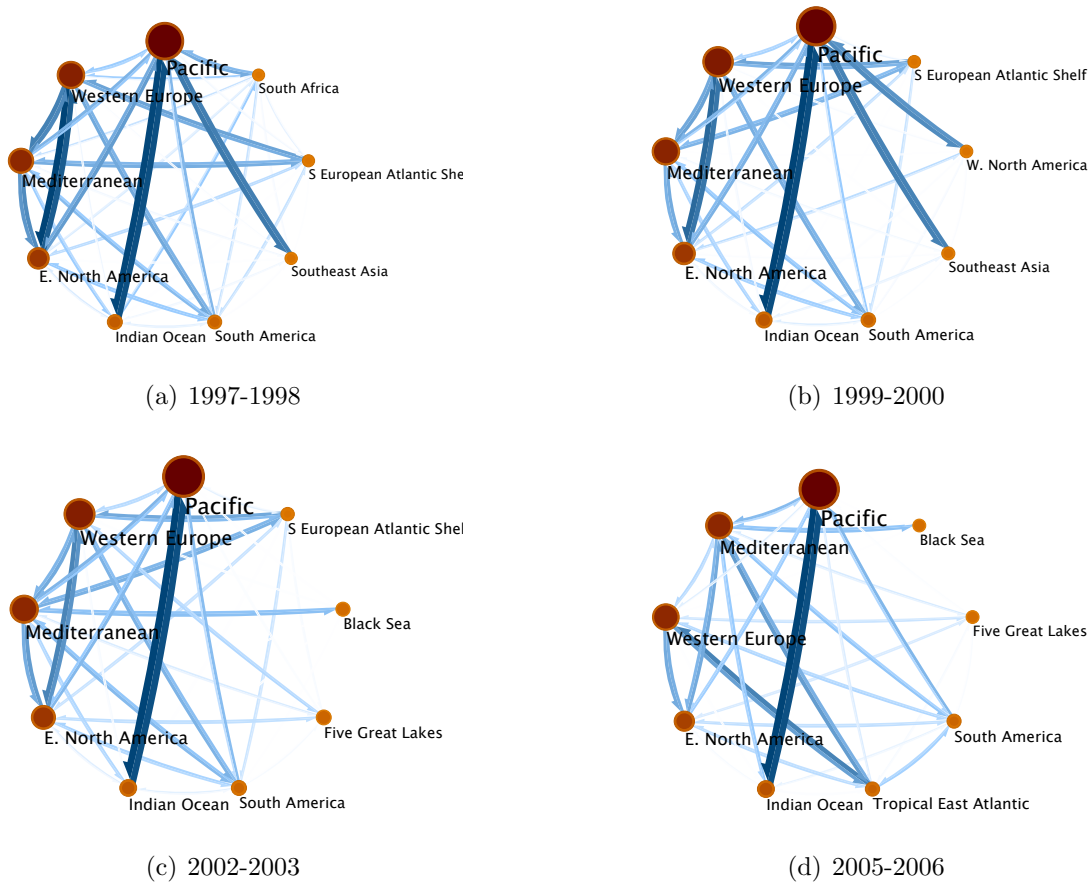

Figure 7.8. **Illustration of inter-cluster and intra-cluster flow.** Here, ratio of darker/lighter region explains the ratio of intra-cluster flow (i.e., flow between ports within a cluster) to inter-cluster flow (i.e., flow between ports belonging to different clusters). Therefore, in major clusters, species exchange among ports within clusters appears to be much higher compared to that of between clusters.

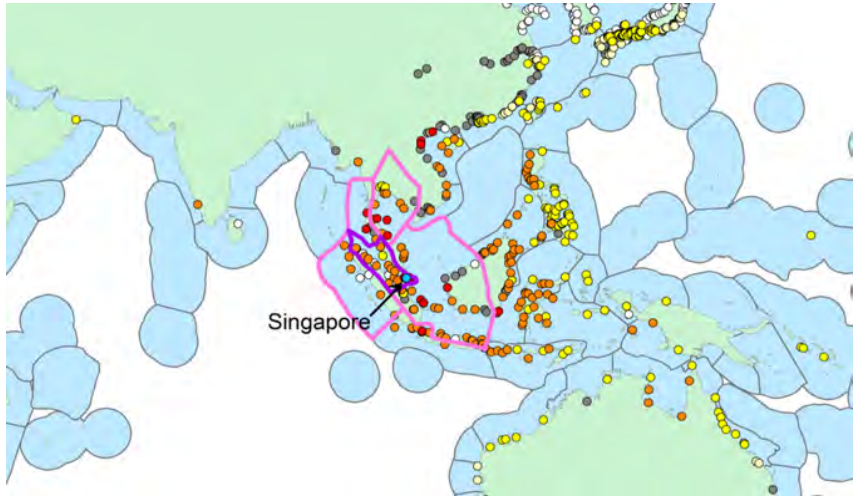

(a) Inner-Cluster Invasion Risk w.r.t. Singapore

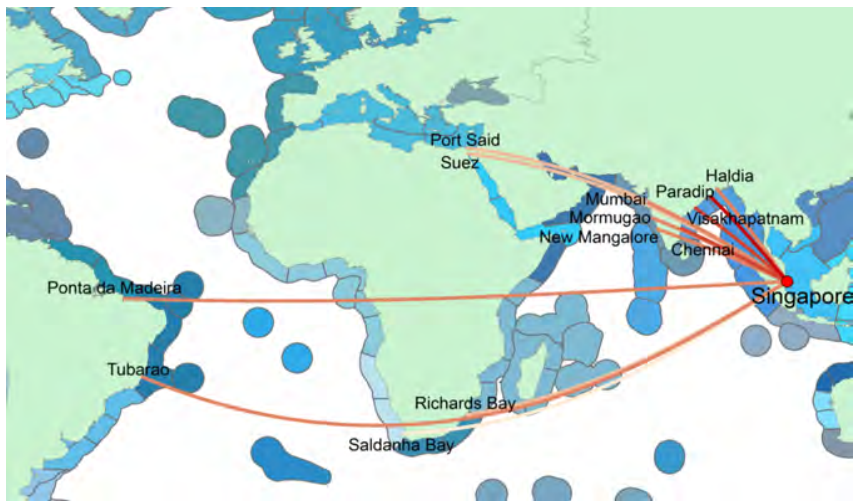

(b) Inter-Cluster Invasion Risk w.r.t. Singapore

Figure 7.9. **NIS invasion risk with respect to Singapore,** where the colors correspond to risk level definitions in Figure 7.6.

CHAPTER 8

SHIPPING NETWORK CONSTRUCTION: A COMPARATIVE STUDY

8.1 Overview

There exists multiple works that utilize network approaches for global shipping analysis. While these studies unanimously chose the same raw data of global ship movements, none of them use exactly the same way to construct the networks. As a result, the network representations for the same data are different. If the network representation influences the subsequent analyses and the ultimate interpretations, then not only the results in these literatures are not comparable to each other, but the network construction step may also be a reason of biased or controversial results.

**Intellectual merit:** In this paper, we conduct a comparative study on how different network construction parameters of the same global shipping data may influence network properties and analysis results. This work in itself does not aim to suggest which approach is "better" than the other. Rather, it is the first step to producing accurate and effective network representations that captures important information in the raw data, and it serves as a reference for researchers in related fields such as road or air traffic network analysis.

**Connections:** This discussion is partially based on the HON representation in Chapter 3. It serves as an extension of the species invasion network modeling in Chapter 7. It can also serve as a guideline for future research of Arctic species invasion in Chapter 9.

**Work status:** This work is accomplished in collaboration with the CoastalSEES research group funded by NSF. It is being prepared for a journal submission.

In this paper, we have illustrated how different network construction parameters of the same global shipping data may influence network properties and analysis results. We observe that the global shipping traffic is imbalanced in terms of directionality, which can only be captured by a directed network. The shipping frequency and capacity is also unevenly distributed; using a weighted network can help preserve such information and distinguish high-traffic connections to low-traffic ones. Explicitly representing indirect linkages in network topologies significantly increases the number of edges in the network, and influences subsequent analysis. Representing higher-order dependencies in the network can help preserve the actual pattern of ship flow in the raw data. The choice of starting year, starting month, and the length of time window all have non-trivial influences on network properties and analysis results.

While we do not aim to suggest a universal guideline to network construction, for the specific task of global shipping representation, we do suggest using a directed and weighted network. If flow dynamics is the research focus, then higher-order network is recommended. The research should specify the starting time and time window as part of the result, better if sensitivity analysis can be provided. For time windows less than one year, we suggest a discussion of seasonality.

8.2  Introduction

Network science has accumulated a powerful toolkit for analysis, from clustering, ranking, to link prediction and anomaly detection. When the network data is not readily available, the first step to performing network analysis is to convert the raw data into a network representation. However, there are various types of networks as discussed in Chapter 2, *does it make a difference in subsequent analyses, if the network construction parameters are different?*

This research is motivated by multiple works that utilize network approaches for global shipping analysis. In Figure **??**, we provide a review of related papers [74–77, 109, 118, 133, 228] from left to right. While these studies unanimously chose the same raw data of global ship movements (provided by Lloyd's List Intelligence[1]), none of them use exactly the same way to construct the networks. Differences in network construction include:

- Directionality
- Weighting mechanism
- Linking mechanism
- First-order / higher-order
- Time window
- Starting year (evolution)
- Starting month (seasonality)
- Coupled network, temporal network, multigraph, hypergraph, ...

Every research follows different paths to construct the network. As a result, the network representations for the same data are different. If the network representation influences the subsequent analyses and the ultimate interpretations, then not only the results in these literatures are not comparable to each other, but the network construction step may also be a reason of biased or controversial results.

In this paper, we conduct a comparative study on how different network construction parameters of the same global shipping data may influence network properties and analysis results. Rather, we first analyze the raw data and extract important features, such as the imbalancedness of shipping traffic, then use those observations to support the choices of the network construction. This work in itself does not aim to suggest which approach is "better" than the other. Rather, it is the first step to producing accurate and effective network representations that captures important information in the raw data, and it serves as a reference for researchers in related fields such as road or air traffic network analysis.

---

[1] https://www.lloydslistintelligence.com/

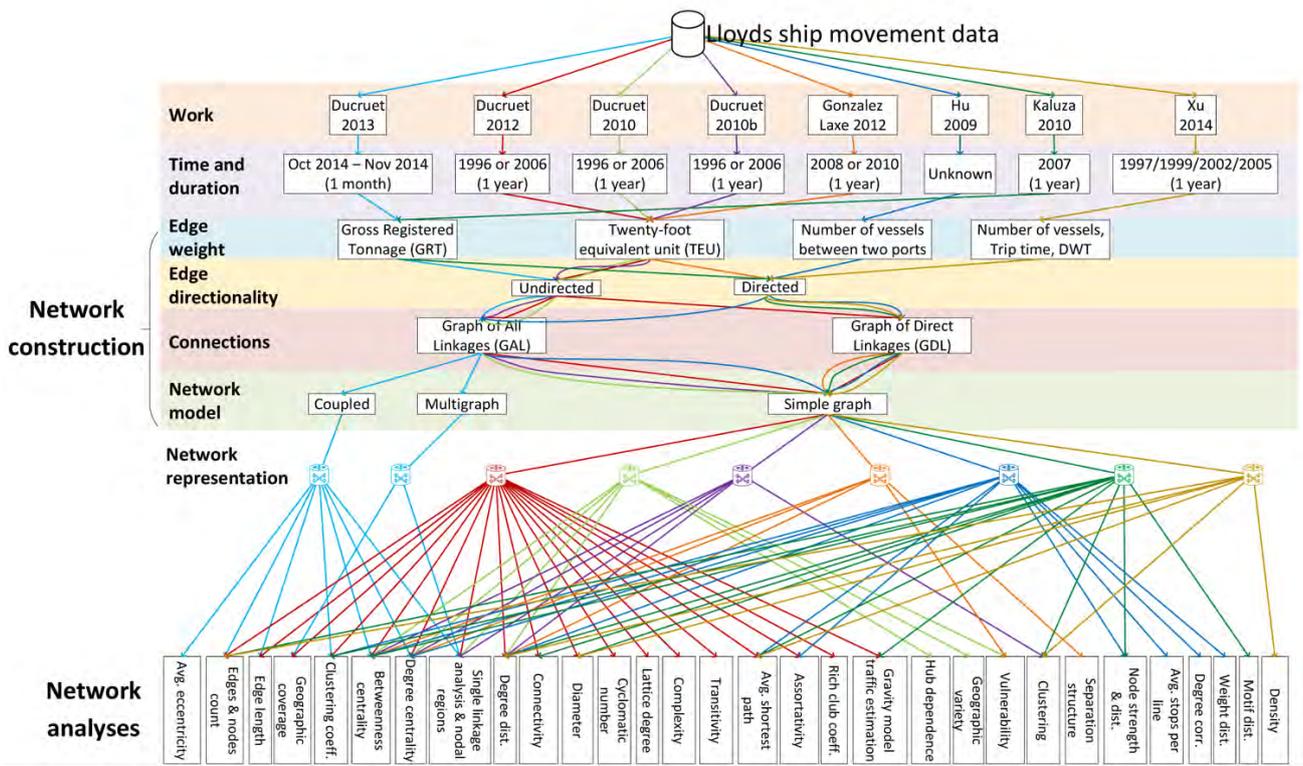

Figure 8.1. Comparison of network construction methods in existing literature for the same global shipping data.

8.3  Preliminaries

In this section, we will review properties of networks. Following sections will conduct comparative analysis of these network properties for different network representations of the same data.

**Definition 8.1.** *Network: A network $G$ is composed of nodes $N$ connected by edges $E$, in the form of $e_{ij} = n_i - n_j$ for undirected networks, and $e_{ij} = n_i \rightarrow n_j$ for directed networks.*

**Definition 8.2.** *In-degree (directed network): how many edges are pointing to the given node. Namely, $d_{in}(i) = |i|$ for $(j \rightarrow i) \in E$*

**Definition 8.3.** *Out-degree (directed network): how many edges are pointing out from the given node. Namely, $d_{out}(i) = |i|$ for $(i \rightarrow j) \in E$*

**Definition 8.4.** *Degree: ow many edges are connected to the given node. Namely, $d(i) = |\{(i - j) \in E\}|$ for undirected network, and $d(i) = d_{in}(i) + d_{out}(i)$ for directed network.*

**Definition 8.5.** *Density: the fraction of edges in the network to all possible edges can be built among the nodes. Namely,*

$$D(G) = \frac{|E|}{|N| \times |N| - 1}$$

*for directed networks, and*

$$D(G) = \frac{2|E|}{|N| \times |N| - 1}$$

*for undirected networks.*

**Definition 8.6.** *Clustering coefficient (generalized, undirected network): The fraction of triangles at the given node to the number of possible triangles among the node* and its neighbors. Namely,

$$c_n = \frac{2T_n}{d_n(d_n-1)}$$

for unweighted networks, where $T_n$ is the number of triangles through n. For weighted networks [186]:

$$c_n = \frac{2T_n}{\sum_{ni \in E} \frac{w_{ui}w_{uj}w_{ij}}{(max(w))^3}}$$

where $w_{ij}$ is the edge weight of $e_{ij}$.

**Definition 8.7.** *Generalized average clustering coefficient: based on the generalized clustering coefficient definition,*

$$C(G) = \frac{1}{|N|} \sum_{n \in G} c_n$$

**Definition 8.8.** *Transitivity (undirected network): the fraction of all possible triangles in the graph. Namely,*

$$Tr(G) = \frac{3 \times \#Triangles}{\#Triads}$$

*in which a triad is three notes sharing at least two edges*

**Definition 8.9.** *Path: A sequence of edges that a node can follow to reach another node. Namely, a path $p_{ij}$ between nodes i and j is $< i, k_1, k_2, k_3, \ldots, k_t, j >$ where $k_* \in N$ and $(i, k_1) \in E$, $(k_1, k_2) \in E$, ... $(k_t, j) \in E$.*

**Definition 8.10.** *Average shortest path: for $i, j \in N$ and $p_{ij}$ exists, shortest path $sp_{ij} = min(p_{ij})$, and average shortest path*

$$asp(G) = \frac{sp_{ij}}{|\{sp_{ij}\}|}$$

**Definition 8.11.** *Diameter: $max(sp_{ij})$*

**Definition 8.12.** *Radius: $min(sp_{ij})$*

**Definition 8.13.** *Connected component: All pairs of nodes in the connected component are connected via paths. Namely, CC is a subgraph of G with $N_{CC} \subseteq N$ and $E_{CC} \subseteq E$, also all $i, j$ in $N_{CC}$ have paths $p_{ij}$*

**Definition 8.14.** *Largest connected component proportion: for all CC of G, largest connected component (LCC) is that with $max(|N_{CC}|)$, and the LCC proportion is*

$$\frac{max(|N_{CC}|)}{N}$$

**Definition 8.15.** *Average degree connectivity coefficient: Whether nodes with higher degrees are connected to nodes with higher degrees or lower degrees. Specifically, the average nearest neighbor degree of node n with degree k is*

$$d_{ann}(n) = \frac{\sum_{i \in \Gamma(n)} d_i}{|\{\Gamma(n)\}|}$$

*where $\Gamma(n)$ is the neighbor of n. The weighted average neighbors degree is defined similarly [23] as*

$$d_{wann}(n) = \frac{1}{s_n} \sum_{i \in \Gamma(n)} w_{ni} k_i$$

*where $s_n$ is the weighted degree of n. The average degree connectivity coefficient is the correlation coefficient between k and the average of $d_{wann}(n_k)$ for nodes $n_k$ with degree k. Intuitively, a positive coefficient means nodes with higher degrees are connected to nodes with higher degrees.*

**Definition 8.16.** *Connected: for undirected networks, connected if $LCC(G) = G$*

*. For directed network, strong connected if every pair of nodes have paths between them; weak connected if the network is not strong connected, but connected when treated as undirected network.*

**Definition 8.17.** *Cyclomatic: number of basic cycles in a network. Namely,*

$$|E| - |N| + |CC|$$

.

**Definition 8.18.** *Flow hierarchy: the fraction of edges not participating in cycles, as defined in [138].*

**Definition 8.19.** *Degree assortativity: whether high-degree nodes are connected to high-degree nodes or low-degree nodes. Namely,*

$$PearsonCorr(d_i, d_j)$$

*for $(i,j) \in E$.*

**Definition 8.20.** *Power-law coefficient: how skewed the distribution of degree is. Specifically, when fitting $a, b$ to*

$$|\{n_k\}| = ak^b$$

*, the coefficient b is the power-law coefficient.*

**Definition 8.21.** *Alpha index: fraction of actual circuits to maximum circuits. Namely,*

$$\alpha(G) = \frac{|E| - |V|}{|V| \times (|V|-1)/2 - (|V|-1)}$$

**Definition 8.22.** *Beta index: edge-to-node ratio. $\beta(G) = |E|/|N|$.*

**Definition 8.23.** *Total length: the sum of lengths of pathways in kilometers.*

**Definition 8.24.** *Eta index: the average lengths (in kms) per edge. Namely, $\eta(G) = L(G)/|E|$.*

**Definition 8.25.** *Theta index: the average traffic at every node. Namely,* $\theta(G) = \sum w/|N|$.

**Definition 8.26.** *Gamma index: ratio of observed links to possible links.*

$$\gamma(G) = \frac{2|E|}{|V| \times (|V|-1)}$$

**Definition 8.27.** *Hub dependence: for a given node, the ratio of the strongest edge that connect to the node to all edges that connect to the node. Namely,*

$$HD(n) = \frac{max(w_{i \to n})}{\sum_i w_{i \to n}}$$

8.4 The Influence of Network Construction Parameters

8.4.1 Edge Directionality

Marine traffic is imbalanced. To begin with, we show that a port that has high a in-degree (number of incoming pathways) does not necessarily have a high out-degree (outgoing pathways), as shown in the ranking of ports by their in-degrees and out-degrees in Table 8.1 . For example, Houston ranks $10^{th}$ for out-degree but $16^{th}$ for in-degree.

If in-degrees and out-degrees are distinguished, we can analyze ports' functionalities from the connectivity perspective. Ports that have higher in-degrees than out-degrees can be seen as "aggregators" (of cargos, resources, etc.); ports have higher out-degrees than in-degrees are "distributors". Figure 8.2 gives the geographical distribution of aggregators versus distributors. The west coast of North America comprises aggregators overwhelmingly, while India is mostly distributors.

The geographical directionality of pathways are imbalanced at ports. In Figure 8.3, ports that have more pathways heading west-bound are colored in red, and

TABLE 8.1

COMPARISON OF THE TOP 20 IN-DEGREE and OUT-DEGREE PORTS

| Ranking | In-degree | Out-degree |
|---:|---|---|
| 1 | Singapore | Singapore |
| 2 | Rotterdam | Rotterdam |
| 3 | Gibraltar | Gibraltar |
| 4 | Antwerp | Antwerp |
| 5 | Las Palmas | Las Palmas |
| 6 | Algeciras | Algeciras |
| 7 | Hamburg | Hamburg |
| 8 | Amsterdam | Ceuta |
| 9 | Ceuta | Amsterdam |
| 10 | Ulsan | Houston |
| 11 | Kaohsiung | Shanghai |
| 12 | Busan | Balboa |
| 13 | Balboa | Hong Kong |
| 14 | Incheon | Ulsan |
| 15 | Hong Kong | St. Petersburg |
| 16 | Houston | Klaipeda |
| 17 | Klaipeda | Busan |
| 18 | Ghent | Kaohsiung |
| 19 | Shanghai | Ghent |
| 20 | Xingang | Gwangyang |

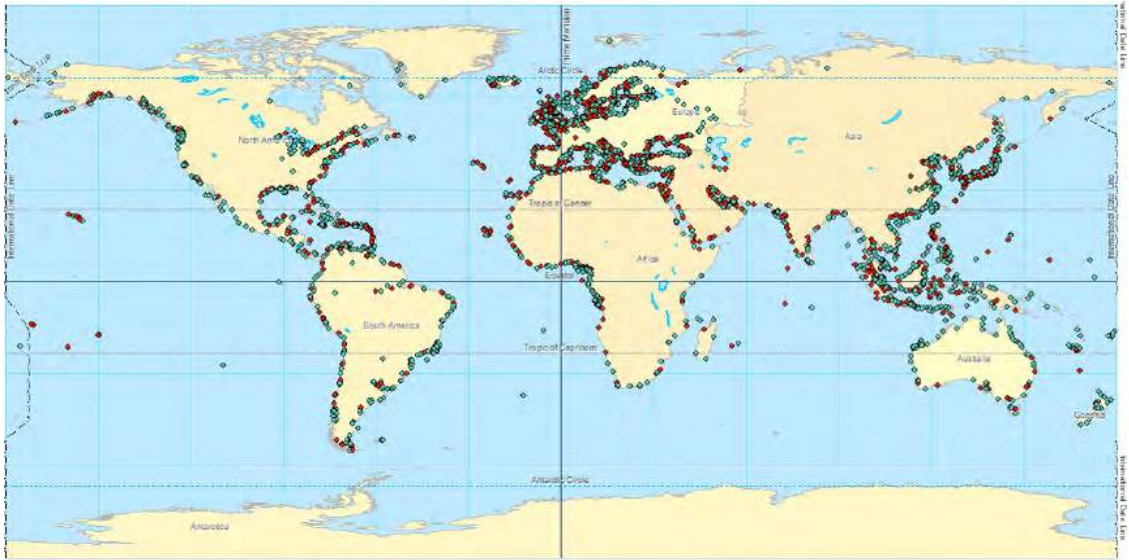

Figure 8.2. Ports with higher out-degree (red) vs ports with higher in-degree (blue).

those with more east-bound pathways are colored in blue. If the geographical directionality is balanced (namely, every port has the same number of east-bound and west-bound shipping pathways), one would expect a map with mixed colors. Instead, we observe distinct regions of colors in Figure 8.3, implying that the direction of circular ship voyages are imbalanced, possibly influence by the directionality of ocean current and the imbalanced demand in cargos. Figure 8.4 further supports the imbalancedness of pathway directionalities. In the figure, ports outside the red boundaries have 2x more voyages in one direction (east or west) than the other.

Edge directionality also influences a variety of network properties, as shown in Table 8.2. All other variables are controlled for (weighted by the number of trips, direct linkage, first-order network, from May 2012 to April 2013.) Notably, for some pair of ports, say $i, j$, there are edges in one direction $i \to j$ but not the other $j \to i$. Even if $j \to i$ never existed in the raw data, in the undirected network, pathways containing $j \to i$ are allowed. That will result in incorrectly smaller average shortest path, diameter, and radius. Also, the number of edges in the directed network is between 1x and 2x of that in the undirected network. All of these observations show that the shipping traffic is not 50/50 in both directions; using the unweighted network will lose important information, or result in biased network properties.

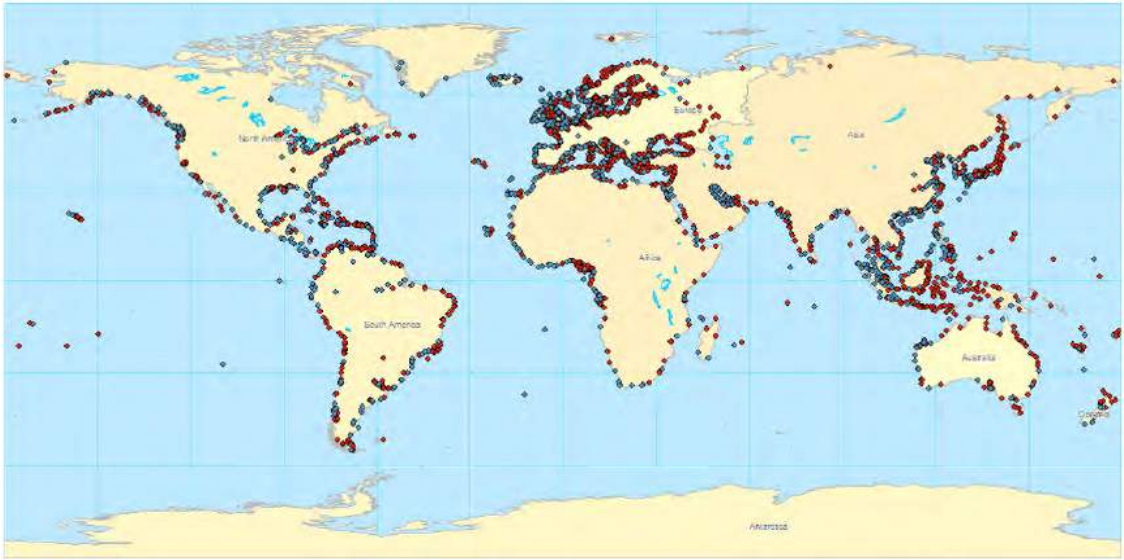

Figure 8.3. Ports with more pathways heading west bound (red) vs more pathways heading east bound (blue).

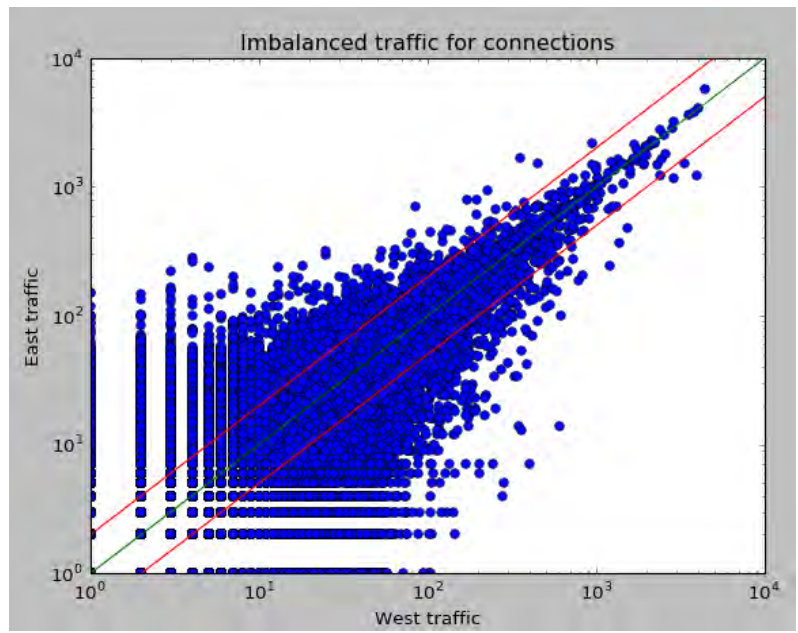

Figure 8.4. For every port, east-bound routes versus west-bound routes. Note that the axes are in log scale. The red boundaries are where the number of east-(west-)bound routes are twice the number of west-(east-)bound routes.

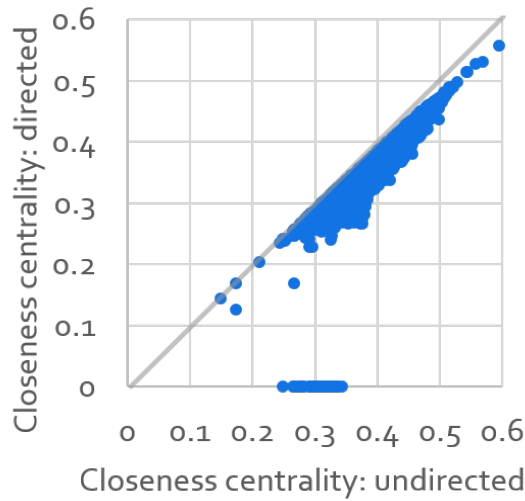

Figure 8.5. Comparing closeness centrality in directed and undirected networks.

Closeness centrality is am important way to estimating the risk of species invasion through global shipping: the higher the closeness centrality, the less steps required to reach other ports. The undirected network results in higher closeness centrality, as in Figure 8.5, because some impossible pathways are deemed possible, thus shortening the shortest paths. As a result, if the undirected network is used, it will give biased estimation of node centrality, as well as the estimated risk of species invasions.The result of ranking by closeness is provided in Table 8.3.

In summary, the global shipping traffic is imbalanced; failing to use the directed network will lead to biases in analysis results.

TABLE 8.2

COMPARING NETWORK PROPERTIES FOR UNDIRECTED
NETWORK AND DIRECTED NETWORK

|  | Undirected |  | Directed |
|---|---:|:---:|---:|
| num_of_nodes | 3595 | = | 3595 |
| **num_of_edges** | **132136** | **<** | **184543** |
| density | 0.02 | > | 0.014 |
| average_degree | 73 | < | 102 |
| highest_degree | 1284 | < | 2051 |
| **avg_shortest_path** | **2.64** | **<** | **2.82** |
| **diameter** | **8** | **<** | **10** |
| **radius** | **4** | **<** | **5** |
| LCG_proportion | 1 | > | 0.984700974 |
| NumConnectedComponents | 1 | < | 56 |
| avg_deg_conn_corr | -0.75 | = | -0.75 |
| beta_index | 36 | < | 51 |
| connected | TRUE |  | weak |
| degree_assortativity | -0.035 | < | -0.03 |
| eta_index | 2901 | > | 2602 |
| power_law | -1.04 | < | -0.93 |
| theta_index | 623 | = | 623 |

TABLE 8.3

PORTS IN UNDIRECTED AND DIRECTED NETWORKS WITH THE HIGHEST CLOSENESS CENTRALITIES

| Ranking | Undirected | Directed |
| --- | --- | --- |
| 1 | Singapore | Singapore |
| 2 | Rotterdam | Rotterdam |
| 3 | Gibraltar | Gibraltar |
| 4 | Las Palmas | Las Palmas |
| 5 | Antwerp | Antwerp |
| 6 | Algeciras | Algeciras |
| 7 | Balboa | Balboa |
| 8 | Hamburg | Hamburg |
| 9 | Amsterdam | Houston |
| 10 | Houston | Ceuta |
| 11 | Ceuta | Amsterdam |
| 12 | Durban | Shanghai |
| 13 | St. Petersburg | St. Petersburg |
| 14 | Shanghai | Durban |
| 15 | Santos | Klaipeda |
| 16 | Santa Cruz de Tenerife | Port Klang |
| 17 | New York | Santos |
| 18 | Port Klang | Flushing |
| 19 | Klaipeda | Bilbao |
| 20 | Bilbao | New Orleans |

### 8.4.2 Edge Weight

We compare three schemes of determining edge weights in the network. (1) Do not assign weights to edges. (2) Assign the number of trips from $i$ to $j$ as the edge weight of $i \to j$. Multiple trips made by the same ship count multiple times. Therefore, port pairs connected with frequent shipping traffic are distinguished from port pairs with occasional traffic. (3) Instead of treating every trip as the same and counting every trip as one, weigh every trip by the Dead Weight Tonnage (DWT) of the corresponding ship, as a relative approximate of the cargo transported. Therefore, pathways with large ships are distinguished with pathways with only small ships.

A comparison of the weighing by # trips and by DWT is provided in Table 8.4. More frequent trips do not necessarily suggest a major shipping pathway. For example, connections between Aomori and Hakodate (two minor cities in Japan) rank $3^{rd}$ and $4^{th}$ when weighted by # trips, but they are not in the top 20 when weighted by DWT. The frequent shipping activities between Aomori and Hakodate are mostly ferries and small fishing boats. On the contrary, the top connections weighted by DWT reflect the size of ports.

Edge weight also influences hub dependence, which is the ratio of the highest traffic pathway of a given port to the total traffic of the port. In other words, a port with high hub dependence has most of its traffic from a single port, and is more vulnerable if the dependent port malfunctions. We compare the list of ports with the least hub dependence (most resilient to malfunctions in other ports) in Table 8.5. For the unweighted network, hub dependence is the reciprocal of node degree; Singapore is the most resilient in under this weighing mechanism. However, when weighted by # trips or DWT, the top is Gibraltar instead of Singapore. The reason is that although Singapore has the most connected pathways, the load is not as evenly distributed among the pathways. The unweighted network representation fails to capture such

TABLE 8.4

STRONGEST 20 EDGES WITH DIFFERENT WEIGHING MECHANISMS

| Rank | By # trips | By DWT |
|---:|---|---|
| 1 | Tanjung Pelepas → Singapore | Port Klang → Singapore |
| 2 | Singapore → Tanjung Pelepas | Singapore → Port Klang |
| 3 | Aomori → Hakodate | Beilun → Yangshan |
| 4 | Hakodate → Aomori | Tanjung Pelepas → Singapore |
| 5 | Yosu → Gwangyang | Hong Kong → Chiwan |
| 6 | Port Klang → Singapore | Singapore → Hong Kong |
| 7 | Singapore → Port Klang | Yangshan → Beilun |
| 8 | Yantai → Dalian | Qianwan → Qingdao |
| 9 | Dalian → Yantai | Dover → Calais |
| 10 | Tokyo → Yokohama | Calais → Dover |
| 11 | Busan → Ulsan | Helsinki → Tallinn |
| 12 | Osaka → Kobe | Holyhead → Dublin |
| 13 | Ulsan → Busan | Dublin → Holyhead |
| 14 | Hong Kong → Macau | Tallinn → Helsinki |
| 15 | Macau → Hong Kong | Yantai → Dalian |
| 16 | Yokohama → Kawasaki | Dalian → Yantai |
| 17 | Busan → Gwangyang | Hong Kong → Yantian |
| 18 | Helsinki → Tallinn | Chiwan → Singapore |
| 19 | Tallinn → Helsinki | Chiwan → Hong Kong |
| 20 | Hong Kong → Chiwan | Kaohsiung → Hong Kong |

information.

The weighing mechanism does not change the network topology (which two ports are connected); therefore, network properties that only depend on topology (such as degree distribution and average shortest path) are not influenced. Nevertheless, network analysis algorithms can yield different results for differently weighted networks. Here we discuss two categories of algorithms, ranking and clustering, in the context of species invasion driven by global shipping.

A typical ranking algorithm is PageRank [161], which uses random walkers (with small probabilities of resets) on the shipping network to simulate the flow or movements among ports, and for each port the algorithm returns the converged relative probabilities of being visited. Since random walker's probability of choosing the next step is proportional to the weights of edges from the current port, the PageRank algorithm is influenced by the weighing mechanism.

In the context of species invasion, PageRank simulates the species being carried by a series of random ships. On the unweighted network, the species flow simulated is only dependent on port connections; on the network weighted by # of trips, ports pairs having more frequent trips in between have stronger flow of species in between; on the network weighted by DWT, the simulation of species flow assumes that a larger ship will carry more species than a smaller ship (e.g., carries more species in the ballast water and creating a higher propagule pressure).

A comparison of PageRank scores is given in Table 8.6. We fix all other parameters (directed network, direct linkage, first-order network, from May 2012 to April 2013). While the top ports Singapore and Rotterdam stays steady, some significant changes follow. Shanghai ranks #20 on the unweighted network, #8 on the network weighted by # trips, and #6 on the network weighted by DWT. Hong Kong also oversees significant increases in ranking from #12 to #4 to #3 when going from unweighted

TABLE 8.5

TOP 20 PORTS WITH THE LEAST HUB DEPENDENCE

| Rank | Unweighted | Weighted by # trips | Weighted by DWT |
|---:|---|---|---|
| 1 | Singapore | Gibraltar | Gibraltar |
| 2 | Rotterdam | Randers | Newcastle(AUS) |
| 3 | Gibraltar | Kolding | Zeytinburnu |
| 4 | Antwerp | Everingen | Bizerta |
| 5 | Las Palmas | Nemrut Bay | Vejle |
| 6 | Algeciras | Mariupol | Singapore |
| 7 | Hamburg | Odda | Vanino |
| 8 | Ceuta | Vejle | Hereke |
| 9 | Amsterdam | Augusta | Augusta |
| 10 | Houston | Newcastle(AUS) | Odda |
| 11 | Shanghai | La Pallice | Ras Tanura |
| 12 | Hong Kong | Tor Bay | Kolding |
| 13 | Balboa | Corunna | Randers |
| 14 | Ulsan | Szczecin | Bonny |
| 15 | St. Petersburg | Aalborg | Odense |
| 16 | Klaipeda | Skive | Kherson |
| 17 | Busan | Cagayan de Oro | Incheon |
| 18 | Kaohsiung | Aviles | Nakskov |
| 19 | Ghent | Kherson | Kalmar |
| 20 | Gwangyang | Mizushima | Mongstad |

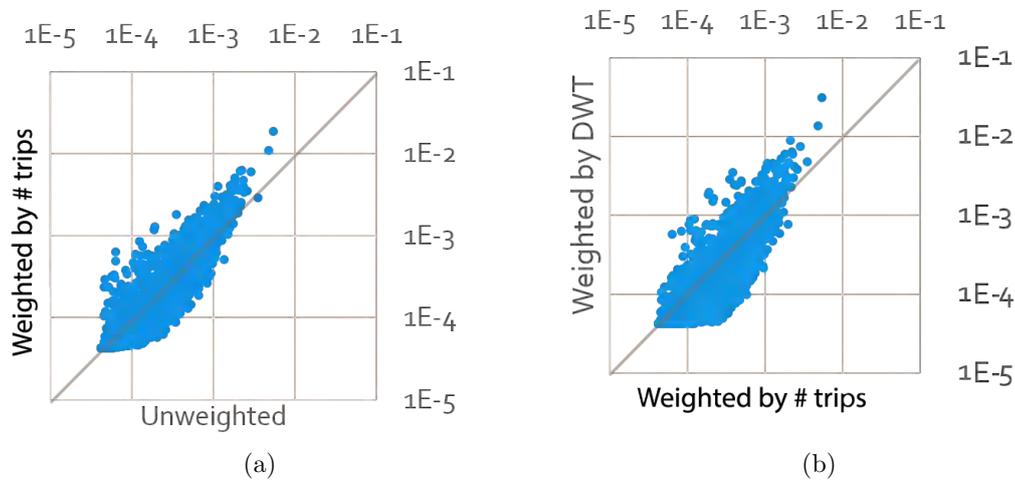

Figure 8.6. **Comparing PageRank scores on networks weighted differently.**

to weighted.

Another comparison is in Figure **??**. Both subfigures show that the weighing mechanisms influences PageRank scores – note that the plots are in log-log scale, so a small deviation from the diagonal still indicates a large difference. In Figure **??**(a), weighing by # trips instead of not weighing the network boosts the PageRank scores of a few ports by a large margin (the scattered dots above the diagonal); the same in Figure **??**(b), that weighing by DWT instead of # trips leads to boosts. A potential reason is latter weighing mechanisms reinforcing certain cycles of shipping routes, thus preventing the flow simulated by PageRank from "leaking" outside the cycles.

A typical clustering algorithm is MapEquation [177], which optimizes the coding length of random walkers' navigation behaviors within and between clusters. The algorithm yields clusters of ports, which are composed of tightly connected ports, but the connections among clusters are relatively sparse. The weighing mechanisms influence the strengths of connections, therefore the clustering results. The clustering

TABLE 8.6

TOP 20 PORTS WITH THE HIGHEST PAGERANK

| Rank | Unweighted | Weighted by # trips | Weighted by DWT |
|---:|---|---|---|
| 1 | Singapore | Singapore | Singapore |
| 2 | Rotterdam | Rotterdam | Rotterdam |
| 3 | Gibraltar | Busan | Hong Kong |
| 4 | Antwerp | Hong Kong | Antwerp |
| 5 | Las Palmas | Antwerp | Busan |
| 6 | Algeciras | Houston | Shanghai |
| 7 | Ulsan | Hamburg | Hamburg |
| 8 | Hamburg | Shanghai | Port Klang |
| 9 | Busan | Surabaya | Gibraltar |
| 10 | Incheon | Jakarta | Kaohsiung |
| 11 | Amsterdam | Port Klang | Beilun |
| 12 | Hong Kong | Piraeus | Bremerhaven |
| 13 | Kaohsiung | Ulsan | Barcelona |
| 14 | Balboa | Yokohama | Piraeus |
| 15 | Ceuta | Kaohsiung | Algeciras |
| 16 | Surabaya | Gwangyang | Xingang |
| 17 | Jakarta | Chiba | Zeebrugge |
| 18 | Houston | Las Palmas | Tanjung Pelepas |
| 19 | Xingang | Algeciras | Yokohama |
| 20 | Shanghai | Tanjung Pelepas | Le Havre |

results on differently weighted networks is in Figure 8.7. On the unweighted network, clustering yields relatively 21 clusters, with 9 of them being non-trivial. When the edges are weighted by # of ships, clusters merge and yields 5 larger clusters. Notably, ports in West Northern America, India and East Southern Africa join the Asia cluster; ports in the Great Lakes and West Southern America join the East Northern America cluster. The merger of clusters can be the result of *frequent* inter-cluster shipping, that are not captured by the unweighted network. When weighted by DWT, there are still five clusters, but the Great Lakes ports separate from East Northern America and join the Northern Europe cluster, and the West Africa ports also join the Northern European cluster.

As a concluding note, the global shipping traffic (shipping frequency and capacity) is unevenly distributed, which can only be captured if the network is weighted. Weighing mechanisms does not influence network properties that only rely on network topology, but influences certain analysis results of network algorithms. Note that by comparing the weighing mechanisms, we are not arguing weighting by DWT is better than by # trips – it is dependent on the network analysis tasks. However, not considering any weight can potentially bias the connections (such as trans-Pacific ties between Asia and West Northern America).

### 8.4.3  Linkage Mechanism

The linkage mechanism answers the following question: does the current step pose influence on multiple steps going forward, or just the next step? Figure 8.8 illustrates the linkage mechanisms. Given the raw ship trajectory A $\to$ B $\to$ C $\to$ D, the conventional approach, direct linkage, considers only the neighboring port calls and creates three edges A $\to$ B, B $\to$ C, and C $\to$ D. However, one may not assume that all cargo from A will go to be, all from B will go to C, and so forth; the cargo from A may as well be carried to D, just going pass B and C. Therefore, the indirect linkage mechanism makes a connection from the current port to every subsequent ports, forming a *clique* for every ship trajectory [109]. A trade-off between direct linkage and indirect linkage is the weighted indirect linkage. It is based on the intuition that neighboring port calls have stronger connection than ports that are distant away. While the method still builds links from the current port to all subsequent ports, distant port calls will be "dampened" [193]. However, how to choose the dampening factor objectively is yet another challenge. In this section, we discuss the direct linkage and indirect linkage mechanisms.

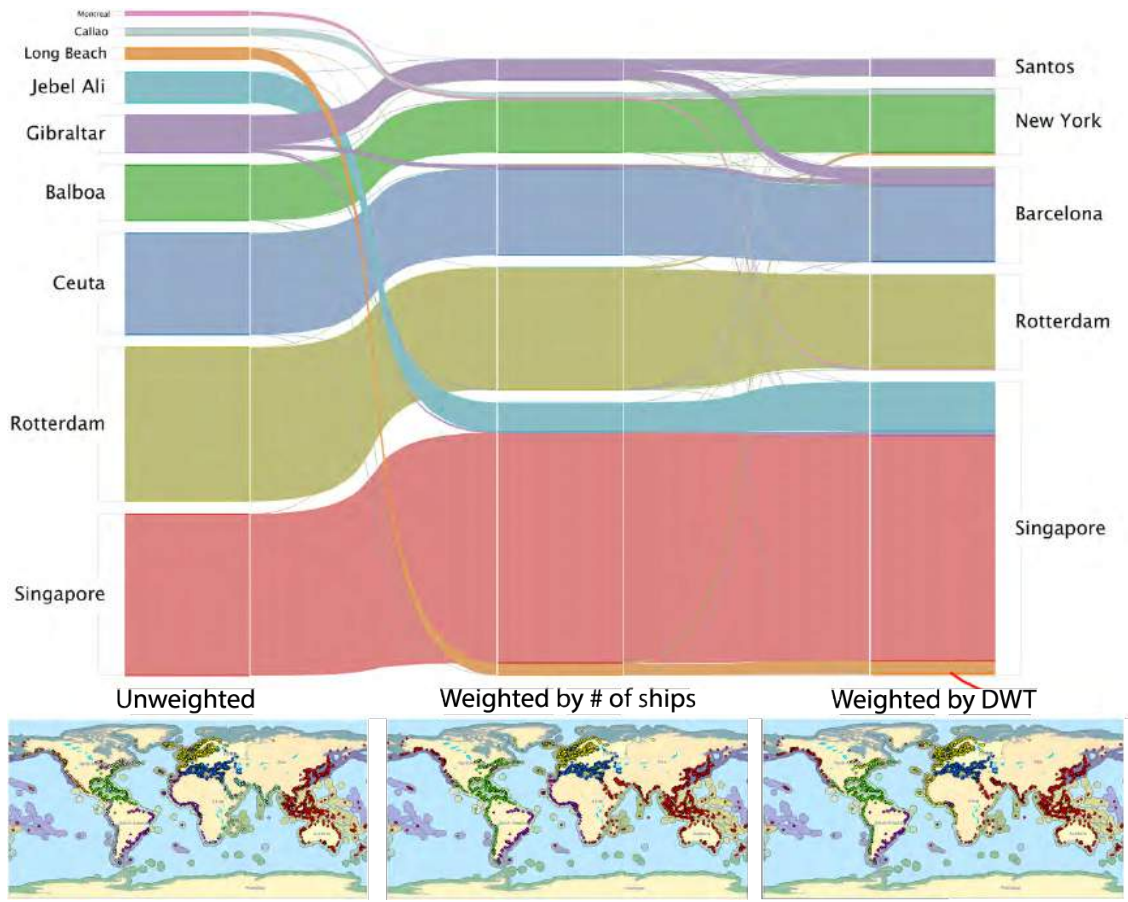

Figure 8.7. Comparing clustering results on differently weighted networks. Above: the alluvial diagram shows the relative flow [177] in each cluster, and illustrates the splits/merges when changing the weighing mechanism. Below: different colors denote the different clusters.

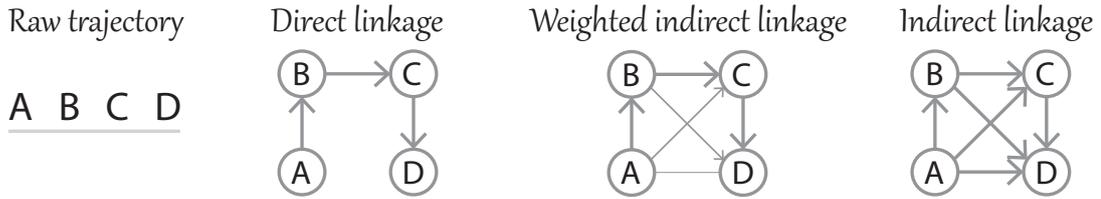

Figure 8.8. Different linkage mechanisms.

A comparison of network properties under different linkage mechanisms is provided in Table 8.7. All other parameters are controlled for (direct network, weighted by # trips, first-order network, from May 2012 to April 2013.) Because under the indirect linkage mechanism, every ship having visited $L$ ports produces a clique with $L(L-1)/2$ edges instead of a path with $L-1$ edges, the overall network will be denser. This is reflected in Table 8.7, where the indirect linkage network has edges 5.6 times that of direct linkage network. The average degree also increases significantly on indirect linkage networks, since indirect connections are made explicit. The abundance of cliques also leads to increased clustering coefficient and transitivity due to the addition of triangles. The average shortest path, diameter and radius all decreased due to the added edges. Note that on the indirect linkage network, the shortest path has a different interpretation: it now represents the number of *distinct ships* to reach

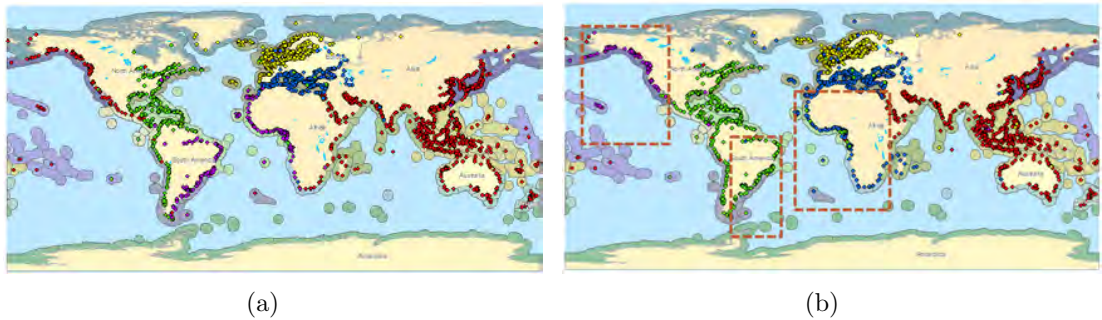

(a)                        (b)

Figure 8.9. **Clustering results on networks with different linkage mechanisms.**

a port, rather than the number of *movements* to reach a port.

Network analysis results are also influenced by linkage mechanism. For example, clustering on indirect linkage in Figure **??** separates West Northern America and Asia, and separates East Southern America and West Africa. A possible reason is that intra-continental links are reinforced, and inter-continental links become relatively weaker on indirect linkage networks.

In brief, the linking mechanism significantly changes the network topology and has profound influence on network properties. Note that the selection of linking mechanism depends on the application context. For example, in the ballast water-driven species invasion context, the direct linkage assumes 100% exchange in ballast water when visiting every port. The indirect linkage assumes the propagule pressure does not matter, and that any amount of species in the residuals of ballast water pose equal harm to the environment. The weighted indirect linkage assumes that the density of species picked up from the current port will be diluted through partial ballast water exchange as the ships sails through multiple ports.

TABLE 8.7

COMPARISON OF NETWORK PROPERTIES UNDER DIFFERENT LINKAGE MECHANISMS

|  | Direct Linkage |  | Indirect Linkage |
| --- | ---: | :---: | ---: |
| num_of_nodes | 3.60E+3 | > | 3.60E+3 |
| **num_of_edges** | **1.32E+5** | **<** | **7.37E+5** |
| **density** | **2.05E-2** | **<** | **1.14E-1** |
| **average_degree** | **7.35E+1** | **<** | **4.10E+2** |
| highest_degree | 1.28E+3 | < | 2.42E+3 |
| **generalized_clustering_coefficient** | **5.48E-1** | **<** | **7.23E-1** |
| **transitivity** | **2.96E-1** | **<** | **4.96E-1** |
| **avg_shortest_path** | **2.65** | **>** | **2.04** |
| **diameter** | **8** | **>** | **5** |
| radius | 4 | > | 3 |
| alpha_index | 1.99E-2 | < | 1.14E-1 |
| avg_deg_conn_corr | -7.54E-1 | > | -7.64E-1 |
| beta_index | 3.68E+1 | < | 2.05E+2 |
| degree_assortativity | -3.59E-2 | > | -3.77E-2 |
| eta_index | 2.90E+3 | < | 5.59E+3 |
| gamma_index | 2.05E-2 | < | 1.14E-1 |
| **power_law** | **-1.05** | **<** | **-5.64E-1** |
| theta_index | 6.23E+2 | < | 3.10E+4 |
| total_length | 3.83E+8 | < | 4.12E+9 |

### 8.4.4 Higher-order Dependencies

Choosing to include higher-order dependencies or only first-order dependencies influences both network properties (such as network density) and analysis results (such as clustering, ranking, anomaly detection), and so on [229]. Please refer to Chapter 3 for detailed discussions.

### 8.4.5 Time Window

Building a network from one week of data versus one year of data, what will be the differences? Figure 8.10 offers a comparative study with time windows, fixing all other parameters (directed network, weighted by # trips, direct linkage, first-order network, observations starting from May 2012). With more observations (from one week to one year), more ports are observed (increased by 1.8x); 52 weeks of observations also yields 20x more diverse shipping routes than a single week, as reflected by the number of edges. The significant increases in edge numbers also leads to denser network and higher average degree. As the network becomes more "connected", the average shortest path, diameter, and radius all decrease; namely, it appears that less steps are required to reach one port from another. Another reflection of the network being more connected is the decrease in the number of connected components, and the growing proportion of the largest connected component.

Clustering result also evolves with longer time windows, as shown in Figure **??**. Given a small time window (14 days), only short and local voyages can be captured; therefore, the algorithm produces three intra-continental clusters. When a larger time window is used (30 days / 60 days), intermediate range-routes start to bind clusters. We observe in Figure **??**(b) the emergence of the cross-continental cluster that binds East Southern America and West Africa, as well as the formation of West Northern America, and the separation of North and South Europe. If the time window is further lengthened (183 days / 365 days), cross continental routes are captured and bind clusters; notably, West Northern America connects with Asia (the cross-Pacific connections).

| | One week | Two weeks | One month | Two months | Six months | One year |
|---|---|---|---|---|---|---|
| num_of_nodes | 2.00E+3 | 2.43E+3 | 2.73E+3 | 2.96E+3 | 3.34E+3 | 3.60E+3 |
| num_of_edges | 9.30E+3 | 2.03E+4 | 4.04E+4 | 6.70E+4 | 1.29E+5 | 1.85E+5 |
| density | 2.32E-3 | 3.45E-3 | 5.41E-3 | 7.63E-3 | 1.16E-2 | 1.43E-2 |
| average_degree | 9.28E+0 | 1.68E+1 | 2.96E+1 | 4.52E+1 | 7.75E+1 | 1.03E+2 |
| highest_degree | 1.56E+2 | 3.45E+2 | 7.09E+2 | 1.06E+3 | 1.63E+3 | 2.05E+3 |
| avg_shortest_path | 6.57E+0 | 4.66E+0 | 3.62E+0 | 3.24E+0 | 2.94E+0 | 2.83E+0 |
| diameter | 2.10E+1 | 1.30E+1 | 9.00E+0 | 8.00E+0 | 1.10E+1 | 1.00E+1 |
| radius | 1.00E+1 | 7.00E+0 | 5.00E+0 | 4.00E+0 | 5.00E+0 | 5.00E+0 |
| LCG_proportion | 7.08E-1 | 8.48E-1 | 9.23E-1 | 9.53E-1 | 9.76E-1 | 9.85E-1 |
| NumConnectedComponents | 5.48E+2 | 3.64E+2 | 2.11E+2 | 1.39E+2 | 8.20E+1 | 5.60E+1 |
| alpha_index | 3.64E-3 | 6.09E-3 | 1.01E-2 | 1.46E-2 | 2.26E-2 | 2.80E-2 |
| avg_deg_conn_corr | 2.08E-1 | 1.07E-1 | -2.76E-1 | -5.54E-1 | -6.91E-1 | -7.57E-1 |
| beta_index | 4.64E+0 | 8.38E+0 | 1.48E+1 | 2.26E+1 | 3.87E+1 | 5.13E+1 |
| connected | no | no | weak | weak | weak | weak |
| cyclomatic | 7.84E+3 | 1.83E+4 | 3.79E+4 | 6.42E+4 | 1.26E+5 | 1.81E+5 |
| degree_assortativity | 8.99E-2 | 4.02E-2 | 1.40E-3 | -1.88E-2 | -2.91E-2 | -3.09E-2 |
| eta_index | 6.01E+2 | 8.74E+2 | 1.37E+3 | 1.84E+3 | 2.35E+3 | 2.60E+3 |
| flow_hierarchy | 7.23E-2 | 2.23E-2 | 6.58E-3 | 2.31E-3 | 7.04E-4 | 2.98E-4 |
| gamma_index | 4.63E-3 | 6.91E-3 | 1.08E-2 | 1.53E-2 | 2.32E-2 | 2.86E-2 |
| power_law | -1.62E+0 | -1.42E+0 | -1.25E+0 | -1.14E+0 | -1.01E+0 | -9.35E-1 |
| theta_index | 1.06E+1 | 2.42E+1 | 5.73E+1 | 1.17E+2 | 3.38E+2 | 6.23E+2 |
| total_length | 5.58E+6 | 1.78E+7 | 5.53E+7 | 1.23E+8 | 3.04E+8 | 4.80E+8 |

Figure 8.10. Comparing network properties given different time windows.

While the time window used for network construction largely depends on data availability, we note the profound influence of time windows on network properties and analysis results. A sensitivity analysis on time window not only makes the results more robust, but also make different research works comparable. Without discussion of time window, arguments such as "the average shortest path of global shipping is 3" become weak.

### 8.4.6 Evolution

Another dimension of temporal consideration is the starting time of observation. The global economy is constantly changing, shaping the global shipping activities. We fix the time window (one year), and change the starting date of the observation, spanning 15 years in 1997, 1999, 2002, 2005, 2008, and 2012. All other parameters are controlled for (directed, weighted by # trips, direct linkage, first-order, 1 year observation, starting from May 1).

A comparative analysis of network properties is in Figure 8.12. The changes in the evolution is non-trivial. The number of nodes (active shipping ports) increased to its peak around 2005, then decreased as of 2012. Meanwhile, the number of edges (diverse shipping routes) increased from 162k to 186k. Notably, degree assortativity increased from -0.054 in 1997 to -0.031; namely, high-degree ports (hub ports such as Singapore) start to connect to more high-degree ports, rather than minor ports.

The starting year also influences network analysis algorithms such as clustering results. Please refer to Chapter 7 for discussion.

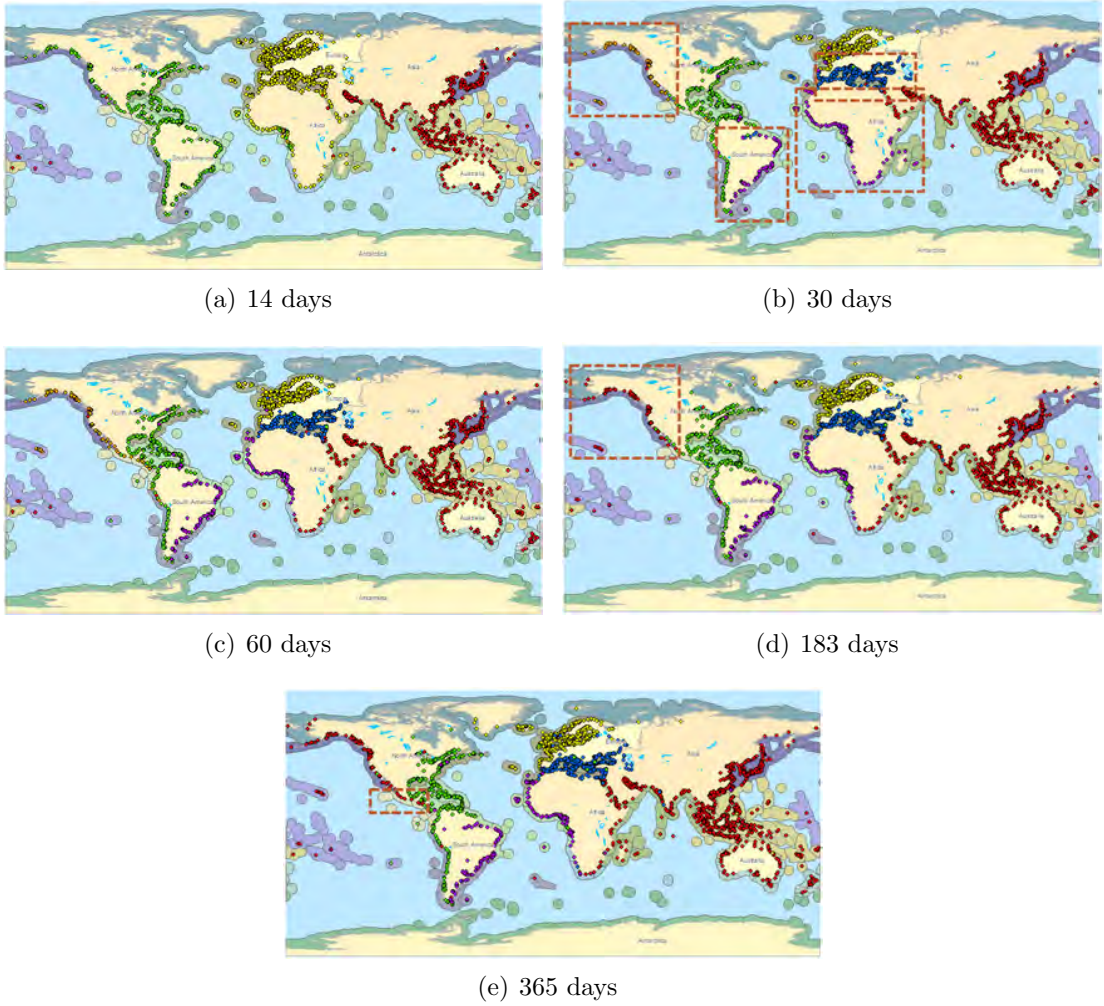

Figure 8.11. **Clustering results on networks constructed from different time windows.**

| | 1997 | 1999 | 2002 | 2005 | 2008 | 2012 |
|---|---|---|---|---|---|---|
| num_of_nodes | 3.52E+3 | 3.59E+3 | 3.85E+3 | 3.87E+3 | 3.86E+3 | 3.60E+3 |
| num_of_edges | 1.61E+5 | 1.66E+5 | 1.61E+5 | 1.64E+5 | 1.65E+5 | 1.85E+5 |
| density | 1.30E-2 | 1.29E-2 | 1.08E-2 | 1.09E-2 | 1.11E-2 | 1.43E-2 |
| average_degree | 9.18E+1 | 9.26E+1 | 8.35E+1 | 8.47E+1 | 8.58E+1 | 1.03E+2 |
| highest_degree | 1.68E+3 | 1.73E+3 | 1.85E+3 | 1.83E+3 | 2.05E+3 | 2.05E+3 |
| avg_shortest_path | 2.86E+0 | 2.86E+0 | 2.91E+0 | 2.93E+0 | 2.92E+0 | 2.83E+0 |
| diameter | 8.00E+0 | 7.00E+0 | 6.00E+0 | 9.00E+0 | 8.00E+0 | 1.00E+1 |
| radius | 4.00E+0 | 4.00E+0 | 4.00E+0 | 4.00E+0 | 4.00E+0 | 5.00E+0 |
| LCG_proportion | 9.83E-1 | 9.81E-1 | 9.74E-1 | 9.72E-1 | 9.81E-1 | 9.85E-1 |
| NumConnectedComponents | 6.10E+1 | 6.80E+1 | 1.02E+2 | 1.09E+2 | 7.60E+1 | 5.60E+1 |
| alpha_index | 2.55E-2 | 2.52E-2 | 2.12E-2 | 2.14E-2 | 2.17E-2 | 2.80E-2 |
| avg_deg_conn_corr | -7.17E-1 | -6.88E-1 | -7.12E-1 | -7.25E-1 | -7.60E-1 | -7.57E-1 |
| beta_index | 4.59E+1 | 4.63E+1 | 4.17E+1 | 4.23E+1 | 4.29E+1 | 5.13E+1 |
| connected | weak | weak | weak | no | weak | weak |
| cyclomatic | 1.58E+5 | 1.63E+5 | 1.57E+5 | 1.60E+5 | 1.62E+5 | 1.81E+5 |
| degree_assortativity | -5.42E-2 | -5.72E-2 | -5.32E-2 | -5.04E-2 | -4.62E-2 | -3.09E-2 |
| eta_index | 2.62E+3 | 2.61E+3 | 2.69E+3 | 2.78E+3 | 2.83E+3 | 2.60E+3 |
| flow_hierarchy | 4.15E-4 | 4.27E-4 | 7.46E-4 | 9.28E-4 | 5.26E-4 | 2.98E-4 |
| gamma_index | 2.61E-2 | 2.58E-2 | 2.17E-2 | 2.19E-2 | 2.23E-2 | 2.86E-2 |
| power_law | -9.18E-1 | -9.43E-1 | -9.84E-1 | -9.78E-1 | -9.65E-1 | -9.35E-1 |
| theta_index | 3.80E+2 | 3.85E+2 | 3.44E+2 | 3.69E+2 | 3.96E+2 | 6.23E+2 |
| total_length | 4.24E+8 | 4.35E+8 | 4.33E+8 | 4.56E+8 | 4.68E+8 | 4.80E+8 |

Figure 8.12. Comparing network properties given different starting years.

### 8.4.7 Seasonality

If the time window is shorter than a full year, the network constructed may be susceptible to seasonality changes. A shipping network constructed based on data in July (summer in Northern Hemisphere when Arctic will be almost ice-free) will be different from that in January (winter in Northern Hemisphere, when the Arctic routes are blocked). A comparison of starting months is provided in Figure 8.13. All other parameters are controlled for (directed, weighted by # trips, direct linkage, first-order, one month starting in 2012/2013). Since the majority of global shipping take place in the Northern Hemisphere, where the sea ice freezes around January, multiple network properties including the number of nodes, edges, density and average degree decrease by up to 10% compared with that in May. If two literatures construct networks base on observations in different months, seasonality is a factor to producing comparable results.

| | May-12 | Jun-12 | Jul-12 | Aug-12 | Sep-12 | Oct-12 | Nov-12 | Dec-12 | Jan-13 | Feb-13 | Mar-13 | Apr-13 |
|---|---|---|---|---|---|---|---|---|---|---|---|---|
| num_of_nodes | 2.73E+3 | 2.77E+3 | 2.73E+3 | 2.73E+3 | 2.76E+3 | 2.77E+3 | 2.74E+3 | 2.71E+3 | 2.66E+3 | 2.65E+3 | 2.69E+3 | 2.76E+3 |
| num_of_edges | 4.04E+4 | 4.05E+4 | 4.04E+4 | 3.93E+4 | 4.01E+4 | 3.97E+4 | 3.89E+4 | 3.77E+4 | 3.70E+4 | 3.76E+4 | 3.85E+4 | 3.98E+4 |
| density | 5.41E-3 | 5.30E-3 | 5.41E-3 | 5.28E-3 | 5.29E-3 | 5.18E-3 | 5.18E-3 | 5.15E-3 | 5.23E-3 | 5.37E-3 | 5.31E-3 | 5.25E-3 |
| average_degree | 2.96E+1 | 2.93E+1 | 2.96E+1 | 2.88E+1 | 2.91E+1 | 2.87E+1 | 2.84E+1 | 2.79E+1 | 2.78E+1 | 2.84E+1 | 2.86E+1 | 2.89E+1 |
| highest_degree | 7.09E+2 | 6.83E+2 | 7.05E+2 | 6.87E+2 | 6.72E+2 | 6.84E+2 | 6.82E+2 | 6.92E+2 | 6.81E+2 | 6.90E+2 | 6.75E+2 | 7.07E+2 |
| avg_shortest_path | 3.62E+0 | 3.64E+0 | 3.61E+0 | 3.65E+0 | 3.66E+0 | 3.65E+0 | 3.66E+0 | 3.69E+0 | 3.71E+0 | 3.67E+0 | 3.72E+0 | 3.69E+0 |
| diameter | 9.00E+0 | 9.00E+0 | 1.20E+1 | 9.00E+0 | 9.00E+0 | 1.00E+1 | 9.00E+0 | 1.20E+1 | 1.00E+1 | 1.00E+1 | 1.00E+1 | 1.00E+1 |
| radius | 5.00E+0 | 5.00E+0 | 6.00E+0 | 5.00E+0 | 5.00E+0 | 5.00E+0 | 5.00E+0 | 6.00E+0 | 5.00E+0 | 5.00E+0 | 5.00E+0 | 5.00E+0 |
| LCG_proportion | 9.23E-1 | 9.11E-1 | 9.20E-1 | 9.14E-1 | 9.22E-1 | 9.17E-1 | 9.06E-1 | 9.12E-1 | 9.03E-1 | 9.12E-1 | 9.15E-1 | 9.17E-1 |
| NumConnectedComponents | 2.11E+2 | 2.47E+2 | 2.19E+2 | 2.35E+2 | 2.15E+2 | 2.30E+2 | 2.57E+2 | 2.32E+2 | 2.49E+2 | 2.22E+2 | 2.28E+2 | 2.29E+2 |
| alpha_index | 1.01E-2 | 9.88E-3 | 1.01E-2 | 9.83E-3 | 9.86E-3 | 9.65E-3 | 9.63E-3 | 9.57E-3 | 9.71E-3 | 1.00E-2 | 9.89E-3 | 9.78E-3 |
| avg_deg_conn_corr | -2.76E-1 | -2.37E-1 | -2.41E-1 | -2.57E-1 | -1.84E-1 | -1.84E-1 | -2.73E-1 | -2.29E-1 | -1.72E-1 | -2.23E-1 | -8.10E-2 | -1.68E-1 |
| beta_index | 1.48E+1 | 1.46E+1 | 1.48E+1 | 1.44E+1 | 1.46E+1 | 1.43E+1 | 1.42E+1 | 1.39E+1 | 1.39E+1 | 1.42E+1 | 1.43E+1 | 1.45E+1 |
| connected | weak | no | no | weak | no | no | weak | no | no | no | no | no |
| cyclomatic | 3.79E+4 | 3.80E+4 | 3.79E+4 | 3.68E+4 | 3.76E+4 | 3.71E+4 | 3.64E+4 | 3.52E+4 | 3.46E+4 | 3.52E+4 | 3.60E+4 | 3.73E+4 |
| degree_assortativity | 1.40E-3 | 3.62E-3 | 3.58E-3 | 3.89E-3 | 1.08E-2 | 1.48E-2 | 9.86E-3 | 1.75E-2 | 1.55E-2 | 6.42E-3 | 1.97E-2 | 1.37E-2 |
| eta_index | 1.37E+3 | 1.34E+3 | 1.37E+3 | 1.33E+3 | 1.33E+3 | 1.36E+3 | 1.37E+3 | 1.37E+3 | 1.34E+3 | 1.35E+3 | 1.35E+3 | 1.35E+3 |
| flow_hierarchy | 6.58E-3 | 8.15E-3 | 7.30E-3 | 8.11E-3 | 7.43E-3 | 7.51E-3 | 8.90E-3 | 7.95E-3 | 8.79E-3 | 7.60E-3 | 7.33E-3 | 7.16E-3 |
| gamma_index | 1.08E-2 | 1.06E-2 | 1.08E-2 | 1.06E-2 | 1.06E-2 | 1.04E-2 | 1.04E-2 | 1.03E-2 | 1.05E-2 | 1.07E-2 | 1.06E-2 | 1.05E-2 |
| power_law | -1.25E+0 | -1.28E+0 | -1.28E+0 | -1.28E+0 | -1.26E+0 | -1.25E+0 | -1.26E+0 | -1.28E+0 | -1.28E+0 | -1.27E+0 | -1.27E+0 | -1.27E+0 |
| theta_index | 5.73E+1 | 5.68E+1 | 5.80E+1 | 5.53E+1 | 5.67E+1 | 5.41E+1 | 5.40E+1 | 5.30E+1 | 5.31E+1 | 5.32E+1 | 5.44E+1 | 5.59E+1 |
| total_length | 5.53E+7 | 5.43E+7 | 5.53E+7 | 5.25E+7 | 5.33E+7 | 5.38E+7 | 5.32E+7 | 5.17E+7 | 4.97E+7 | 5.07E+7 | 5.18E+7 | 5.37E+7 |

Figure 8.13. Comparing network properties given different months.

## 8.5 Discussions

In this paper, we have illustrated how different network construction parameters of the same global shipping data may influence network properties and analysis results. We observe that the global shipping traffic is imbalanced in terms of directionality, which can only be captured by a directed network. The shipping frequency and capacity is also unevenly distributed; using a weighted network can help preserve such information and distinguish high-traffic connections to low-traffic ones. Explicitly representing indirect linkages in network topologies significantly increases the number of edges in the network, and influences subsequent analysis. Representing higher-order dependencies in the network can help preserve the actual pattern of ship flow in the raw data. The choice of starting year, starting month, and the length of time window all have non-trivial influences on network properties and analysis results.

While we do not aim to suggest a universal guideline to network construction, for the specific task of global shipping representation, we do suggest using a directed and weighted network. If flow dynamics is the research focus, then higher-order network is recommended. The research should specify the starting time and time window as part of the result, better if sensitivity analysis can be provided. For time windows less than one year, we suggest a discussion of seasonality.

This work has its limitation in that there is not yet a quantitative measurement of the "quality" of network; that question in itself is non-trivial, and largely depends on the context of application. This work is the first step to producing accurate and effective network representations that captures important information in the raw data. The mining of patterns from the raw data, combined with domain knowledge, is the ultimate guideline of choosing the parameters for now. There remains other aspects worthy of discussion; for example, how to determine the weights for the weighted indirect linkage approach given a specific application, what are the influences of modeling as hypergraphs, temporal networks, coupled networks, and so on. The analysis pro-cess serves as a reference for related research, such as road or air traffic network analysis, information diffusion on social networks, human interactions, and so on.

CHAPTER 9

NETWORK MODELING AND PROJECTION OF SHIP-BORNE SPECIES INTRODUCTION AND DISPERSAL IN THE ARCTIC

9.1 Overview

Rapid climate change has sweeping implications for the economic, environment and policy landscape of the Arctic region. Climate change will lead to sea ice loss, increased economic and political focus on the region, and a dramatic increase in Arctic shipping activity. As a result, the risk of harmful marine species being introduced into this critical region will increase unless additional management steps are implemented.

**Intellectual merits:** Using big data sets about shipping, ecoregions, and environmental conditions, we leverage network analysis and data mining techniques to assess, visualize, and project ballast water mediated species introductions into the Arctic and ship-mediated dispersal of non-native species within the Arctic. We identify high-risk connections between Arctic and non-Arctic ports that could be sources of invasive species. Using higher-order network analysis, we further distinguish critical shipping routes that facilitate species dispersal within the Arctic. Our decadal projections of current trends reveal the emergence of shipping hubs in the Arctic, and suggest that the cumulative risk of species introduction is increasing and becoming more concentrated at these emergent shipping hubs. The risk assessment and projection framework proposed in this paper could inform risk-based assessment and management of ship-borne invasive species in the Arctic.

**Connections:** This work combines the HON algorithm in Chapter 3, the first-order network analysis at the global scale in Chapter 7, the discussion of network construction algorithms in Chapter 8, the SF-HON input in the HON+ algorithm in Chapter 4, and the visualization of HONVis in Chapter 5.

**Work status:**  This work is accomplished in collaboration with the CoastalSEES research group funded by NSF. It is under review at Nature Communications.

## 9.2 Introduction

Global trade and transportation networks can introduce non-native species to ecosystems via land, sea, and air as evidenced by thousands of reported cases [34, 125, 139, 180]. As trade and transportation volumes increase, so does the potential for human-assisted species introduction via these mechanisms [44, 62, 111, 180, 193]. When introduced to a compatible environment, a species can become established, with a subset of established species becoming invasive, i.e., threatening to the economy, environment or human health [137]. The World Wildlife Fund estimates that between 2004 and 2009, aquatic invasive species caused at least 50 billion dollars of damage to fisheries, aquaculture, water supply systems, industrial infrastructure and harbours [25]. Established invasive species are in many cases impossible or expensive to eradicate, and eradication efforts can harm native species [25, 26]. The difficulty of eradication coupled with the potential economic impact of invasive species underscores the high value of prevention [26]. Improved management of invasive species, including prevention strategies, is a primary goal of multiple international agreements including the Convention on Biological Diversity. A detailed understanding of species introduction potential through transportation networks [25, 26] is an important prerequisite for effective prevention.

The global shipping network is the dominant vector for the unintentional translocation of species to new ecosystems [151], typically through ballast water discharges and biofouling (i.e., organisms attached to the surfaces of ships) [26, 46, 61, 73, 193] Prerequisites for ship-borne species invasion include that the species survive transportation, reproduce in the new environment, and spread [26, 193, 228]. These processes are in turn influenced by multiple factors including ship type, voyage duration, ballast water uptake / discharge volume and location, environmental differences between the source and destination ports, and the environmental tolerance of the organism [149, 193, 223, 228]. Therefore, assessing species invasion risk is a complex task involving data sets of many types and from many sources.

Aquatic invasive species are a widely-recognized threat to Arctic ecosystems [61]. The complex interplay between climate change, shipping activities, and environmental conditions in the Arctic adds dimensionality to the risk evaluation [26, 61, 228]. Furthermore, climate change has the potential to melt Arctic sea ice, leading to increased shipping, changes in shipping patterns, and increased human activity and interest in the Arctic [144]. In recent years the number of ships traveling through the Arctic has increased 20% annually, leading to an associated increase in species introduction risk [148]. Climate change may expand the range of invasive species by altering the temperature and salinity of ports, allowing species to survive in locations they could not previously [223]. Literature has focused on the increase in shipping and the impact of climate change on environmental conditions. Here we build on these results to assess how these processes interact to influence Arctic species invasion.

The complex process of ship mediated species introduction, the economic importance of global shipping ($\tilde{9}0\%$ of global trade [?]), the interactions between stakeholders and nations within the Arctic, and the uncertain effects of climate change mean that addressing the issue of aquatic invasive species in the Arctic is complex, important and urgent [137, 228]. In the May 2017 Fairbanks Declaration [60], the eight member countries of the Arctic Council endorsed an action plan to reduce the impact of invasive species in the Arctic, emphasizing the importance of shipping as a primary pathway of species introductions [192]. Here we employ network analysis and data mining techniques to assess, visualize, and project aquatic species introduction into and dispersal within the Arctic via shipping. Without such analyses, it will be impossible to prioritize surveillance, prevention, and other management efforts among ports and routes, which will be necessary to achieve the goals laid out by the Arctic Council [192].

We accomplish these goals by building a risk assessment and projection framework tailored for shipping into and within the Arctic. We use the best available global data sets on ship movements [120] (Lloyd's List Intelligence (LLI), ballast water discharge data [156], biogeographical data [2, 205], and environmental data (temperature, salinity) [11, 120, 136]. We focus on ballast water-meditated invasions, but the framework can readily be extended to hull fouling [228]. Building on previous modeling frameworks to assess the relative risk of introduction posed by ships [59, 193, 228, 229], we develop a novel approach for evaluating the current and future relative risk of species dispersal into and within the Arctic posed by ballast water discharges and analyze the different components of risk.

Our analysis includes the following components. First, we use a first-order network of all voyages that originate outside the Arctic and end in the Arctic to estimate the relative risk of species introduction based on shipping frequency, ship size, ship type, trip duration, and ballast water exchange patterns. Second, we use the same network analysis to estimate the risk of establishment of introduced species based on the similarity of the environment in the origin and destination ports. Third, we combine these two components of risk in a visualization to identify those ports of origin that pose a high risk of delivering non-native species to Arctic destination ports with an environment likely to foster species establishment. Fourth, we shift our attention from identifying the Arctic ports at highest risk of initial introduction and establishment to the relative risk of subsequent ship-driven dispersal among ports within the Arctic. We compare dispersal risk from a first-order network to that estimated from a higher-order network30 which incorporates the dependency of a ship's next destination on the origin of its previous voyage. Fifth, to identify the relative risk for different Arctic ports of receiving two species of concern, the soft-shell clam (*Mya arenaria*) and the red king crab (*Paralithodes camtschaticus*), we apply a global first-order network analysis that combines introduction into the Arctic with dispersal within the Arctic, and includes the potential for stepping stone invasion (species using an Arctic port at risk of species introduction as the stepping stone to disperse to other uninvaded Arctic ports). Finally, under the assumption that current climate-change driven shipping trends continue, we use the global first-order network to project the potential for ship-mediated species introduction and dispersal in the Arctic over the next decade.

Our analyses are novel and important for a number of reasons: (1) we distinguish between the risk of species introduction into the Arctic from ship-driven dispersal within the Arctic; (2) we illustrate the potential importance of modeling ship-borne invasions with a higher-order network; (3) we employ an innovative approach to assessing the risk of establishment based on environmental tolerance; (4) we provide the first Arctic-wide projection of future ship-borne species invasion risk, including two case studies of immediate relevance to management efforts. Our results and framework provide a foundation for risk-based prioritization among ports for surveillance and efforts to prevent biological invasions of the Arctic.

## 9.3 Results

### 9.3.1 Species Introduction to the Arctic

Using the Arctic conservation area boundary defined by the Arctic Council, we identify 310 Arctic and 7,187 non-Arctic ports within the LLI data (Table 9.1). Here we focus on the species introduction pathways (purple pathways in Figure 9.1): the subset of 3,902 pathways directly connecting Arctic ports to non-Arctic ports along which voyages were recorded during 1997–2013 (see Methods for details). Since species are only considered potentially invasive if they have been introduced to an area by human activity, we did not consider species transported from neighboring ecoregions as non-native because they could disperse to these locations naturally. Removing pathways from neighboring ecoregions further reduce the number of intro-duction pathways we analyzed to 2,874.

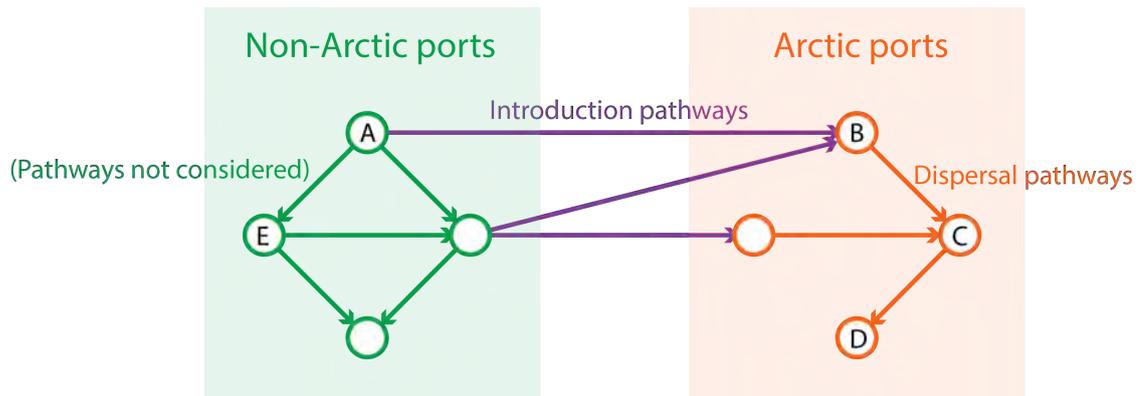

Figure 9.1. Illustration of species introduction pathways (from non-Arctic port to Arctic port) and dispersal pathways (from Arctic port to Arctic port).

TABLE 9.1

SUMMARY STATISTICS AND THE EVOLUTION OF SHIPPING

OBSERVED

|  |  | 1997-1998 | 1999-2000 | 2002-2003 | 2005-2006 | 2008-2009 | 2012-2013 | All years |
|---|---|---|---|---|---|---|---|---|
| Introduction to the Arctic | Recipient ports | 104 | 112 | 102 | 97 | 88 | 78 | 183 |
| | Direct pathways | 1,109 | 1,078 | 1,067 | 1,155 | 966 | 1,171 | 3,902 |
| | Hub with the most incoming pathways | Murmansk 110 | Murmansk 117 | Murmansk 135 | Murmansk 167 | Murmansk 116 | Murmansk 158 | Murmansk 344 |
| | Power-law distr. coeff. | -0.64 | -0.73 | -0.71 | -0.61 | -0.64 | -0.51 | -0.58 |
| | Voyages | 3,522 | 3,136 | 3,633 | 4,676 | 3,877 | 5,454 | 24,298 |
| | Sum DWT | $3.65 \times 10^7$ | $3.21 \times 10^7$ | $3.79 \times 10^7$ | $5.70 \times 10^7$ | $5.52 \times 10^7$ | $7.07 \times 10^7$ | $2.89108$ |
| | DWT per voyage | $1.04 \times 10^4$ | $1.02 \times 10^4$ | $1.04 \times 10^4$ | $1.22 \times 10^4$ | $1.42 \times 10^4$ | $1.30 \times 10^4$ | $1.19 \times 10^4$ |
| Dispersal in the Arctic | Recipient ports | 89 | 83 | 82 | 100 | 73 | 74 | 168 |
| | Direct pathways | 414 | 318 | 354 | 348 | 304 | 522 | 1,269 |
| | Hub with the most incoming pathways | Murmansk 18 | Murmansk 13 | Tromso 21 | Tromso 18 | Tromso 22 | Reykjavik 31 | Tromso 45 |
| | Power-law distr. coeff. | -1.17 | -1.09 | -1.14 | -1.3 | -1.11 | -0.73 | -1.02 |
| | Voyages | 2,579 | 1,856 | 3,271 | 3,246 | 3,404 | 9,710 | 24,066 |
| | Sum DWT | $1.15 \times 10^7$ | $8.93 \times 10^6$ | $1.57 \times 10^7$ | $2.00 \times 10^7$ | $1.77 \times 10^7$ | $5.75 \times 10^7$ | $1.31 \times 10^8$ |
| | DWT per voyage | $4.47 \times 10^3$ | $4.81 \times 10^3$ | $4.79 \times 10^3$ | $6.16 \times 10^3$ | $5.21 \times 10^3$ | $5.92 \times 10^3$ | $5.46 \times 10^3$ |

We denote the relative risk of species introduction from a non-Arctic port $i$ to Arctic port $j$ as $P_{i \to j}$; it is not an absolute probability of species introduction, but a robust metric of the relative risk of species introduction determined by shipping frequency, ship size, ship type, ballast water exchange, and likelihood of species survival in ballast (see Methods). Many pathways connect the Arctic to distant ports in Australia, South America, and Africa (Figure 9.2). A subset of pathways (red in Figure 9.2) constitute high-risk pathways (as indicated by $P_{i \to j}$) that connect Arctic ports with non-Arctic ports. High-risk pathways are concentrated in Northwestern Europe (e.g., the major ports of Rotterdam, Hamburg, and Amsterdam) and are connected by numerous voyages to the Arctic ports of Narvik in Norway and Murmansk in Russia (Table 9.2). For each port in the Arctic we aggregate the risk of invasion as $P_j = 1 - \prod_i (1 - P_{i \to j})$, denoted by node sizes (Figure 9.2). Of course, ports associated with high-risk pathways (such as Murmansk) demonstrate high aggregated risks; however, even some ports such as Afognak and Kodiak (both in Alaska) associated with low-risk pathways demonstrate high aggregated risks because of the substantial number of different ports to which they are connected (Figure 9.2).

Given that species with narrow temperature and salinity tolerances may not survive translocation to environmentally dissimilar ports, we next investigated the influence of species' sensitivity to environmental change on the potential for invasion along different introduction pathways. To simplify the analysis, we categorized species into six groups that reflect different environmental tolerances to temperature and salinity [228], based on empirically-estimated long term thermal tolerances of marine invertebrate taxa [175]. Only species with tolerance to a wide range of both temperature and salinity can survive translocation along pathways with large environmental differences between source and destination port (Figure 9.3a, light blue). In contrast, many species can survive translocation along pathways that connect environmen-

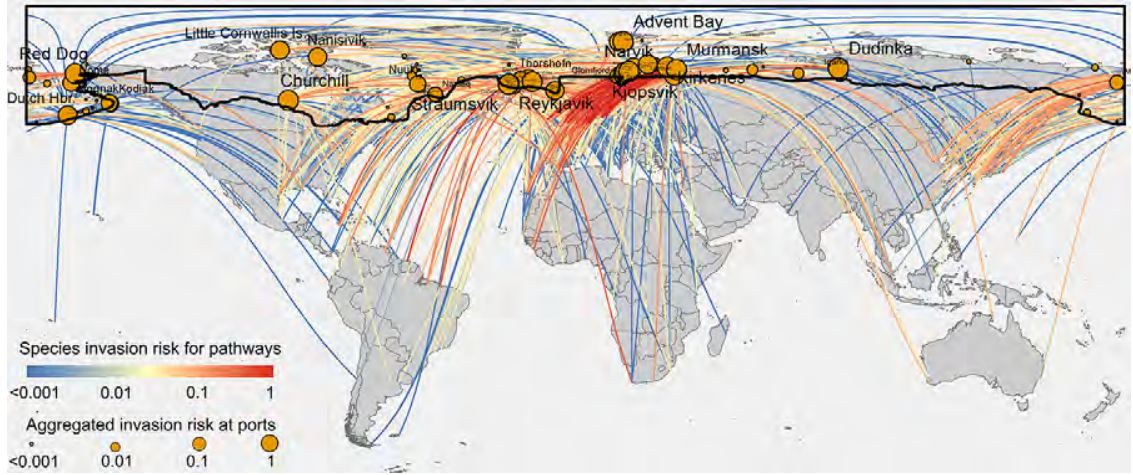

Figure 9.2. A global overview of species introduction pathways into the Arctic via shipping. Colors of links indicate the relative risk of invasion $P_{i\to j}$ from non-Arctic port $i$ to Arctic port $j$. Sizes of nodes indicate the risk of invasion to that port $P_j$ aggregated over all voyages into that port. The black outline delineates the Arctic boundary as defined in text.

TABLE 9.2

TOP 10 POTENTIAL SPECIES INTRODUCTION AND DISPERSAL PATHWAYS RANKED BY $P_{i\to j}$

| Rank | Introduction pathway (non-Arctic → Arctic) | Dispersal pathway (Arctic → Arctic) |
| --- | --- | --- |
| 1 | Bremen, DEU → Narvik, NOR | Murmansk, RUS → Dudinka, RUS |
| 2 | Rotterdam, NLD → Murmansk, RUS | Dudinka, RUS → Murmansk, RUS |
| 3 | Hamburg, DEU → Narvik, NOR | Murmansk, RUS → Glomfjord, NOR |
| 4 | Rotterdam, NLD → Narvik, NOR | Kandalaksha, RUS → Murmansk, RUS |
| 5 | Hamburg, DEU → Murmansk, RUS | Hammerfest, NOR → Tromso, NOR |
| 6 | Amsterdam, NLD → Murmansk, RUS | Harstad, NOR → Tromso, NOR |
| 7 | Dunkirk, FRA → Narvik, NOR | Leirpollen, NOR → Grundartangi, ISL |
| 8 | Ymuiden, NLD → Narvik, NOR | Tromso, NOR → Hammerfest, NOR |
| 9 | Amsterdam, NLD → Narvik, NOR | Tromso, NOR → Bodo, NOR |
| 10 | Ghent, BEL →urmansk, RUS | Murmansk, RUS → Vitino, RUS |

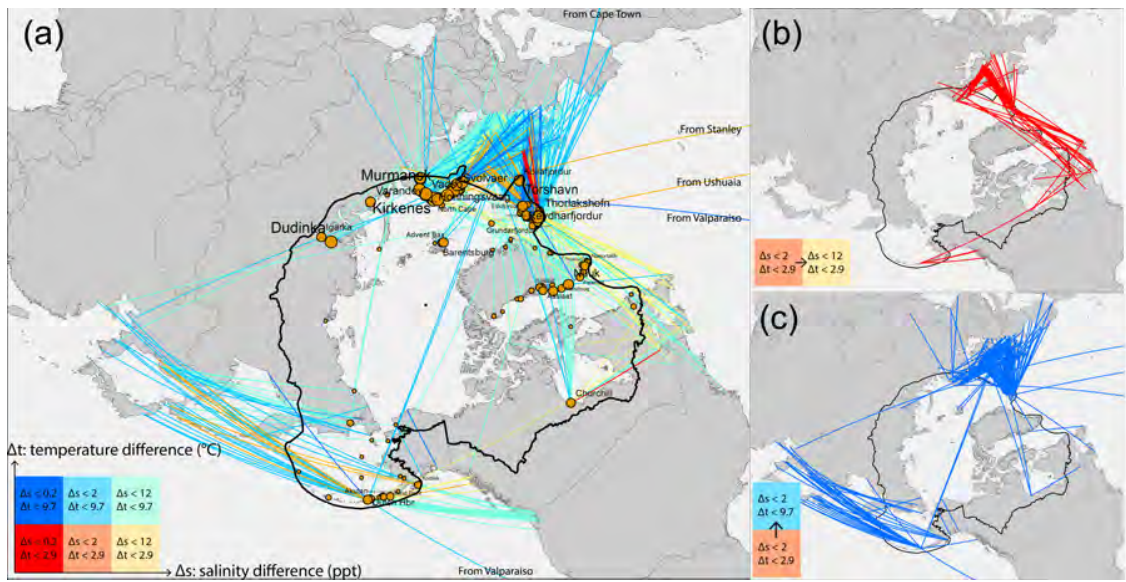

Figure 9.3. The influence of species' sensitivity to environmental change on introduction pathways. (a) Species introduction pathways from shipping originating outside the Arctic, organized by environmental tolerance groups, with link colors indicating the temperature and salinity differences between ports. Only the species with high tolerance to temperature and salinity changes are likely to survive the light blue pathways, whereas most species are likely to survive the dark red pathways. (b) Additional pathways that link ports with substantial differences in salinity (differences indicated in key). (c) Additional pathways that link ports with substantial differences in temperature (differences indicated in key).

tally similar ports (Figure 9.3a, dark red), like those from Port Alfred (Canada) to Churchill (Canada), and from Seaham (UK) to Akranes (Iceland).

9.3.2 Risk of Species Establishment in the Arctic

We further investigate the distinct roles played by species' temperature and salinity tolerance in determining potential invasion pathways. For example, species with a tolerance of salinity differences up to 12 ppt rather than 2 ppt will have access to additional invasion pathways as shown in Figure 9.3b, particularly from low salinity ports in East Canada to high salinity ports in Iceland. Meanwhile, species with a tolerance of temperature differences up to 9.7 °C rather than 2.9 °C will have access to additional invasion pathways as shown in Figure 9.3c, notably from North East Asia to Alaska. Such observations are important when devising targeted control strategies for species with certain environmental tolerances. Figure 9.4 compares species introduction networks available to species with different tolerances to temperature and salinity; the top introduction pathways in each of the six tolerance groups are given in Table 9.3.

### 9.3.3 Ship-borne Species Dispersal within the Arctic

Here we focus on species dispersal pathways (orange pathways in Figure 9.1): the subset of 1,269 Arctic first-order (direct) shipping routes along which voyages between ports within the Arctic were recorded in the LLI data (Figure 9.5). Several high-risk connections exist between ports in Northern Europe, including those connecting Murmansk in Russia and Tromso in Norway (Table 9.2). The aggregated risks of invasion at ports in Norway and Iceland are particularly high, due to the many strong dispersal pathways connecting to them.

The analyses above assume that a ship's next destination is independent of its previous destinations. However, this can be an oversimplification: our data show that a destination depends not only on a ship's current location, but also on its previous locations [229]. Such higher-order dependencies in ship movements imply higher-order patterns in species flow, which are not captured in a first-order network like those illustrated above (Figure 9.5). We model species flows as a network to highlight the connectivity among Arctic ports: nodes represent ports, links that connect nodes represent the species flow pathways, and links' weights (strengths) represent the relative risks of pathways. Complex movement patterns – for example ships (species) traveling from Tromso to Murmansk are more likely to then travel to

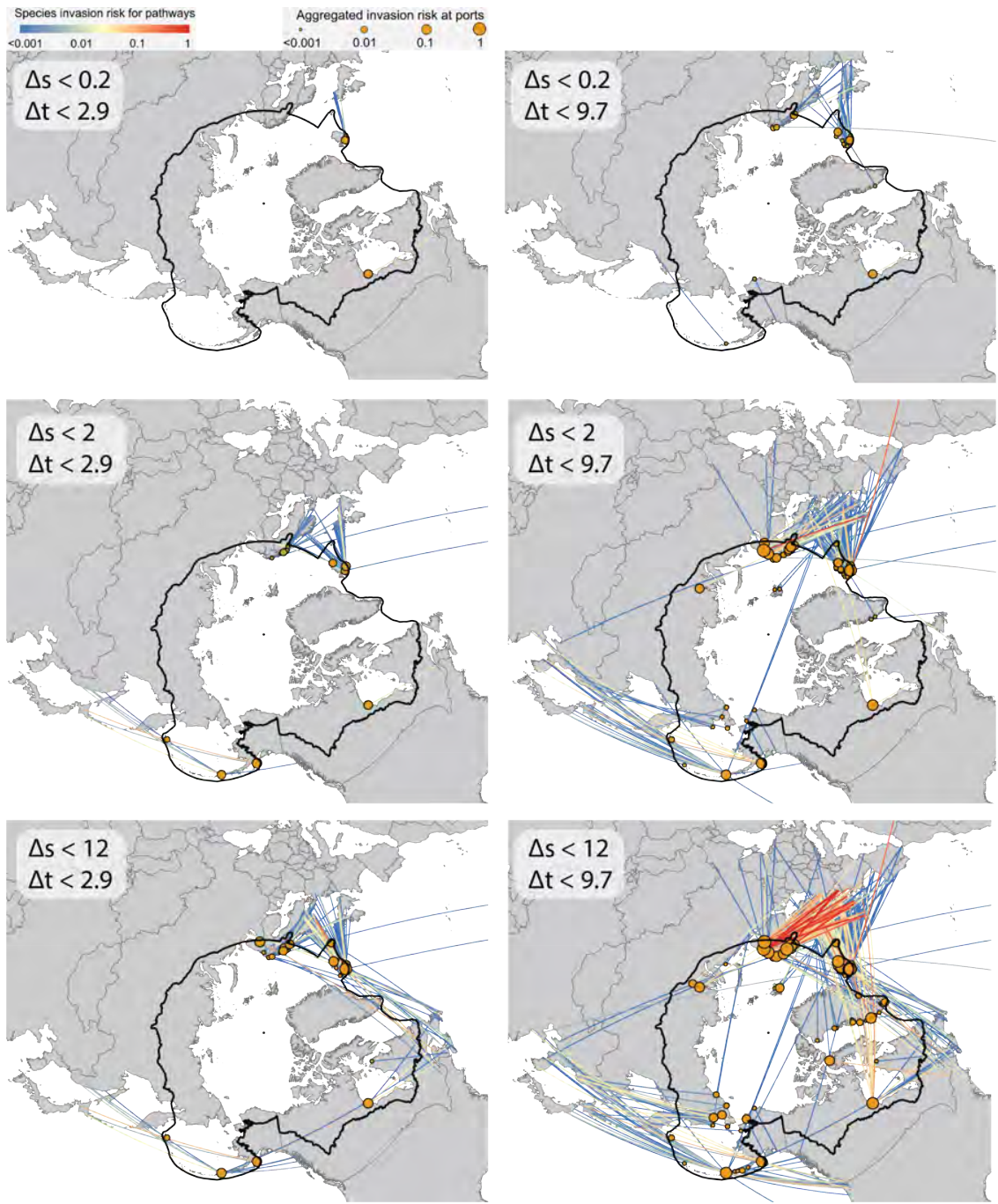

Figure 9.4. For species in different environmental tolerance groups, lines denote species invasion pathways, line colors the probability of invasion $P_{i \to j}$, node sizes the aggregated risk of invasion $P_j$ at ports.

TABLE 9.3

TOP FIVE INTRODUCTION PATHWAYS FOR ENVIRONMENTAL TOLERANCE GROUPS

| Environmental tolerance groups | Introduction pathway |
|---|---|
| Salinity tolerance <0.2 ppt<br>Temperature tolerance <2.9 | Port Alfred, CAN → Churchill, CAN |
| | Seaham, GBR → Akranes, ISL |
| | Blyth, GBR → Straumsvik, ISL |
| | Aberdeen, GBR → Straumsvik, ISL |
| | Fraserburgh, GBR → Vestmannaeyjar, ISL |
| Salinity tolerance <0.2 ppt<br>Temperature tolerance <9.7 | Port Alfred, CAN → Churchill, CAN |
| | Buckie, GBR → Eskifjordur, ISL |
| | Seaham, GBR → Akranes, ISL |
| | Cork, IRL → Glomfjord, NOR |
| | Kirkwall, GBR → Reykjavik, ISL |
| Salinity tolerance <2 ppt<br>Temperature tolerance <2.9 | Tomakomai, JPN → Afognak, USA |
| | Port Alfred, CAN → Churchill, CAN |
| | Kushiro, JPN → Dutch Harbor, USA |
| | Immingham, GBR → Vestmannaeyjar, ISL |
| | Archangel, RUS → Seydhisfjordur, ISL |
| Salinity tolerance <2 ppt<br>Temperature tolerance <9.7 | Aughinish Island, IRL → Murmansk, RUS |
| | Cape Town, ZAF → Straumsvik, ISL |
| | Hunterston, GBR → Kirkenes, NOR |
| | Immingham, GBR → Reykjavik, ISL |
| | Port Talbot, GBR → Kirkenes, NOR |
| Salinity tolerance <12 ppt<br>Temperature tolerance <2.9 | Stravanger, NOR → Hafnarfjordur, ISL |
| | Dalhousie, CAN → Kandalaksha, RUS |
| | Haugesund, NOR → Hafnarfjordur, ISL |
| | Tomakomai, JPN → Afognak, USA |
| | Seven Islands, → CAN Grundartangi, ISL |
| Salinity tolerance <12 ppt<br>Temperature tolerance <9.7 | Rotterdam, NLD → Murmansk, RUS |
| | Rotterdam, NLD → Narvik, NOR |
| | Amsterdam, NLD → Murmansk, RUS |
| | Amsterdam, NLD → Narvik, NOR |
| | Ghent, BEL → Murmansk, RUS |

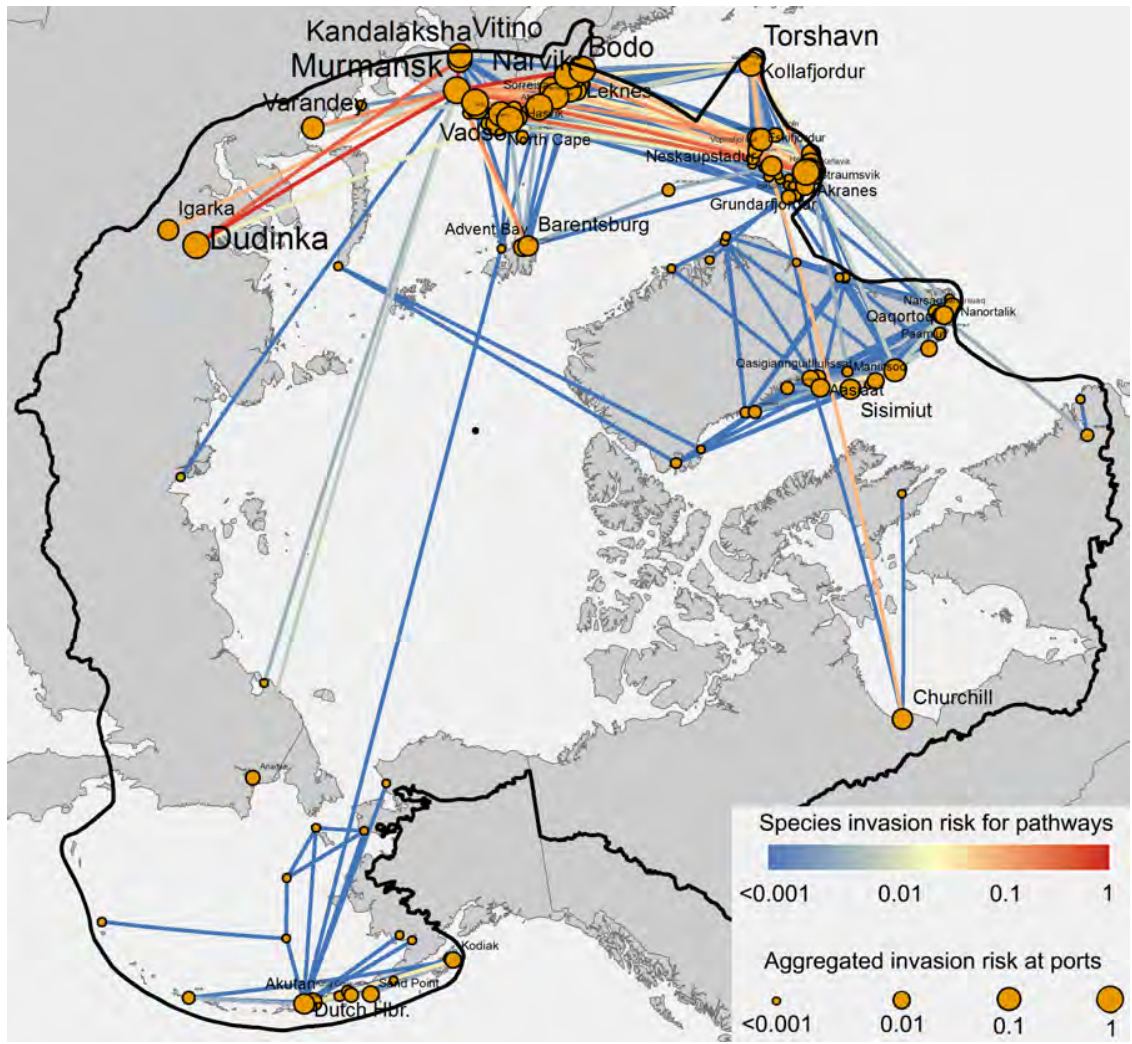

Figure 9.5. Species dispersal pathways within the Arctic. Colors of links indicate the relative risk of invasion. Sizes of nodes indicate the aggregated invasion risks for direct intra-Arctic species dispersal.

Bodo – are hidden in the layout in Figure 9.5, but are directly observable through the higher-order network (Figure 9.6a).

Figure 9.6b shows the Arctic SF-HON (see Methods on the construction of SF-HON). In this layout, ports are placed closer to each other if they have stronger species dispersal pathways between them (using layout developed by Yifan Hu [108]), thus highlighting the connections among ports rather than their geographic distributions. Using this layout, ports in the SF-HON are automatically grouped into five distinct clusters (determined by the clustering algorithm developed by Blondel et al. [36]), which largely self-organize by region. The inter-cluster connections are much weaker than intra-cluster connections; specifically, the Greenland cluster has only a few weak connections to the Icelandic cluster, and there are no significant dispersal pathways ($P_{i \to j} > 0.001$) between the Alaskan cluster and other clusters. These observations highlight opportunities to devise species control strategies targeting the loose connections between groups (such as those between Greenland and Iceland), to effectively prevent (or slow down) species dispersal from one cluster to another.

We simulate species flow in multiple steps, including potential flow between ports that are not directly connected. Inspired by the PageRank algorithm [161], we conceptualize species flow as a random walk process, which randomly follows pathways present in the SF-HON (hitchhiking on random ships traveling known pathways), and has a small probability of being "reset" to a random port (hitchhiking on ships traveling unobserved routes). The random walk result is reflected in Figure 9.6 as node sizes, which indicates the relative probability that a species will be transported to that port by randomly flowing through the SF-HON. Note that the random walk results presented here, which incorporate indirect species flows through multiple steps of ship movements, differ from the aggregated risk associated with single ship movements (a side-by-side comparison is provided in Table 9.4).

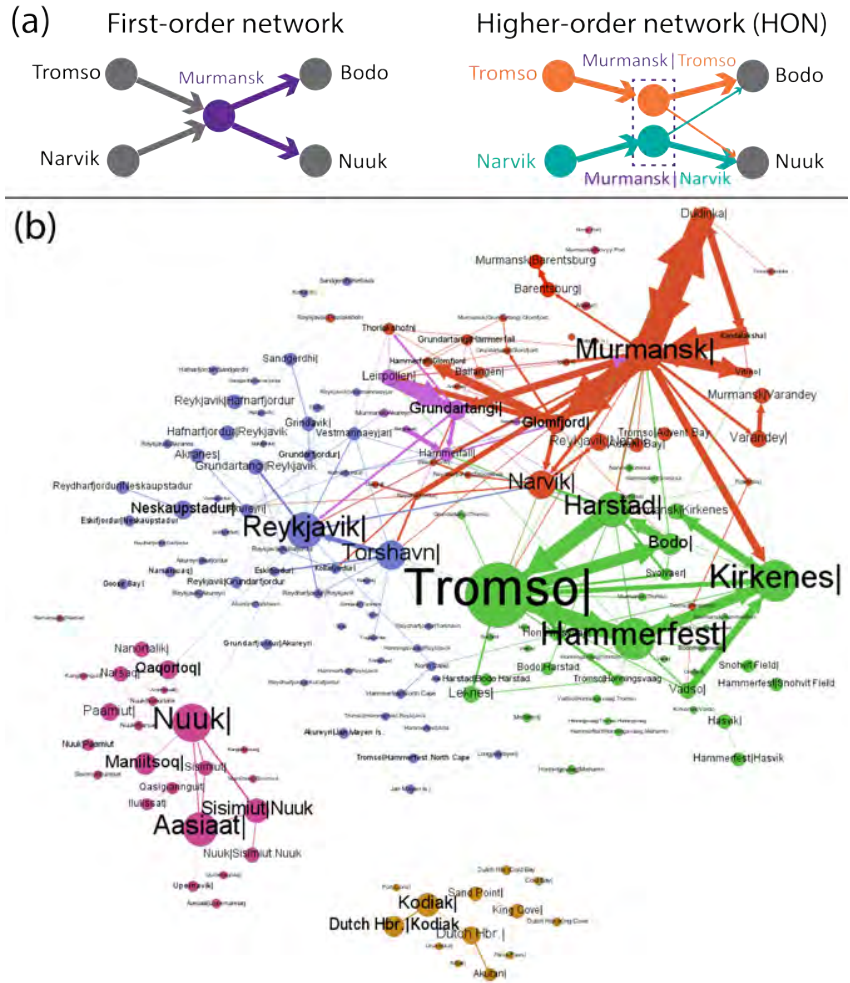

Figure 9.6. Species flow higher-order network in the Arctic. (a) Example of species flow represented as networks. Ship movements and species flow can depend on multiple previous steps, for example, where a ship will go from Murmansk is influenced by from where the ship came to Murmansk. Such higher-order dependencies are ignored by the first-order network, but can be captured by the higher-order network. (b) Species flow higher-order network (SF-HON) in the Arctic. Nodes represent ports (with labels in the form of [$CurrentPort$]|[$OptionalPreviousPorts$]), edges represent non-trivial species flow pathways (with $P_{i \to j} < 0.001$), and edge weights are species flow probabilities $P_{i \to j}$. Nodes closer to each other have stronger connections. Clusters of ports tightly coupled by species flows are distinguished by colors. Multiple nodes with the same [$CurrentPort$] represent the same physical location but with different previous locations. The size of nodes represents the relative probability that species end up at the given port by randomly flowing through the SF-HON.

TABLE 9.4

PORTS WITH THE HIGHEST INTRA-ARCTIC INVASION RISKS

| Rank | Risk of single-step direct invasion | Risk of multi-step indirect invasion |
|---|---|---|
| 1 | Murmansk, RUS | Tromso, NOR |
| 2 | Tromso, NOR | Reykjavik, ISL |
| 3 | Dudinka, RUS | Murmansk, RUS |
| 4 | Glomfjord, NOR | Hammerfest, NOR |
| 5 | Hammerfest, NOR | Nuuk, GRL |
| 6 | Kirkenes, NOR | Kirkenes, NOR |
| 7 | Grundartangi, ISL | Harstad, NOR |
| 8 | Harstad, NOR | Dutch Harbor, USA |
| 9 | Hammerfall, NOR | Grundartangi, ISL |
| 10 | Bodo, NOR | Aasiaat, GRL |

Left: aggregated risk associated with single ship movements obtained in the conventional network. Right: indirect species flows through multiple steps of ship movements estimated using random walks on SF-HON.

Take Reykjavik in Iceland as an example, the aggregated single-step risk for Reykjavik only ranks $14^{th}$ among all Arctic ports, since the pathways pointing to it are weak ($max(P_{\bullet \to Reykjavik}) = 0.14$ compared to $max(P_{\bullet \to Murmansk}) = 0.70$); however, Reykjavik ranks $3^{rd}$ for multiple-step risk of invasion, and is the central port in the Iceland cluster, therefore species at other ports in that cluster have a high probability of eventually flowing to the topologically highly connected port of Reykjavik. The case for Reykjavik as an invasion hotspot is supported by recent reports of several ship-borne non-native species establishing themselves in southwest Iceland [214]. Another example supporting the multi-step invasion risk rankings is the presence of at least 8 cryptogenic NIS in Dutch Harbor Alaska [181], a port that ranks $8th$ in multi-step invasion risk but only $24^{th}$ in the single-step risk. Unfortunately, existing survey reports are lacking for most Arctic ports, preventing a more rigorous test of the single-step and multi-step invasion risk rankings. Additionally, since new species may take years or decades to establish and become detectable [182], current survey data may not present the true picture of recent introductions. Therefore, we recommend that when devising management strategies, the single step-based ranking is suitable for short-term policies focusing on direct invasions, and the multiple step-based ranking is a better guideline for long-term strategies.

### 9.3.4 Case Studies for Soft-shell Clam and Red King Crab

We also conducted a risk assessment for specific species of interest to demonstrate how this framework could be used to develop species-specific management strategies. We conduct case studies on the soft-shell clam due to its wide distribution, broad environmental tolerances, and potential for (and possible history of) transport in ballast water [83]; and red king crab, due to its more compact distribution, narrow environmental tolerances, potential for ballast water transport during its pelagic phase, and concern about its impact on native species [117].

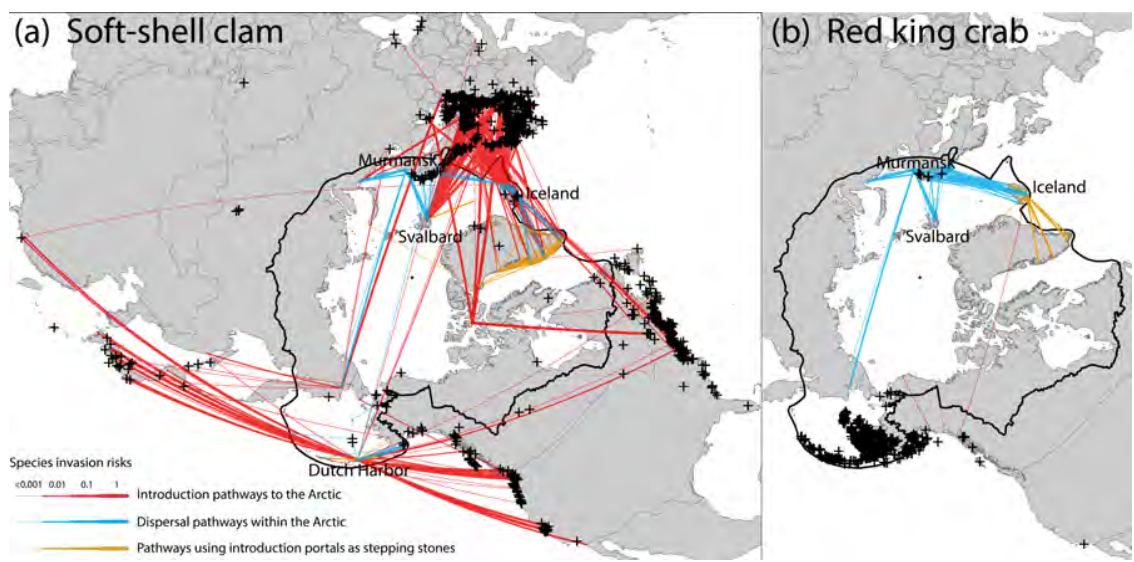

Figure 9.7. Case studies for soft-shell clam and red king crab. The known distributions of (a) Soft-shell clam and (b) red king crab (marked with "+" in each subfigure), and the potential stepping stone pathways of invasion in the Arctic based on their potential introductions by the shipping network and each species' environmental tolerances. Primary introduction pathways are red with subsequent introductions (dispersal) in blue (secondary) and yellow (tertiary). The width of links indicates the strength of connection by shipping.

Soft-shell clam is widely distributed across Arctic and non-Arctic regions (Figure Figure 9.7a), and can exist under a wide range of temperature and salinity conditions (2–28°C, 1.5–35 PSU) [13]. The potential introduction pathways (denoted in red) are mainly from Northern Europe to Iceland and Svalbard, and from Japan and the west coast of North America to Dutch Harbor in Alaska, USA (also see Figure 9.8 and Table 9.5). Since soft-shell clams already exist at Murmansk in Russia, there are no introduction pathways connected to Murmansk despite it being a highly active Arctic port. Meanwhile, Murmansk may play a key role in the dispersal of soft-shell clams within the Arctic (Figure 9.7a, thick blue pathways). We further discuss a special case of species dispersal: the stepping stone process [12, 73, 85]. Species introduction from Port A to Port B can lead to further dispersal from Port B to Port C; we call the pathway from Port B to Port C as the stepping-stone pathway (illustrated in Figure 9.1). Europe–Greenland will result in stepping stone ports in Greenland, which are in turn connected to multiple ports across West Greenland connections (Figure 9.7a, yellow pathways). Thus, if soft-shell clams are introduced to Greenland through Europe, the stepping stone effect will facilitate its dispersal across West Greenland. The development of surveillance and management strategies should consider such stepping stone pathways to effectively prevent the spread of soft-shell clams.

Red king crab is less widely distributed across the Arctic and non-Arctic regions (Figure 9.7b), and exists in a more limited range of conditions (-2–18°C and 28–35 PSU) [54, 142]. While there are only three weak introduction pathways into the Arctic, if red king crabs are introduced from Skagway in USA to Reykjavik in Iceland, Reykjavik could become the source of many stepping stone invasions to Icelandic and Greenlandic ports (Figure 9.7b, yellow paths). Therefore, high value could be realized in preventing species introduction from Skagway to Reykjavik and/or eradicating any incipient invasion at Reykjavik to prevent Reykjavik from becoming a stepping stone port. Likewise, a management focus on the small area between Murmansk and Norway (Figure 9.7b, blue paths; also see Figure 9.8 and Table 9.5) could have high value in preventing the dispersal of red king crabs from entering the protected areas of northeastern Greenland.

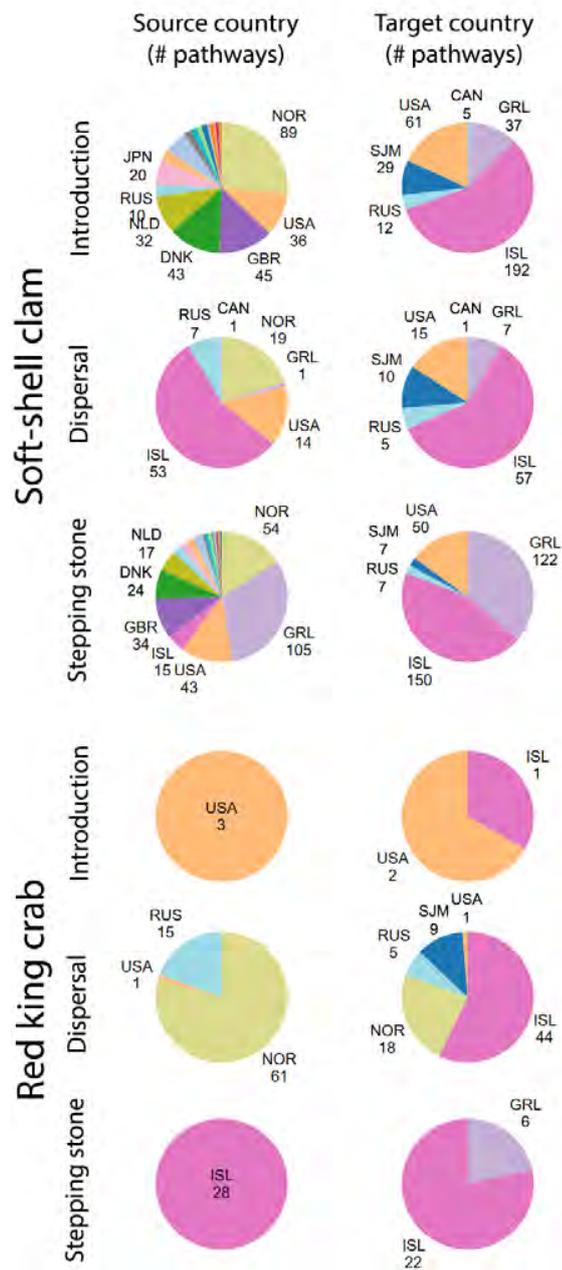

Figure 9.8. Source and target countries for introduction, dispersal, and stepping stone pathways for soft-shell clams and red king crabs.

TABLE 9.5

TOP FIVE SPECIES-SPECIFIC INTRODUCTION, DISPERSAL, AND
STEPPING STONE PATHWAYS

| *Soft-shell clam* | | |
|---|---|---|
| Introduction | Dispersal | Stepping stone |
| Brevik, NOR → Reydharfjordur, ISL | Murmansk, RUS → Varandey, RUS | Nuuk, GRL → Sisimiut, GRL |
| Europoort, NLD → Advent Bay, SJM | Murmansk, RUS → Barentsburg, SJM | Aasiaat, GRL → Nuuk, GRL |
| Europoort, NLD → Barentsburg, SJM | Straumsvik, ISL → Reydharfjordur, ISL | Sisimiut, GRL → Nuuk, GRL |
| Rotterdam, NLD → Reydharfjordur, ISL | Kodiak, USA → Dutch Hbr., USA | Akutan, USA → Dutch Hbr., USA |
| Fukuyama, JPN → Dutch Hbr., USA | Reykjavik, ISL → Reydharfjordur, ISL | Sisimiut, GRL → Aasiaat, GRL |

| *Red king crab* | | |
|---|---|---|
| Introduction | Dispersal | Stepping stone |
| Cordova, USA → Wainwright, USA | Tromso, NOR → Bodo, NOR | Reykjavik, ISL → Reydharfjordur, ISL |
| Valdez, USA → Wainwright, USA | Murmansk, RUS → Varandey, RUS | Reykjavik, ISL → Vestmannaeyjar, ISL |
| Skagway, USA → Reykjavik, ISL | Narvik, NOR → Reykjavik, ISL | Reykjavik, ISL → Qaqortoq, GRL |
| | Murmansk, RUS → Barentsburg, SJM | Reykjavik, ISL → Akranes, ISL |
| | Tromso, NOR → Leknes, NOR | Reykjavik, ISL → Nuuk, GRL |

### 9.3.5  Projection of Risk

We conduct a predictive analysis of ship-borne species invasion risks in the Arctic based on the evolution of shipping patterns observed from 1997 to 2012 (Figure 9.9). We use a linear model for this task, given the sparsity of available data, because it requires the least number of variables and assumptions. Across the 15-year observational span, in spite of ever-growing shipping activities in the Arctic, the number of distinct introduction and dispersal pathways only slightly increased; meanwhile, the number of recipient ports for introduction and dispersal pathways decreased at a rate of 2.03 and 0.85 ports per year respectively, indicating that Arctic shipping traffic (that was originally more distributed) is being "rewired" to a few hub ports. The relationship between the number of incoming pathways $p$ and the number of ports with $p$ incoming pathways $f_p$ follows a power-law distribution $f_p = ap^b$. The power-law coefficient $b$ increased from -0.64 to -0.51 over the 15 years, indicating a longer tailed distribution and the emergence of highly connected hubs.

Compared with the slight growth in the number of distinct pathways, the number of ship voyages (indicating shipping intensity) increased at a higher rate (55% more incoming traffic; 277% surge in intra-Arctic traffic) over the 15 years. Besides ship voyages, the sum of dead weight tonnage (DWT) of ships (a crude estimate of overall propagule pressure) and the average DWT per voyage (indicating the average ship size) has also increased rapidly. These observations all point to an increased risk of ship mediated species invasion and dispersal in the Arctic, which is verified by a significant trend toward increase in the average invasion risk per introduction pathway ($p = 0.006, r^2 = 0.88$), and a less significant trend toward increase in the average risk per dispersal pathway ($p = 0.238, r^2 = 0.32$), likely a result of there being comparatively less intra-Arctic shipping.

We also simulated how the observed emergence of hubs influences the aggregated risks of invasion at ports. To this end, we adapt the idea of the preferential attach-ment mechanism from the BarabásiAlbert network model [20]: based on the species invasion network in 2012, each pathway has a small probability of being rewired to a new Arctic port to simulate changes in shipping routes. The pathway being rewired has a higher likelihood of connecting to a port that has more connections–a hub port like Murmansk–than to a random minor Arctic port (details in the Methods section). We combine this projected topological evolution of pathways with the projected increase in invasion risk per pathway and compute the aggregated introduction and dispersal risks for Arctic ports in 2027.

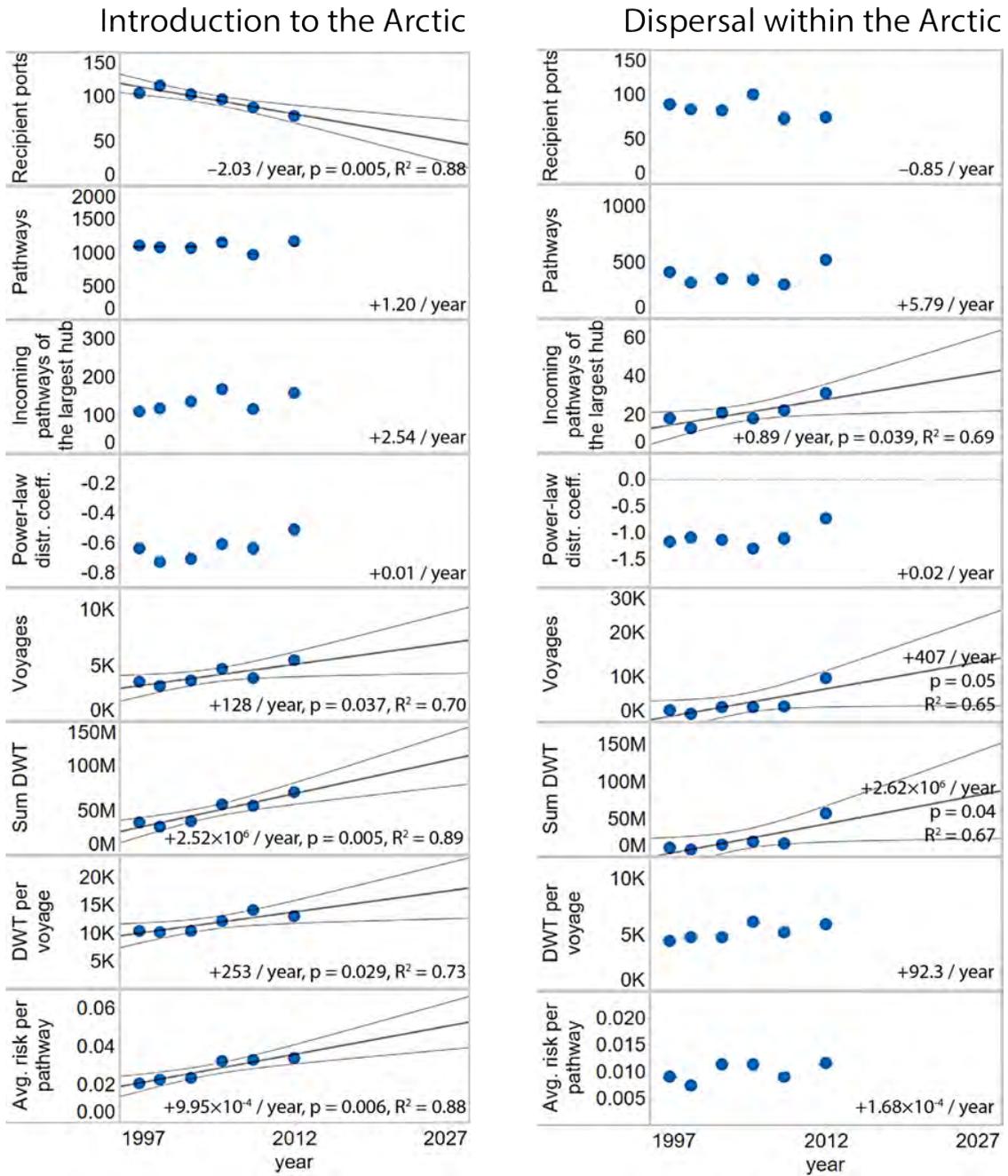

Figure 9.9. The evolution and the projection of shipping activities and species invasion risks in the Arctic. Left column: species introduction pathways from non-Arctic ports to Arctic ports; right column: species dispersal pathways within the Arctic. 95% confidence intervals for projections are given for regressions that have $p < 0.05$.

Over both the observed and simulated 15-year periods (1997–2012 and 2012–2027), the number of ports at risk has consistently decreased, most evidently in Greenland. The decreasing number of low-risk ports suggests that these minor ports are becoming inactive, and their flows are being redirected to hub ports; meanwhile, aggregated risk is increasing at the hub ports due to the rewiring of pathways to hubs, and the overall increase in risks per pathway. Geographically, the ports most at risk in the near future center around three locations: Northwest Russia (Murmansk) and Norway (Tromso), Iceland (Reykjavik, etc.), and Alaska (Dutch Harbor, etc.). Thus, management efforts in these locations are likely to be the most cost-effective to protect the entire Arctic.

## 9.4 Discussion

### 9.4.1 Key Observations

Our work presents the most comprehensive assessment and projection available of ballast water mediated Arctic species invasion risk. Ours is the first attempt to integrate big data sets from shipping voyages, ballasting records, environmental data on ports, and the current distribution of marine species. While future improvements in data sets and modeling methods will refine subsequent efforts, the current re-sults provide sound guidance as to which ports to prioritize for management efforts, given limited resources and the Arctic Council's goal of protecting the Arctic region from invasion [60]. The UN's International Maritime Organization (IMO) 2004 International Convention for the Control and Management of Ships' Ballast Water and Sediments will come into force in September 2017, and will provide a uniform international policy framework for reducing ballast-mediated invasions. Management mandated by the IMO convention could be augmented by geographically targeted efforts coordinated among Arctic Council member nations to protect the Arctic from threats revealed by our analysis, including risks not previously recognized.

Our analysis of species introduction pathways highlights the potential for species introductions originating from distant locations normally thought of as being disconnected from the Arctic, including Australia, South America, and Africa, particularly by species that are tolerant to a wide range of environmental conditions. It also reveals high-risk introduction pathways densely distributed in Northwestern Europe (the Murmansk–Narvik region). We further demonstrated how the same route poses a different level of risk for different species, depending on the average and range of environmental tolerances. Thus, our analyses can be used at a general, route-specific level or applied more specifically to species of concern based on knowledge of species-specific environmental tolerances.

Our analysis of species dispersal by shipping within the Arctic leverages the higher-order network to capture the influence on species flow of path-dependent ship movement patterns. The higher-order network analysis likely produces more accurate long-term species flow estimates with random walks. The higher-order network approach reveals clusters of ports that are loosely connected to each other (e.g., a single route connects Alaska to the other port clusters), highlighting opportunities to effectively prevent or slow down species dispersal within the Arctic by improved management on a small number of inter-cluster routes.

The case studies for soft-shell clam and red king crab – species with quite different distributions and environmental tolerances – highlight our framework's ability to provide insight into specific scenarios. We assess the role of the stepping stone effect whereby species introduction can occur through an intermediate port, and identify ports at which management could prevent stepping stone invasions. For example, targeted management on a single route could prevent Reykjavik from acting as a stepping stone for red king crab.

Our projection of species introduction risk presents a first glimpse into the future risk posed to the pan-Arctic by ballast-mediated species introduction. Our projection of current trends indicates that shipping intensity (number of trips, average or aggregated ship capacities) will continue increasing, along with the average invasion risk per pathway. While the number of pathways will likely remain steady, the pathways will be rerouted through Arctic shipping hubs such as Murmansk, resulting in three high-risk regions (Northwest Russia, Iceland, and Alaska). Increased management of ports in those regions would therefore have a disproportionate positive impact on risk across the entire Arctic. Overall the current results could aid in the development of effective Arctic invasive species management policies. As summarized above, our risk assessment framework, and the specific applications to soft-shell clam and red king crab, identified shipping routes where ballast water management will have the greatest impact on the overall threat of species invasion. Given finite resources and the especially challenging Arctic environment, information like that presented here can help guide the placement of resources for surveillance, prevention, and other species management strategies. Place-based strategies that decision-makers could consider include: (1) prioritizing locations for surveillance efforts, including new eDNA and other genetic-based methods, to inform early detection and rapid response efforts [137]; (2) choosing the location for on-shore ballast water treatment facilities for high-risk regions or ports that are becoming hubs (e.g., Murmansk and Narvik where many high risk pathways connect) [145]; (3) more stringent management policies on the highest risk routes and ports, including routes through inter-cluster connections in the Arctic, and ports (e.g., Reykjavik) with high potential for fostering many stepping stone invasions; and (4) developing site-based and/or mobile equipment and protocols for rapid response control efforts when an incipient invasion is discovered.

### 9.4.2 Opportunities for Future Improvement and Application

We believe that our approach can be refined in the future following improvements in data streams, and the incorporation of new data streams that capture other aspects of rapidly changing conditions in the Arctic. Although the LLI data set is the most comprehensive data set available on global shipping, it has limited observations in the northern regions of Canada. The NBIC ballast water data set, from which we parameterized the ballast water discharge model, is limited to the U.S. The GBIF species occurrence data is limited by the scope of biological monitoring efforts and is, no doubt, incomplete. All underlying phenomena are dynamic and potentially interact. Given melting Arctic sea ice and other warming temperatures, temperature and salinity in ports and other coastal environments may change rapidly, underscoring the need for more frequent and widespread environmental monitoring and real-time incorporation in global datasets could be very important in future analyses. Some of the key biological relationships–the likelihood of species establishment as a function of the volume of ballast discharge or discharge frequency or concentration of organisms in discharge–could be improved with more empirical research but are likely to remain highly contingent and uncertain. Thus, our analyses do not attempt to compute the absolute magnitude of invasion risks; rather, we use available data to estimate relative risks, which we believe are far more robust than any estimates of absolute risk.

Future research could couple ballast water discharge monitoring, species occur-rence monitoring, and long-term monitoring of the invasion processes to complement existing data. More comprehensive information about port types, port usage patterns and ships would allow for the implementation of machine learning approaches to modeling ballast water discharge and species invasion risks, getting closer to robust estimates of absolute risk in future.

Finally, our risk assessment and prediction framework can be extended in multiple ways. The risk of invasion from biofouling risk is thought to be commensurate to that from ballast discharges [209], but is currently poorly studied and has not received policy attention like ballast-mediated invasions. However, we have taken initial steps toward modeling biofouling mediated invasions [228], and when more data become available on biofouling patterns on different surfaces of different ship types, trip speed and detailed route trajectories (the satellite-based Automatic Identification System shows promise), our framework for assessing ballast water mediated invasion risk could be adapted to assess biofouling risk. The higher-order network approach can also extend beyond the Arctic to investigate species flows at the global scale. Our framework could be paired with species monitoring efforts, to produce regions of interest to guide species monitoring, and validate our model framework using observed changes in species distributions.

## 9.5 Methods

### 9.5.1 Data Sets and Preprocessing

We utilized global ship movement data for the years 1997, 1999, 2002, 2005, 2008, and 2012 (starting on May $1^{st}$ of these years and ending on Apr $30^{th}$ of the following years) from the Lloyd's List Intelligence (LLI). This data set was organized by individual voyages, totaling 12,723,028 records across the six years. The data also included unique vessel identifiers, vessel type (150 categories), gross weight ton-nage, dead weight tonnage, vessel departure and arrival port, departure and arrival dates, and miscellaneous information not used in our analysis. These records were screened for quality (duplicate ports of call made consecutively by the same ship were combined and records with ambiguous destinations were ignored), yielding 9,569,619 ship movements. The data was further subset to include only voyages to and between Arctic ports, herein defined as ports located within the boundary laid out by the Arctic Council's Conservation of Arctic Flora and Fauna working group [192], yielding 48,364 voyages through 3,902 introduction pathways from non-Arctic ports to Arctic ports, and 4,715 voyages through 1,269 dispersal pathways within the Arctic.

Ballast water discharge records for the years 2004 to 2016 for ships completing foreign and domestic voyages in Alaska were collected from the U.S. National Ballast Water Information Clearinghouse (NBIC) [156], totaling 4,926 records. This data set included information about vessel type (9 categories), ballast water discharge and gross weight tonnage. Records with missing information, zero discharges, or where recorded ballast water discharge exceeded recorded ballast water capacity were removed, leaving 1,280 valid records. Since ships sailing to and through the Arctic face unique climate conditions and have distinct patterns (frequencies and amounts) of ballast water update/discharge, we subset the NBIC data to include only voyages to and between Arctic ports, and used that subset in our estimates of ballast water discharge patterns.

Georeferenced records of species occurrences for soft-shell clams and red king crabs were obtained from the Global Biodiversity Information Facility (GBIF) [212] based on the observation as of March 2017. Temperature and salinity tolerance data for of soft-shell clams and red king crabs were obtained from literature sources [54, 117, 142, 206]. Information about annual average water temperature and salinity conditions were obtained from the Global Ports Database [120] wherever possible, and complemented with the World Ocean Atlas [11] by looking up the surface water condi-tions closest to ports' locations. Ecoregion data was obtained from Marine Ecoregion of the World (MEOW) [205] and Freshwater Ecoregion of the World (FEOW) [1].

9.5.2  Calculation of Invasion Risks

The relative risk of invasive species introduction was first calculated for every ship movement in the Lloyd's data, then aggregated for every pathway. Inspired by the introduction risk equation from Seebens et al. [193], for a ship $s$ making the trip $t$ from port $i$ to port $j$ which took $\Delta_{i \to j}^{(t)}$ days and discharged $D_{i \to j}^{(t)}$ ballast water, the relative risk of invasion for this trip is:

$$P_{i \to j}^{(t)} = (1 - e^{-\lambda D_{i \to j}^{(t)}})e^{-\mu \Delta_{i \to j}^{(t)}}$$

based on the intuition that species has higher chance of being transported through the trip if there was larger amount of ballast water discharge, or if the trip was short increasing the probability of species survival in the ballast. The duration $\Delta_{i \to j}^{(t)}$ of trip $t$ was taken from the Lloyd's data, and the daily species mortality rate is available from the Lloyd's data, and the daily species mortality rate $\mu = 0.02$ was chosen based on the work of Seebens et al. [193]. The species introduction potential per volume of discharge parameter $\lambda$ was given as $\lambda = 3.22 \times 10^{-6}$ based on Xu et al. [228], so that $P_{i \to j}^{(t)}$ is 0.8 when ballast discharge volume is $500,000 m^3$ and trip duration is zero. The volume of ballast water $D_{i \to j}^{(t)}$ translocated by trip $t$ made by a ship with type $k$ and gross weight tonnage $GWT$ is estimated using a modified version of the approach of Seebens et al. [193]:

$$D_{i \to j}^{(t)} = Z_k W_{GWT}$$

where $Z_k$ is the fraction of non-zero releases for ship type $k$, and $W_{GWT}$ is the estimated discharge in metric tons for a ship with gross weight tonnage GWT. The

TABLE 9.6

THE FRACTION OF NON-ZERO RELEASES (Z)

| Vessel type | Z |
|---|---|
| Bulker | 0.94 |
| Reefer | 0.38 |
| General cargo | 0.28 |
| Unknown | 0.23 |
| RoRo | 0.20 |
| Other | 0.14 |
| Tanker | 0.11 |
| Container | 0.06 |
| Passenger | 0.05 |

150 ship types in the LLI data were mapped to the 9 types present in the NBIC data and given ship type k from the 9 types in the NBIC data, the ballast water discharge frequency $Z_k$ is computed based on the NBIC data, yielding the mapping of $k \to Z_k$ in Table 9.6. We estimated $W_{GWT}$, by removing the zero discharge records from the NBIC data, randomly splitting the data into training (70%) and testing (30%) data sets, and fitting a random forest regression to the training data in R. The random forest regression predicted $W_{GWT}$ as a function of ship type and gross weight tonnage and was validated using the testing set yielding an $R^2$ of 0.93 (Figure 9.10).

We computed the invasion risk $P_{i \to j}^{(t)}$ based on ship size, ship type, and trip duration. Assuming $P_{i \to j}^{(t)}$ are independent, then the aggregated probability of invasion for a pathway $i \to j$ is:

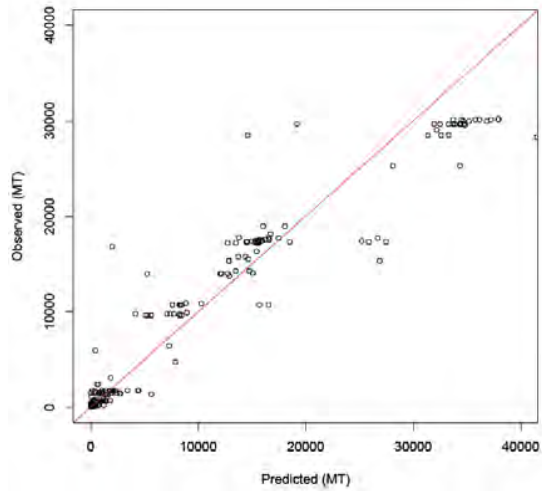

Figure 9.10. Observed ballast discharges versus predicted ballast discharges, showing prediction performance of the ballast discharge modeling. 1:1 regression line plotted for reference in red.

$$P_{i \to j} = 1 - \prod_t (1 - P_{i \to j}^{(t)})$$

and the aggregated probability of invasion for a given target port $T$ is:

$$P_T = 1 - \prod_S (1 - P_{S \to T})$$

9.5.3    Species-specific Case Studies

Given a species of interest (soft-shell clam or red king crab), we start with pathways that have observed ship movements in the Lloyds data. We first filter the pathways based on the GBIF species occurrence data: for each pathway, species observations must already exist within 200km of the source port, but not yet exist within 200km of the target port. Using the annual average temperature and salinity as recorded in the Global Ports Database and the World Ocean Atlas, these poten-tial introduction, dispersal, and stepping stone pathways are filtered to include those with target ports within the environmental tolerance for a given species (2 - 28 °C and 1.5 - 35 PSU for soft-shell clam and -2 - 18 °C and 28 - 35 PSU for red king crab) [54, 117, 142, 206].

### 9.5.4 Projecting the Risk of Invasion per Pathway

The observed general increase in invasion risk is a result of multiple trends including higher shipping frequency, larger ship capacity, shorter voyage times and so on. Moreover, the inter-dependencies of these underlying factors are not directly clear. Instead of predicting these underlying factors separately and aggregating them assuming independence, to make the least assumptions, we use the single depended variable: per-pathway species invasion risk (the bottom two subfigures in Figure 9.9) for projection.

### 9.5.5 Simulating the Topological Evolution of Pathways

Given the introduction or dispersal pathways in 2012, we repeat the following rewiring procedure: randomly choose a pathway, keep the strength of connection, and rewire the pathway to connect to a new target port (which cannot be the current source port); the probability of each port being chosen as the new target port is proportional to the number of incoming pathways. Therefore, the rewiring procedure is more likely to attach a pathway to a hub port that is already highly connected, resulting in a decrease of number of recipient ports. The rewiring process is repeated until the projected number of recipient ports has been reached (50 ports for introduction pathways in 2027, and 64 ports for dispersal pathways in 2027). Finally, the resulting power-law coefficient is tested to see if it deviates more than 5% from the projected power-law coefficient (in 2027 -0.398 for introduction pathways and -0.615 for dispersal pathways). If this is the case then the simulation is restarted. Therefore, the simulation guarantees the resulting re-wiring matches the projected number of recipient ports and the power-law coefficient in Figure 9.9. Despite the stochastic nature of this simulation, given the constraints on the two parameters, the resulting aggregated risks at ports in 2027 (the map in Figure 9.11) is stable.

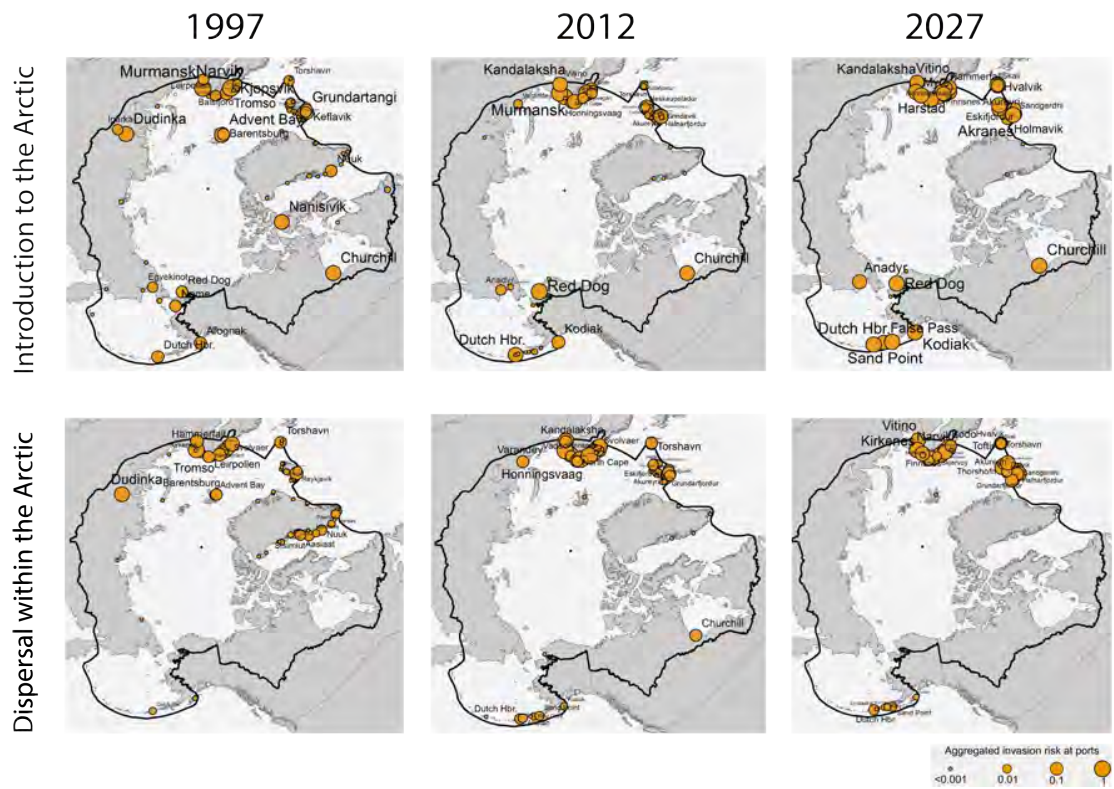

Figure 9.11. The projected introduction and dispersal risks for Arctic ports in 2027, along with 1997 and 2012 for reference.

# CHAPTER 10

# DETECTING ANOMALIES IN SEQUENTIAL DATA WITH HON

## 10.1 Overview

A major branch of anomaly detection methods rely on dynamic networks: raw sequence data is first converted to a series of networks, then critical change points are identified in the evolving network structure. However, existing approaches use first-order networks (FONs) to represent the underlying raw data, which may lose important higher-order sequence patterns, making higher-order anomalies undetectable in subsequent analysis.

**Intellectual merit:** By replacing FONs with higher-order networks (HONs), we show that existing anomaly detection algorithms can better capture higher-order anomalies that may otherwise be ignored. We construct a large-scale synthetic dataset with 11 billion shipping movements that verifies the effectiveness of the proposed method in capturing variable orders of anomalies. We also test our method with real-world taxi trajectory data, showing the ability of proposed method in amplifying anomaly signals when higher-order anomalous patterns exist.

**Connections:** This research is an application based on the HON representation in Chapter 3 and 4. It can serve as a module for the visualization framework in Chapter 5.

**Work status:** This work is accomplished in collaboration with the U.S. Army Research Lab Network Science Collaborative Technology Alliance (ARL NS-CTA). Prof. Bruno Ribeiro from Purdue University also contributed to the formalization of the problem. It is currently under review at ICDM 2017.

## 10.2 Introduction

Anomalies are valuable to detect in the ever-increasing flow of data, as they are indicative of deviations that may deem immediate attention and intervention. City managers detect anomalies in urban traffic to respond proactively to incidents; football coaches detect anomalies in football passing patterns to identify changes in opponent's strategies; banks detect anomalies in signatures to identify fraudulent signatures; governments detect anomalies to identify terrorist activities. For all these real-world applications, it is critical that anomaly detection algorithms do not leave anomalous signals undetected.

Nevertheless, certain types of anomalies are more difficult to identify than others. This is especially true when modeling a large complex system as a network, where the anomalies manifest themselves in the network dynamics. A comprehensive review of detecting change points in the evolving networks is presented in [7]. However, all approaches in this direction use the first-order network (FON) to represent the underlying raw data (such as trajectories or event sequences). As shown in the related work, FON can lose important trajectory information inherent in the raw data [229], and thus any anomaly detection can lead to incorrect outcomes. If the network representation loses important movement pattern information in the first place, the subsequent anomaly detection algorithms that rely on accurate network representations may not capture changes in higher-order movements.

**Example.** Fig. 10.1 illustrates the challenge of detecting such anomalies. Given four vehicle trajectories spanning two days as the input, conventionally, a traffic network is built for each day, with the nodes representing locations and edges representing the traffic between locations. This dynamic traffic network is then monitored; a change in the network topology indicates an anomaly in traffic patterns. According to the original trajectories, in Day 1, where a vehicle goes from $c$ does not depend on where the vehicle comes to $c$. In Day 2, however, all vehicles coming from $a$ to $c$ go to $d$, and all coming from $b$ to $c$ go to $e$; namely, second-order movement patterns have emerged. Since the first-order (pairwise) traffic of $a \to c$, $b \to c$, $c \to d$ and $c \to e$ remain the same in Day 1 and Day 2, the first-order traffic network in the two days are exactly the same; therefore, network-based anomaly detection algorithms cannot capture the emergence of this second-order anomaly by monitoring the first-order traffic network.

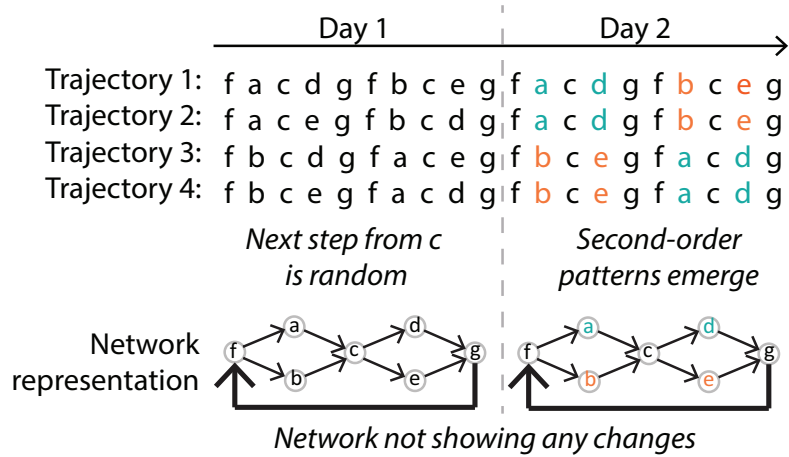

Figure 10.1. Higher-order anomalies not captured by the conventional network-based anomaly detection methods.

**Contributions.** We propose a high-order network approach that can effectively capture higher-order anomalies. By replacing the first-order network (FON) with the higher-order network (HON) in the dynamic network representation of raw trajectory data, we show existing anomaly detection algorithms can capture higher-order anomalies that may otherwise be ignored. We evaluate the effectiveness of the proposed method on detecting anomalies in sequences of variable dependency orders through a large-scale synthetic data with 11 billion movements, and show various scenarios where our method captures high-order dependency changes when conventional methods using FONs fail. Finally, we apply the proposed method on a real-world taxi trajectory data, showing its ability in amplifying higher-order anomaly signals.

## 10.3 Related Work

**Anomaly detection in sequential data.** There are two categories of sequential data anomaly detection tasks. Suppose the input data has trajectories of multiple cars, one direction is to identify the few cars with trajectories significantly different than others (revealing traffic violation, dangerous driving, etc.); this direction has been studied with sequence similarities [47, 203], sliding windows [224], Markovian methods [146, 176, 208], HMM methods [131], and so on. The other direction is to identify the times when most cars suddenly change their movement behaviors (revealing traffic incidents, major events, etc.). It is also known as "event detection", which has been studied with sliding windows [121–123] and dynamic networks [7]. A more comprehensive review is in [48]. Our study focuses on identifying anomalous time windows rather than individual trajectories, thus relate closely with the second category.

**Anomaly detection in dynamic networks.** Unlike the task of detecting anomalous nodes and edges in a single static network, anomaly detection in dynamic networks [7, 48] uses multiple snapshots of networks to represent the interactions of interest (such as interacting molecules [173], elements in frames of videos [185], flow of invasive species [228], etc.), then identifies the time when the network topology shows significant changes. Our work follows this general framework, with innovations in network constructions.

## 10.4 Methods

**Definition.** The procedure of a network-based anomaly detection method takes as the input the sequential data $\mathcal{S} = [S_1, S_2, \ldots, S_T]$ divided into $T$ time windows, namely $t \in [1, T]$. In each time window, the sequential data is represented with a network, i.e., $S_i \to G_i$, yielding a dynamic network $\mathcal{G} = [G_1, G_2, \ldots, G_T]$. The dynamic network $\mathcal{G}$ is then used to find the changing point(s) $c \in [1, T]$ when $G_{c-1}$ and $G_c$ are significantly different. The changing points (as the outputs) correspond to the anomalous events in the raw data.

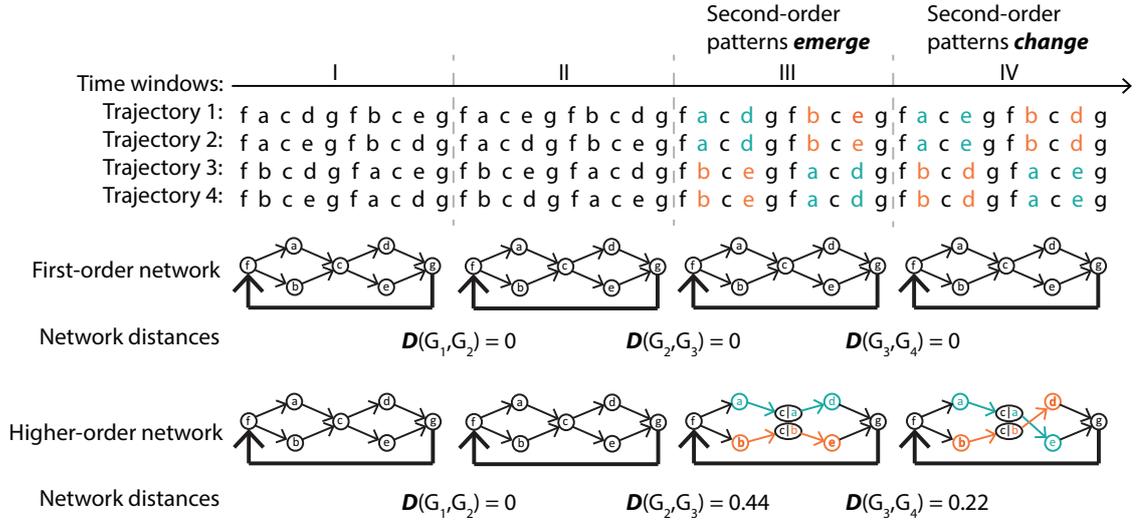

Figure 10.2. Comparing anomaly detection based on the first-order dynamic network and the higher-order dynamic network.

The basic idea behind dynamic network anomaly detection is to apply a network distance metric $\mathcal{D}$ to measure the degree of changes in neighboring networks $d_t = \mathcal{D}(G_t, G_{t+1})$, then the problem is reduced to finding the anomalies in the time series of $[d_1, d_2, \ldots, d_{T-1}]$.

We propose to use, as foundation, the higher-order network (HON) [229] $\mathcal{G}$ to represent the underlying sequential data $\mathcal{S}$ for the $S_i \to G_i$ step. HON can capture higher-order dependencies in the raw data, and reflect such information in network topologies via higher-order nodes and edges. By using HON, existing anomaly detection algorithms can discover *higher-order anomalies* that may otherwise remain undetected. We introduce several innovations to the original HON algorithm, notating it as HON+, especially in making it parameter free and scalable. This scalability is critical for applicability to large networks.

**Example.** Fig. 10.2 illustrates higher-order anomalies, and how HON can help discover such anomalies. Suppose there are four trajectories in the raw data. In time window I, the traffic going through $c$ randomly goes to $d$ or $e$, regardless whether the traffic came to $c$ from $a$ or $b$. Given these observations, a FON and a HON can be built; the HON in this case is identical to the FON, since the movements are random and there are no higher-order dependencies. In time window II, the movements are still random, with 50% traffic from $c$ going to $d$ and another 50% going to $e$. Therefore, both FON and HON remain the same as time window I, showing network distances of 0.

In time window III, second-order patterns *emerge*: all traffic from $a$ to $c$ goes to $d$, and all traffic from $b$ to $c$ goes to $e$. Since the aggregated traffic from $c$ to $d$ and $e$ remains the same, FONs remain *exactly the same*. As a result, the network distance to the previous time window is 0, missing this newly emerged pattern. Conversely, HON uses additional higher-order nodes and edges to capture higher-order dependencies. Because the traffic distribution from $a \to c$ changed from 50% $d$ and 50% $e$ to 100% $d$, the HON algorithm creates a new node $c|a$ (representing $c$ given the last step being $a$), and the path $a \to c \to d$ now becomes $a \to c|a \to d$. Similarly, another node $c|b$ was created. Therefore, the emergence of the second-order pattern in the raw data is reflected by the non-trivial changes in the topology of HON. Due to the complete changes in four out of the nine edges, the network distance $\mathcal{D}(G_2, G_3) = 0.44 > 0$, successfully captures this *higher-order anomaly*, i.e., we detect a significant change in higher-order movement patterns.

In time window IV, the second-order movement pattern *changes*: all traffic from $a$ to $c$ now goes to $e$, and all traffic from $b$ to $c$ now goes to $d$. Since the aggregated traffic from $c$ to $d$ and $e$ remains the same, FON remains the same, again failing to reflect the changes. However, HON captures the changes by connecting $a \to c|a \to e$ and $b \to c|b \to d$. The complete changes in two out of nine edges results in a network distance of 0.22, again capturing the changes in higher-order movement patterns.

As for the network distance metric $\mathcal{D}$, we use a simple weight distance [199], defined as

$$\mathcal{D}(G, H) = \frac{\sum\limits_{u,v \in V} \frac{|w_E^G(u,v) - w_E^H(u,v)|}{max(w_E^G(u,v) - w_E^H(u,v))}}{|E_G \cup E_H|}$$

with $w$ being the edge weights and $|E|$ being the total number of edges. Note that the proposed HON-based anomaly detection framework can also use other distance metrics designed for weighted directed networks, such as MCS Weight [199], Graph Edit Distance for weighted networks [87], Modality [126], Entropy and Spectral methods [170].

## 10.5 Results

### 10.5.1 Large-scale Synthetic Data

#### 10.5.1.1 Data Preparation

We first use synthetic data with known higher-order anomalies to test the effectiveness of the HON-based anomaly detection method. For a comprehensive test, we aim to synthesize input trajectories with variable orders of movement patterns. We fulfill this task by starting from the basic case, then gradually adding or changing higher-order movement rules, and see if the proposed method can successfully identify these anomalies.

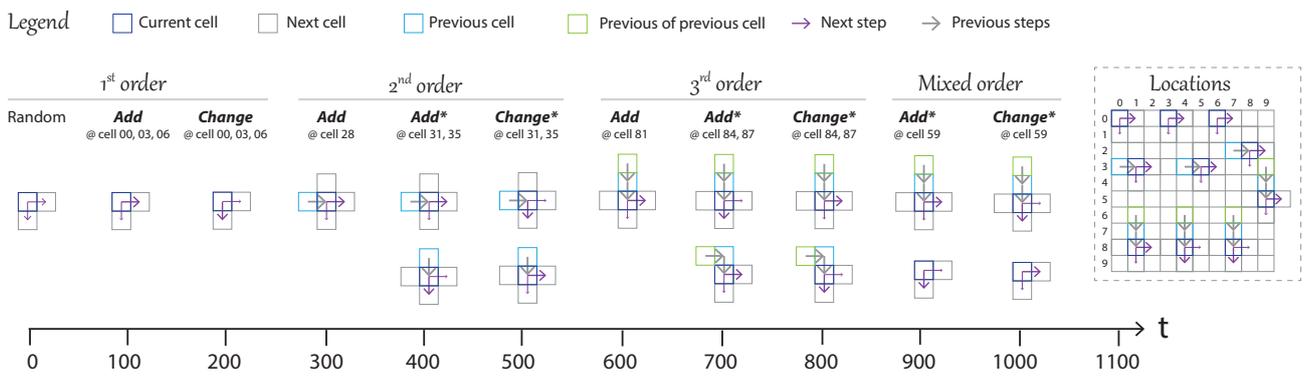

Figure 10.3. Construction of synthetic data. How variable orders of movement patterns are synthesized.

The full process to synthesize the input trajectories is illustrated in Fig. 10.3. To begin with, we assume there are 100,000 ships moving on a 10x10 grid, with cells numbered from 00 to 99 (the "locations" inset on the right of Fig. 10.3). At each time stamp, every ship moves 100 steps, resulting in 10,000,000 movements. For each of the following 11 cases, we maintain the movement simulation rules for 100 time windows. In total, we generate 11,000,000,000 movements for the subsequent anomaly detection task.

**Initial random movement case.** At $t = [0, 99]$, each ship can move to the neighboring cell in each step, either moving right or moving down with 50/50 chances, with wrapping (moving right on the rightmost cell will end up at the leftmost cell in the same row, and moving down on the bottom cell will end up at the top cell in the same column).

**Emergence of first-order dependency.** At $t = [100, 199]$, we impose the following first-order rule of movement: all ships coming to cell 00, 03 and 06 will have 90% chance of moving to the right and 10% chance of moving down in the next step. The locations of these dependency rules are highlighted in the "locations" grid. Since the traffic between pairs of cells (00–01, 00–10, 03–04, 03–13, 06–07 and 06–16) will change, both FON and HON should be able to reflect this change in pairwise connection at $t = 100$.

**Change of first-order dependency.** At $t = [200, 299]$, we change the existing first-order movement rules: all ships coming to cell 00, 03 and 06 will now have 90% chance of moving down in the next step, and 10% chance of moving right. This change at $t = 200$ should also be reflected in both FON and HON.

**Emergence of second-order dependency.** At $t = [300, 399]$, we keep the previous first-order rules, and impose a new second-order rule: all ships coming from cell 27 to 28 will have 90% chance of moving to the right in the next step, and 10% chance of moving down (also highlighted in "locations"). This change at $t = 300$ should be reflected in HON; since it also influences the first-order traffic (traffic of $27 \to 28 \to 29/38$ changes from 1:1 to 7:3), FON should show minor changes.

**Emergence of complementary second-order dependencies.** At $t = [400, 499]$, we keep the previous first-order and second-order rules, and impose a pair of new second-order rules: (1) all ships coming from cell 30 to 31 (and 34 to 35) will have 90% chance of moving to the right in the next step, and 10% chance of moving down; (2) all ships coming from cell 21 to 31 (and 25 to 35) will have 90% chance of moving down, and 10% chance of moving right (also highlighted in "locations"). The combined effect of these two complementary second-order dependencies is that the first-order traffic from cell 31 and 35 remain unchanged. Therefore, this change at $t = 400$ will *not* be expected to incur any changes in FON, but will introduce new higher-order nodes and edges in HON.

**Change of complementary second-order dependencies.** At $t = [500, 599]$, we keep the previous first-order and second-order rules, and flip the rules for the complementary second-order dependencies: (1) all ships coming from cell 30 to 31 (and 34 to 35) will have 90% chance of moving down, and 10% chance of moving right; (2) all ships coming from cell 21 to 31 (and 25 to 35) will have 90% chance of moving right, and 10% chance of moving down. This change at $t = 500$ shall be reflected in HON, but shall *not* be expected to incur any changes in FON.

**Emergence of third-order dependency.** At $t = [600, 699]$, we keep the previous first-order and second-order rules, and impose a new third-order rule: all ships coming from cell 61 through 71 to 81 will have 90% chance of moving to the right in the next step, and 10% chance of moving down (also highlighted in the "locations" grid). This change at $t = 600$ should be reflected in HON; since it influences the first-order traffic slightly (from 1:1 to 3:2), FON will show minor changes.

**Emergence of complementary third-order dependencies.** At $t = [700, 799]$, we keep the previous first-order, second-order, and third-order rules, and impose a pair of new third-order rules: (1) all ships coming from cell 64 through 74 to 84 (and 67 through 77 to 87) will have 90% chance of moving to the right in the next step, and 10% chance of moving down; (2) all ships coming from 73 through 74 to 84 (and 76 through 77 to 87) will have 90% chance of moving down, and 10% chance of moving right (also highlighted in the "locations" grid). The combined effect of these two complementary third-order dependencies is that the first-order traffic from cell 84 and 87 will not change. This change at $t = 700$ should be captured by HON but not FON.

**Change of complementary third-order dependencies.** At $t = [800, 899]$, we keep the previous first-order, second-order, and third-order rules, and flip the rules for the complementary third-order dependencies. Again, this change at $t = 800$ should be captured by HON but not FON.

**Emergence of complementary mixed-order dependency.** At $t = [900, 999]$, we keep the previous first-order, second-order, and third rules, and impose a new third-order rule and a first-order rule: (1) all ships coming from cell 39 through 49 to 59 will have 90% chance of moving to the right in the next step, and 10% chance of moving down; (2) all ships at cell 59 will have 11/30 chance of moving right and 19/30 chance of moving down (also highlighted in the "locations" grid). The combined effect of the complementary third-order and first-order dependencies is that the first-order traffic from cell 84 and 87 will not change. This change at $t = 900$ should be captured by HON but not FON.

**Change of complementary mixed-order dependency.** At $t = [1000, 1099]$, we keep the previous first-order, second-order, and third rules, and flip the rules for the mixed-order dependencies. This change at $t = 1000$ should be captured by HON but not FON.

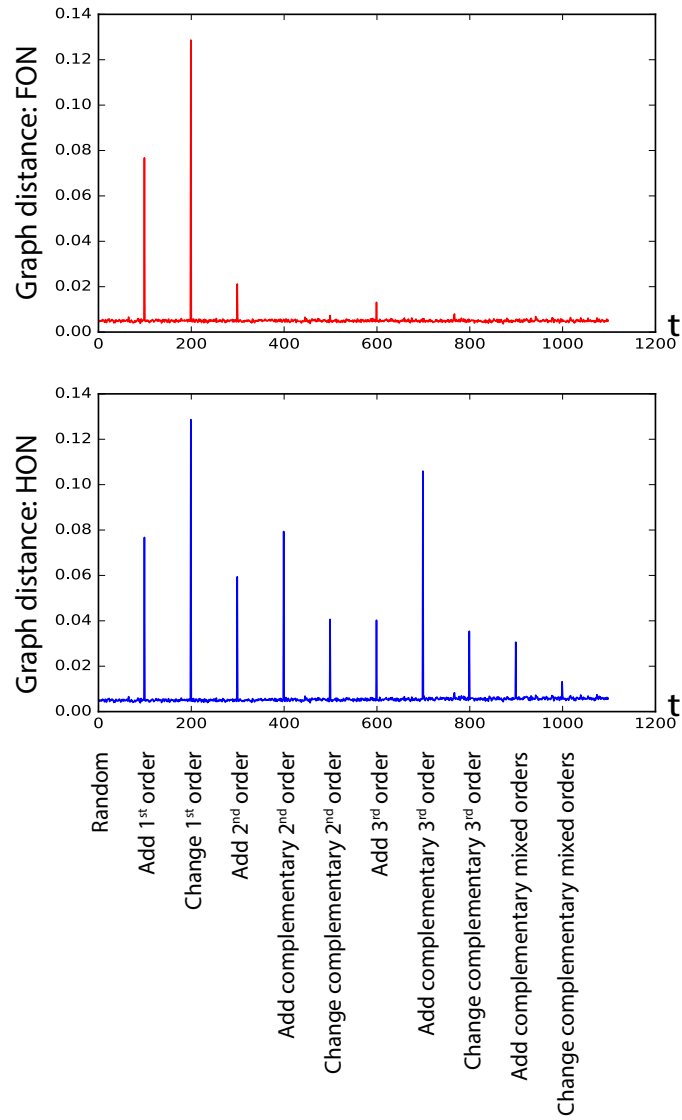

Figure 10.4. Graph distances on dynamic FON and HON.

10.5.1.2   Anomaly Detection Results

We compare anomaly detection results using the dynamic network of FON and HON representations. The graph distances of each neighboring time window is shown in Fig. 10.4. We observe that both FON and HON can capture the addition and changes in first-order movement patterns ($t = 100, t = 200$). As expected, the addition of second-order ($t = 300$) and third-order rules ($t = 600$) also slightly changes the first-order traffic. Therefore, FON does reflect those changes, but the spikes incurred are not as significant as when changes are made directly to first-order rules. Meanwhile, HON shows strong signals for the addition of first-order and second-order rules, due to the addition of higher-order nodes and edges. Finally, the addition and changes of complementary rules of second order, third order, and mixed orders do not incur any changes in FON, since they do not change the first-order traffic. On the contrary, HON produces strong signals to these changes in complementary rules.

In brief, the large synthetic data with known dependencies supports the effectiveness of HON in identifying higher-order anomalies. If FON is used as the dynamic network for anomaly detection, it will yield weak signal or no signal for higher-order anomalies.

10.5.2   Real-world Porto Taxi Data

10.5.2.1   Data Preparation

The data we used was the ECML/PKDD 2015 challenge data[1], which contains one year (Jul. 1, 2013 to Jun. 30, 2014) of all the 442 taxi GPS trajectories in Porto, Portugal. The coordinates of each taxi was collected at every 15 seconds. To discretize the geolocation data into reasonable points of interests that are repre-sentative of population density, we mapped all coordinates to the nearest 41 police stations (Fig. 10.6(a)). We remove duplicate POIs in neighboring time windows for each trajectory. We take every week as a single time window (in order not to compare weekdays to weekends), yielding 52 time windows, each containing 442 trajectories of POIs. The trajectories in each week is used to construct the FON and HON traffic networks. The HON+ algorithm proposed in the Methods Section is used (with the optional parameters $ThresholdMultipler = 5$ and $MinSupport = 10$ to remove rare observations).

---

[1] http://www.geolink.pt/ecmlpkdd2015-challenge/dataset.html

### 10.5.3 Observations with FON and HON

Given the 52 networks for both FON and HON, we compute the graph distances for neighboring time windows, as in Fig. 10.5(a). While the trend of HON resembles that of FON, the graph distances between Week 44 (also Week 2) and the neighboring weeks are particularly more significant in HON than that in FON. Such differences are also indicated in the histograms of graph distances in Fig. 10.5(b) and (c), where the orange circles highlight the same anomalous signals, which is more significant on HON than on HON.

We focus on the case of Week 44 to understand why HON produces stronger signal than FON at this time window. We notice that Porto's second most important festival, "Burning of the Ribbons", lasts from May 4 to May 11 in 2014 and falls within Week 44 of our study. The festival involves parades, road closures, and is popular among tourists, which could be the underlying reason to the changes in taxis' movement patterns. After plotting the traffic HON of Week 43 and Week 44 in Fig. 10.6, we notice that multiple higher-order nodes and edges emerge in Week 44, indicating the emergence of higher-order traffic patterns. The newly emerged higher-order patterns correspond to police stations labeled from 9 to 14, which is where the event's main venue (Queimdromo in the City Park) and participating universities

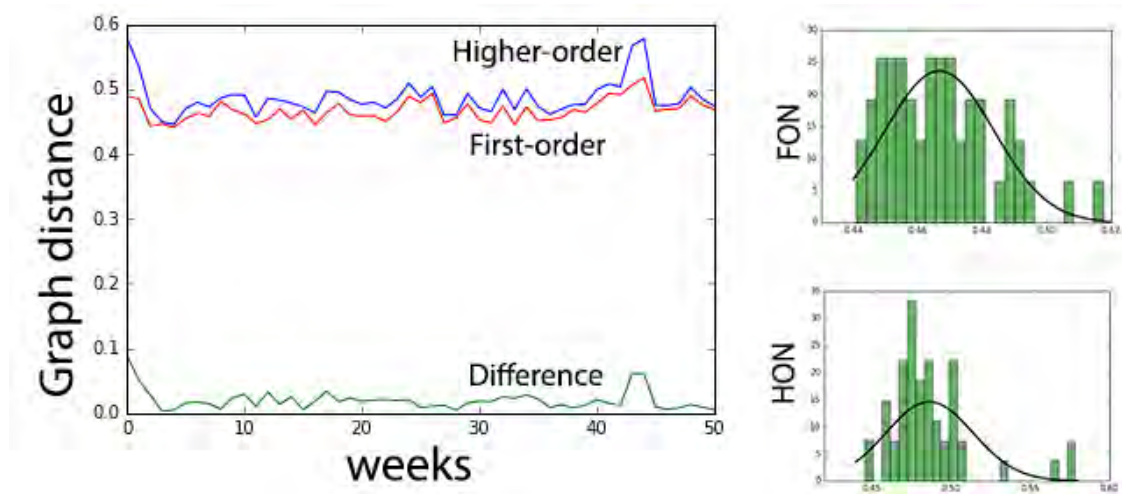

Figure 10.5. (**a**) Anomaly detection results on the dynamic network of FON and HON. (**b**) and (**c**) HON amplifies the anomalous traffic patterns.

locate at[2].

Recall that if there is a change in higher-order movement patterns that incurs changes in first-order movement patterns, HON can capture both, but FON can only capture the aggregated first-order traffic changes. In such cases, the bigger changes in HON graph distances can *amplify* the anomalous signal, and making anomalies that contain higher-order anomalies easier to be detected.

## 10.6 Discussion

This section presents a new network-based anomaly detection approach that is capable of detecting higher-order anomalies in sequences that cannot be captured by first-order networks (FONs). By replacing the FON with a high-order network (HON) in the dynamic network representation of raw sequence data, existing anomaly detection algorithms can capture changes in higher-order movement patterns that may have been ignored otherwise. With a 11 billion-movement synthetic data, we comprehensively test the effectiveness of HON in capturing anomalies with variable orders of dependencies. In real-world data HONs also show the ability to amplify anomaly signals.

---

[2] http://www.maiahoje.pt/noticias/ler-noticia.php?noticia=577

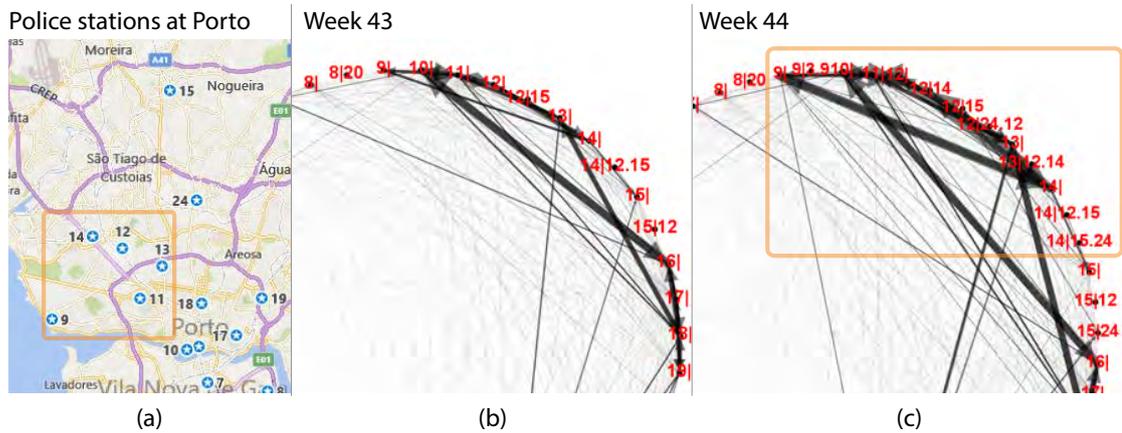

Figure 10.6. (**a**) Labeling of police stations in urban areas of Porto. (**b**) and (**c**) the emergence of higher-order traffic patterns in Week 44 ("Burning of the Ribbons" festival) captured by HON, corresponding to the highlighted region in (a).

There are multiple directions for future improvements. First, one can apply the proposed method to analyze more real-world data sets, such as football ball-passing data. With signature data, our method may also serve as an indicator of fraud detection. For terrorist activities or network hacking records, our method can help identify attacks and higher-order anomalies that may otherwise be ignored by first-order methods. Second, one can try other graph distance metrics, including those discussed in [170], and structure-based metrics [7] that factor in changes of clustering or ranking results, and local properties on the network. Finally, it would be of practical importance to integrate this module into the HON visualization framework HoNVis [211] to facilitate interactive exploration and decision making.

PART III

DIFFUSION DYNAMICS ON IMPLICIT SOCIAL NETWORKS

# CHAPTER 11

# MINING FEATURES ASSOCIATED WITH EFFECTIVE TWEETS

## 11.1 Overview

What tweet features are associated with higher effectiveness in tweets? The answer to this question is the basis for varieties of tasks including the analysis of advertising campaigns, the prediction of user engagement, and the extraction of signals for automated trading.

**Intellectual merits:** Through the mining of 122 million engagements of 2.5 million original tweets, we present a systematic review of tweet time, entities, composition, and user account features. We show that the relationship between various features and tweeting effectiveness is non-linear; for example, tweets that use a few hashtags have higher effectiveness than using no or too many hashtags. This research closely relates to various industrial applications that are based on tweet features.

**Connections:** The features discussed in this work serves as a foundation of the diffusion of retail traders' attention analysis on Twitter in Chapter 12.

**Work status:** This research is the joint effort with Prof. Nitesh Chawla. It has been accepted at Advances in Social Network Analysis and Mining (ASONAM) 2017.

## 11.2 Introduction

With 316 million monthly active users and 500 million tweets sent per day [217], Twitter is the most successful microblogging service, and is often compared with news media [129]. Every tweet (a message of at most 140 characters posted on Twitter) has the potential to reach millions of Twitter users, and is essentially a tiny advertisement (delivering information, message, idea or knowledge). An interesting fact is that a tweet posted by someone with more followers does not necessarily make a bigger impact [45]; instead, the effectiveness of a tweet is about the engagement rate that the tweet receives among the followers and beyond. The question is: *what features are associated with an effective tweet?*

Twitter is a gold mine for tracking and predicting the diffusion of information and public opinions. Stock trading platforms such as TD Ameritrade have recently integrated live Twitter feeds into their web trading interface, displaying company-related tweets to investors in real-time [9]. Tweets closely correlates with retail traders' attention [49, 88], and can move stock prices [50]. Similarly, politicians hire analysts to understand public opinions on Twitter and what campaigns will be effective [42, 57, 132, 202, 216], governments monitor tweets to detect emerging events [58, 200, 225], and so on. A systematic understanding of features that are closely related to effective tweeting is a first step to extracting valuable for building predictive models.

The study of what features are associated with a "viral" tweet has attracted much attention in the recent years, mostly with regard to the *retweeting* behavior. There has been in-depth analysis on certain dimensions of tweets, such as sentiment analysis [3, 16, 128, 168], Tsur *et al.* [215] on multiple features of hashtags, Naveed *et al.* [155] on tweet content such as emoticons, Dabeer *et al.* [65] and Liu *et al.* [135] on timing of tweets, and so on. The correlation between these features and the number of retweets has been studied using conventional statistical methods such as Principle

Component Analysis [207], generalized linear model [116] and so on. Furthermore, retweet prediction is studied using both statistical and machine learning methods [51, 230, 232]. Related fields such as social influence [18, 45], information diffusion and network analysis on social networks have also used retweet dynamics [5, 19, 72, 231].

The first problem is that although *favorites and replies* account for 47% of user engagements (in our data of 122 million engagements), in previous studies they are rarely factored into the effectiveness of a tweet. The second problem is the *assumption of linear relationship* between tweet features and tweet effectiveness, as seen in the prediction of user engagement [49, 207]. Such linear models can only represent either "longer tweets are more effective" or "shorter tweets are more effective", but cannot capture non-linear relationships.

We present a systematic review and analysis of factors associated with the effectiveness of a tweet:

- We factor all three forms of engagements (retweets, favorites, and replies) for a more comprehensive measurement of a tweet's effectiveness, and provide comparisons with previous work.

- We discuss if the posting time, entities (hashtags, pictures, etc.), the composition (length, sentiment, etc.), and account features have association with tweeting effectiveness. We also analyze new features that have not been analyzed before, such as embedding videos and gifs in tweets and the usage of third-party tools.

- We show that the relationship between various features and tweeting effectiveness is non-linear; for example, using a certain number of hashtags is more effective than using no or too many hashtags. The non-linear correlations suggest important design considerations toward accurate prediction of tweeting effectiveness.

## 11.3 Materials and Methods

### 11.3.1 Data Preparation

There are two major differences in data collection compared with existing work. While related studies [116, 202, 207] use Twitter's REST API to crawl users' historical tweets (which is limited to at most 3,200 most recent tweets from any account), we used the Streaming API[1] to continuously collect tweets in real-time, yielding a more comprehensive corpus of tweets from Nov. 1, 2013 to Apr. 30, 2015 (18 months in total). The Streaming API yields 121,772,646 user engagements (retweets, favorites, and replies) on 2,452,120 original tweets posted by the 258 accounts we monitored. We have made the tweet and user IDs publicly available at http://dx.doi.org/10.6084/m9.figshare.1548304[2]

Another major difference is that we chose to monitor fewer accounts but collect their continuous tweeting history, as opposed to related works [57, 167, 207] that used the "sample" endpoint of public Streaming API which yields a random 1% sample of overall tweets. Although the "sample" endpoint returns tweets from all Twitter users, the random 1% sample is too sparse to reflect the exact user engagement [153]. For example, when a tweet receives 100 retweets, the "sample" endpoint may only capture the $20^{th}$, drastically underestimating user engagement. We collected tweets posted by 258 active Twitter accounts in four categories: 65 official Twitter accounts of major media outlets (e.g., @nytimes, @WSJ), 138 official Twitter accounts of S&P500 companies (e.g., @Starbucks, @Walmart), 24 CEOs of big companies (e.g., @elonmusk, @WesternUnionCEO), and 31 famous investors (e.g., @Carl_C_Icahn, @christine_benz). These accounts are manually verified as the official accounts, and are actively posting tweets during our observation.

---

[1] https://dev.twitter.com/streaming/overview

[2] This data set is used for the following analysis; all times are in U.S. Eastern Time unless otherwise specified.

### 11.3.2 Tweeting Effectiveness

We define the *effectiveness* $E$ of a tweet $t$ composed by user $u$ as:

$$E(t, u) = \frac{R(t) + F(t) + C(t)}{N(u)} \quad (11.1)$$

where $R(t)$ is the number of retweets, $F(t)$ the number of favorites, $C(t)$ the number of comments (replies), respectively — these three combine as the overall user engagement. On the denominator, $N(u)$ is the number of followers. This definition of a tweet's effectiveness follows the intuition that **(i)** given the same number of followers, a tweet receiving more user engagement is more effective, and **(ii)** given the same overall user engagement, the tweet posted to fewer followers is more effective. This definition implies that having more followers does not indicate the user's tweets are more effective, as shown in Fig. 11.1(a). For example, @Disney has nearly the same number of followers as @WholeFoods, but the average effectiveness of tweets posted by @WholeFoods is 471 times lower than that of @Disney.

By including favorites and replies, our definition of tweet engagement aims at providing a more comprehensive measurement of tweeting effectiveness, and follows the official definition of engagement[3]. Our approach differs from[51, 116, 207, 230, 232] which focus on retweets, [65] which uses the timeliness of retweets a tweet receives, and [184] which considers retweets and replies but not favorites. Our motivation is that the number of retweets is insufficient for reflecting the overall user engagement; the proportion of retweets in all three forms of user engagement is not constant in different tweets, as shown in Fig. 11.1(b). For example, a tweet like "How was your last experience flying with American Airlines?" is more likely to receive replies rather than retweets. In brief, the number of retweets is insufficient for reflecting the overall user engagement.

---

[3]https://business.twitter.com/basics/how-to-create-a-twitter-content-strategy

## 11.4 Results

### 11.4.1 Time to tweet

**Weekends correlate with higher engagement.** Tweets posted in different days of week correlate with different effectiveness. As shown in Fig. 11.2, tweets posted at weekends (Saturdays and Sundays) generally attract more interest than those posted on weekdays; in particular, tweets posted on Sundays are relatively 30% more effective than those posted on Thursdays. A possible reason to the high engagement rate during weekends is people staying at home for longer hours and spending more time on Twitter. It is worth noting that the accounts we monitor post only half as many tweets at weekends compared to weekdays.

**Hour of tweet matters.** Previous study [184] suggests that posting tweets during the daytime is generally more effective than posting at night. However, after breaking down by different types of accounts, our research shows that for accounts owned by media companies, tweeting at night is associated with higher effectiveness (40% more than during daytime), as in Fig. 11.2. Moreover, for media outlet accounts, the peak engagement rate hours align with the "prime-time" of television (7 PM – 11 PM across Continental U.S.) [14], which is known to attract the most viewers. This could be the joined effect of (1) people having more time on their smart phones after work and (2) an increasing number of TV programs are adopting Twitter as a channel to interact with their audiences in real-time. Nevertheless, media outlet accounts post twice in working hours (9 AM – 5 PM), mismatching the peak engagement hours at night. On the contrary, Twitter accounts of S&P500 companies have higher engagement during working hours (9 AM – 5 PM), and that aligns well with their tweeting volumes.

**Tweeting frequently correlates with lower effectiveness.** In the log-log scale plot of Fig. 11.4, posting ten times more frequently associates with about ten times lower effectiveness for every tweet. In particular, S&P500 companies tend to flood users' timelines, and have lower tweeting effectiveness in general (the purple × marks on bottom right, such as @WholeFoods, @Walmart, and @ChipotleTweets); CEOs of big companies tweet only occasionally, but every word of them counts (the blue ○ marks on the top left, such as @dkhos and @Donahoe_John). A compelling inference is: *the overall user engagement per day is not influenced by tweeting frequency.*

11.4.2 Entities in Tweets

**Non-linear correlation between #URLs and effectiveness.** Fig. 11.5-(a) shows that having a few URLs in the tweet is associated with higher effectiveness; in particular, having three URLs are associated with eight times the effectiveness compared to tweets without URLs. Media outlets have higher adoption rates to URLs (Fig. 11.5-(b)), by putting the headline in the tweet and including a few URLs to the full story. On the other hand, tweets are not simply "the more URLs the more effective". Having too many (four or more) URLs take a significant proportion of valuable 140 characters in the twee; if the idea is not clearly conveyed by the tweet itself, readers are less likely to engage with the tweet. For example, the following tweet receives no engagement at all:

> "Kiev unrest: LIVE: http://t.co/YoGbFPeM3E PHOTOS: http://t.co/fjRznDUAyF Why? http://t.co/EUEO4SgRN3 http://t.co/9S1hPFNaf2" — @ABC

**Non-linear relationship between #hashtags and effectiveness.** As shown in Fig. 11.6-(a), having no more than 10 hashtags is associated with 2–3 times higher effectiveness than having no hashtags. Having a few hashtags not only clearly identifies the topic of the tweet, but also makes the tweet more discoverable via searching, browsing tweets of the same topic, and clicking on the "trending" dock; these multiple channels of exposure can be the reason of higher effectiveness. Meanwhile, tweets having more than ten hashtags are hard to read and are associated with lower effectiveness.

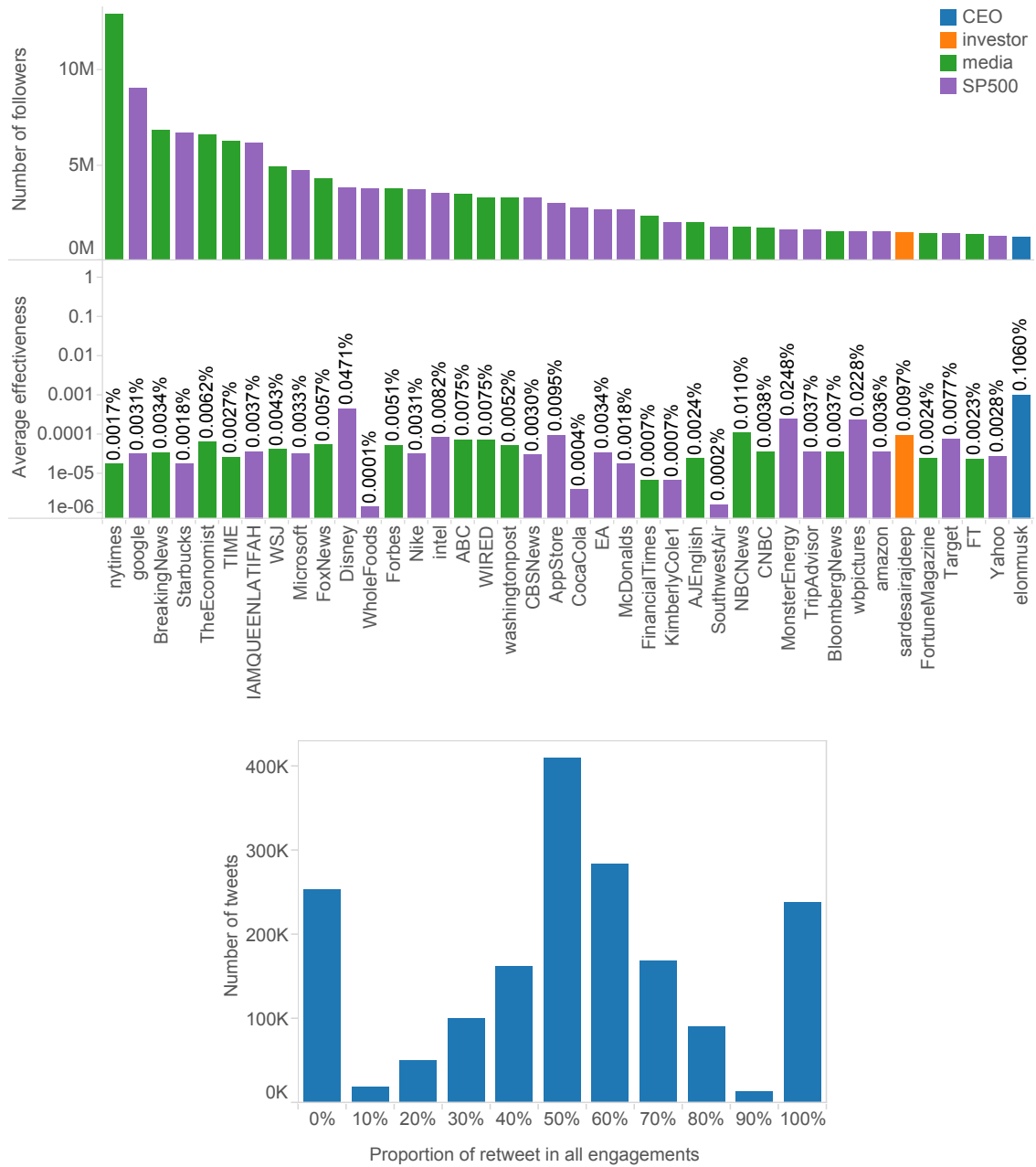

Figure 11.1. **(a)** Tweeting effectiveness for 40 accounts with the most followers in our data. For accounts with similar number of followers (e.g., @Disney and @WholeFoods), their tweeting effectiveness can differ by hundreds of times (note that effectiveness is shown in log-scale). **(b)** For every tweet, the proportion of retweet in all engagements (retweets, favorites, replies), aggregated in histogram.

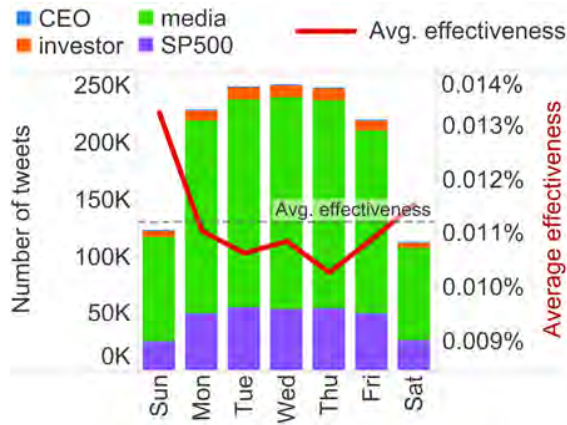

Figure 11.2. **Tweeting effectiveness versus different days of week, and the actual tweets per day, for non-reply tweets.** Weekends are the best time to engage followers, but only half as many tweets are posted at weekends.

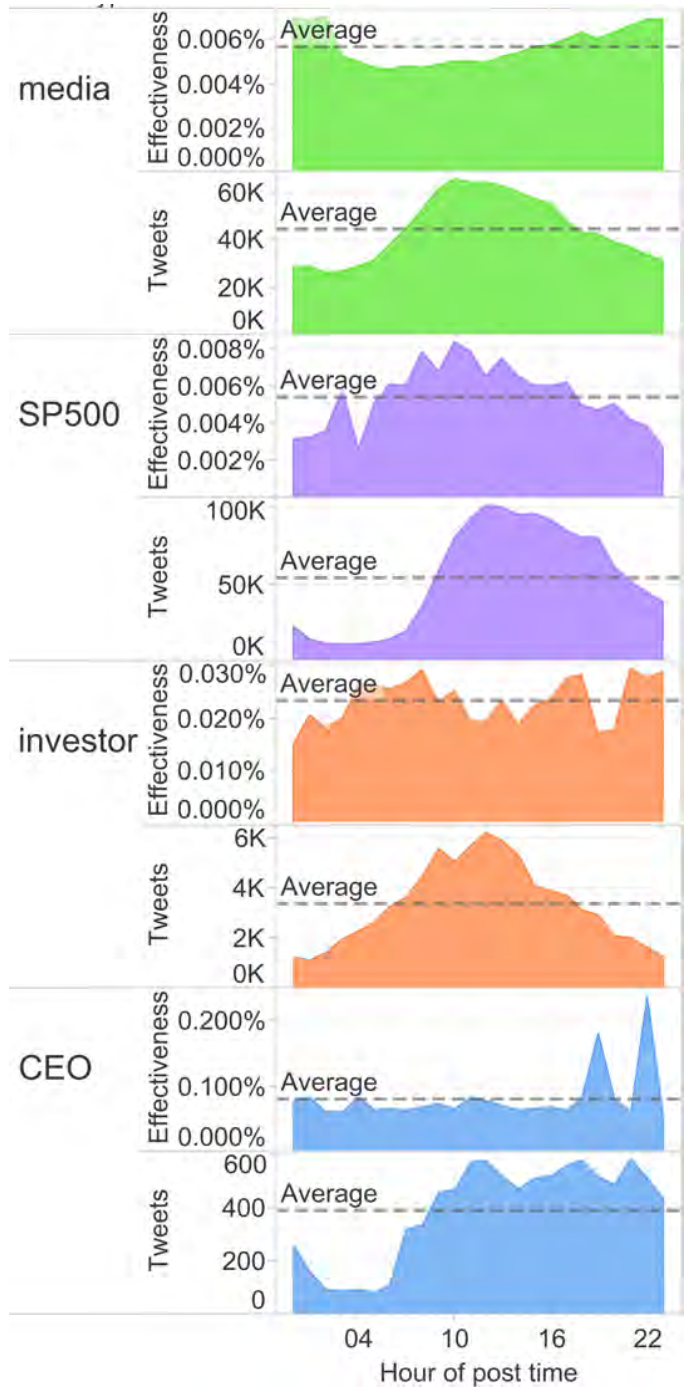

Figure 11.3. **Tweeting effectiveness versus the number of actual tweets posted in different hours of day.** Hour of post time is in U.S. Eastern Time.

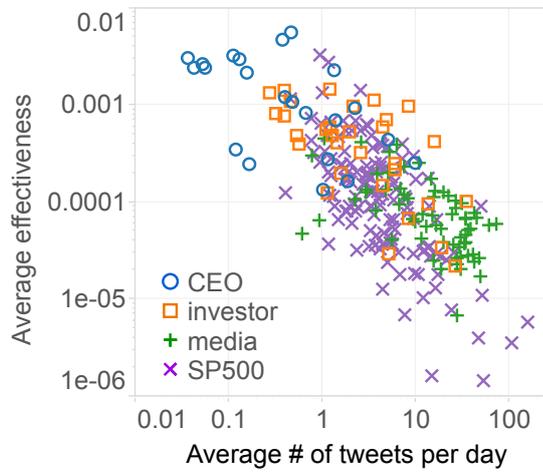

Figure 11.4. **Tweeting effectiveness versus frequency of tweeting.** Frequent tweeting is associated with low tweeting effectiveness.

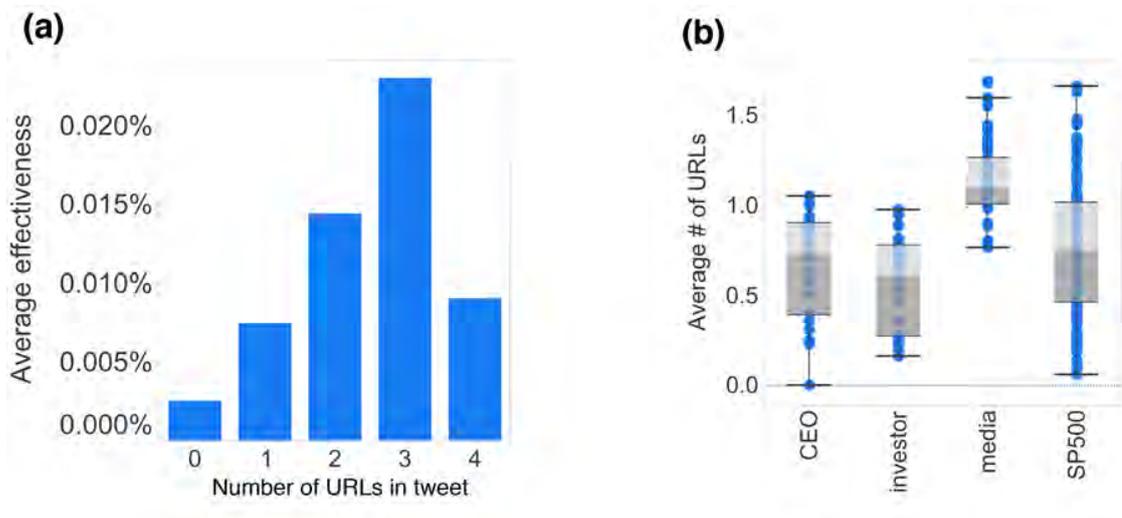

Figure 11.5. **Tweeting effectiveness versus number of links in tweet.** (a) Having three URLs in tweet is associated with the highest effectiveness. (b) News media have higher adoption rate to URLs.

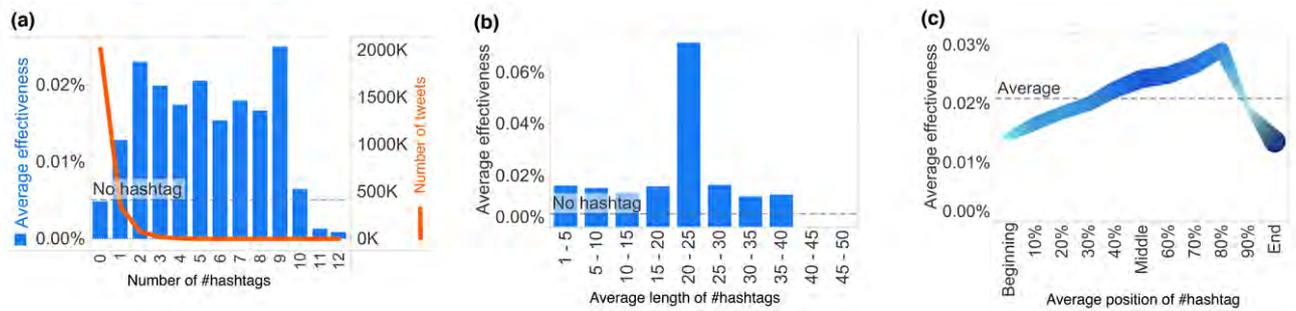

Figure 11.6. **Tweeting effectiveness versus hashtags.** (a) Having hashtags is positively associated with influence; 82% of tweets in our observation do not have hashtags. (b) Non-linear correlation between the length of hashtag and effectiveness. Succinct but descriptive hashtags with 20–25 characters are associated with the highest influence. (c) Tweets having hashtags in the middle are associated with higher effectiveness. We observe that hashtags are most frequently placed at the end of tweets (line width shows the relative number of tweets); such tweets show 40% less effectiveness than average.

**Hashtags having 20 – 25 characters associate with higher effectiveness.**
Because hashtags cannot include spaces, #AnOverwhelminglyLongHashtagThatIsBeyondFourtyCharacters is hard to interpret. Fig. 11.6-(b) suggests having hashtags with more than 40 characters is worse than having no hashtag at all, while 20 – 25 characters are the most effective. The most effective tweets in our observation have hashtags such as #FrenchToastCrunchIsBack by @GeneralMills and #GiveAChildABreakfast by @KelloggsUK, both being succinct and descriptive.

**Tweets have hashtags in the middle demonstrate higher effectiveness.**
The position of hashtag is an important feature as well. As shown in Fig. 11.6-(c), blending the hashtag with text rather than specifically mentioning it at the beginning or the end of correlate with higher effectiveness. Blending hashtags with text does not disrupt the flow of reading and produces less fragmented words, thus potentially more welcomed by readers. However, as the line width in Fig. 11.6-(c) showing the relative number of tweets, a large proportion of tweets have hashtags appended at the end of tweets, and are 40% less effective as average.

**Tweets with $symbol are less effective.** A less known way to enrich tweets is prefixing "$" to stock ticker symbols (e.g., $AAPL), which has similar usage and function as hashtags. About 1% of tweets have symbols in our data; we observe using symbols in tweets correlate negatively with the effectiveness of tweets in general, as shown in Fig. 11.7(a).

**@Mentioning has non-linear relationship with tweeting effectiveness.**
As in Fig. 11.7(b), mentioning a few influential accounts associate with improved effectiveness, but mentioning more than eight accounts correlate with low effectiveness (the tweet is spam or looks like spam.) To illustrate the underlying dynamics of mentioning, below is a tweet posted by @CampbellsChunky, who has 10 thousand followers:

"@Seahawks fans- head to @Safeway & find @CampbellsChunky to snap

a #selfie w/ @RSherman_25 using @blippar #mamasboy"

This tweet only received two retweets initially. But after @Seahawks (which has 900 thousand followers) retweeted the message, the tweet was flooded by hundreds of retweets and favorites from fans of @Seahawks.

**@Replies do not go far.** @Mentions and @replies both have the form of "@ScreenName", but @replies are created by clicking the "reply" button below the tweet, and the @OriginalPoster is automatically mentioned at the beginning of tweet. Although replies are publicly viewable and retweetable, *"people will only see others' @replies in their home timeline if they are following both the sender and recipient of the @reply"*, according to Twitter[4]; namely, a reply to a specific user has a much smaller group of audiences and consequentially have lower effectiveness. Our observation shows that a reply is 11 times less effective than a non-reply on average.

**A picture is worth a thousand words, but not gifs and videos.** While only 17% tweets in our data embed pictures, tweets having a picture is generally more favored (4.5 times more effective) than those that don't, as in Fig. 11.8. The positive correlation is stronger for S&P500 accounts, and less as strong for media accounts. On the contrary, gifs and videos do not always correlate with higher effectiveness in tweets – it depends on the type of original account. In our observation, media and S&P500 accounts show higher effectiveness when including gifs and videos, but lower for CEOs and investors. We observe that gifs and videos by CEOs and investors are generally more casual and personal, while the followers who could be expecting more serious business information. In our data, adoption rates for gifs (about 1/1,000) and videos (1/10,000) are are still relatively low.

---

[4]https://support.twitter.com/articles/14023-what-are-replies-and-mentions

11.4.3   Composition of Tweets

**Effective tweets are either succinct or long.**[5] Previous study in [184] suggests that shorter tweets are more powerful than longer ones, but by using a finer granularity in the tweet length, we discover the relationship is again non-linear, that very short and very long tweets are equally powerful, and more effective than mid-length tweets. As shown in Fig. 11.7(c), succinct tweets that have less than 20 characters or descriptive tweets that have more than 115 characters have higher than average effectiveness. A succinct tweet such as

> "#justdoit" — by @Nike

does great job reinforcing the brand's image and shows high engagement. Descriptive, long tweets such as

> "Downside risk for Scottish independence is virtually zero. The upside is enormous. Don't let fear deprive you of opportunity. #indyref" — by @maxkeiser

also successfully engages the followers, by using three short sentences of a similar meaning, one reinforcing another.

However, as the width of line in Fig. 11.7(c) showing the relative number of tweets in our observation, the majority of tweets are neither succinct nor descriptive (between 20 and 115 characters), correlating with lower than average effectiveness. There is also an interesting "dent" between 90 and 110 characters, where tweeting effectiveness goes down and hit a local low around 100 characters; another smaller "dent" happens starting at 120 characters. To understand the reason of the dents, we use darkness of color to show the likelihood of a tweet embedding a photo. We see a considerable decrease of the likelihood of embedding photos for tweets beyond 90 characters, another decrease beyond 120 characters. This is because 22–23 characters are reserved for each photo or link, for tweets with more than 95 characters it is not possible to have a photo and a link (and most tweets choose link over photo); and for tweets with more than 118 characters, it is impossible to include either a photo or a link. As shown in previous sections that photos and links are positively correlated to tweeting effectiveness, the inability to embed photos or links can be the underlying reason to the two "dents". Nevertheless, it does not change the overall trend that succinct tweets and long tweets are more effective.

---

[5]We are interested in the plain text in tweet: we exclude the number of characters taken up by links (22 characters each link, 23 characters for https links, including photos and other media contents, as in https://support.twitter.com/articles/78124-posting-links-in-a-tweet).

**Exclamation marks and question marks are useful features.** Exclamation marks (!) and question marks (?) are punctuations expressing feelings; in Twitter, they get more attention from readers and correlate with higher effectiveness. Tweets that have at least one exclamation mark is 2.3 times more effective than those that do not have. A tweet like

> "Attention NYC #Directioners! Were giving you the chance to attend a private live concert with @OneDirection: http://t.co/KOLaBv1ucb #1D" — by @HasbroNews

uses the exclamation mark early to grab followers' attention, and shows high effectiveness. Similarly, tweets that have at least one question mark is 25% more effective than those do not have.

**Words in effective tweets.** With at most 140 characters, the choice of words can make a difference. By analyzing the most frequently used 100 words (combining tense & plural variants, excluding stop words) in non-reply tweets, we observe that tweets containing certain words are associated with considerably higher/lower effectiveness, shown in Table 11.1.

Notably, the keyword "retweet" is associated with 14 times higher effectiveness. A typical tweet containing the keyword "retweet"

> "Retweet if you're excited for the year's best bout #MayweatherMaid-

TABLE 11.1

WORDS THAT CORRELATE WITH MAJOR CHANGES IN EFFECTIVENESS

| Positive correlation | | Negative correlation | |
|---|---:|---|---:|
| Keyword | Effectiveness | Keyword | Effectiveness |
| retweet | +1394% | stock | −62% |
| win | +327% | deal | −50% |
| happy | +216% | loss | −46% |
| apologize | +185% | sales | −46% |
| thank | +165% | wrong | −43% |
| appreciate | +145% | weekday | −43% |
| worse | +99% | china | −41% |
| rt | +94% | google | −38% |
| hashtag | +94% | apple | −36% |
| favorite | +93% | obama | −35% |
| welcome | +78% | issue | −28% |
| great | +68% | say | −25% |
| location | +62% | report | −20% |
| awesome | +55% | business | −18% |
| check | +54% | lose | −17% |

ana2! http://t.co/VHeXFZFHwo" — by @DIRECTV

explicitly asks the followers to retweet the message, and is associated with high effectiveness. The short form "rt", also demonstrates a positive correlation (94% more effective), although not as strong as spelling out the word "retweet". Similarly, having the keyword "favorite" is associated with 93% higher effectiveness.

Other observations include that good news (e.g., "win") has more engagement than bad news (e.g., "lose"). Effective tweets generally contain positive words (e.g., "happy", "thank", "appreciate", "welcome", "great", "awesome"). Meanwhile, stock related words (e.g., "stock", "deal", "loss", "sales", "business") have less engagement, which may be the result that investors not wanting to retweet valuable news to other uninformed investors. Quoting (e.g., "say", "report") rather than directly bring up the message correlate with lower effectiveness. Last but not least, tweets discussing popular topics (e.g., "china", "google", "apple", "obama") are associated with lower than average effectiveness – if every user has fixed engagement in a given topic, but many more sources are talking about the same topic at the same time, the user engagement for every tweet is diluted.

**Positive tweets have higher effectiveness.** To understand how positive words and negative words collectively associate with effectiveness of tweets, we do a simple but effective sentiment analysis: we define the *sentiment score* of a tweet as the number of positive words minus the number of negative words (using the lexicon made available by Liu *et al.* [107, 134]), so that a higher sentiment score means there are more positive words than negative words in the tweet, and vice versa.

In Fig. 11.9(a) we see that tweets with sentiment scores further away from 0 generally have higher effectiveness, which indicates that tweets that take clear positions (and possibly express strong feelings) are more effective than neutral tweets. In particular, tweets with positive sentiment scores are more effective than those with negative scores. A typical tweet with positive sentiment score is:

> "Incredibly proud of our outstanding team at @NewsRadio930 WBEN Buffalo. Extraordinary 24/7 in-depth coverage helping the city cope."
> — by @DavidFieldETM

This tweet uses five positive words to deliver the message, and demonstrates high engagement rate. In our observation, however, more than half of tweets have sentiment scores between -1 and 1 (with no or equal positive / negative words) and demonstrate lower effectiveness.

As a comparison with the result of [116], in which the authors claim that negative tweets are the most likely to be retweeted, our analysis is based on the overall user engagement, which factors retweet, favorite, and replies into the measurement, which can be the reason of the different observation. Our observation can be cross-validated by the examples given in Table 11.1 where the top words that improve effectiveness are generally positive words, and the changes of effectiveness (the percentages) correlated with using positive words are generally higher than using negative words.

**Third-party platforms $\neq$ higher effectiveness.** There exists numerous third-party platforms for composing and publishing tweets, with advanced features such as bulk uploading messages, collaborative composition, and so on. However, as shown in Fig. 11.9(b), although the majority of the tweets in this research are through third-party platforms, tweets published via the official platforms have the best effectiveness.

Twitter Ads helps tweets achieve the highest average effectiveness, but it would be an unfair comparison with other platforms, because Twitter Ads allow content to be delivered to targeted users by inserting the tweet to their home timelines, even if they do not follow the account. Tweets posted through iPhone have more than twice of the average effectiveness. A possible reason is that when significant events happen and first-hand information needs to be published, the easiest and perhaps the only available way to tweet is using the smartphone at hand.

### 11.4.4 Account Features

**Most account features do not associate with tweeting effectiveness.** There are multiple ways for increasing Twitter account's creditworthiness (e.g., getting the account officially verified) and improving the account's appearance (e.g., changing profile backgrounds). However, we discover that most, if not all, account features do not demonstrate significant correlation with the effectiveness of tweets. Such non-significant features are as follows:

- **Friendship:** followers count, friends count, user favorites count, user listed count
- **Name and description:** length of screen name, length of user name, length of user description, user name has number, screen name has number

- **Profile customization:** using customized profile, using customized profile image, using background image
- **Location:** geo enabled, time zone
- **Verification:** officially verified by Twitter

**Tweets posted by newer users correlate with higher effectiveness.** Early adopters tend to make a bigger impact in fields such as journalism and academia, but to our surprise, it is not the case in Twitter. According to Fig. 11.9(c), newly created accounts on Twitter generally correlate with higher effectiveness. For example, @Carl_C_Icahn and @jpmorgan joined Twitter about only two years prior to the study, but their tweeting effectiveness are about 70 times higher than accounts created eight years prior to the study, such as @Starbucks and @BBCBusiness. Although @BBCBusiness has about 1 million followers, which is 5 times more than @Carl_C_Icahn, the overall impact of @BBCBusiness makes is 14 times smaller than @Carl_C_Icahn.

## 11.5 Discussion

In this section, we presented a systematic analysis of features that associate with a tweet's effectiveness; namely, how the posting time, tweet entities, composition, and account features associate with the effectiveness of tweets. We incorporated favorites and replies in addition to retweets for a more comprehensive reflection of user engagement. We showed agreements of discoveries between our discoveries and previous work (for example, tweets posted in weekends are more effective), and also some discrepancies (for example, long tweets are equally effective as short tweets). We further discussed new features that have not been analyzed before in terms of effectiveness of tweets, such as embedding videos and gifs in tweets and using third-party tools to compose and publish tweets.

What we observed were not simply linear relationships like "the more hashtags the better"; rather, our analyses revealed the non-linear relationship between various tweet features and the effectiveness of a tweet. For tasks that rely on tweet features, such as user engagement prediction, marketing campaign analysis, and automated trading, the non-linear relationship suggests important design considerations against simple linear models.

Note that by showing correlation between these features and tweeting effectiveness, we do not argue causality, nor claim that a tweet will be effective if it contains features that correlate with high effectiveness. Nevertheless, on the large corpus of data we collected, features revealed by this study can serve as a reference to those who aim to improve their tweeting effectiveness. Our discussion of multiple Twitter features can also inspire feature mining in other social network platforms such as Facebook, Instagram, and so on.

There are several directions to extend this research. First is to extend data collection to other types of users, such as celebrities, politicians, government & organizations, and so on. The current approach has reached its maximum rate of collection on a single machine; extension in data collection will require multiple machines with distinct IP addresses. Second is to account for more factors in the definition of engagement, such as link click through rates, impressions, and so on. Such information can also help weigh the relative importance of retweet, favorites, and replies, and calibrate the measurement of user engagement. This direction will be feasible once Twitter opens access such information. Third is to incorporate more sophisticated sentiment analysis algorithms.

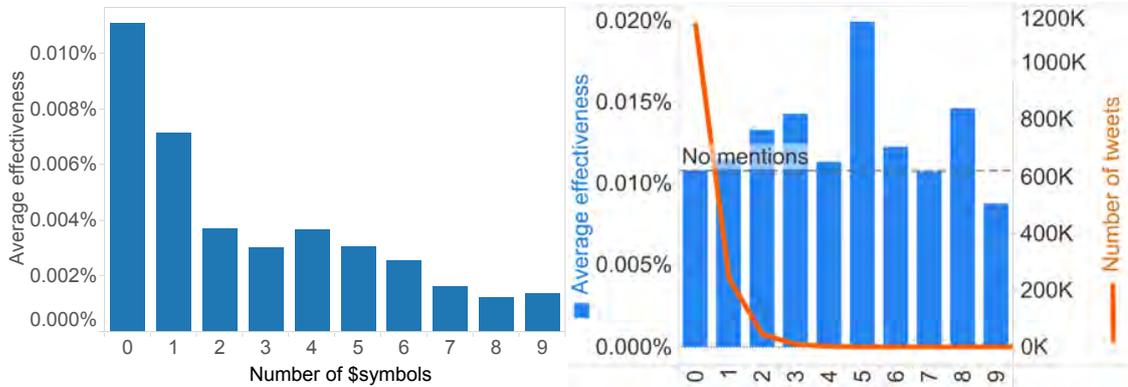

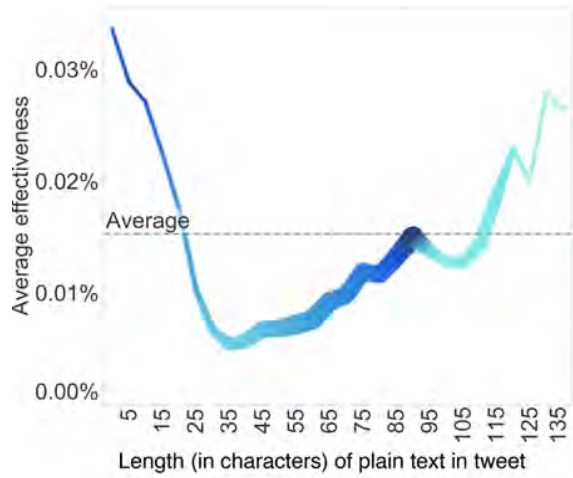

Figure 11.7. **(a)** The usage of symbols negatively correlates with effectiveness. **(b)** Mentioning a few influential accounts correlate with higher effectiveness. As the line indicating the actual number of tweets, 79% of tweets do not mention anyone. **(c)** Either very succinct tweets that are under 20 characters or very long tweets that exceed 115 characters have higher than average effectiveness. Line width represents the amount of tweet, showing a large proportion of tweets are neither long nor short and have lower than average effectiveness. Darkness of color shows the likelihood of a tweet embedding a photo.

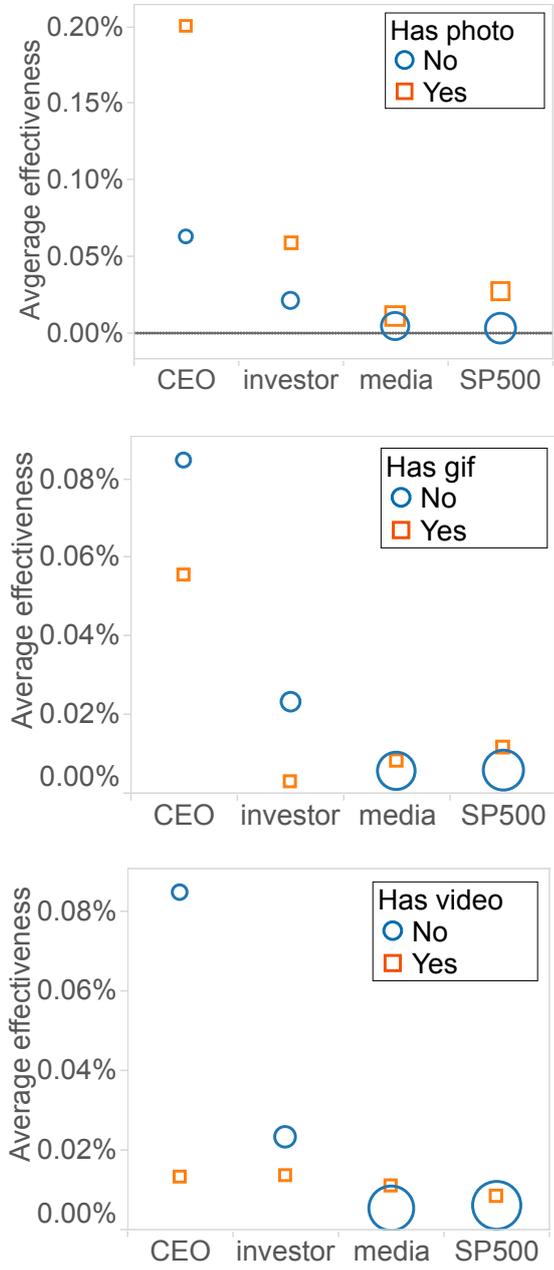

Figure 11.8. **Tweeting effectiveness versus embedding pictures / gifs / videos or not.** Symbol size indicates relative number of tweets in log scale.

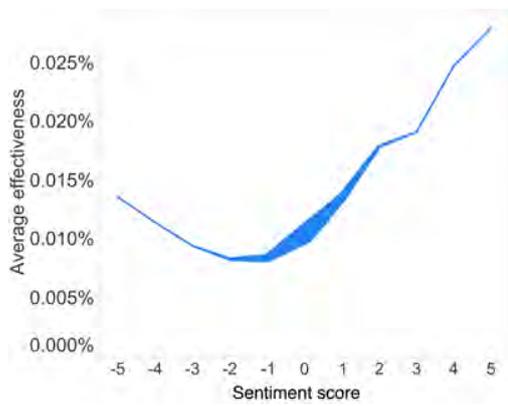

(a) Sentiment

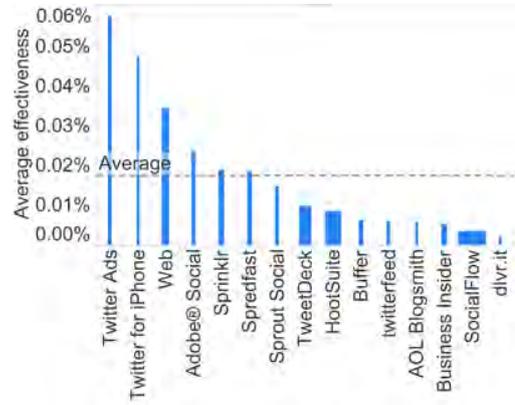

(b) Platform

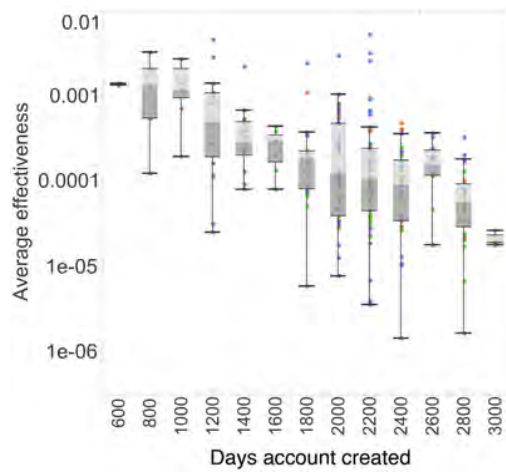

(c) Age of account

Figure 11.9. **(a)** Tweets with many positive words or negative words expressing strong feelings tend to have higher effectiveness. Line width denote the relative number of tweets. **(b)** Tweeting effectiveness versus the platform being used. Bar width denotes the relative number of tweets sent through the platform in our data. **(c)** Accounts newly created on Twitter generally demonstrate higher tweeting effectiveness.

# CHAPTER 12

# DIFFUSION ON IMPLICIT TWITTER NETWORK

## 12.1 Overview

**Intellectual merits:** We track information diffusion in real time by monitoring how the news is tweeted and retweeted on Twitter. We find that news diffusion is highly correlated with intraday trading, especially for retail trading. News diffusion leads to a lower bid-ask spread and price pressure on the news day that is completely reverted the next day. The result is robust when we employ an instrumental variables approach. Our results show that information diffusion via Twitter does not incorporate new information into the stock price. Rather, Twitter spreads stale news, albeit at a much higher speed than traditional media.

**Connections:** This work is based on the features extracted from Chapter 11.

**Work status:** This work [49] is joint effort with Zhi Da, Mao Ye and Nitesh Chawla. It is currently under review at Journal of Management Science.

*We were witnessing another explosion of technological innovations that facilitate interpersonal communication, consisting of e-mail and chat rooms, and after 2000 these led to social media These new and effective media for interactive (if not face-to-face) communication may have the effect of expanding yet again the interpersonal contagion of ideas. They may have allowed enthusiasm for the market to spread much more widely than it would otherwise have. Certainly we are still learning how to regulate the use of these new media in the public interest.*

*Robert Shiller*
*Irrational Exuberance, Third Edition (2015)*

## 12.2 Introduction

In the past, investors learned the news through reading newspapers, watching television, or by word-of-mouth. The advent of social media fundamentally changed how information is produced and disseminated in financial markets [80]. In this paper, we focus on the role of social media on information dissemination. Take Twitter as an example; people receive short electronic messages known as "tweets" through the Twitter accounts they follow, and they can re-transmit the message by retweeting. Does information diffusion through social media affect investors' trading behavior? If so, what is the impact of such information diffusion on return and liquidity? We provide the first analysis of the impact of information diffusion in social media on the stock market by tracking the history of tweets and their retweets.

Under the semi-strong form of market efficiency, information diffusion through Twitter should be inconsequential. The information in tweets is considered to be stale or becomes stale almost instantly after the initial tweet. Rational traders should not react to tweets and/or retweets, and the transmission of news via Twitter should not affect trading volume, return, and/or liquidity.

The information contained in tweets and retweets is public by default, and public information can affect trading behavior, return, and/or liquidity through two channels: the new information channel and the stale news channel. Under the new information channel, public news is not incorporated into prices until investors pay attention [164]. Hirshleifer, Lim, and Teoh [101], DellaVigna and Pollet [68], and Peng and Xiong [165] find that investors' limited attention causes underreactions to news, that is, the slow adjustment of price to fundamental value. The information channel implies a permanent adjustment of prices, although the particular path of adjustment depends on whether traders also learn from the price [98, 140], and the existence of other types of traders may cause overshooting and partial reversal [105]. Hong, Hong, and Ungureanu [104] predict that return increases with the speed of information diffusion. Their model also implies that liquidity decreases with the speed of the diffusion, because faster diffusion increases the number of informed traders relative to uninformed traders, thereby the level of information asymmetry.

Under the stale news channel, tweets and retweets affect trading, return, and liquidity if investors overreact to stale news. Ho and Michaely [102], Huberman and Regev [110], Gilbert [89], and Tetlock [213] find that the stock market reacts to previously published news, suggesting that relevant information is neglected at the time of the previous news. Under the stale news channel, the irrational exuberance generated through social media can temporally move the price away from fundamentals. As stale news only generates noise trading, liquidity increases with the speed of diffusion.

We find that news diffusion through Twitter is associated with price pressure and then reversal, along with an increase in liquidity. We begin our analysis by constructing a unique dynamic measure of information diffusion. The construction of the measure starts from tracking the increase in the number of retweets as well as the number of followers reached by the retweets over time. The empirical cumulative distribution function (CDF) of the information diffusion is the fraction of retweets or the followers reached by retweets in a given interval (e.g., 10 minutes) relative to a terminal time (e.g., one hour or three hours). For example, if the tweet reaches 200,000 users in the first 10 minutes and 1 million users in an hour, the information diffusion speed in the first 10 minutes is 0.2. This CDF function is essential in theoretical models on information diffusion [63, 104, 105, 140], and we provide the first direct proxy for the CDF function over time.

We then define trading intensity as the CDF function of trading volume. We find a strong correlation between information diffusion speed and trading intensity, even after controlling for information diffusion through traditional media outlets, time of day, recent turnover and volatility, past and contemporaneous returns, and various fixed effects. A 1% increase in information diffusion speed is associated with a 0.14% increase in trading intensity. The number increases to 0.21% for retail trading intensity, and to 0.36% for 20% of retail investors with the largest stock holdings.

We also find a strong positive contemporaneous relationship between information diffusion and stock returns. The more Twitter users a tweet reaches, the higher the stock returns on that day (from 10 minutes after the tweet to market close). However, we find that the higher returns are completely reverted the next day. The price overshooting and reversal are driven mostly by smaller stocks. Qualitatively, our results are similar to price pressure led by retail attention in Barber and Odean [22], Tetlock [213], and Da, Engelberg, and Gao [64], but the whole cycle of price pressure and reversal transpires at a much higher speed. A researcher using low-frequency data would find that social media has no impact on stock returns. Information transmitted via social media reduces the cost of paying attention, which expedites but does not eliminate the retail-induced price pressure and its reversal. A further support for the price pressure interpretation is the greater decrease in the bid-ask spread for stocks whose tweets reach more users. The spread decrease is consistent with lower adverse selection risk as retail trading picks up.

Certainly, it is possible that higher return leads to faster retweets, or both returns and information diffusion are driven by the same unobserved factors. Fortunately, the field of computer science discovers a number of predictors of diffusion speed that do not directly predict volume, return, and liquidity [51, 116, 167, 207]. For example, the network structure affects information diffusion speed. A tweet retweeted by users with more followers in the first 10 minutes will diffuse more rapidly afterwards. We also find that a tweet that comes from an active Twitter account generating many new tweets per day diffuses slowly, consistent with the investor distraction hypothesis proposed by Hirshleifer, Lim, and Teoh [101]. We find that Tweets with pictures or hashtags diffuse more rapidly, and that a tweet with a URL link diffuses more slowly. One interpretation is that Tweets with pictures or hashtags attracts attention, but it takes time to read the linked article in URL. Since pictures, hashtags and URLs are independent of future returns and not directly related to trading or valuation, we use them to instrument our diffusion measure.

We multiply the predicted information diffusion rate using these instruments by the number of Twitter users a tweet reaches within 10 minutes. This generates the same price overshooting and reversal pattern. Surprisingly, the total number of Twitter users a tweet can reach within 10 minutes alone does not have predictive power on the return and liquidity. Therefore, it is the information diffusion speed that is predictive of the return and liquidity, which again highlights the importance of having a dynamic, not static measure of information diffusion.

We then check the robustness of our results using an out-of-sample exercise. We use only data from the first six months of our sample period (2013/11 – 2014/04) to run the predictive regression and then apply the regression coefficients to the next six months (2014/05 – 2014/10). The predicted information diffusion rate is therefore free of forward-looking bias and can be computed in real-time. We then detect positive price pressure and the subsequent reversal of the retweeting effects for the second half of our sample (2014/05 – 2014/10).

The rest of the paper is organized as follows. In Section 1, we describe the data. In Section 2, we propose the measure of information diffusion and examine its relation with trading, return, and liquidity. In Section 3, we provide a robustness check of the results using an instrumental variable approach. We conclude in Section 4.

## 12.3   Data Description

Using Twitter's Streaming APIs, we track the history of tweets and retweets of 78 major media outlets (e.g., @WSJ and @CNBC), 56 active Twitter accounts of S&P 1,500 CEOs and CFOs (e.g., Elon Musk of Tesla Motors), (Chen, Hwang, and Liu, 2013), and 143 Twitter accounts of S&P 500 companies (e.g., @TysonFoods). We focus on news from trustworthy outlets to avoid potential noise or even rumors from social media. We captured all original tweets posted by any of these 277 accounts and their retweets from November 1, 2013 through October 31, 2014.

Table 12.1 reports the summary statistics for these 277 accounts. The 78 Twitter accounts from media outlets tend to have more followers, with a mean of 888,545 followers and a median of 100,446 followers. For example, @nytimes has more than 11 million followers and @WSJ has more than 4 million followers. The 143 official Twitter accounts of S&P 500 companies also have many followers, with a mean of 601,931 followers and a median of 125,521 followers. Both @Google and @Starbucks have more than 5 million followers apiece. Firm CEOs and CFOs have fewer followers; the mean is 54,576 followers and the median is only 621 followers. @ericschmidt, @RalphLauren, and @MichaelDell attract the most followers (779K, 672K, and 628K, respectively).

Table 12.1 also reports that the average number of years since inception is 5.7 for media outlet accounts, 5.3 for company accounts, and 4.3 for CEO/CFO accounts. Twitter accounts by media outlets are the most active with almost 7,488 tweets per year per account, followed by S&P 500 company accounts (3,334 tweets per year per account). The Twitter accounts of CEOs and CFOs are the least active, with only

TABLE 12.1

SUMMARY STATISTICS ON TWITTER ACCOUNTS IN THE SAMPLE

|  | N | Mean | Std dev | Q1 | Median | Q3 |
|---|---|---|---|---|---|---|
| **# of Followers** | | | | | | |
| CEO/CFO | 56 | 54,576 | 173,192 | 167 | 621 | 7034 |
| Media | 78 | 888,545 | 1,789,724 | 17,802 | 100,446 | 923,497 |
| SP500 | 143 | 601,931 | 1,222,074 | 42,249 | 125,521 | 467,134 |
| **Years Since Inception** | | | | | | |
| CEO/CFO | 56 | 4.3 | 1.8 | 3.1 | 4.5 | 5.4 |
| Media | 78 | 5.7 | 1.7 | 5.1 | 5.6 | 6.6 |
| SP500 | 143 | 5.3 | 1.7 | 4.9 | 5.4 | 6.0 |
| **Tweets Per Year** | | | | | | |
| CEO/CFO | 56 | 264 | 703 | 6 | 45 | 186 |
| Media | 78 | 7,488 | 6,312 | 2,157 | 6,076 | 11,821 |
| SP500 | 143 | 3,334 | 6,987 | 788 | 1,413 | 2,985 |

This table reports summary statistics on the 277 Twitter accounts we monitored from 2013/11 to 2014/10. They include 78 major media outlets (e.g., @WSJ and @CNBC), 56 active accounts of S&P 1,500 CEOs and CFOs (e.g., Elon Musk), and 143 official Twitter accounts of S&P 500 companies (e.g., @TysonFoods).

264 tweets per year per account.

To identify potentially influential tweets, we apply the following filters to the tweets:

1. Having been retweeted more than 50 times.

2. Having been posted during extended trading hours (4:00 to 20:00 ET).

3. Mentioning at least one company that is in the Russell 3000 index.

4. If multiple events happened to the same company, we pick the one with the most retweets. If multiple tweets about the same event are captured, we pick the one that was sent the earliest.

TABLE 12.2

EXAMPLE OF TWEETS IN OUR SAMPLE

| Date | Source | Tweet | Ticker |
|---|---|---|---|
| 11/12/2013 | @WSJ | AirWatch expresses interest in buying service division of Blackberry: http://t.co/R9vTFvfHkD | BBRY |
| 11/14/2013 | @FordTrucks | @Ford?F-150 EcoBoost hits 400,000 sales, saving 45 million gallons of gas annually: http://t.co/xYRgWVGoph?#BuiltFordTough | F |
| 11/22/2013 | @paradimeshift | Western Union and tradition bank wire transfers are dead! 11/23/13 $147 Million transferred for 37 CENTS! #bitcoin | WU |
| 12/09/2013 | @ABC | Just in: American Airlines/US Airways merger complete says company - @ABCaviation | AAL |
| 12/19/2013 | @DavidJBarger | Very cool @JetBlue's SJU Team welcomed N903JB, our first A321, "Bigger Brighter Bluer" to the airline! http://t.co/IU7JFJt9Y4 | JBLU |
| 01/09/2014 | @EMCcorp | Congratulations to David Goulden - new CEO of #EMC. Joe Tucci will continue as Chairman & CEO of EMC Corporation http://t.co/no4P9BYOwT | EMC |
| 01/29/2014 | @BreakingNews | Facebook earnings: Q4 EPS $0.31 ex-items v. $0.27 estimate; revenues $2.59 billion v. $2.33 billion estimate - @CNBC http://t.co/sNqDbtfyzv | FB |
| 02/05/2014 | @ReutersBiz | Twitter reports revenue of $243 million, up 116 percent year-over-year | TWTR |
| 02/19/2014 | @businessinsider | TESLA EXPECTS 55% VEHICLE DELIVERY GROWTH IN 2014 http://t.co/aXQZAqHd0z | TSLA |
| 03/04/2014 | @CNET | 2015 Lamborghini Huracan debuts with Nvidia-powered digital dashboard http://t.co/j7bvnt9JuH http://t.co/XlfBKsU85Q | NVDA |

TABLE 12.3

SUMMARY STATISTICS OF FIRMS IN OUR SAMPLE

|                 | Mkt Cap (M$) | Turnover | Volatility | IO    |
|-----------------|-------------:|---------:|-----------:|------:|
| Mean            | 136,668      | 4.20     | 0.022      | 0.577 |
| Median          | 85,186       | 2.05     | 0.016      | 0.602 |
| Std dev         | 144,755      | 4.80     | 0.019      | 0.175 |
| CRSP percentile | 89.9         | 62.5     | 50.5       | 80.0  |

We carry out steps (1) and (2) automatically using a computer script, and we perform steps (3) and (4) manually (e.g., distinguishing the mentions of the tech company "Apple" from the fruit "Apple"; identifying different tweets that discuss the same topic, etc.). The selection process leaves us with 1,261 tweets. These tweets originate from 115 Twitter accounts and cover 178 distinct stocks. Table 12.2 contains a few sample tweets, which cover a wide range of news (mergers and acquisitions, earnings announcements, product launches, independent research, etc.).

Of the 115 distinct Twitter accounts, @WSJ generates the most tweets in our sample (270), followed by @Forbes (129), and @CNBC (83). Of the 178 distinct stocks, Apple (AAPL) appears most frequently (92 times) in tweets, followed by Facebook (FB, 88 times), Google (GOOG, 82 times), Twitter (TWTR, 81 times), Microsoft (MSFT, 67 times), and Tesla (TSLA, 64 times). Table 12.3 presents summary statistics on the stocks in our sample. The average stock size is at the 90th percentile of the CRSP universe. The average institutional ownership is also large, at 57.7%, corresponding to the 80th percentile of the CRSP universe. The volatility of the average stock in our sample is similar to that of an average stock in the CRSP universe but has higher turnover.

In Panel A of Figure 12.1, we plot the number of retweets over time during the first hour after an original tweet, with each time interval representing 10 minutes. The median tweet in our sample will be retweeted 68 times by the end of the first hour. The small number of 68 retweets, however, reaches 3 million people, because many accounts that retweet the news also have a large number of followers (Panel B of Figure 12.1).

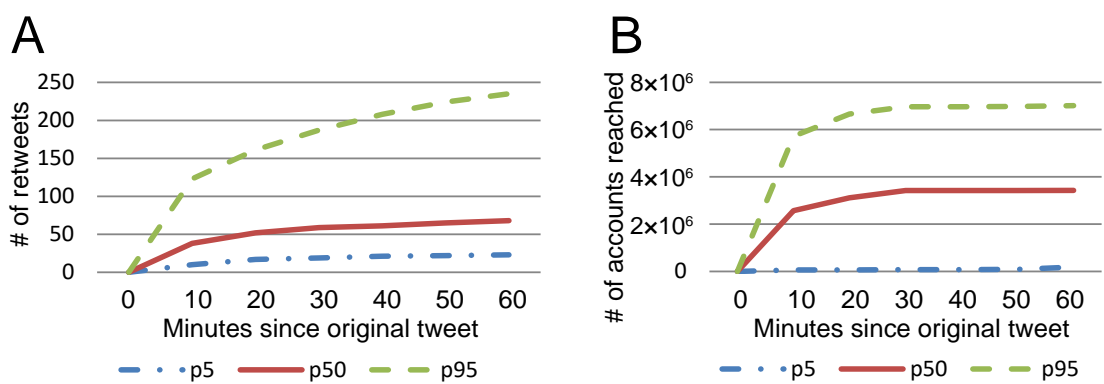

Figure 12.1. In Panel A, we plot the total number of retweets during the first hour after the original tweet, for the median case, for the $5^{th}$ percentile, and for the $95^{th}$ percentile. In Panel B, we also account for the number of followers of each Twitter account that posts the original tweet or the retweet. As a result, the number measures the number of potential users the tweet can reach in the first hour. Each time interval represents 10 minutes.

We use the NYSE Daily Trade and Quote (DTAQ) database to construct the complete National Best Bid and Offer (NBBO). The DTAQ provides two files that contain official NBBO quotes. If a single exchange has both the best bid and the best offer, then the official NBBO will be recorded in the DTAQ Quotes File. Otherwise, the NBBO quotes will be recorded in the DTAQ NBBO file. Following the procedure proposed by Holden and Jacobsen (2014), we combined the NBBO quotes from both files to construct the complete official NBBO. We then compute bid-ask spreads and intraday returns using bid-ask midpoints.

As market-wide intraday retail trading volume data are not directly available, we use the trading volume from the Trade Report Facility (TRF) as a proxy for the market-wide retail trading volume. The results are supplemented by a proprietary dataset on retail trading from TD Ameritrade (TDA). The market-wide proxy is constructed based on the empirical finding of Battalio, Corwin, and Jennings (2015) that non-direct limit and market orders are seldom routed to public exchanges but are often internalized by broker-dealers. Therefore, we use TRF volume (exchange symbol D in the TAQ dataset) as our proxy for market-wide retail trading. This measure has two limitations. First, TRF volume also contains volume from dark pools [130]. Second, Battalio, Corwin, and Jennings [24] find that some retail brokers route orders to public exchanges, including TDA. Therefore, we supplement our market-wide proxy of retail trading with a proprietary dataset from TDA. This dataset includes 331 million de-identified transactions made by 2.8 million clients from June 1, 2010 to June 10, 2014.

We use Ravenpack data to control for news coverage on other media outlets. Following Hafez [95], we only keep news events with a novelty score of 100 and relevance score above 75. For each tweet in our sample, we then count the amount of news on the same stock on the day of the tweet up to the time of the tweet and also trace how this news count changes after the tweet. This allows us to measure both the amount of news coverage in other media outlets and how this coverage changes over time.

## 12.4 Regression Analysis

In this section, we first define our measure of the information diffusion speed. Next, we examine two questions: (1) How does the information diffusion speed affect trading intensity, especially among retail investors? (2) How does the information diffusion speed affect asset prices and stock liquidity?

### 12.4.1 Definitions of Information Diffusion Speed and Trading Intensity

The driver of the asset-pricing dynamics in information diffusion models is the proportion of agents who know the information earlier than others, which is characterized by the CDF function [104]. Yet the empirical literature does not include a dynamic proxy for the CDF function. In this paper, we provide the first direct proxy for the CDF function to fill this void.

For each tweet in our sample, we calculate the total number of its retweets within the first hour. Next, we divide the hour into six 10-minute intervals and calculate the number of retweets in the 10-, 20-, 30-, 40-, 50-, and 60-minute intervals relative to the total of retweets within the first hour. By construction, this number is 1 after 60 minutes. A fast diffusion of information implies a quick convergence to 1 over time. For simplicity, we classify a diffusion process as fast if more than 60% of total first-hour retweets occur in the first 10 minutes; we classify a diffusion process as slow if less than 40% of total first-hour retweets occur in the first 10-minute interval. The result is robust under other specifications. Panel A of Figure 12.2 presents the average information diffusion speed of the fast and slow diffusion in our sample.

Similarly, trading intensity is also a normalized measure. We divide the cumulative volume in the first 10, 20, 30, 40, 50, and 60 minutes by the total volume in the first hour. Panel B of Figure 12.2 presents the average trading intensity of the fast and slow information diffusion in our sample.

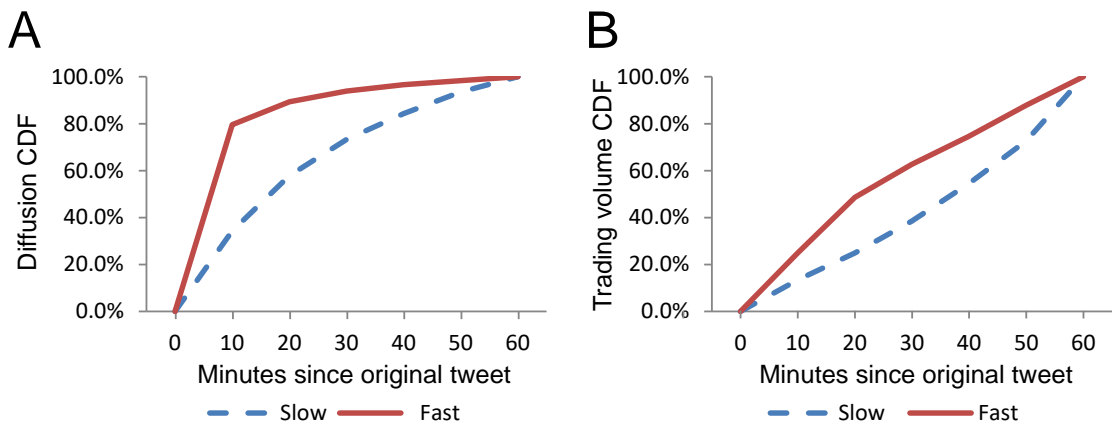

Figure 12.2. In Panels A and B, we plot the cumulative numbers of retweets and trading volumes for each of the six 10-minute intervals during the first hour following a tweet. Both variables are normalized by their totals during the first hour, so the plot resembles a cumulative distribution function (CDF). Rapid diffusion occurs when more than 60% of total first-hour retweets occur in the first 10 minutes; slow diffusion occurs when less than 40% of total first-hour retweets occur in the first 10 minutes.

12.4.2  Trading Intensity Increases with Information Diffusion Speed

We next examine how the information diffusion speed is related to trading intensity. We discuss the results for aggregated volume and then retail volume.

12.4.2.1  Information Diffusion Speed and Total Volume

Figure 12.2 shows that 25.0% of the first-hour trades take place in the first 10 minutes after a tweet for the fast diffusion case. In contrast, only 13.4% of the first-hour trading takes place in the first 10 minutes for the slow diffusion case.

To overcome any potential omitted variable bias, we control for other factors that can drive the correlation between diffusion rate and trading intensity. For example, both retweets and trading intensity may have intraday seasonality, and we control for this seasonality using time dummies. Also, extreme returns immediately following a tweet could trigger both retweets and trading. In addition, the information diffusion rate may be correlated with news coverage for the same stock in other media outlets.

We control these variables in Table 12.4. The dependent variable, the percentage of first-hour total trading that occurs in the first 10 minutes following a tweet, measures trading intensity. The main independent variable, diffusion, measures the percentage of first-hour retweets that occurs in the first 10 minutes. Other control variables include a dummy variable equal to 1 if the tweet takes place before 9:30 ET (pre-market); a dummy variable equal to 1 if the tweet takes place between 12:30 and 16:00 ET (afternoon); a dummy variable equal to 1 if the tweet takes place after 16:00 ET (post-market); log market capitalization (size); turnover (turn); daily return volatility in the past 30 days (volatility); book-to-market ratio (bm); absolute stock returns over the market in the past hour (abs past 1h ret); absolute stock returns over the market in the first 10 minutes after the tweet (abs 10m ret); log number of media coverage on the same day but prior to the tweet (media); and percentage of first-hour media coverage for the stock that occurs in the first 10 minutes after the tweet (media_diffusion). We include stock and Twitter account fixed effects in our regression. The standard error is clustered by ticker. The sample covers 1,261 tweets during one year from 2013/11 to 2014/10.

Table 12.4 reports the regression results for trading volume in TAQ. This regression links retweets to trading during the first hour after a tweet. The dependent variable is the percentage of first-hour trading that occurs in the first 10 minutes. The main independent variable, diffusion, measures the percentage of first-hour retweets that occur in the first 10 minutes. Other control variables include pre-market (a dummy variable equal to 1 if the tweet takes place before 9:30 ET); afternoon (a dummy variable equal to 1 if the tweet takes place between 12:30 and 16:00 ET); post-market (a dummy variable equal to 1 if the tweet takes place after 16:00 ET);

TABLE 12.4

DIFFUSION SPEED AND TRADING INTENSITY: TOTAL TRADING VOLUME

|  | (1) | (2) | (3) | (4) |
|---|---|---|---|---|
| intercept | 0.02 | 0.11*** | 0.01 | -0.04 |
|  | (0.65) | (3.27) | (0.12) | (-0.42) |
| diffusion | 0.30*** | 0.17*** | 0.14** | 0.14** |
|  | (5.13) | (3.07) | (2.25) | (2.27) |
| pre-Market |  | -0.12*** | -0.12*** | -0.12*** |
|  |  | (-9.88) | (-7.70) | (-7.93) |
| afternoon |  | -0.02 | -0.01 | -0.01 |
|  |  | (-1.58) | (-1.02) | (-0.95) |
| post-Market |  | 0.07*** | 0.07*** | 0.07*** |
|  |  | (3.67) | (2.86) | (2.96) |
| size |  |  | 0.00 | 0.01* |
|  |  |  | (1.00) | (1.71) |
| turn |  |  | 0.00 | 0.00 |
|  |  |  | (-0.37) | (0.47) |
| volatility |  |  | 1.43*** | 1.51*** |
|  |  |  | (10.09) | (10.49) |
| bm |  |  | 0.00 | 0.00 |
|  |  |  | (-0.53) | (-0.43) |
| abs past 1 hr ret |  |  | -0.46*** | -0.39** |
|  |  |  | (-2.41) | (-2.07) |
| abs 10m ret |  |  | 2.71*** | 2.89*** |
|  |  |  | (3.10) | (3.10) |
| media |  |  |  | -0.01 |
|  |  |  |  | (-1.35) |
| media diffusion |  |  |  | -0.03 |
|  |  |  |  | (-1.13) |
| stock FE | Y | Y | Y | Y |
| account FE | Y | Y | Y | Y |
| $R^2$ | 0.031 | 0.123 | 0.134 | 0.138 |

size (log market capitalization); turn (turnover); volatility (daily returns volatility in the past 30 days); bm (book-to-market ratio); abs past 1h ret (absolute stock returns over the market in the past hour); abs 10m ret (absolute stock returns over the market in the first 10 minutes after the tweet); media (log number of media coverage on the same day but prior to the tweet); media diffusion (percentage of first-hour media coverage occurs in the first 10 minutes). We include stock and Twitter account fixed effects. Standard errors are clustered by ticker. The sample covers 1,261 tweets from 2013/11 to 2014/10. T-statistics are in parentheses. ***, **, and * indicate statistical significance at the 1%, 5%, and 10% levels, respectively.

In Table 12.4, the results confirm the strong unconditional correlation between diffusion rate and trading intensity observed in Figure 12.2. A 1% increase in the diffusion rate leads to a 0.3% increase in trading intensity, with a t-value of 5.13. Once we control for time of day in column (2), the effect attenuates to 0.17% but is still highly significant. The number reduces to 0.14 but remains significant (t-value = 2.25) after we control for size, turnover, volatility, and book-to-market in Column (3).

Surprisingly, the results in column (4) show that neither the amount of news coverage in other media outlets nor its diffusion rate drives the trading intensity. After controlling for the information diffusion in Twitter, the coefficient of media coverage is negative, although not statistically significant. Plus, the inclusion of news coverage in other media and media diffusion has no impact on the coefficient before Twitter, which remains at 0.14, with a t-value of 2.27. These results are consistent with the Shiller's [195] conjecture that conventional mediaprint media, television, and radiohave limited ability to generate active behavior. We provide evidence to support Shiller's [195] claim that interpersonal and interactive communications have the most powerful impact on behavior. To the best of our knowledge, there is no horse race between these two ways of spreading news in the literature other than a survey conducted by Pound and Shiller [196]. Our paper provides direct empirical evidence that the attention and action of investors is more stimulated by interactive communications.

#### 12.4.2.2  Diffusion Speed and Retail Volume

We find an even stronger link between diffusion speed and trading intensity for retail investors. While retail investor trading volume is not directly available, we compute trading intensity using only TRF volume and use it as the dependent variable in the regressions.

Table 12.5 reports the regression results for trading volume from TRFs (exchange symbol D from the TAQ dataset). We find that a 1% increase in information diffusion speed is associated with a 0.21% increase in retail trading intensity, after controlling for other factors (Column 4). Therefore, the link between information diffusion speed and retail trading intensity is 50% stronger for retail investors than for all investors.

We then take advantage of a unique brokerage account dataset from TDA that includes 331 million transactions made by 2.8 million clients from June 1, 2010 to June 10, 2014. The data have been provided by TDA through an academic data collaborative agreement. Individual clients are not identified in the data, although demographic characteristics such as age and gender are included for each anonymous ID. We are also able to track the history of trading from an individual through this unique ID. While trades in the TDA data represent a subset of all trades, it is a relatively clean subset of retail trading.

We merge our tweet sample with TDA brokerage-account-level transaction data for the overlapping period from 2013/11 to 2014/06. We focus only on stock trades from "individual" accounts in TDA. Since investors at TDA rarely trade during after-hour sessions, we focus on tweets during market trading hours (9:30 to 16:00 ET). We

TABLE 12.5

DIFFUSION SPEED AND TRADING INTENSITY: TRF TRADING VOLUME

|  | (1) | (2) | (3) | (4) |
|---|---|---|---|---|
| *intercept* | 0.12*** | 0.24*** | 0.33** | 0.35** |
|  | (2.64) | (5.95) | (2.37) | (2.38) |
| *diffusion* | 0.44*** | 0.25*** | 0.21** | 0.21** |
|  | (5.84) | (3.69) | (2.91) | (2.89) |
| *pre-Market* |  | -0.24*** | -0.22*** | -0.22*** |
|  |  | (-11.81) | (-8.23) | (-8.12) |
| *afternoon* |  | -0.02 | -0.02 | -0.02 |
|  |  | (-1.51) | (-1.17) | (-1.15) |
| *post-Market* |  | 0.14*** | 0.14*** | 0.14*** |
|  |  | (4.89) | (3.81) | (3.67) |
| *size* |  |  | 0.00 | -0.01 |
|  |  |  | (-0.61) | (-0.73) |
| *turn* |  |  | 0.00 | 0.00 |
|  |  |  | (1.29) | (1.30) |
| *volatility* |  |  | -1.87 | -1.99 |
|  |  |  | (-1.19) | (-1.23) |
| *bm* |  |  | 0.00 | 0.00 |
|  |  |  | (1.49) | (1.48) |
| *abs past 1 hr ret* |  |  | -0.81*** | -0.79** |
|  |  |  | (-3.35) | (-3.29) |
| *abs 10m ret* |  |  | 2.21*** | 2.22*** |
|  |  |  | (1.97) | (1.93) |
| *media* |  |  |  | 0.01 |
|  |  |  |  | (0.63) |
| *media diffusion* |  |  |  | -0.03 |
|  |  |  |  | (-0.72) |
| *stock FE* | Y | Y | Y | Y |
| *account FE* | Y | Y | Y | Y |
| $R^2$ | 0.036 | 0.215 | 0.248 | 0.249 |

TABLE 12.6

TD AMERITRADE BROKERAGE ACCOUNT DATA

| Across 35,443 accounts | | | | | |
|---|---|---|---|---|---|
| *Avg char* | Mean | Std dev | Q1 | Median | Q3 |
| *Age* | 48.7 | 14.3 | 38 | 48 | 58 |
| *Stock holding ($)* | 78,063 | 167,449 | 3,481 | 20,146 | 74,288 |
| *Net trade* | 0.153 | 1.023 | -1 | 0.167 | 1 |
| *Trade intensity* | 20.00% | 36.60% | 0.00% | 0.00% | 25.00% |
| *Across 331 tweets* | | | | | |
| *Avg char* | Mean | Std dev | Q1 | Median | Q3 |
| *# of accounts* | 194.4 | 288.9 | 19 | 69 | 242 |
| *% of Male* | 72.10% | 12.20% | 67.90% | 71.70% | 76.20% |
| *Age* | 51 | 5 | 48.8 | 50.4 | 52.9 |
| *Stock holding ($)* | 91,004 | 51,164 | 68,843 | 84,461 | 102,741 |
| *Net trade* | 0.06 | 0.43 | -0.202 | 0.045 | 0.333 |
| *Trade intensity* | 20.40% | 17.10% | 10.20% | 18.50% | 26.10% |

examine only accounts that trade the corresponding stock at least once during the first hours after a tweet. Altogether, our selection criteria result in a merged dataset that contains 331 tweets and trades from 35,443 individual TDA accounts.

Table 12.6 provides summary statistics for our merged sample. We merge our tweet sample with the TD Ameritrade (TDA) brokerage-account-level transaction data during the overlapping period from 2013/11 to 2014/06. We focus on tweets during market hours and retail accounts that trade the corresponding stock at least once during the first three hours after a tweet. The merged sample contains 331 distinct tweets and 35,443 distinct TDA accounts. To compute net trades, one buy (sell) is counted as 1 (-1). Trade intensity is again measured as the percentage of first-hour trading that occurs in the first 10 minutes.

The average individual TDA account holder is 49 years old in 2014. The first age quartile is 38 and the third quartile is 58. Their median stock holdings with TDA are worth $20,000 with 25% holding stocks worth more than $74,000. When they trade during the first hour after a tweet, they are more likely to buy. Both the mean and median of the net trade variable are positive (with $t-value > 5.00$). This finding provides direct support for Barber and Odean (2008), who argue that retail attention leads to positive price pressure on average since retail investors are less likely to short. On average, 20% of all first-hour trades take place during the first 10 minutes following the tweet.

The bottom half of Table 12.6 reports the summary statistics for the 331 tweets in our merged sample. On average, we observe trades from almost 200 individual accounts following a tweet. Seventy-two percent of the trades come from male account holders. The average account holder is 51 years old, holding about $91,000 in stocks, and is more likely to buy stocks (rather than sell them). Twenty percent of all first-hour trades take place during the first 10 minutes following a tweet.

We repeat the regressions in Table 12.4 for our merged sample and report the results in Table 12.7. We measure trade intensity using (I) all trades from TAQ; (II) all TDA trades; (III) all TDA trades of female investors; (IV) all TDA trades of young investors ($age < 35$); and (V) all TDA trades of "rich" investors (with stock holdings $> \$100K$). We include the same set of control variables as those used in the regressions for Table 12.4. Since we focus on tweets during the normal trading hour, the pre-market and post-market dummies drop out. Other control variables include afternoon (a dummy variable equal to 1 if the tweet takes place between 12:30

TABLE 12.7

THE LINK BETWEEN DIFFUSION SPEED AND TRADING

INTENSITY FOR TD AMERITRADE DATA

|  | All Trades (I) | TDA Trades (II) | TDA Female (III) | TDA Young (IV) | TDA Rich (V) |
|---|---|---|---|---|---|
| *intercept* | -0.06 | -0.03 | -0.07 | -0.09 | 0.3 |
|  | (-0.44) | (-0.12) | (-0.24) | (-0.19) | -0.94 |
| *diffusion* | 0.16*** | 0.23** | 0.27** | 0.17 | 0.36*** |
|  | -2.78 | -1.99 | -1.99 | -0.97 | -2.69 |
| *afternoon* | -0.01 | 0.01 | 0.04 | 0 | 0.06** |
|  | (-0.50) | -0.57 | -1.4 | (-0.04) | -2.01 |
| *size* | 0.01 | 0 | 0 | 0.01 | -0.01 |
|  | -1.16 | -0.31 | -0.32 | -0.23 | (-0.84) |
| *turn* | 0 | 0 | 0 | 0 | 0 |
|  | -0.84 | -0.91 | (-0.25) | (-0.77) | -1.48 |
| *volatility* | -0.19 | -1.99 | 1.4 | 8.45 | -8.36* |
|  | (-0.10) | (-0.55) | -0.25 | -1 | (-1.69) |
| *bm* | 0.00* | 0.00** | 0 | 0 | 0.00* |
|  | -1.93 | -2.46 | -0.52 | (-0.40) | -1.74 |
| *abs past 1h ret* | -0.17 | -0.12 | 0.55 | -1.43 | 2.08 |
|  | (-0.15) | (-0.09) | -0.3 | (-0.91) | -0.86 |
| *abs 10m ret* | 3.35 | 7.71* | 6.84* | 6.78 | 2.88 |
|  | -1.38 | -1.7 | -1.71 | -1.36 | -0.9 |
| *media* | 0 | 0.01 | -0.02 | 0 | 0.01 |
|  | -0.39 | -0.42 | (-0.85) | -0.33 | -0.36 |
| *media diffusion* | -0.05*** | -0.03 | -0.04 | -0.12*** | 0.02 |
|  | (-2.86) | (-0.85) | (-1.08) | (-2.95) | -0.39 |
| *stock FE* | Y | Y | Y | Y | Y |
| *account FE* | Y | Y | Y | Y | Y |
| $R^2$ | 0.081 | 0.079 | 0.064 | 0.054 | 0.092 |

and 16:00 ET); post-market (a dummy variable equal to 1 if the tweet takes place after 16:00 ET); size (log market capitalization); turn (turnover); volatility (daily returns volatility in the past 30 days); bm (book-to-market ratio); abs past 1h ret (absolute stock returns over the market in the past hour); abs 10m ret (absolute stock returns over the market in the first 10 minutes after the tweet); media (log number of media coverage on the same day but prior to the tweet); media_diffusion (percentage of first-hour media coverage occurs in the first 10 minutes). We include stock and Twitter account fixed effects. Standard errors are clustered by ticker. T-statistics are in parentheses. ***, **, and * indicate statistical significance at the 1%, 5%, and 10% levels, respectively.

Column I of the Table 12.7 confirms the strong correlation between diffusion speed and trading intensity measured using all trades in our merged sample of 331 tweets. A 1% increase in the information diffusion rate leads to a significant 0.16% increase in trading intensity. The coefficient of 0.16 is higher than the corresponding coefficient of 0.14 using the larger 1,261 tweet sample (see Table 12.4), possibly because we focus on trading hours.

The results of regression II suggest a much stronger link between information diffusion speed and retail trading intensity measured using all TDA trades. The coefficient for the information diffusion variable increases from 0.16 to 0.23. The results of regression III suggest an even stronger link between information diffusion speed and retail trading intensity among female investors, who account for less than 30% of all TDA investors in our sample.

The strongest link between information diffusion speed and retail trading intensity we find in our sample is among the 20% of TDA investors with the largest stock holdings. For them, a 1% increase in the information diffusion rate leads to a 0.36% increase in trading intensity. This is not surprising as traders with higher investments in stocks should be more attentive to financial news and thus react more quickly to that news.

Surprisingly, the weakest link between diffusion speed and retail trading intensity we find is among the 20% of TDA investors who are younger than age 35, who should be more frequent users of Twitter. For them, a 1% increase in the information diffusion rate leads to only a 0.17% increase in trading intensity, for two possible reasons. First, younger investors typically have fewer financial resources and therefore fewer investment assets and lower investment value, which means they may be constrained by commissions or other fixed transaction costs. Attention is one such cost. Investors with less valuable investments may have a weaker incentive to follow a particular firm. Second, compared with average TDA investors in our sample, who are close to retirement age, younger investors have to focus more on work during trading hours and thus may not trade immediately after a tweet.

Overall, the TDA data we examine provide direct support for the concept that diffusion speed measured using retweets is more strongly related to retail trading.

### 12.4.3 Diffusion Speed, Stock Returns, and Liquidity

Insofar as diffusion speed measured using retweets relates to trading intensity, we next examine how it affects prices and dollar bid-ask spreads. We measure contemporaneous stock returns (in excess of the market) from 10 minutes after an initial tweet until the end of the day (CAR%[10m, close d0]). We skip the first 10 minutes after the tweet to avoid mechanical correlation in the next section. We also measure stock returns (in excess of the market) on the next trading day (CAR%[close d0, close d1]). We then examine the change in stock liquidity as the average dollar bid-ask spread during the three hours after a tweet minus the average dollar bid-ask spread during the hour before the tweet.

We measure information diffusion speed using the total number of Twitter users the tweet can reach after three hours (diffusion_3hr). If an influential Twitter user with 5,000 followers retweets, the number of Twitter users the tweet can reach will increase by 5,000. Focusing on a three-hour horizon after a tweet makes the measure comparable across tweets. We find similar results when we measure the level of retail attention until the end of the day, as most of the retweets take place during the first three hours after the initial tweet.

We then regress contemporaneous stock returns, stock returns the next day, and the dollar spread change on information diffusion speed in panel regressions. To avoid the mechanical effect that more breaking news leads to both higher returns and diffusion speed, we control for both stock returns over the market in the first 10 minutes after a tweet (10m ret) and absolute value of 10-minute return (abs 10m ret). As a result, the return is measured relative to the return over the first ten minutes after a tweet. We also control for stock returns over the market in the past hour (past 1h ret) and absolute value of past hour returns (abs past 1h ret). The variable media measures the log number of media coverage on the same day and up to three hours after the tweet. Other control variables include pre-market, afternoon, post-market, size, turn, volatility, and bm.

Table 12.8 reports the regression results from the full sample for all 1,261 tweets from 2013/11 to 2014/10. The dependent variables are stock returns (in excess of the market and by percentage) 10 minutes after a tweet until the end of the day (CAR%[10m, close d0]); stock returns (in excess of the market and by percentage) on the next trading day (CAR%[close d0, close d1]); and the change in the average dollar spread from the one hour before the tweet to the one hour after. The main independent variable is diffusion_3hr, which measures the log number of users the tweet can potentially reach three hours after the tweet. Other control variables include pre-market (a dummy variable equal to 1 if the tweet takes place before 9:30 ET); afternoon (a dummy variable equal to 1 if the tweet takes place between 12:30

TABLE 12.8

RETAIL ATTENTIONS, STOCK RETURNS, AND CHANGES IN
DOLLAR SPREAD FOR ALL STOCKS

|  | CAR% [10m, close d0] | CAR% [close d0, close d1] | Spread_chg |
|---|---|---|---|
| *intercept* | -4.28*** | -5.16*** | 0.04 |
|  | (-2,68) | -2.85 | -0.07 |
| *diffusion_3hr* | 0.21*** | -0.21*** | -0.05* |
|  | -3.16 | (-2.74) | (-1.84) |
| *Pre-market* | 0.13 | -0.23 | -0.78*** |
|  | -0.56 | (-1.00) | (-9.52) |
| *afternoon* | -0.03 | -0.25 | 0.34*** |
|  | (-0.16) | (-0.93) | -7.32 |
| *post-market* | 0.02 | -0.12 | 0.23*** |
|  | -0.11 | -0.51 | -4.62 |
| *size* | 0.07 | -0.02 | 0.02 |
|  | -1.02 | (-0.31) | -1 |
| *turn* | 0 | 0.00* | 0.00* |
|  | -0.15 | -1.85 | (-1.84) |
| *volatility* | -17.75 | -137.13** | 13.70** |
|  | (-0.30) | (-2.29) | -2.21 |
| *bm* | 0.01 | 0.06** | -0.01 |
|  | -0.29 | -2.11 | (-1.62) |
| *past 1h ret* | 14.09 | -6.53 | -1.61 |
|  | (-0.45) | (-1.07) | (-1.27) |
| *abs past 1h ret* | 16.01 | 1.68 | -1.47 |
|  | -1.52 | -0.21 | (-0.88) |
| *10 ret* | -6.11 | -38.37* | -4.81 |
|  | (-0.29) | (-1.67) | (-1.92) |
| *abs 10m ret* | 64.81*** | 20.61 | 2.25 |
|  | -3.15 | -0.65 | -0.64 |
| *media* | -0.02 | -0.02 | -0.01 |
|  | (-0.21) | (-0.16) | (-0.48) |
| *isbreaking* | -0.2 | -0.88*** | -0.06 |
|  | (-0.95) | (-2.87) | (-0.67) |
| *stock FE* | Y | Y | Y |
| *account FE* | Y | Y | Y |
|  | 0.117 | 0.081 | 0.341 |

and 16:00 ET); post-market (a dummy variable equal to 1 if the tweet takes place after 16:00 ET); size (log market capitalization); turn (turnover); volatility (daily returns volatility in the past 30 days); bm (book-to-market ratio); past 1h ret (stock returns over the market in the past hour); abs past 1h ret (absolute value of past 1h ret); 10m ret (stock returns over the market in the first 10 minutes after the tweet); abs 10m ret (absolute value of 10m ret); media (log number of media coverage on the same day and up to three hours after the tweet); Isbreaking (a dummy variable equal to 1 if the tweet contains "breaking"). We include stock and Twitter account fixed effects. Standard errors are clustered by ticker.

We observe a positive and significant association between information diffusion speed and contemporaneous returns. A one-standard-deviation increase in our retail attention measures (diffusion_3hr) leads to a 23 bps increase in contemporaneous-day returns. Interestingly, the higher returns that result due to a tweet revert completely the next day. Such temporary price overshooting and subsequent reversal is consistent with the stale news channel documented by Tetlock [213]. This finding is also consistent with the retail attention mechanism documented by Da, Engelberg, and Gao [64].

The economic magnitude of the price pressure is similar to that in Da, Engelberg, and Gao [64]. Yet the cycle of price pressure and reversal ends at a much higher speed in Twitter. In Da, Engelberg, and Gao [64], the price pressure and reversal cycle lasts for two weeks, but this cycle lasts for less than a day in our sample. Researchers using low-frequency data may not be able to detect the price pressure and reversal cycle. Taken together, our results are consistent with the hypothesis that tweets spread stale news among investors, which generate price pressure followed by reversal. The Twitter platform, however, speeds up the decimation of stale news.

We also find that a one-standard-deviation increase in the information diffusion speed decreases the bid-ask spread by 5.45 bps points (1.09*0.05), which provides further support for the stale news channel. This result is in contrast to Hong, Hong and Ungureanu [104], who find that fast information diffusion increases the number of informed investors relative to the number of uninformed investors, which reduces liquidity. Under the stale news channel, noise trading is generated through information diffusion, which can increase liquidity.

The return and liquidity results are much more pronounced among stocks with a market capitalization that is below the median. Table 12.9 reports the results for tweets on firms with market capitalization below the median of all stocks in our sample. A one-standard-deviation increase in diffusion measures (diffusion_3hr) leads to a much higher 46 bps increase in contemporaneous-day returns. Again, the price pressure completely reverted the day following a tweet. Table 12.9 also shows an even greater decrease in the bid-ask spread after a tweet concerning smaller stocks.

## 12.5 Instrumental Variables Approach

In the previous section, we establish a correlation between information diffusion speed and trading, return, and liquidity. Yet, correlation does not necessarily imply causality. Fortunately, researchers in computer science have developed advanced machine-learning techniques that can be employed to predict information cascades on large social networks. A number of these predictors do not have a direct relation with trading, return, and liquidity. We use these instrumental variables to first generate a predictive value for the information diffusion speed, and then examine whether it can be used to forecast trading, return, and liquidity. We use ordinary least squares (OLS) to predict the diffusion speed so that our method is similar to the two-stage least squares (2SLS) regression. More sophisticated machine-learning techniques such as support vector machine, neural networks, and decision tree-based algorithms provide stronger statistical power, and are available upon request.

TABLE 12.9

RETAIL ATTENTIONS, STOCK RETURNS, & CHANGE IN DOLLAR SPREAD FOR SMALL STOCKS

|  | CAR% [10m, close d0] | CAR% [close d0, close d1] | Spread_chg |
|---|---:|---:|---:|
| *intercept* | -7.59*** | 7.93*** | 0.7 |
|  | (-2.52) | -2.07 | -0.86 |
| *diffusion_3hr* | 0.39*** | -0.38*** | -0.06* |
|  | -3.24 | (-2.99) | (-2.03) |
| *Pre-market* | 0.33 | -0.63 | -0.89*** |
|  | -0.79 | (-1.48) | (-9.01) |
| *afternoon* | 0.01 | -0.28 | 0.36*** |
|  | -0.04 | (-0.57) | -6.8 |
| *post-market* | 0.13 | -0.51 | 0.21*** |
|  | -0.43 | (-1.09) | -3.92 |
| *size* | 0.1 | -0.01 | 0.01 |
|  | -0.67 | (-0.08) | -0.2 |
| *turn* | 0 | 0.00* | 0.00* |
|  | -0.08 | -1.72 | (-1.08) |
| *volatility* | -13.09 | -150.53** | 9.19** |
|  | (-0.20) | (-2.19) | -1.26 |
| *bm* | 0.01 | 0.07** | -0.01 |
|  | -0.42 | -2.28 | (-1.54) |
| *past 1h ret* | 6.35 | -7.88 | -1.76 |
|  | (-0.67) | (-1.25) | (-1.45) |
| *abs past 1h ret* | 17.44 | 5 | 0.82 |
|  | -1.56 | -0.58 | (-0.50) |
| *10 ret* | -5.94 | -44.17* | -4.91 |
|  | (-0.25) | (-1.80) | (-1.92) |
| *abs 10m ret* | 67.05*** | 22.47 | 2.91 |
|  | -3.01 | -0.66 | -0.81 |
| *media* | 0.01 | 0.05 | -0.04 |
|  | -0.08 | (-0.28) | (-1.15) |
| *isbreaking* | -0.24 | -1.46*** | -0.09 |
|  | (-0.56) | (-2.51) | (-0.77) |
| *stock FE* | Y | Y | Y |
| *account FE* | Y | Y | Y |
|  | 0.148 | 0.113 | 0.37 |

The dependent variable of interest is the future information diffusion rate on Twitter. Specifically, the growth rate is calculated as log(diffusion_3hr) log(diffusion_10m), where diffusion_10m and diffusion_3hr are the number of users a tweet potentially reaches after 10 minutes and after three hours, respectively.

There are two types of variables that can be employed to predict information diffusion speed. Suh et al. [207], Petrovic, Osborne and Lavrenko [167], and Jenders, Kasneci and Naumann [116], among others, rely mostly on the content and source of a tweet, while Cheng et al. [51] suggest that how information diffuses in the first few minutes after a tweet (also known as temporal features) and the characteristics of people who have retweeted are also crucial factors in predicting information cascades.

For the content of a tweet, we include in the regression a dummy variable equal to 1 if the tweet contains a picture (HasPicture), a dummy variable equal to 1 if the tweet contains hashtags (HasHashtags), and a dummy variable equal to 1 if the tweet contains URL links (HasURLs). Tweets with pictures or hashtags should disseminate faster because they typically grab users' attention. On the other hand, a tweet with a URL link should diffuse more slowly as it takes time to read the linked material. Certainly, we cannot fully rule out the possibility that a tweet with a picture or hashtag is more important, whereas news with URL is less important. Yet we suggest that these three variables should not have a direct effect on trading, return, and liquidity other than through the information diffusion rate.

For the temporal features and characteristics of people, we first find the Twitter account for the last five retweets before the 10-minute cut-off time. The variable log(# of followers of recent retweeters) is defined as the log of the total number of followers for these five Twitter accounts. Intuitively, a retweet from an account with more followers tends to disseminate faster. As the network of followers is established well before news is tweeted, it is unlikely to be affected directly by trading, return, and liquidity. Motivated by the investor distraction hypothesis put forth by Hirshleifer,

Lim and Teoh [101], we also include the average daily number of tweets sent by a Twitter account (Total # of tweets).

The regressions for the results in Table 12.10 include the inverse of the average time lapse between the five most recent retweets before the 10-minute cut-off time (Speed of recent retweets), the hour of the tweet (Hour), a dummy variable equal to 1 if the tweet is sent from the West Coast (IsWest), and a dummy variable equal to 1 if the tweet is sent by a CEO (IsCEO). The results show that tweets with pictures or hashtags increase the information diffusion speed, whereas inclusion of a URL decreases information diffusion speed. If an initial tweet is retweeted quickly by users with more followers, the retweet will disseminate more quickly. Tweets sent from a Twitter account generating more tweets per day will disseminate slower, consistent with the "driven-to-distraction" hypothesis of Hirshleifer, Lim, and Teoh [101]. In addition, if recent retweets are posted in rapid-fire fashion, the initial tweet will disseminate faster. The predictive power of these temporal features is consistent with the findings in Cheng et al. [51].

Our predictive variables are all observable during the first ten minutes after an initial tweet, and thus are independent of future returns measured after 10 minutes. In addition, most of these variables are not directly related to the value and liquidity of a stock. We therefore use them to instrument our information diffusion speed.

Specifically, we first compute the predicted attention diffusion rate from the regression. We then multiply the diffusion rate by the total number of Twitter users a tweet can reach after 10 minutes and use this product in our analysis. Intuitively, this product measures the expected number of users the tweet can reach using the information set available 10 minutes after the initial tweet. We then link the predictive retail attention to contemporaneous and future stock returns using the same panel regressions that we use for Table 12.8.

TABLE 12.10

DIFFUSION PREDICTION

| Variable | CAR% |
|---|---|
| | [10m, close d0] |
| *Intercept* | -4.15*** |
| | (-10.87) |
| *Total # of tweets* | -0.01*** |
| | (-3.71) |
| *Log (# of followers of recent re-tweeters)* | 0.02** |
| | -2.17 |
| *Speed of recent retweets* | 4.31*** |
| | -5.16 |
| *Hour* | -0.05*** |
| | (-4.95) |
| *Iswest* | 0.75*** |
| | -5.8 |
| *IsCEO* | 1.24** |
| | 2.35 |
| *HasPicture* | 0.63*** |
| | (6.28 |
| *HasURLs* | -0.37** |
| | (-2.52) |
| *HasHashtags* | 0.41** |
| | 2.46 |
| $R^2$ | 0.19 |

TABLE 12.11

PREDICTED RETAIL ATTENTION AND STOCK RETURN PREDICTIONS USING PREDICTED DIFFUSION

|  | CAR% [10m, close d0] | CAR% [close d0, close d1] |
|---|---|---|
| *Intercept* | -4.20*** | 4.62 |
|  | (-2.61) | -2.56 |
| *Predicted diff* | 0.20*** | -0.17** |
|  | -3.14 | -2.22 |
| *Pre-market* | 0.211 | -0.23 |
|  | -0.94 | (-0.95) |
| *afternoon* | -0.02 | -0.27 |
|  | (-0.11) | (-1.04) |
| *post-market* | 0.02 | -0.06 |
|  | -0.11 | (-0.26) |
| *size* | 0.08 | -0.03 |
|  | -1.07 | (-0.31) |
| *turn* | 0 | 0.00* |
|  | -0.2 | -1.88 |
| *volatility* | -15.69 | -139.40** |
|  | (-0.26) | (-2.31) |
| *bm* | 0 | 0.06** |
|  | -0.07 | -2.1 |
| *past 1h ret* | -3.95 | -38.31* |
|  | (-0.19) | (-1.65) |
| *abs past 1h ret* | -2.46 | -6.01 |
|  | (-0.28) | (-0.98) |
| *10 ret* | 16.74 | 1.63 |
|  | -1.59 | -0.02 |
| *abs 10m ret* | 62.56*** | 20.26 |
|  | -3.04 | -0.63 |
| *media* | -0.04 | -0.06 |
|  | (-0.44) | (-0.49) |
| *isbreaking* | -0.23 | -0.91*** |
|  | (-1.09) | (-2.95) |
| *stock FE* | Y | Y |
| *account FE* | Y | Y |
| $R^2$ | 0.126 | 0.081 |

We report the predicted retail attention and stock return predictions in Table 12.11. The dependent variables are stock returns (in excess of the market and by percentage) 10 minutes after a tweet until the end of the day (CAR%[10m, close d0]) and stock returns (in excess of the market and by percentage) on the next trading day (CAR%[close d0, close d1. The main independent variable is predicted diff, which measures the log number of users the tweet is predicted to reach three hours after the tweet. It is computed by summing the log number of Twitter users the tweet reaches after 10 minutes (diffusion_10m) and the predicted log growth rate from the regression shown in Table 12.8. We include stock and Twitter account fixed effects. Standard errors are clustered by ticker. The sample covers all 1,261 tweets from 2013/11 to 2014/10. T-statistics are in parentheses. ***, **, and * indicate statistical significance at the 1When we include the predicted diffusion rate in the regression, the results in Table 12.11 exhibit the same price overshooting and reversal pattern as in Table 12.8.

Interestingly, we do not detect this pattern when using only the total number of Twitter users the tweet can reach after 10 minutes in the regression for the results in Table 12.12. The main independent variable is simply diffusion_10m. Other control variables include pre-market (a dummy variable equal to 1 if the tweet takes place before 9:30 ET); afternoon (a dummy variable equal to 1 if the tweet takes place between 12:30 and 16:00 ET); post-market (a dummy variable equal to 1 if the tweet takes place after 16:00 ET); size (log market capitalization); turn (turnover); volatility (daily returns volatility in the past 30 days); bm (book-to-market ratio); past 1h ret (stock returns over the market in the past hour); abs past 1h ret (absolute value of past 1h ret); 10m ret (stock returns over the market in the first 10 minutes after the tweet); abs 10m ret (absolute value of 10m ret); media (log number of media coverage on the same day and up to three hours after the tweet); Isbreaking (a dummy variable equal to 1 if the tweet contains "breaking"). In other words, the predicted information diffusion speed 10 minutes after a tweet is crucial for measuring the actual information diffusion speed.

Finally, we conduct a predictive out-of-sample exercise. We use data from the first six months of our sample period (2013/11 – 2014/04) to run the predictive regression and then apply the regression coefficients to the next six months (2014/05 – 2014/10) in computing the diffusion rate. The predicted retail attention measure is therefore free of forward-looking bias and can be computed in real time. We then link the predictive retail attention to contemporaneous and future stock returns using only the second half of our sample period.

In Table 12.13 we perform out-of-sample prediction on retail attention and stock returns. We break our one-year sample period (2013/11-2014/10) into an in-sample period (2013/11-2014/04) and an out-of-sample period (2014/05-2014/10). We estimate the predictive regression from Table 12.8 during the in-sample period only. We then take the estimated coefficients and apply them to the out-of-sample period to compute predicted diff. In other words, predicted diff is observable 10 minutes after a tweet. We then link predicted diff to future returns in the out-of-sample period. The dependent variables are stock returns (in excess of the market and by percentage) 10 minutes after the tweet until the end of the day (CAR%[10m, close d0]) and stock returns (in excess of the market and by percentage) on the next trading day (CAR%[close d0, close d1. Other control variables include pre-market (a dummy variable equal to 1 if the tweet takes place before 9:30 ET); afternoon (a dummy variable equal to 1 if the tweet takes place between 12:30 and 16:00 ET); post-market (a dummy variable equal to 1 if the tweet takes place after 16:00 ET); size (log market capitalization); turn (turnover); volatility (daily returns volatility in the past 30 days); bm (book-to-market ratio); past 1h ret (stock returns over the market in the past hour); abs past 1h ret (absolute value of past 1h ret); 10m ret (stock returns

TABLE 12.12

PREDICTED RETAIL ATTENTION AND STOCK RETURN PREDICTIONS

|  | CAR% [10m, close d0] | CAR% [close d0, close d1] |
|---|---|---|
| *Intercept* | -3.64** | 2.48 |
|  | (-2.18) | -1.25 |
| *Predicted diff* | 0.11 | -0.04 |
|  | -1.62 | (-0.44) |
| *Pre-market* | 0.04 | -0.17 |
|  | -0.15 | (-0.71) |
| *afternoon* | -0.12 | -17 |
|  | (-0.54) | (-0.62) |
| *post-market* | 0.07 | -0.05 |
|  | -0.42 | (-0.18) |
| *size* | 0.12 | -0.01 |
|  | -1.5 | (-0.07) |
| *turn* | 0 | 0.00* |
|  | -0.3 | -1.9 |
| *volatility* | -15.12 | -129.72** |
|  | (-0.27) | (-2.20) |
| *bm* | 0.01 | 0.04 |
|  | -0.44 | -1.5 |
| *past 1h ret* | -9.49 | -40.37* |
|  | (-0.38) | (-1.81) |
| *abs past 1h ret* | -10 | -8.1 |
|  | (-1.08) | (-1.31) |
| *10 ret* | 9.36 | 4.08 |
|  | -0.8 | -0.56 |
| *abs 10m ret* | 74.37*** | 23.46 |
|  | -2.89 | -0.81 |
| *media* | -0.07 | -0.16 |
|  | (-0.74) | (-1.45) |
| *isbreaking* | -0.22 | -0.95*** |
|  | (-0.80) | (-2.88) |
| *stock FE* | Y | Y |
| *account FE* | Y | Y |
| $R^2$ | 0.082 | 0.069 |

over the market in the first 10 minutes after the tweet); abs 10m ret (absolute value of 10m ret); media (log number of media coverage on the same day and up to three hours after the tweet); Isbreaking (a dummy variable equal to 1 if the tweet contains "breaking"). We include stock and Twitter account fixed effects. Standard errors are clustered by ticker. The regression uses tweets during the out-of-sample period from 2014/05 to 2014/10. T-statistics are in parentheses. ***, **, and * indicate statistical significance at the 1%, 5%, and 10% levels, respectively. The results in Table 12.13 show that the predicted diffusion speed forecasts the price pressure and subsequent reversal out-of-sample, thus providing even stronger evidence for the stale news hypothesis.

## 12.6 Related Work

Our paper contributes to the burgeoning literature on social media. The unique feature of social media is the dynamic and interactive way information is distributed from user to user. This dynamic feature differentiates social media from other information dispersal methods [195]. Yet most researchers treat Twitter as a traditional information distribution outlet. Sprenger and Welpe [216], Bollen, Mao and Zeng [39], Brown [41], Mittal and Goel [150], Rao and Srivastava [195], Nann, Krauss and Schoder [154], and Oliveira et al. [159] apply text analysis to generate new sentiment measures. On the other hand, Blankespoor, Miller and White [35], Bhagwat and Burch [32], and Chen, Hwang, and Liu [50] consider social media as alternative way to release information, which reduces information asymmetry, improves stock liquidity, and attracts more investors. Under these two approach, only the initial information release matters. A notable exception is Giannini, Irvine and Shu [88], who use Twitter to track the change in investor disagreement, but they do not track the spread of information across agents. To the best of our knowledge, we are the

TABLE 12.13

PREDICTED RETAIL ATTENTION AND STOCK RETURNS:

OUT-OF-SAMPLE PREDICTIONS

|  | CAR% [10m, close d0] | CAR% [close d0, close d1] |
|---|---|---|
| *Intercept* | -0.93 | 4.17* |
|  | (-0.48) | -1.77 |
| *Predicted diff* | 0.15* | -0.22** |
|  | -1.88 | (-2.21) |
| *pre-market* | 0.56* | -0.44** |
|  | -1.94 | (-1.35) |
| *afternoon* | 0.14 | -0.69** |
|  | -0.67 | (-2.17) |
| *post-market* | 0.15 | -0.4 |
|  | -0.87 | (-1.22) |
| *size* | -0.06 | 0.07 |
|  | -0.75 | -0.63 |
| *turn* | 0 | 0.00** |
|  | -0.41 | -2.07 |
| *volatility* | -41.89 | -159.92** |
|  | (-0.54) | (-2.24) |
| *bm* | 0.01 | 0.05 |
|  | -0.26 | -0.9 |
| *past 1h ret* | 29.07 | -121.99*** |
|  | -0.67 | (-2.86) |
| *abs past 1h ret* | -4.08 | -1.49 |
|  | (-0.41) | (-0.22) |
| *10 ret* | 14.52 | 1.53 |
|  | -1.27 | -0.16 |
| *abs 10m ret* | 76.43 | 99.81* |
|  | -1.5 | -1.85 |
| *media* | 0.04 | 0.01 |
|  | -0.39 | -0.04 |
| *Isbreaking* | -0.17 | -1.18*** |
|  | (-0.70) | (-2.85) |
| *stock FE* | Y | Y |
| *account FE* | Y | Y |
| $R^2$ | 0.134 | 0.187 |

first to analyze Twitter's dynamic features of information diffusion using a financial perspective. This dynamic feature, in turn, allows us to contribute to two other lines of literature: information diffusion and social interaction.

The key variable in the theoretical information diffusion literature is the portion of investors who know the information relative to the portion who do not. The change in the portion over time serves as the main driver for volume, return, and liquidity [63, 98, 104, 105, 140]. Yet the existing proxies for information diffusion are all static. By tracking tweets and retweets over time, we contribute to the literature by proposing the first dynamic proxy for information diffusion.

Word-of-mouth communication has been shown to affect market outcome. In the theoretical literature, the information exchange can happen through a network of friends [160], or through information percolation, a random meeting of agents without a network [78, 79]. A fundamental challenge for empirical work is to document word-of-mouth communication, in which one agent receives a piece of information from another agent, and then transmits it to a third agent. There are indirect proxies of word-of-mouth communication such as physical proximity [40, 106, 113], sociability [113], ownership [197], common schooling [56], coworkers [197], and correlated stock trades [160]. One challenge of word-of-mouth communication literature is to differentiate it with homophily, in which investors act alike because they share similar backgrounds, not because they share information. To the best of our knowledge, documentation of word-of-mouth communication only exists as anecdotes through criminal investigations on insider trading [4, 195] or Ponzi schemes [174]. The primary drawback of this approach is generality, because the information diffusion in illegal activity can be substantially different from normal information diffusion [4]. Tweets and retweets provide the first direct proxy of word-of-mouth communication under normal conditions. Besides finding direct evidence that word-of-mouth communication affects stock return and liquidity, we also provide a direct proxy of information diffusion. This proxy may be valuable in direct empirical tests on theories of information percolation and network effects.

Studies document that limited investor attention affects trading behavior, return, and liquidity. Limited investor attention can lead to underreaction to actual news [101, 164], but it can also lead to overreaction to stale news [89, 213].

Our empirical results show that Twitter mainly serves the second function. We also find that the attention generated by tweets and retweets has a large impact on retail traders. In this sense, our paper contributes to the literature on retail attention [22, 64].

We do not argue that social media makes stock prices less efficient, because price pressure and reversal have been documented for the era before Twitter. Instead, we maintain that information dispersal through Twitter speeds up the price pressure and reversal. The cycle of price pressure and reversal usually lasts for several days or even weeks [22, 64], but in our paper price pressure accumulates within a day and completely reverses the next day. Our speeding up finding provides supportive evidence for [195], who finds that the invention of the telephone sped up word-of-mouth communication in last century. Apparently, Twitter also speeds up the word-of-mouth communication because it is fast and interactive.

12.7 Conclusion

In this paper, we track the diffusion of news by monitoring how such news is tweeted and retweeted on Twitter. We find the diffusion speed to be highly correlated with intraday retail trading patterns. The resulting retail attention leads to lower bid-ask spreads and positive price pressure on the news day, but these effects are completely reverted the next day. The amount of retail attention that news generates on Twitter can be predicted using characteristics of users, accounts, and tweets. The fact that predicted retail attention generates similar results helps to alleviate concerns about reverse causality and endogeneity. Taken together, we show that the role of Twitter is to spread stale news. Twitter generates price pressure and reversal, albeit at a much faster speed than the cycle generated by traditional media. This finding sheds some light on the question raised by Shiller (2015) on the impact of social media on financial markets.

More broadly, we are among the first to construct a dynamic and direct measure of information diffusion and word-of-mouth communication. This measure can be applied in a number of ways to test the implications of information diffusion or social network theory. For example, we can test the differential impacts of learning from trading and learning from information diffusion. The advent of social media provides a unique opportunity for researchers to construct new measures that is hard to obtain using traditional media. Using our measure to test the implications of economic theory could prove very fruitful.

CHAPTER 13

CONCLUSION AND FUTURE DIRECTIONS

13.1 Summary of Contributions

In this dissertation, we aim to make one concrete step toward building a connection between between raw data and network representations. We achieve this by proposing the higher-order network, including its mathematical model, efficient construction algorithm and open-source software implementation, the first visualization software toolkit, a tutorial that demonstrates a real-world scenario of constructing HON from raw data. We demonstrate HON's applications in clustering, ranking, anomaly detection and other dynamic processes on networks. With the ever-increasing complexity of dynamic systems and the higher-order interactions data derived from such systems, HON points out a new direction in network analysis that effectively utilize higher-order signals for pattern discovery and supporting decision making.

This dissertation contains three inter-connected parts, being the theoretical foundations of HON, real-world applications of HON, and dynamic processes on implicit networks. Specifically, Part I first reviews diverse types of data, how these data had been converted to network representations, and what information had been lost during the conversion. It shows that sequential data, temporal data, time series data and other higher-order data types do not have effective network representations. Then the higher-order network, that is able to preserve higher-order dependencies in the network structure accurately and compactly, is proposed. It demonstrates the effectiveness of HON, as well as its influence on network analysis results such as clustering and ranking. Improvements to make the HON construction algorithm scalable and parameter-free are introduced, and methods to extend HON's input to diffusion data, temporal data, and others are summarized. This dissertation then introduces the HONVis software toolkit, which is the first visualization and interactive exploration software for HON. Finally, this part concludes with a tutorial that demonstrates the full process of building HON from raw data using the Python implementation of HON.

Part II focuses on real-world applications of representing big data as networks. It first provides a network model of the species invasion network at the global scale, and demonstrates how network analysis such as clustering can provide insights to targeted species control and support decision making. It then questions the conventional network construction approach that makes multiple implicit assumptions. It provides a comparative study on the construction of shipping network models, and show how these different network representations influence network analysis results. This part then focuses on the Arctic region which is overseeing rapid increases in shipping traffic and rising risk of invasions. It discusses species introduction and dispersal in the Arctic using HON, reveals the emergence of Arctic shipping hubs and stepping-stone species introduction ports. It projects the invasion risk in the Arctic on the decadal scale through a combination of regression and dynamic network evolution modeling. Finally, this part reveals a striking case when conventional network-based anomaly detection methods may fail on higher-order data, and proposes the HON-based anomaly detection framework that effectively identifies and amplifies higher-order anomalous signals.

Part III reviews dynamic processes on implicit social networks. The mining of features associated with effective tweets reveals non-linear relationship between multiple tweet features and user engagement rate, suggesting that machine learning models should not simply assume linear relationships such as "longer tweets are more likely to be more influential". These features serve as the input of the research on retail attentions reflected by information diffusion on Twitter. It shows that information diffusion via Twitter does not incorporate new information into the stock price. Rather, Twitter spreads stale news, albeit at a much higher speed than traditional media.

## 13.2 Future Directions

### 13.2.1 Low Hanging Fruits

**HON algorithm.** While HON has become the state-of-the-art method in this direction, there are still opportunities. For example, besides KL-divergence, how does other metrics of distance measurements (such as cosine similarities) influence the results? Besides random walk-based simulation of movements, what are other evaluation metrics of HON's accuracies?

**Influence on other network analyses.** Besides clustering, ranking, anomaly detection, can the higher-order topologies of HON serve as a feature for link prediction? What is HON's influence on diffusion processes other than Twitter?

**HON implementation.** While HON's code is easily parallelizable, it is less intuitive for the HON+ algorithm. How to parallelize the implementation to scale up HON's ability for even bigger data and online computation? A Python package that can be installed through package management software is also desirable. As for visualization, besides making the tool accessible through web browsers, it is desirable to incorporate the temporal dimension for exploring the evolution of higher-order dependencies, the anomaly detection module for visualizing higher-order anomalies, and so on.

**Interpretations of network properties.** How does HON influence basic network properties, such as degree distributions, betweenness centralities, average shortest paths, and many others? More importantly, how to interpret the changes in these network properties, given that HON may use multiple higher-order nodes to represent the same physical location?

### 13.2.2 Promising Directions

**Integrating HON with other network models.** Can we combine HON and heterogeneous networks to capture richer types of entities on the same network? Can we combine HON with temporal networks to incorporate link activation intervals on HON? What is the connection between HON and other multi-layered network representations in capturing higher-order patterns, such as Edler et al's approach by combining network structures that yield similar dynamics [81], Flavio et al's approach by combining multiple layers of first-order networks [84], Benson et al's approach on higher-order network topologies [30], Scholtes's approach through temporal networks [188], and so on.

**Providing features for network-based machine learning algorithms.** Machine learning algorithms that are based on network structures may also benefit from the information incorporated in the higher-order structures of HON. For example, PageRank on a network of words has been used for keyword extraction in text mining; since HON can help improve the accuracy of the implicit random walkers, can it improve the keyword extraction result? What if HON is combined with representation learning such as Deepwalk [166] or Node2Vec [93]?

To facilitate the exploration of these directions for the network science and data mining community, we have made code, video demo and papers are made available at `http://www.HigherOrderNetwork.com`